\pdfoutput=1
\documentclass[a4paper, 12pt]{report} 
\usepackage[colorlinks=true, linkcolor=blue, citecolor=blue, urlcolor=blue, plainpages=false, pdfpagelabels]{hyperref} 
\usepackage{graphicx}

\usepackage{amsmath,amssymb}
\usepackage{tikz}
\usepackage{amsthm}
\newtheorem{principle}{Principle}

\usepackage{longtable}
\usepackage{multirow}
\usepackage{array}
\usepackage{microtype}

\newtheoremstyle{myaxiom}
  {} % Space above
  {} % Space below
  {\itshape} % Body font
  {} % Indent amount
  {\bfseries} % Theorem head font
  {:} % Punctuation after theorem head
  {.5em} % Space after theorem head
  {} % Theorem head spec
\theoremstyle{myaxiom}

\usepackage[intoc]{nomencl} % Load the package, [intoc] adds to ToC
\makenomenclature % Initialize the system
\usepackage{subcaption} % Moved here to load after nomencl/tocbasic
\usepackage{etoolbox} % Load this package as well
\renewcommand\nomgroup[1]{%
  \item[\bfseries % Make group headers bold
  \ifstrequal{#1}{R}{Roman Symbols}{% Standard Latin letters
  \ifstrequal{#1}{G}{Greek Symbols}{% Greek letters
  \ifstrequal{#1}{S}{Subscripts and Superscripts}{% If you define subscripts separately
  \ifstrequal{#1}{O}{Operators and Mathematical Symbols}{% Nabla, partial derivatives etc.
  \ifstrequal{#1}{A}{Acronyms}{% Any acronyms used
  {}}}}}}% % Default empty group if none match
]}

\title{The Mechanics of Macroscopic Electrodynamics}
\author{Bela Schulte Westhoff} % Add other required info like Department, University, Advisor(s)
\date{\today \\[2ex] \large\textit{Manuscript Draft}} % Using \large\textit makes it stand out a bit

\begin{document}

% --- FRONT MATTER ---

% 1. Title Page
\hypersetup{pageanchor=false}
\maketitle 
\hypersetup{pageanchor=true}
% Some templates use \begin{titlepage}...\end{titlepage} instead or handle it automatically

% --- Optional: Copyright Page ---
% \cleardoublepage % Ensures it starts on a right-hand page if using book class/twoside option
% \thispagestyle{empty} % Usually no page number
% \vspace*{\fill}
% \begin{center}
% Copyright \copyright\ \the\year\ Your Name \\ All rights reserved.
% \end{center}
% \vspace*{\fill}
% \clearpage

% --- Set page style for front matter (often plain) and numbering (roman) ---
\pagestyle{plain} % Or as required by guidelines
\pagenumbering{roman} 
% \setcounter{page}{ii} % Adjust if title page counts as 'i'

% 2. Abstract
\cleardoublepage % Start on a new right-hand page (adjust if not needed)

\begin{abstract}
\noindent
Classical Electrodynamics is typically regarded as a closed chapter of physics---an established framework of the 19th and early 20th centuries. However, the exact definition of its primary mechanical quantities in ponderable media remains the subject of a century-long debate. The standard macroscopic description admits multiple, competing localizations of force density, momentum, and field energy (Minkowski, Abraham, Einstein--Laub, variational forms, and others). The prevailing view treats these alternatives as formally equivalent---different ``splits'' of the same total balance, reconciled by postulating appropriate material counterparts.

This monograph challenges the consensus that this ambiguity is merely a matter of convention. We argue that global conservation alone does not settle the question of physical coupling. A macroscopic formulation must also be mechanically coherent in the strict kinematic sense that power transfer is the work rate of the same force acting on the same mass target. We formalize this requirement as the \emph{Force--Energy Consistency Criterion} (FECC)---a ``Kinematic Lock'' (\(P=\mathbf f\!\cdot\!\mathbf v\))---and use it to perform a systematic mechanical audit of standard macroscopic tensors. Under this scrutiny, the Macroscopic Vacuum (Lorentz) Formulation emerges as the sole mechanically consistent description of the total transfer of energy and momentum from field to matter. The internal distribution of this energy is shown to be macroscopically indeterminate, requiring valid microstructural models to resolve.

To explain why a vacuum-form tensor can govern macroscopic matter, the book reconstructs the theory from the microscopic Lorentz baseline by reinterpreting spatial averaging as \emph{spectral filtering}. This perspective identifies a universal \emph{host interface} that routes electromagnetic energy into three macroscopic destinations: mechanical work, irreversible heat, and reversible high-frequency storage (binding). The resulting framework exposes a structural isomorphism in which thermodynamics, continuum mechanics, and electrodynamics appear not as independent fields, but as coupled spectral projections of a unified microscopic reality.
\end{abstract}

\addcontentsline{toc}{chapter}{Abstract} % Manually add Abstract to ToC if needed

% --- PREFACE ---

% --- END OF PREFACE ---
% 3. Optional: Dedication
% \cleardoublepage
% \thispagestyle{empty} % Often no page number
% \vspace*{\fill}
% \begin{center}
%     \textit{Dedication text...} 
% \end{center}
% \vspace*{\fill}
% \addcontentsline{toc}{chapter}{Dedication} % Optional: add to ToC

% 5. Table of Contents
\cleardoublepage
\tableofcontents

% 6. Optional: List of Figures
% \cleardoublepage
% \listoffigures
% \addcontentsline{toc}{chapter}{List of Figures} 

% 7. Optional: List of Tables
% \cleardoublepage
% \listoftables
% \addcontentsline{toc}{chapter}{List of Tables} 

% 8. Optional: Nomenclature / List of Symbols
% \cleardoublepage
% \chapter*{Nomenclature} 
% \addcontentsline{toc}{chapter}{Nomenclature} 
% % Use appropriate package (nomencl, glossaries) or a simple tabular environment
% % \begin{tabular}{ll} ... \end{tabular}
% ...

% --- MAIN BODY ---
% --- Nomenclature ---
\cleardoublepage
\printnomenclature
% If you didn't use \usepackage[intoc]{nomencl}, uncomment the next line:
% \addcontentsline{toc}{chapter}{Nomenclature}

\cleardoublepage
\chapter*{Nomenclature}
\addcontentsline{toc}{chapter}{Nomenclature}

\setlength{\emergencystretch}{3em} % Helps with line breaking in narrow columns
\begin{longtable}{@{} c @{\quad} l @{\quad} p{0.6\textwidth} @{}}

% --- Headers and Footers for Longtable ---
\multicolumn{3}{l}{\textbf{Roman Symbols -- Fields and Potentials}} \\
\hline \textbf{Symbol} & \textbf{Units} & \textbf{Description} \\ \hline \endfirsthead
% Header for continued pages
\multicolumn{3}{l}{\textbf{Nomenclature} -- \textit{Continued from previous page}} \\ \hline \textbf{Symbol} & \textbf{Units} & \textbf{Description} \\ \hline \endhead
% Footer for all pages except the last
\hline \multicolumn{3}{r}{\textit{Continued on next page}} \\ \endfoot
% Footer for the last page
\hline \endlastfoot

%--------------------------------------------------------------------
% --- A. FOUNDATIONAL FRAMEWORKS & CORE CONCEPTS ---
%--------------------------------------------------------------------
\multicolumn{3}{l}{\textit{1. Microscopic (Ground Truth) Quantities -- Chapter~\ref{chap:FreeCharges}}} \\
$\mathbf{e}, \mathbf{b}$ & V/m, T & The true, unaveraged microscopic electric and magnetic fields. \\
$\rho_{\text{micro}}, \mathbf{j}_{\text{micro}}$ & C/m$^3$, A/m$^2$ & Total microscopic charge and current densities. \\
$\mathbf{f}_{\text{em}}$ & N/m$^3$ & The true microscopic Lorentz force density, $\rho_{\text{micro}}\mathbf{e} + \mathbf{j}_{\text{micro}}\times\mathbf{b}$. \\
$T_{\text{em}}^{\mu\nu}$ & J/m$^3$ & The fundamental energy-momentum tensor of the microscopic field. \\
$\mathbf{f}_{\text{host}}, P_{\text{host}}$ & N/m$^3$, W/m$^3$ & Microscopic Host Force and Power; mediate all non-EM interactions between Active Sources and the Host Medium (Chapter~\ref{chap:HostInterface}). \\[1ex]

\multicolumn{3}{l}{\textit{2. The Macroscopic Formulation of this Work -- Chapter~\ref{chap:HostInterface}}} \\
$\mathbf{E}, \mathbf{B}$ & V/m, T & Macroscopic fields, defined as the spatial average $\langle\mathbf{e}\rangle$ and $\langle\mathbf{b}\rangle$. \\
$\rho_{\text{total}}, \mathbf{J}_{\text{total}}$ & C/m$^3$, A/m$^2$ & Total macroscopic charge and current densities, $\langle\rho_{\text{micro}}\rangle, \langle\mathbf{j}_{\text{micro}}\rangle$. \\
$T^{\mu\nu}_{EM}$ & J/m$^3$ & The \textbf{vacuum-form} energy-momentum tensor, expressed with macroscopic fields. It is central to the formulation developed in this work. \\
$\mathbf{f}_{Lorentz}$ & N/m$^3$ & Total macroscopic Lorentz force density, $\rho_{\text{total}}\mathbf{E} + \mathbf{J}_{\text{total}}\times\mathbf{B}$. \\
$\mathbf{J}_{\text{total}} \cdot \mathbf{E}$ & W/m$^3$ & Total power density transferred from the electromagnetic field to all charge carriers. \\[1ex]

\multicolumn{3}{l}{\textit{3. Alternative Historical Formulations}} \\
$u_{M}, P_{\text{matter},M}$ & & Energy density and postulated power exchange in the \textbf{Minkowski} formulation. \\
$u_{A}, P_{\text{matter},A}$ & & Energy density and postulated power exchange in the \textbf{Abraham} formulation. \\
$u_{conv}$ & J/m$^3$ & ``Conventional'' energy density from standard textbook formulations (see Chapter~\ref{chap:Critique_ArbitrarySplit}). \\
$\mathbf{f}_{KH}$ & N/m$^3$ & Korteweg-Helmholtz force, typically derived from an energy functional like $u_{conv}$. \\[2ex]

%--------------------------------------------------------------------
% --- B. PRIMARY PHYSICAL QUANTITIES ---
%--------------------------------------------------------------------
\multicolumn{3}{l}{\textbf{B. Primary Physical Quantities}} \\ \hline
\multicolumn{3}{l}{\textit{1. Fields, Potentials, and Fluxes}} \\
$\phi, \mathbf{A}$ & V, T$\cdot$m & Macroscopic Scalar and Vector Potentials ($\mathbf{E} = -\nabla\phi - \partial_t\mathbf{A}$, $\mathbf{B} = \nabla\times\mathbf{A}$). \\
$u_{EM}, \mathbf{g}_{EM}$ & J/m$^3$, Ns/m$^3$ & Energy density and momentum density of the consistent (vacuum-form) framework. \\
$\mathbf{S}_{EM}, \mathbf{T}_{EM}$ & W/m$^2$, N/m$^2$ & Poynting vector and Maxwell stress tensor of the consistent framework. \\
$\mathbf{D}, \mathbf{H}$ & C/m$^2$, A/m & Auxiliary fields, $\mathbf{D} \equiv \varepsilon_0\mathbf{E} + \mathbf{P}$ and $\mathbf{H} \equiv \mathbf{B}/\mu_0 - \mathbf{M}$. Defined as mathematical conveniences for simplifying source terms. \\
$\mathbf{P}, \mathbf{M}$ & C/m$^2$, A/m & Macroscopic polarization and magnetization, $\langle\mathbf{p}\rangle$ and $\langle\mathbf{m}\rangle$. \\
$\mathbf{p}, \mathbf{m}$ & C/m$^2$, A/m & Microscopic polarization and magnetization density fields. \\
$F^{\mu\nu}, G^{\mu\nu}$ & & Relativistic field tensor and its dual. \\[1ex]

\multicolumn{3}{l}{\textit{2. Sources and Material Parameters}} \\
$\rho_{f}, \mathbf{J}_{f}$ & C/m$^3$, A/m$^2$ & Macroscopic free charge and current densities. \\
$\rho_{b}, \mathbf{J}_{b}$ & C/m$^3$, A/m$^2$ & Macroscopic bound charge ($-\nabla\cdot\mathbf{P}$) and current ($\partial\mathbf{P}/\partial t + \nabla \times \mathbf{M}$). \\
$\sigma$ & S/m & Electrical conductivity. \\
$\chi_e, \chi_m$ & 1 & Electric and Magnetic Susceptibilities. \\
$\varepsilon, \mu$ & F/m, H/m & Permittivity and Permeability of the medium (Effective parameters). \\
$n$ & 1 & Refractive index ($n = \sqrt{\varepsilon\mu/\varepsilon_0\mu_0}$). \\
$\kappa$ & W/(m$\cdot$K) & Thermal conductivity. \\
$N$ & 1/m$^3$ & Macroscopic number density of particles. \\
$J^{\mu}_{f}, J^{\mu}_{b}, J^{\mu}_{\text{total}}$ & A/m$^2$ & Free, bound, and total 4-current densities. \\
$Z, Z_0$ & $\Omega$ & Impedance and Characteristic Impedance of free space ($\sqrt{\mu_0/\varepsilon_0}$). \\[1ex]

\multicolumn{3}{l}{\textit{3. Forces \& Power}} \\
$P_{\text{diss}}, P_{\text{mech}}$ & W/m$^3$ & Power density dissipated as heat and power density converted to mechanical work. Both are components of the total host power, $-P_{\text{host}}$. \\
$\mathbf{f}_{\text{pragmatic}}$ & N/m$^3$ & Complete pragmatic force density approximation; explicitly not energy-consistent (Chapter~\ref{chap:PointDipole_Audit}). \\
$\mathbf{f}_{\text{dipole}}$ & N/m$^3$ & Pragmatic model for the force density on bound matter, forming the core of $\mathbf{f}_{\text{pragmatic}}$. \\
$\mathbf{f}_{\text{Kelvin}}$ & N/m$^3$ & Kelvin force, a simplified baseline case of $\mathbf{f}_{\text{dipole}}$. \\
$\mathbf{f}_Q$ & N/m$^3$ & Quantum Constraint Force (Poincaré Stress) maintaining the stability of the mesoscopic dipole (Chapter~\ref{chap:ForceDensity}). \\
$\mathbf{\Phi}$ & N/m$^3$ & Force Discrepancy Vector representing the anomalous force in the Minkowski formulation (Chapter~\ref{chap:Macro_Critique}). \\
$\Phi_E$ & W/m$^3$ & Energy Discrepancy Scalar representing the anomalous power in the Minkowski formulation (Chapter~\ref{chap:Macro_Critique}). \\
$\mathbf{p}_{dp}, \mathbf{m}_{dp}$ & C$\cdot$m, A$\cdot$m$^2$ & Point electric and magnetic dipole moments used in pragmatic models. \\
$\mathbf{f}_{\text{other}}, P_{\text{other}}$ & N/m$^3$, W/m$^3$ & Conceptual non-EM force and power densities (e.g., from a battery) used in general balance equations. \\[1ex]

\multicolumn{3}{l}{\textit{4. Kinematics, Mechanics \& Thermodynamics}} \\
$t, \mathbf{r}$ & s, m & Time and Position vector. \\
$\mathbf{x}$ & m & Position vector (alternative to $\mathbf{r}$). \\
$\mathbf{v}$ & m/s & Velocity vector. \\
$\beta, \gamma$ & 1 & Normalized velocity ($\mathbf{v}/c$) and Lorentz factor ($1/\sqrt{1-\beta^2}$). \\
$\omega, \mathbf{k}$ & rad/s, 1/m & Angular frequency and Wave vector. \\
$T$ & K & Temperature. \\
$k_B$ & J/K & Boltzmann constant. \\
$d_f$ & 1 & Number of degrees of freedom per particle. \\
$\rho_m, \mathbf{g}_{\text{mech}}$ & kg/m$^3$, kg/(m$^2\cdot$s) & Mass density and mechanical momentum density of charge carriers. \\
$u_{\text{mech}}, \mathbf{T}_{\text{kin}}$ & J/m$^3$, N/m$^2$ & Kinetic energy density and kinetic stress tensor of charge carriers. \\
$u_{spin}$ & J/m$^3$ & Conceptual energy of the ``spin battery'' used to model constant magnetic moments (Sec.~\ref{sec:Spin}). \\
$\mathbf{v}_{\text{bulk}}, \mathbf{v}_{\text{carrier}}$ & m/s & Bulk velocity of a medium and the effective velocity of charge carriers. \\[2ex]

%--------------------------------------------------------------------
% --- C. SUBSCRIPTS & SUPERSCRIPTS ---
%--------------------------------------------------------------------
\multicolumn{3}{l}{\textbf{C. Subscripts \& Superscripts}} \\ \hline
\textit{micro} & & Denotes a true, unaveraged microscopic quantity (e.g., $\mathbf{e}, \rho_{\text{micro}}$). \\
\textit{EM} & & Denotes a quantity from the consistent, vacuum-form formulation (e.g., $T^{\mu\nu}_{EM}$). \\
\textit{M}, \textit{A}, \textit{EL} & & Denotes a quantity from the Minkowski, Abraham, or Einstein-Laub formulation. \\
\textit{conv} & & Denotes a quantity from the conventional textbook formulation (e.g., $u_{conv}$). \\
\textit{f}, \textit{b}, \textit{total} & & Distinguishes between free, bound, and total sources. \\
\textit{mech}, \textit{kin} & & Denotes a mechanical or kinetic quantity. \\
\textit{host} & & Denotes an interaction with the non-EM Host Medium (Chapter~\ref{chap:HostInterface}). \\
\textit{pragmatic}, \textit{eff} & & Denotes an approximate quantity from the pragmatic models (Chapter~\ref{chap:PointDipole_Audit}). \\
$s$ & & Index denoting a specific particle species (e.g., electrons, ions). \\[2ex]

%--------------------------------------------------------------------
% --- D. MATHEMATICAL SYMBOLS & CONSTANTS ---
%--------------------------------------------------------------------
\multicolumn{3}{l}{\textbf{D. Mathematical Symbols \& Constants}} \\ \hline
$\varepsilon_0, \mu_0$ & F/m, H/m & Permittivity and permeability of free space. \\
$c$ & m/s & Speed of light in vacuum, $1/\sqrt{\varepsilon_0\mu_0}$. \\
$\hbar$ & J$\cdot$s & Reduced Planck constant. \\
$\langle \dots \rangle$ & & Spatial averaging operator (equivalent to a low-pass filter). \\
$\nabla, \partial_t, \partial_\mu$ & & Standard differential operators (del, time derivative, 4-gradient). \\
$\delta(\mathbf{r})$ & 1/m$^3$ & Dirac delta function. \\
$g^{\mu\nu}, \eta^{\mu\nu}$ & & Metric tensor (Minkowski signature typically $+---$). \\
\end{longtable}

% --- MAIN BODY ---
\cleardoublepage 
% --- Switch page numbering to Arabic ---
\pagenumbering{arabic} 
\setcounter{page}{1} % Reset page counter to 1 for the main body

\chapter*{Prologue: Classical Physics --- The Physics of the Human Scale}
\addcontentsline{toc}{chapter}{Prologue: Classical Physics --- The Physics of the Human Scale}
\label{chap:Prologue}

This monograph investigates the foundations of classical physics at the scale where we live, measure, and build.
While its technical focus is macroscopic electrodynamics in matter---and its coupling to thermodynamics and continuum mechanics---its deeper aim is to expose the ``hidden machinery'' that classical physics routinely uses but rarely names:
the macroscopic world is a \emph{filtered} description of microscopic reality.

Before we write down tensors, constitutive laws, or mechanical audits, we must step back from the blackboard to ask a different kind of question.
Not \emph{how} the equations work, but \emph{why} they matter.
Why, in an age of Quantum Field Theory and General Relativity, do we still rely on the classical description of the world?
Why do engineers, experimentalists, and even theorists continue to speak in the language of forces, stresses, heat, and work?

The common answer is that classical physics is a convenient approximation.
This is true, but it is superficial.
It hides a more fundamental question: approximation \emph{to what}, and \emph{for whom}?

\bigskip

Modern fundamental physics is, in a precise sense, the \textbf{Physics of Extremes}.
To reach the most elementary description, we have pushed observation outward and downward:
toward the \emph{infinitesimal}, where quantum fluctuations and the structure of matter appear,
and toward the \emph{infinite}, where gravity shapes spacetime and the architecture of the cosmos becomes visible.
This pursuit is among the great intellectual achievements of humanity.

Yet, the world in which we actually think and act sits between these extremes.
It is the world of the cup and the circuit, the motor and the muscle, the fluid and the solid; the world in which we experience warmth, pressure, sound, color, friction, and fatigue.
The fundamental descriptions of particles and curvature do not speak this language directly.
Not because they are wrong, but because they are too \emph{fine-grained}: their ``signal'' contains vastly more information than our lives can receive, process, or use.

Macroscopic classical physics is, therefore, not merely a computational shortcut.
It is the \textbf{Physics of the Human Scale}:
the disciplined language of what can be resolved by finite observers equipped with finite instruments.
It is the theory of the world as it is \emph{accessible}.

\section*{The Mechanism: The Observer as a Filter}

A single cup of coffee contains on the order of $10^{23}$ particles.
Microscopically, it is a storm of electromagnetic interactions and stochastic molecular motion.
Yet this is not the world we inhabit.
We perceive \emph{temperature}, \emph{color}, \emph{aroma}, \emph{viscosity}.
These qualities are not layers added to the ``real'' physics; they are the stable outputs of a massive \emph{spectral compression}.

The human observer does not merely \emph{apply} a filter to reality; the human observer \emph{is} the filter.
Our senses are physical devices defined by limited bandwidth, finite resolution, and finite integration time.
They cannot track microscopic detail.
They integrate, average, damp, and discard.
The macroscopic world is what remains when microscopic complexity is projected onto the axes of human perception.

This is not a philosophical slogan; it is a physical fact regarding measurement.

\begin{description}
    \item[\textsc{Hearing}:]
    The ear is a damped mechanical oscillator coupled to the acoustic field.
    It does not report Brownian motion molecule-by-molecule.
    It resonates with coherent pressure variations in a narrow band and suppresses incoherent fluctuations as ``noise.''
    In this sense, \emph{silence is not the absence of energy}; it is the success of a filter.

    \item[\textsc{Sight}:]
    The eye is a chemical integrator.
    It converts the discrete arrival of billions of photons into continuous sensations of brightness and color through temporal averaging.
    The world does not present itself to us as photon counts; it presents itself as a stable macroscopic field.

    \item[\textsc{Touch}:]
    The skin reports ``solidity'' even though, microscopically, matter is almost entirely vacuum.
    What we feel as rigidity is the macroscopic outcome of electromagnetic constraints and collective response.
    We do not perceive electron--electron repulsion directly; we perceive its filtered, mechanical consequence.
\end{description}

Everything we distinguish---the plant, the ocean, the machine, the symphony---is a product of this filtering.
To perceive the microscopic reality directly would not be enlightenment; it would be overwhelming noise.
Even the experiments we conduct to explore the \textsc{Physics of Extremes} are ultimately mediated through this interface.
The ``illusion'' of the continuous world is not a lie; it is the only truth we can inhabit.

\bigskip

\section*{The Price of Simplicity}

This process of spectral filtering is what renders classical physics possible.
By suppressing microscopic stochasticity, it generates stable variables and predictive laws.
Temperature, elasticity, viscosity, permittivity: these are not microscopic primitives.
They are the \emph{compressed summaries} of vast, hidden degrees of freedom.

Yet, the same operation that grants simplicity imposes a structural cost.
When information is discarded, the mapping from micro to macro becomes non-invertible.
Different microscopic configurations can result in the identical macroscopic state.
The description becomes \emph{underdetermined}:
multiple internal mechanisms may produce the exact same external measurements.

This indeterminacy is not merely a philosophical nuisance; it is the source of a practical crisis.
It explains how a century-old debate can persist within an otherwise deterministic framework.
In matter, electrodynamics couples to mechanics and thermodynamics, but the macroscopic fields alone lack the information to uniquely localize that coupling.

The Abraham--Minkowski controversy is the archetypal symptom of this ambiguity.
At stake is not a matter of aesthetic preference for one tensor over another.
At stake is the most basic question of mechanical reality:
\emph{Where} is the momentum?
\emph{Where} does the force act?
\emph{How} does field energy partition into heat, deformation, and work?
If the macroscopic theory loses track of the specific kinematic carriers, it may remain globally consistent while becoming locally opaque.

\bigskip
\section*{The Objectives of this Synthesis}

The purpose of this monograph is to treat classical physics as a single, mechanically coherent architecture rather than a collection of adjacent subjects.
To explore this, the work proposes two foundational commitments.

\begin{description}
    \item[\textsc{I. Mechanical Closure is Non-Negotiable.}]
    Force and power are not independent accounting choices.
    They are kinematically linked: power is strictly the projection of a force onto the velocity of a definite mass target ($P = \mathbf{f} \cdot \mathbf{v}$).
    Any macroscopic formulation that permits energy transfer without strictly identifying its kinematic recipient is mechanically incomplete.
    A central tool of this work is to re-impose this closure as a strict criterion for validity.

    \item[\textsc{II. The Macroscopic Must be Derived, Not Postulated.}]
    Macroscopic equations appear not as a secondary layer of physics superimposed upon the microscopic one.
    They are the low-frequency signal that remains after high-frequency structure has been filtered out.
    Once this is accepted, thermodynamics, continuum mechanics, and macroscopic electrodynamics reveal themselves as \emph{distinct spectral projections} of one underlying microscopic reality.
    Their couplings are not mysterious add-ons; they are the unavoidable routing of energy and momentum across scales.
\end{description}

\bigskip

The aim is a fundamental shift in perspective.
Instead of treating macroscopic electrodynamics as a closed formalism that merely requires the correct constitutive parameters, we treat it as an \emph{interface}---an effective description that must remain faithful to the mechanics of the system it compresses.

This explains why the text insists on what may seem, at first glance, like a philosophical demand: to keep energy transfer rigorously tied to kinematic mass.
Without that demand, the set of mathematically consistent theories is infinite.
With it, the space of physically admissible descriptions becomes sharply constrained.

\bigskip

\section*{Why It Matters}

The importance of this project extends beyond settling a historical controversy.
A mechanically explicit macroscopic theory is vital wherever fields drive matter:
in precision metrology, optical manipulation, soft-matter electromechanics, and the study of high-field dielectrics.
It is essential for the design of devices that deliberately exploit the coupling between electromagnetic, mechanical, and thermal domains.

More fundamentally, it matters because classical physics is the language in which we \emph{certify reality}.
It is the regime in which we define what ``work,'' ``force,'' and ``energy'' actually mean---not as mathematical abstractions,
but as quantities tied to measurement and kinematics.
If the macroscopic interface is ambiguous, our interpretation of every experiment relying on that interface becomes ambiguous with it.

\bigskip

\section*{A Note to the Reader}

This prologue frames classical physics as the ``Physics of the Human Scale'' not to romanticize it,
but to emphasize its true status: it is the rigorous description of a world inherently bandwidth-limited.
The continuous macroscopic world is not a lie.
It is the only coherent world available to finite observers.
By understanding the mechanics of this emergence, we do not diminish the reality of the experience; we secure the language to describe it.

\bigskip

\noindent
With this perspective established, we proceed to the technical inquiry.
This work constitutes an attempt to:
examine how macroscopic electrodynamics emerges from the microscopic vacuum,
understand why its standard formulations have historically diverged,
and demonstrate what is recovered when the filtered description is forced to remain mechanically closed.

\chapter{Introduction: The Macroscopic Ambiguity}
\label{chap:Introduction}

\section{The Challenge of Coherence}
\label{sec:Intro_Coherence}

In the vacuum, Classical Electrodynamics stands as a complete structure.
Maxwell's equations provide a description of the electromagnetic field that is rigorous, deterministic, and universally accepted.
However, when the focus shifts from the clarity of the vacuum into the complexity of ponderable media, the blueprint becomes structurally ambiguous.

For over a century, the physics of continuous media has been characterized by an unresolved structural question.
When addressing a fundamental mechanical inquiry---\textit{Where exactly does the force act inside a polarized dielectric?} or \textit{How is momentum conserved when a magnet moves?}---one is presented with a variety of competing maps. 
Standard textbooks typically present a specific solution, often the Minkowski energy and momentum definitions, as the effective description of reality.
In these texts, the debate regarding force density is generally treated as a settled matter of bookkeeping.

The ``modern resolution,'' famously articulated by Penfield and Haus~\cite{PenfieldHaus1967}, posits that this ambiguity is inherent to the theory.
This view argues that \textit{all} formulations (Minkowski, Abraham, Chu, etc.) can be correct, provided one postulates the appropriate ``material counterpart'' tensor to balance the conservation laws.

This monograph conducts a structural re-examination of this view. 
It indicates that while the modern consensus is mathematically robust, Classical Electrodynamics may not be a closed chapter in this regard.
While the consensus frames the choice of force law as a matter of convention, the analysis presented herein suggests that different definitions of energy, force density, and momentum represent distinct physical descriptions.
The "hypothesis of equivalence" (or "arbitrary split") is identified as resting on a \textbf{category error}---a topological mismatch between the mathematical definition of momentum and the physical reality of mass.

The requisite link is a simple kinematic lock, formalised here as the \textit{force-energy consistency criterion} (FECC).
This criterion follows directly from Newton's Second Law ($\mathbf{f} = m \mathbf{a}$).
Because force is defined as the time-rate of change of momentum of a massive body, any transfer of energy ($\Delta U$) via a force must be kinematically coupled to the velocity ($\mathbf{v}$) of that specific mass target.

It is argued that this mechanical link has been obscured in electrodynamics for the last 120 years.
The inquiry was, in this view, not concluded---the link was simply decoupled in the formalism.
This analysis indicates that if this kinematic lock is restored, the structural ambiguity is addressed, suggesting that the macroscopic vacuum energy-momentum-stress tensor provides the consistent description of the closed system.

\section{A Survey of Formal Divergence}
\label{sec:Intro_Crisis}

The perception of Classical Electrodynamics as a settled theory is challenged by a systematic lack of consensus regarding its most basic operational outputs: electromagnetic momentum and energy in matter.
The following survey is not merely a historical review; it shows that various physical hypotheses regarding the energy-momentum tensor have coexisted for over a century.
The persistence of these competing formalisms reveals a deep structural ambiguity at the heart of the subject.

\subsection{Energy-Momentum-Stress Tensors: The Formalism of Divergence}

The structural disagreement between the formulations is fundamental.
It represents a conflicting landscape of hypotheses, proposing different definitions for the core elements of the theory: the electromagnetic stress tensor (momentum flux), the electromagnetic momentum density, the energy density, and the energy flux (Poynting vector).
The Energy-Momentum-Stress Tensor formally unifies two separate conservation laws: a vector equation governing momentum and a scalar equation governing energy.

\textbf{Momentum Balance (Spatial Components):}
The spatial divergence of the tensor corresponds to the force density.
It describes how and where the electromagnetic field couples to the mechanical domain:
\begin{equation}
    \mathbf{f} = -\frac{\partial \mathbf{g}}{\partial t} - \nabla \cdot \mathbf{T}
\end{equation}
Physically, this interprets the force density $\mathbf{f}$ as the local sink of electromagnetic momentum.
Whenever the divergence of the momentum flux is not balanced by the time variation of the momentum density:
\begin{equation}
    -\frac{\partial \mathbf{g}}{\partial t} - \nabla \cdot \mathbf{T} \neq 0
\end{equation}
there is a coupling to the mechanical domain.
The loss of electromagnetic momentum manifests as a gain in local mechanical momentum.

\textbf{Energy Balance (Temporal Component):}
The temporal component represents the power balance.
It describes how and where the field couples to the energy domain:
\begin{equation}
    P = -\frac{\partial u}{\partial t} - \nabla \cdot \mathbf{S}
\end{equation}
This scalar equation follows the same conceptual logic.
Wherever there is a local sink of electromagnetic energy:
\begin{equation}
    -\frac{\partial u}{\partial t} - \nabla \cdot \mathbf{S} \neq 0
\end{equation}
energy leaves the electromagnetic domain and is transferred into the mechanical domain as power density $P$.

These four elements---momentum density $\mathbf{g}$, stress tensor $\mathbf{T}$, energy density $u$, and energy flux $\mathbf{S}$---define the complete tensor.
They thereby define the force density and power density predictions of the theory:
\begin{equation}
\boxed{
\underbrace{
\begin{pmatrix}
f_x & f_y & f_z & P
\end{pmatrix}
}_{\text{Mechanical Sink (Force/Power)}}
= -
\begin{pmatrix}
\partial_t & \partial_x & \partial_y & \partial_z
\end{pmatrix}
\cdot
\underbrace{
\begin{pmatrix}
g_x & g_y & g_z & u \\
T_{xx} & T_{yx} & T_{zx} & S_x \\
T_{xy} & T_{yy} & T_{zy} & S_y \\
T_{xz} & T_{yz} & T_{zz} & S_z
\end{pmatrix}
}_{\text{Electromagnetic Source (Field Tensor)}}
}
\end{equation}

\subsubsection{The Minkowski Tensor~\cite{Minkowski1908}}
Proposed as the natural covariant extension of macroscopic electrodynamics:
\begin{itemize}
    \item \textbf{Energy Density:} $u_M = \frac{1}{2} (\mathbf{E} \cdot \mathbf{D} + \mathbf{B} \cdot \mathbf{H})$
    \item \textbf{Momentum Density:} $\mathbf{g}_M = \mathbf{D} \times \mathbf{B}$
    \item \textbf{Energy Flux (Poynting Vector):} $\mathbf{S}_M = \mathbf{E} \times \mathbf{H}$
    \item \textbf{Stress Tensor:} $\mathbf{T}_M = \mathbf{E} \otimes \mathbf{D} + \mathbf{H} \otimes \mathbf{B} - u_M \mathbf{I}$
\end{itemize}
While mathematically elegant, the Minkowski tensor is asymmetric ($T^{ij} \neq T^{ji}$), leading to well-known difficulties in conserving angular momentum without introducing compensating material torque terms.

\subsubsection{The Abraham Tensor}
Abraham argued for symmetry, modifying the stress tensor and momentum density while retaining the Minkowski energy density:
\begin{itemize}
    \item \textbf{Energy Density:} $u_A = \frac{1}{2} (\mathbf{E} \cdot \mathbf{D} + \mathbf{B} \cdot \mathbf{H})$
    \item \textbf{Momentum Density:} $\mathbf{g}_A = \frac{1}{c^2}(\mathbf{E} \times \mathbf{H})$
    \item \textbf{Energy Flux (Poynting Vector):} $\mathbf{S}_A = \mathbf{E} \times \mathbf{H}$
    \item \textbf{Stress Tensor:} $\mathbf{T}_A = \frac{1}{2} \left( \mathbf{E} \otimes \mathbf{D} + \mathbf{D} \otimes \mathbf{E} + \mathbf{H} \otimes \mathbf{B} + \mathbf{B} \otimes \mathbf{H} \right) - u_A \mathbf{I}$
\end{itemize}

\subsubsection{The Vacuum Formulation (Amperian)~\cite{PenfieldHaus1967}}
This framework adheres strictly to the microscopic definitions, positing that the field retains its vacuum structure even within matter:
\begin{itemize}
    \item \textbf{Energy Density:} $u_0 = \frac{1}{2} (\varepsilon_0 E^2 + \mu_0^{-1} B^2)$
    \item \textbf{Momentum Density:} $\mathbf{g}_0 = \varepsilon_0(\mathbf{E} \times \mathbf{B})$
    \item \textbf{Energy Flux (Poynting Vector):} $\mathbf{S}_0 = \mu_0^{-1}(\mathbf{E} \times \mathbf{B})$
    \item \textbf{Stress Tensor:} $\mathbf{T}_0 = \varepsilon_0 \mathbf{E} \otimes \mathbf{E} + \mu_0^{-1} \mathbf{B} \otimes \mathbf{B} - u_0 \mathbf{I}$
\end{itemize}

\subsubsection{The Scope of Formal Divergence}
The formalisms above represent only the most prominent positions.
The literature contains numerous other proposals, including the kinematic formulation of \textbf{Chu}~\cite{FanoChuAdler1960}, the symmetric tensor of \textbf{Boffi}~\cite{Boffi1957}, and the statistical mechanical derivation of \textbf{de Groot and Suttorp}~\cite{deGroot1972}.

\subsection{Divergence in Force Density}
The different tensor formulations predict different force densities.
Crucially, these are not merely alternative mathematical representations of the same physics; they represent distinct physical hypotheses regarding the \textit{localization} of interaction.
This divergence is not limited to the choice of stress tensor.
The literature offers a broader spectrum of force density proposals, often derived from specific microscopic models or thermodynamic arguments.
This list is by no means exhaustive.
 
One prominent alternative is the \textbf{Einstein-Laub Force Density}, derived from a microscopic model of averaged dipoles~\cite{EinsteinLaub1908}:
\begin{equation*}
    \mathbf{f}_{EL} = (\mathbf{P} \cdot \nabla)\mathbf{E} + \mu_0(\mathbf{M} \cdot \nabla)\mathbf{H} + \frac{\partial\mathbf{P}}{\partial t} \times \mu_0\mathbf{H} + \mu_0\mathbf{E} \times \frac{\partial\mathbf{M}}{\partial t}
\end{equation*}
Another is the Kelvin force, often applied in quasistatic scenarios~\cite{Zangwill2013, Melcher1981}:
\begin{equation*}
    \mathbf{f}_{\text{Kelvin}} = (\mathbf{P} \cdot \nabla)\mathbf{E} + (\mathbf{M} \cdot \nabla)\mathbf{B}
\end{equation*}

\subsection[Energy-Derived Forces: Korteweg-Helmholtz]{Energy-Derived Forces: The\\ Korteweg-Helmholtz Approach}
Beyond direct force postulation, a parallel and prominent tradition in the literature derives mechanical effects from energy-based variational principles.
Complementing the direct force expressions, the influential \textbf{Korteweg-Helmholtz} (KH) approach derives force through such a method: calculating the gradient of a pre-defined energy functional.
Championed by figures such as Stratton and Landau \& Lifshitz, this method yields a distinct force density known for its simplicity in linear media~\cite{Stratton1941, LandauLifshitzVol8}:
\begin{equation*}
    \mathbf{f}_{KH} = - \frac{1}{2} E^2 \nabla\varepsilon - \frac{1}{2} H^2 \nabla\mu
\end{equation*}

\subsection{Divergence in Energy Flux}
This ambiguity extends to the definition of energy flux.
Just as the force density varies, the partitioning of energy between the field and the material host varies radically depending on the chosen formalism.
It is worth noting that the formulations listed within the tensors above are not exhaustive; consistent thermodynamic treatments often yield results that differ subtly from the canonical tensors.

Standard texts such as \cite{Jackson1999} and Stratton \cite{Stratton1941} often present the "conventional" Poynting theorem energy density as:
\begin{equation}
    u_{conv} = \int \left( \mathbf{E} \cdot d\mathbf{D} + \mathbf{H} \cdot d\mathbf{B} \right)
\end{equation}
This definition leads to a power balance equation:
\begin{equation}
     P_{conv} = \mathbf{j}_f \cdot \mathbf{E} = -\nabla \cdot (\mathbf{E} \times \mathbf{H}) - \left( \mathbf{E} \cdot \frac{\partial \mathbf{D}}{\partial t} + \mathbf{H} \cdot \frac{\partial \mathbf{B}}{\partial t} \right)
    \label{eq:intro_conventional_poynting_def}
\end{equation}
It is important to note that this thermodynamic energy definition differs subtly but significantly from the fixed energy densities $u_M$ and $u_A$ proposed by Minkowski and Abraham.

\subsection{The Undetermined System}
The Abraham-Minkowski controversy, historically focused on the definition of momentum density ($\mathbf{g}_A$ vs. $\mathbf{g}_M$), is often treated as an isolated anomaly.
However, viewed through the lens of this survey, it appears as the most visible symptom of a broader structural indeterminacy.
The theory, in its standard formulation, lacks precise definitions for its fundamental mechanical outputs.
There is no agreed-upon map for where the force acts or how the momentum flows.
As the list of competing formalisms illustrates, ambiguity extends to every mechanical output of the theory: force, power, momentum flux, and energy flux.

The existence of such structural ambiguity in a classical, deterministic theory suggests that the standard axioms of macroscopic electrodynamics, while necessary, are by themselves \textbf{insufficient} to uniquely determine the mechanical state of matter.
This suggests that the system is mathematically under-constrained.
Consequently, the resolution to these paradoxes cannot be found by simply re-analyzing Maxwell's equations or appealing to convention.
It motivates the introduction of a rigorous external constraint—a mechanical anchor capable of distinguishing the physical path from the mathematical possibilities.
This manuscript proposes that this constraint is found in the axiomatic laws of mechanics.

\section{The Landscape of Ambiguity}

How does the literature deal with this ambiguity?
A survey of the most prominent textbooks defining the modern consensus reveals a spectrum of approaches, ranging from specific endorsements to principled agnosticism.

\paragraph{Zangwill}
In \textit{Modern Electrodynamics} (2012), Zangwill addresses the controversy by adopting a pragmatic consensus: he treats the competing formalisms not as mutually exclusive theories, but as descriptions of distinct physical quantities.
He identifies the Minkowski vector ($\mathbf{D} \times \mathbf{B}$) as the ``canonical'' pseudomomentum relevant to wave propagation \textit{in} a medium, while retaining the Abraham vector ($\mathbf{E} \times \mathbf{H}/c^2$) to describe the ``kinetic'' momentum of the field itself.
For the mechanics of continuous media, he utilizes the Minkowski stress tensor (derived via the variational principle), demonstrating that it yields the standard Helmholtz force density ($\mathbf{f} = \rho \mathbf{E} + \mathbf{j} \times \mathbf{B} - \frac{1}{2}E^2 \nabla \varepsilon$) required to describe electrostriction and internal stresses in fluids.
He argues that neither formalism is complete on its own and that consistency is restored only when the stress tensor of the material medium is explicitly included.
Notably, Zangwill restricts his analysis to the Abraham-Minkowski duality; he does not consider or present alternative historical formulations such as the Einstein-Laub, Chu, Boffi, or Amperian tensors.

\paragraph{Jackson}
In \textit{Classical Electrodynamics} (3rd ed., 1999), Jackson approaches the subject by emphasizing that the division of the total system's momentum into ``electromagnetic'' and ``mechanical'' components is largely a matter of definition and convention.
While he acknowledges the symmetry arguments favoring the Abraham vector ($\mathbf{E} \times \mathbf{H}/c^2$)---specifically for satisfying the conservation of angular momentum---he primarily utilizes the Minkowski stress tensor for macroscopic problems involving continuous media.
Through a thermodynamic analysis of virtual displacements, Jackson derives the Helmholtz force density ($\mathbf{f} = \rho \mathbf{E} + \mathbf{J} \times \mathbf{B} - \frac{1}{2}E^2 \nabla \varepsilon - \frac{1}{2}H^2 \nabla \mu$), which he presents as the correct formulation for calculating bulk forces, explicitly accounting for electrostriction and magnetostriction in fluids.
He concludes that while the Minkowski momentum ($\mathbf{D} \times \mathbf{B}$) is often more convenient for describing wave propagation in a medium (pseudomomentum), the ambiguities of the controversy disappear only when the total momentum of the closed system is evaluated.
Consistent with the standard canonical approach, Jackson focuses his rigorous derivations on the Maxwell-Lorentz macroscopic theory and does not provide an in-depth treatment of alternative historical formulations such as the Chu, Einstein-Laub, or Amperian tensors.

\paragraph{Stratton}
In \textit{Electromagnetic Theory} (1941), Stratton anchors the theory in thermodynamics rather than arbitrary tensor postulation.
He explicitly denies the ``essential physical reality'' of the stress tensor components themselves, treating them strictly as mathematical auxiliaries.

Regarding force density, Stratton explicitly rejects the Amperian (or Livens) formulation --- which models forces via effective bound charge ($\rho_{pol}\mathbf{E}$) and current --- arguing that it is ``manifestly incorrect'' for deformable media because it neglects the work associated with elastic strain.
Instead, he utilizes the Principle of Virtual Work to derive the Helmholtz force density, which uniquely incorporates the electrostriction term $\frac{1}{2} \nabla [ E^2 \tau (\partial \varepsilon/\partial \tau) ]$.

Regarding momentum, Stratton discards the Minkowski formulation ($\mathbf{D} \times \mathbf{B}$) entirely.
He argues that its resulting asymmetric stress tensor violates the theorem of the conservation of angular momentum.
Consequently, he adopts the Abraham-von Laue hypothesis ($\mathbf{g} = \mathbf{E} \times \mathbf{H}/c^2$) as the unique solution satisfying relativistic symmetry and the theorem of the center of energy.
Notably, Stratton frames the choice as a binary between the mathematically correct (Abraham) and the physically inconsistent (Minkowski).

\paragraph{Griffiths}
In \textit{Introduction to Electrodynamics} (5th ed., 2024), Griffiths addresses the ambiguity by adhering strictly to the microscopic (vacuum) formulation, effectively bypassing the macroscopic tensor controversy.
He defines the electromagnetic energy density as $u = \frac{1}{2}(\varepsilon_0 E^2 + \frac{1}{\mu_0} B^2)$ and the energy flux as the standard Poynting vector $\mathbf{S} = \frac{1}{\mu_0}(\mathbf{E} \times \mathbf{B})$.

He does not discuss or mention the Abraham or Minkowski tensors, nor does he present the competing macroscopic formulations involving $\mathbf{D}$ and $\mathbf{H}$.
Instead, he explicitly defines the electromagnetic force density $\mathbf{f}$ strictly as the Lorentz force density acting on charge and current distributions:
\begin{equation}
    \mathbf{f} = \rho\mathbf{E} + \mathbf{J} \times \mathbf{B}
\end{equation}
He does not consider the Korteweg-Helmholtz, Kelvin, or Einstein-Laub force densities.
While he calculates forces on dielectrics and dipoles, he derives these from the energy budget or the Lorentz force on bound charges, without referencing these alternative stress-tensor hypotheses.

\paragraph{Landau and Lifshitz}
In \textit{Electrodynamics of Continuous Media} (Vol. 8), Landau and Lifshitz bypass the ambiguity of postulating a stress tensor by grounding their formulation entirely in thermodynamics.
Rather than selecting between competing definitions, they derive the stress tensor using a variational principle applied to the total free energy of the system (matter plus field).
This approach yields the Korteweg-Helmholtz force density ($\mathbf{f} = \nabla [\frac{1}{2} E^2 \rho (\partial \varepsilon / \partial \rho)] - \frac{1}{2} E^2 \nabla \varepsilon$), which they present as the definitive description of the mechanical state, distinguishing themselves by rigorously including the electrostriction terms often omitted in elementary treatments.
They argue that attempting to separate the ``field'' energy from the ``material'' energy is physically meaningless in a continuous medium; consequently, the stress tensor is treated not as an independent physical hypothesis but as a derived quantity required to satisfy thermodynamic equilibrium.
In this rigorous derivation, they do not engage with the Abraham-Minkowski momentum controversy, nor do they present or discuss alternative historical formulations such as the Einstein-Laub, Chu, or Amperian tensors.

\paragraph{Eringen and Maugin}
In \textit{Electrodynamics of Continua} (1990) \cite{EringenMaugin1990}, Eringen and Maugin address the formal divergence not by selecting a preferred energy momentum stress tensor, but by constructing the macroscopic theory from the bottom up via the statistical averaging of microscopic Maxwell-Lorentz electron theory.
Rather than postulating a macroscopic axiom, they derive the electromagnetic force density explicitly from a microscopic model of bound charges and dipoles, yielding a definitive source term that accounts for dynamic polarization and magnetization interactions similar to the Einstein-Laub force.
They fundamentally argue that the separation of the total momentum flux into ``electromagnetic'' and ``mechanical'' components is mathematically non-unique; as demonstrated rigorously by Maugin in \cite{Maugin1980}, competing formulations (such as Abraham, Minkowski, or their own Grot-Eringen tensor) are physically equivalent descriptions distinguishable only by the choice of independent thermodynamic variables.
Consequently, they bypass the controversy by grounding the theory in thermodynamics: they utilize a total energy balance law and define material response via a Helmholtz free energy functional, which rigorously couples field and deformation without requiring an arbitrary isolation of field energy from material energy.

\paragraph{Wald}
In \textit{Advanced Classical Electromagnetism} (2022), Wald frames the formal divergence by identifying the \textbf{microscopic Lorentz framework} as the sole fundamental reality.
He asserts that the ``true'' electromagnetic stress-energy tensor is uniquely determined by the \textbf{vacuum formulation} of microscopic electrodynamics.

Regarding the \textbf{Abraham-Minkowski con\-tro\-versy}, Wald explicitly adopts a stance of \textbf{agnosticism}, asserting that the dispute is ill-posed.
He states that ``neither formula represents the true electromagnetic momentum density'' inside matter.
The macroscopic theory, derived by averaging, is thus not fundamental.
He declines to select a macroscopic tensor, arguing that it will always be a simplification of reality---an averaged system where the high-gradient atomic field interactions are filtered out, and the fundamental reality of interaction is obscured.

Regarding force density, Wald bypasses the postulation of macroscopic stress tensors entirely.
He derives the macroscopic force density directly from the microscopic averaging of the fundamental Lorentz force on point charges.
This yields results similar to the Kelvin force density (specifically the $(\mathbf{P} \cdot \nabla)\mathbf{E}$ term) as the correct description of the mechanical state, rejecting formulations not grounded in the microscopic Lorentz theory.

\paragraph{de Groot and Suttorp}
In \textit{Foundations of Electrodynamics} (1972) \cite{deGroot1972}, de Groot and Suttorp provide perhaps the most rigorous statistical mechanical resolution to the divergence.
Sharing the fundamental epistemological stance of Wald, Eringen, and Maugin, they explicitly reject the idea that macroscopic laws can be postulated independently.
Instead, they assert that the microscopic Maxwell-Lorentz framework represents the only underlying reality, and that macroscopic tensors must be derived strictly via the statistical averaging of microscopic motion.

Their approach distinguishes itself by introducing an intermediate atomic level, where point charges are first grouped into stable aggregates defined by multipole moments before statistical averaging is applied.
This derivation yields a macroscopic force density that rigorously validates the Kelvin force structure (specifically the $(\nabla \mathbf{E}) \cdot \mathbf{P}$ interaction term) and the Amperian magnetic force.
 
Crucially, their statistical decomposition resolves the Abraham-Minkowski controversy by strictly separating field and matter contributions.
They demonstrate that terms involving particle properties (mass, velocity, internal energy) naturally fall into the mechanical pressure tensor (specifically into kinetic and correlation pressure terms).
Consequently, they reject both the Minkowski ($\mathbf{D} \times \mathbf{B}$) and Abraham ($\varepsilon_0 \mathbf{E} \times \mathbf{H}$) momenta in favor of the Vacuum formulation, identifying the true electromagnetic momentum density solely as $\varepsilon_0 \mathbf{E} \times \mathbf{B}$.
They further derive the energy flux as the standard Poynting vector $\mathbf{E} \times \mathbf{H}$, noting that the divergence-free nature of the energy balance allows for this specific choice without contradicting the vacuum momentum density.

\paragraph{Penfield \& Haus}
In \textit{Electrodynamics of Moving Media} (1967), Penfield and Haus attribute the historical controversy to the arbitrary split of the total system's energy-momentum into electromagnetic and mechanical subsystems.
They demonstrate that the Minkowski, Abraham, and Chu formulations are physically equivalent, provided that each arbitrary field definition is paired with its thermodynamically consistent material model.
Consequently, they argue that structural ambiguity disappears only when the closed system is evaluated as a whole.
To resolve the indeterminacy in force prediction, they discard the reliance on postulated stress tensors and instead employ the Principle of Virtual Power.
Through this variational method, they derive a unique, formulation-independent force density that recovers distinct historical expressions based on the material's complexity: they identify the Kelvin force as the correct description for a model of isolated, non-interacting dipoles, while the Helmholtz force is required for a continuous dielectric fluid to account for electrostrictive particle interactions.
While acknowledging the validity of all consistent formulations, they advocate for the Chu formulation as the most robust framework, as it simplifies the arbitrary split by treating polarization and magnetization strictly as material sources rather than flux constituents.

\paragraph{Brevik}
In his comprehensive review \textit{Experiments in Phenomenological Elec\-tro\-dynamics} (1979), Brevik asserts that phenomenological electrodynamics offers no unique prescription for the separation of the total energy-momentum tensor into field and matter parts.
Despite this formal ambiguity, he identifies the Helmholtz force density---derived via the thermodynamic variational principle (Korteweg-Helmholtz)---as the single consistent theory capable of explaining all examined experiments across frequency domains.
He concludes that while the Minkowski tensor is valid and convenient for describing momentum transport in optics, the Abraham force term is physically real and necessary to explain low-frequency mechanical oscillations (such as in the Walker-Lahoz experiment).

\subsection{Summary of the Modern Consensus}

The history of this subject is not characterized by the eventual triumph of a single macroscopic formulation.
Rather, as detailed in the comprehensive modern review by Pfeifer et al.~\cite{Pfeifer2007}, the debate has largely resolved into a consensus of \textbf{formal equivalence}.
Crucially, this consensus has flourished in the absence of a diagnostic criterion capable of physically validating the tensor itself.

This modern perspective, foundational to most contemporary textbooks, adopts the view articulated by Penfield and Haus~\cite{PenfieldHaus1967}: that the division of the total energy-momentum tensor into ``electromagnetic'' and ``mechanical'' parts is not a physical observable, but an \textbf{arbitrary split}.
Under this paradigm, the search for a unique electromagnetic tensor is considered ill-posed.
Any algebraically consistent formulation (Minkowski, Abraham, Chu, etc.) is accepted as valid, provided it is accompanied by the corresponding material subsystems required to satisfy global conservation laws.
The modern resolution is thus one of convention rather than discovery.

\section{Thesis: A Structural Argument}
\label{sec:Intro_Approach}

The following analysis re-examines the consensus. A structural analysis of the ``Arbitrary Split'' is conducted, suggesting that Classical Electrodynamics may not be a closed chapter in this regard.

We argue that different definitions of energy, force density, and momentum represent \textbf{mutually exclusive physical hypotheses}. The decision to assign a specific momentum density to the field is not merely a bookkeeping choice; it is a physical hypothesis about where the energy resides and how it moves. The ``Arbitrary Split'' is therefore argued to be an insufficient physical hypothesis: while multiple algebraic formulations are possible, the FECC implies they cannot all be mechanically true simultaneously. The description of physical reality required a specific determination.

\subsection{Reframing the Divergence}

The persistence of this controversy for over a century is suggested to stem from the lack of a diagnostic tool capable of auditing the tensors themselves.
The historical focus has centered on \textit{Global Conservation Laws}, where it is well established that one can choose \textbf{any} electromagnetic tensor provided one postulates a corresponding ``material counterpart'' to balance the total budget.
This ``Total System'' approach, while necessary, is argued to be insufficiently constrained: as long as the coupling between field and matter is treated as a free variable, there are infinite ways to partition the energy while satisfying global conservation.

The consensus championed by Penfield, Haus, and others is identified as relying on a classification divergence: it conflates \textit{algebraic equivalence} (balancing the equation) with \textit{physical identity} (validating the terms).
Postulating ``effective'' momentum terms to balance an asymmetric tensor achieves global conservation, but it obscures the local mechanical reality.
To resolve this, criteria usage stricter than global conservation---one that enforces the correct local definition of work---is proposed.
The fundamental constraint of mechanics is therefore re-imposed: the \textit{force-energy consistency criterion} (FECC).

The definitions of force density ($\mathbf{f}$) and power density ($P$) are argued to be not independent choices subject to convention; they are rigidly coupled by the velocity of the massive charge carriers ($\mathbf{v}$) via the identity:
\begin{equation}
    P \equiv \mathbf{f} \cdot \mathbf{v}
\end{equation}
Crucially, $\mathbf{v}$ refers strictly to the kinematic velocity of well defined mass targets.
By enforcing this ``kinematic lock,'' the ``freedom'' to choose between tensors is argued to vanish.
One is not free to choose the momentum.
The mass has a velocity, and the field has a force.
If $P \neq \mathbf{f} \cdot \mathbf{v}$, the theory is mathematically consistent but suggested to be mechanically ungrounded.

\subsection{The Hierarchy of Uniqueness}
\label{sec:Intro_Hierarchy}

Central to this thesis is the establishment of a clear boundary between what is physically unique and what is structurally indeterminate. We propose that the resolution of the tensor controversy requires a strict "Separation of Powers" in macroscopic electrodynamics, organized into a clear hierarchy.

\paragraph{Level 1: Microscopic Uniqueness (Ground Truth).}
At the fundamental level, the coupling is unique. The Lorentz force and the interaction $\mathbf{j}\cdot\mathbf{e}$ provide the absolute ground truth. The \textit{Force-Energy Consistency Criterion} (FECC) applies strictly at the carrier level.

\paragraph{Level 2: Macroscopic Uniqueness of Exchange (Input Budget).}
This constitutes \textbf{Stage 1: Exchange / Supply}. Here, we determine the energy-momentum transferred from the electromagnetic field into the material system.
After filtering, it is demonstrated that what remains invariant (and therefore unique) is the \textit{total} EM$\to$matter input.
This is identified as the domain of the \textbf{Macroscopic Vacuum Tensor}. It is proposed as the unique master equation for the input ledger. It is argued that historical tensors (Minkowski, Abraham) fail here because they conflate the stages, packaging allocation assumptions into the exchange tensor.

\paragraph{Level 3: Macroscopic Indeterminacy of Allocation (Internal Routing).}
This constitutes \textbf{Stage 2: Allocation / Demand}. Here, we determine how the received energy-momentum is distributed within matter---into bulk motion, deformation, heat, and reversible storage.
We propose that this allocation \textbf{cannot}, in general, be inferred from macroscopic fields alone, because filtering deletes the microstructural topology. It requires constitutive and microstructural closure.
Thus, it is argued that while the \textit{input budget} is unique, the \textit{internal split} is dependent on the microstructure.

\subsection{The Mechanical Synthesis: The Unified Matrix}

The analysis aligns with the epistemological stance of Wald, de Groot, Suttorp, and Eringen: the microscopic Lorentz theory represents the sole fundamental framework of electrodynamics.
Macroscopic laws are not independent truths but are derived simplifications obtained via averaging.
However, a distinct methodological perspective on this procedure is introduced.
The analysis proposes that spatial averaging should be understood rigorously as a \textit{spectral filter}---the mathematical realization of the observer constraints discussed in the Prologue.
Mathematically, convolution with an averaging kernel is identical to a low-pass filter in the spatial frequency domain.

This shift in perspective transforms the problem from one of ``approximation'' to one of \textit{information compression}.
By applying this filter, the microscopic electromechanical reality is subdivided into two distinct domains: the coherent low-frequency ``Signal'' and the incoherent high-frequency ``Noise.''
Crucially, the analysis suggests that the ``Noise'' is not random; it is kinematically structured.
By decomposing the microscopic motion of positive and negative charges, three distinct kinematic modes are identified, each corresponding to a major discipline of classical physics:

\begin{itemize}
    \item \textbf{Thermodynamics\\ (Unordered Motion):} The physics of high-frequency, incoherent particle velocity (Heat).
    \item \textbf{Continuum Mechanics (Ordered Motion):} The physics of low-frequency, coherent motion of neutral aggregates (Deformation).
    \item \textbf{Electrodynamics (Relative Motion):} The physics of the relative velocity between ion and electron populations (Current).
\end{itemize}

This leads to the framework of the Mechanical Synthesis.
Macroscopic ``Material Response'' parameters---temperature, permittivity, elasticity, and viscosity---are interpreted as \textit{emergent spectral coefficients} of the underlying high-frequency microscopic reality.
They represent the integrated energy of the high-frequency interactions that have been filtered out of the macroscopic description.
Temperature is the compression of incoherent kinetic energy; elasticity is the compression of high-frequency binding fields.
Thus, the three pillars of classical physics are conceptually unified as spectral projections of the same underlying mechanical reality.

\subsection{New Methodological Tools}

In summary, two major tools are advanced to the debate:

\begin{itemize}
    \item \textbf{The Diagnostic Tool (Top-Down):} The \textit{force-energy consistency criterion} (FECC). The analysis replaces the flexible ``Global Conservation'' standard with a rigid ``Local Topology'' standard, testing whether the force density strictly drives the kinematic mass velocity ($P = \mathbf{f} \cdot \mathbf{v}$). In other words, the force proposed by a tensor formulation must match the energy output defined by that very same tensor.
    \item \textbf{The Derivation Method (Bottom-Up):} The \textit{spectral filtering perspective}. The analysis replaces the concept of ``Averaging'' (approximation) with ``Filtering'' (architecture), treating macroscopic laws as the low-frequency signal of the microscopic vacuum.
\end{itemize}

By strictly applying these tools, the ambiguity of the macroscopic energy-momentum-stress tensor is aimed to be clarified.

\section{Structure of the Argument}
\label{sec:Intro_Methodology}

The argument of this manuscript is constructed as a systematic analysis, proceeding in seven logical stages: the establishment of the ground truth, the derivation of the macroscopic filter, the application of that filter to the three canonical domains of interaction, the analysis of historical models, and the final mechanical synthesis.

\paragraph{Part \ref{part:Micro}: The Microscopic Foundation}
The rigorous baseline is established by adopting the microscopic Lorentz theory as the fundamental ground truth.
At this level, it is argued that the definitions of force and momentum are not matters of convention but are rigidly constrained by the \textit{Force-Energy Consistency Criterion} (FECC).
The theorem of mutual exclusivity is derived, indicating that the competing historical tensors (Minkowski, Abraham) are not merely alternative "viewpoints" but mutually exclusive physical hypotheses.

\paragraph{Part \ref{part:MacroscopicFilter}: The Bottom-Up Construction}
This methodological shift---from "averaging" to "information compression"---suggests that the classical domains of thermodynamics, continuum mechanics, and electrodynamics are not independent disciplines.
We propose that they are emergent spectral projections of the same underlying kinematic reality, coupled via the \textit{host interface}.

\paragraph{Part \ref{part:ElectricResponse}: The Electric Response}
This diagnostic framework is applied to the statics and dynamics of charge separation.
The "stored energy" of polarization is traced to its physical location in the variance of microscopic binding fields, aiming to resolve the energetic paradoxes of dielectrics.

\paragraph{Part \ref{part:MagneticResponse}: The Magnetic Response}
The analysis turns to the dynamics of rotational currents.
Electron spin is modeled as an active superconducting loop in an effort to resolve the "Artifactual Work" paradox.
The "Ampere Topology" is argued to be the mechanically consistent description of magnetic force.

\paragraph{Part \ref{part:MovingMatter}: Moving Matter and Relativistic Sources}
The analysis extends to the relativistic regime, proposing that the phenomenology of moving matter is a geometric consequence of the relativistic Source Tensor.

\paragraph{Part \ref{part:NewPart6}: The Analysis of Macroscopic Models}
The historical formulations (Minkowski, Abraham, Korteweg-Helmholtz) are subjected to a rigorous examination using the \textit{Force-Energy Consistency Criterion} (FECC).
The analysis aims to highlight the mechanical inconsistencies, artifactual forces, and topological filtering inherent in these bundled energy functionals.

\paragraph{Part \ref{part:Synthesis}: The Mechanical Synthesis}
The argument concludes by synthesizing these findings into a unified matrix of classical physics.
The macroscopic ambiguity is addressed in favor of the \textit{Principle of Physical Exclusivity}, identifying the vacuum-form tensor as the description consistent with the proposed FECC.

\subsection{Detailed Roadmap}

The argument unfolds across the following structure:

\begin{itemize}
    \item \textbf{Part I: Microscopic Electrodynamics}
    \begin{itemize}
        \item Chapter \ref{chap:part1_intro}: Introduction to Part I
        \item Chapter \ref{chap:FreeCharges}: The Physical Foundation: Microscopic Interactions
        \item Chapter \ref{chap:Critique_ArbitrarySplit}: The Arbitrary Split: A Microscopic Analysis
        \item Chapter \ref{chap:Experiment}: Experiment and the Interpretation of Evidence
    \end{itemize}
    
    \item \textbf{Part II: The Macroscopic Filter}
    \begin{itemize}
        \item Chapter \ref{chap:Part2_Intro}: Introduction: The Bridge of Scale
        \item Chapter \ref{chap:ThermoAndContinuum}: Thermodynamics and Continuum Mechanics
        \item Chapter \ref{chap:HostInterface}: The Host Interface: The Macroscopic Transmission
    \end{itemize}
    
    \item \textbf{Part III: The Electric Response}
    \begin{itemize}
        \item Chapter \ref{chap:IntroPart3}: Introduction to Part III
        \item Chapter \ref{chap:Conductors}: Free Charges on Conductors
        \item Chapter \ref{chap:Dielectrics}: Dielectrics: Binding Energy
        \item Chapter \ref{chap:ForceDensity}: Force Density: The Hierarchy of Deformation
        \item Chapter \ref{chap:BudgetInvariance}: Invariance of Total Momentum and Energy Budget
    \end{itemize}

    \item \textbf{Part IV: The Magnetic Response}
    \begin{itemize}
        \item Chapter \ref{chap:IntroPart4}: Introduction to Part IV
        \item Chapter \ref{chap:RotationalCurrents}: Free Rotational Currents
        \item Chapter \ref{chap:Spin}: Spin
        \item Chapter \ref{chap:Magnetic_Binding_Energy}: Magnetics: Binding Energy
        \item Chapter \ref{chap:MagneticForce}: Force Density in Magnetics: The Ampere Topology
    \end{itemize}

    \item \textbf{Part V: Moving Matter and Relativistic Sources}
    \begin{itemize}
        \item Chapter \ref{chap:Moving_Matter}: The Physics of Moving Sources
    \end{itemize}

    \item \textbf{Part VI: The Analysis of Macroscopic Models}
    \begin{itemize}
        \item Chapter \ref{ch:review_vacuum_framework}: The Vacuum Framework: The Standard of Judgment
        \item Chapter \ref{chap:Macro_Critique}: The Mechanical Audit of the Minkowski Tensor
        \item Chapter \ref{chap:Abraham_Audit}: The Focus on Symmetry: The Abraham Audit
        \item Chapter \ref{chap:PointDipole_Audit}: The Focus on Deformation: The Point-Dipole Audit
        \item Chapter \ref{chap:VariationalMethod}: The Variational Circularity: Force, Energy, and Mass
        \item Chapter \ref{chap:Part6_Conclusion}: Conclusion: The Resolution of the Paradox
    \end{itemize}

    \item \textbf{Part VII: The Mechanical Synthesis}
    \begin{itemize}
        \item Chapter \ref{chap:ReviewPart1to6}: Review of Parts I-VI
        \item Chapter \ref{chap:MechanicalUnification}: The Mechanical Unification
        \item Chapter \ref{chap:Conclusion}: Conclusion: The Mechanical Synthesis
    \end{itemize}
\end{itemize}

\section{The Emergent Result: A Return to Simplicity}

The foundational ambiguity of macroscopic electrodynamics is addressed herein through a systematic, first-principles reconstruction.
By enforcing the \textit{Force-Energy Consistency Criterion} (FECC), the indeterminacy of the energy-momentum tensor is constrained.
The analysis yields the \textit{Principle of Physical Exclusivity}: The system is argued to be mechanically determined, implying a unique path for the flow of energy.
A systematic simplification is suggested to emerge from this analysis.
The mechanically consistent formulation of macroscopic electrodynamics is argued to be the vacuum formulation---mirroring the exact microscopic formulation of the electromagnetic tensor.

Moreover, by deriving the macroscopic law through a filtering perspective, a structural isomorphism of classical physics is proposed.
The domains of continuum mechanics, thermodynamics, and macroscopic electromagnetism are conceptually unified as filtered views of the shared underlying microscopic reality.
Thus, the synthesis offered implies not the invention of new physics, but the rigorous audit of the mechanics of the standard theory.

\part{Part I: Microscopic Electrodynamics}
\label{part:Micro}

\chapter{Introduction to Part I}
\label{chap:part1_intro}

The primary objective of this book is the derivation of a consistent macroscopic theory of electrodynamics. To achieve this, we propose a "ladder of abstraction"—a hierarchical framework that rigorously separates the analysis into distinct layers of information resolution. This structural organization allows us to navigate from the fundamental quantum reality to the engineering continuum without losing logical consistency.

We define three distinct stages of analysis:

\begin{enumerate}
    \item \textbf{Stage 1 (Quantum Baseline):} Quantum Electrodynamics (QED). The domain of probabilistic wavefunctions, uncertainty, and fundamental microscopic interactions.
    \item \textbf{Stage 2 (Classical Microscopic Baseline):} The Lorentz Theory. The domain of deterministic trajectories, continuous fields, and discrete charge carriers interaction in a vacuum.
    \item \textbf{Stage 3 (Macroscopic Model):} The Continuum Theory. The domain of averaged fields, smooth media, and observable engineering variables.
\end{enumerate}

Part I of this book focuses on the definition of \textbf{Stage 2}. It establishes the deterministic microscopic baseline that serves as the input signal for the macroscopic filter derived in Part II.

\section{The Microscopic Baseline}
\label{chap:microscopic_baseline}

This chapter defines the physical interactions and modeling choices that constitute Stage 2: the interaction of charges and ions in the vacuum. This baseline serves as the deterministic input signal for the macroscopic averaging process. We adopt the standard \textit{classical microscopic model} (the Lorentz Theory) as our starting point \cite{Zangwill2013, Wald_AdvancedEM, deGroot1972}. This represents the \textit{first simplification} in our descent down the "ladder of abstraction," replacing the probabilistic QED model with a deterministic classical framework.

\section{The General Approach: Toward a Macroscopic Framework}

The fundamental description of nature is provided by Quantum Electrodynamics (QED), which represents our current understanding of the microscopic world. However, for the purposes of engineering and macroscopic analysis, this description offers a signal of effectively infinite bandwidth—a spectrum of computational complexity that is unusable for macroscopic predictions. To navigate the macroscopic world, this signal must be \textit{compressed}.

This work presents a microscopic deterministic ground truth which is strictly a simplification of the underlying quantum reality. The transition from Stage 1 to Stage 2 involves a fundamental shift from a probabilistic description to a deterministic one: instead of a particle cloud defined by a wavefunction, the model begins with deterministic microscopic charge carriers.

To achieve this consistency, a stabilizing mechanism must be introduced. Classical electrodynamics alone cannot explain why atoms do not collapse (Earnshaw's Theorem) or why discrete charges do not explode due to self-repulsion. To prevent these instabilities without solving the Schrödinger equation, \textit{quantum constraint forces} are introduced. These act as boundary conditions within the deterministic classical theory, maintaining the stability of the charge carriers. The stability of matter is thus treated not as a dynamic variable to be solved, but as a \textbf{pre-existing structural constraint}—the "passive stage" upon which the classical dynamics play out.

\subsection{Methodological Consistency: Intentional Simplification}

It is crucial to understand that these quantum boundary conditions are not merely convenient approximations but represent an \textit{intentional modeling choice} driven by the overarching goal of this work: the derivation of a macroscopic theory.

Macroscopic physics is, by definition, a low-resolution model of reality; its utility lies precisely in its suppression of high-frequency quantum information. Thus, the replacement of the probabilistic quantum cloud with a deterministic "Extended Body" is the \textbf{intentional simplification of the input signal}. This deterministic simplification, stabilized by passive quantum forces, constitutes the \textbf{structural justification} for the existence of the classical model itself. The utility of the classical domain lies precisely in its ability to filter out internal quantum complexity to resolve the deterministic trajectory of the center of mass.

The goal is not to replace QED with a new microscopic theory of the electron or ion. Rather, the goal is to simplify QED behavior sufficiently to serve as a tractable input for the macroscopic filter. The validity of the final goal—the macroscopic theory—will not be judged by the literal truth of its microscopic baseline, but by the rigorous internal consistency of the resulting user interface and its ability to predict macroscopic experiments.

\section{Modeling Choices: The Vacuum and the Charge}

Microscopically, we make specific modeling choices regarding the nature of the charge carriers—electrons and ions—to ensure the theory remains mathematically regular and suitable for macroscopic averaging.

\subsection{The Continuum vs. The Point}

Electrons and ions are not modeled as mathematical points. Instead, effective charge carriers are treated as very small but continuous distributions of charge. This "Extended Body" approach is grounded in two primary rationales:

\begin{enumerate}
    \item \textbf{The Macroscopic Goal.} The ultimate objective is a macroscopic theory. The transition from the microscopic to the macroscopic domain is physically an averaging process—conceptually, a convolution of the microscopic theory with an averaging volume. Thus, the macroscopic theory is ultimately a continuous theory, characterized by densities rather than point charges. Further, as the derivation will demonstrate, the subsequent averaging process renders the precise geometric details of this input irrelevant. Whether the electron is modeled as a sphere, a shell, or a cloud, the emergent macroscopic variables (charge density, polarization) are mathematically identical. Therefore, the choice to use densities instead of point charges at the microscopic level is methodologically consistent. With the goal of a macroscopic theory in mind, it is rational to model the charges microscopically as continuous distributions from the outset.
    
    \item \textbf{Mathematical Regularity.} By avoiding the point charge model, we avoid the well-known self-energy singularities and divergences associated with point sources (see Wald \cite{Wald_AdvancedEM}). This allows for a rigorous mathematical treatment of the fields without resorting to ad-hoc renormalization techniques within the classical framework.
\end{enumerate}

While literal rigidity is problematic in the context of special relativity, it is adopted here strictly as a \textbf{kinematic idealization}, valid in the limit where the observation scale $L$ is much larger than the particle radius $R$. The "Extended Body" model serves as a mathematical modeling choice facilitating the development of the macroscopic theory, rather than a claim about fundamental reality.

\subsection{The Resolution Limit (The Pixel Size)}

The spatial resolution of the baseline is defined at the Compton wavelength of the electron ($\lambda_c = h/mc \approx 2.43 \times 10^{-12}$ m). As noted by Zangwill \cite{Zangwill2013}, empirical evidence suggests that the vacuum Maxwell equations remain valid interaction laws down to this scale. This is taken as the "pixel size" of the microscopic image.

This resolution scale necessitates one further act of data compression: the \textbf{simplification of the nucleus}. The internal structure of the atomic nucleus (quarks, gluons, and strong nuclear forces) operates at the femtometer scale ($10^{-15}$ m), far below our resolution limit. Therefore, this complexity is compressed into a single "unitary entity": the \textbf{ion}. The nucleus is treated not as a dynamic quantum system, but as a stable, positively charged entity characterized solely by its position, mass, and total charge. The "how" of nuclear stability is effectively omitted to establish a manageable input for the electromagnetic theory.

\chapter{The Physical Foundation: Microscopic Interactions}
\label{chap:FreeCharges}

\section{The Two-Domain Architecture}
\label{sec:Micro_Architecture}

The rigorous physical description begins by conceptually separating microscopic reality into two interacting domains:
\begin{enumerate}
    \item \textbf{The Mechanical Domain (The Signal Source):} The domain of mass and its mechanical momentum.
    \item \textbf{The Electromagnetic Domain (The Carrier):} The domain of charge, current, and the electromagnetic fields.
\end{enumerate}

A distinction is made between two fundamental species ($s$):
\begin{enumerate}
    \item \textit{Electrons} ($s=-$): The light, mobile charge carriers.
    \item \textit{Ions} ($s=+$): The tightly bound atomic nuclei and core electrons, treated here as single, massive, relatively inert entities.
\end{enumerate}

Each species is characterized by an invariant mass $m_s$ and a charge $q_s$. The complete dynamic state is defined fundamentally by the position $\mathbf{x}_{s,i}(t)$ and velocity $\mathbf{v}_{s,i}(t)$ of every particle $i$.

\subsection{The Kinematic Map: Centroids and Form Factors}

As justified in the preceding chapter, the particles are modeled microscopically not as ideal point charges, but as extended bodies with a finite, albeit small, distribution.

The Dirac delta function is utilized to map the discrete coordinates of the particle \textit{centroids} into continuous field variables. While the particles are physically modeled as Extended Bodies, mathematically the distributions are treated as singularities for the purposes of derivation, a standard practice in rigorous treatments \cite{Wald_AdvancedEM}. This yields two fundamental kinematic fields for each species:

\begin{itemize}
    \item The \textit{Microscopic Centroid Density} $n_s(\mathbf{x}, t)$:
    \begin{equation}
        n_s(\mathbf{x}, t) = \sum_{i=1}^{N_s} \delta(\mathbf{x} - \mathbf{x}_{s,i}(t)).
        \label{eq:micro_number_density}
    \end{equation}
    \item The \textit{Microscopic Centroid Flux} $\boldsymbol{\Gamma}_s(\mathbf{x}, t)$:
    \begin{equation}
        \boldsymbol{\Gamma}_s(\mathbf{x}, t) = \sum_{i=1}^{N_s} \mathbf{v}_{s,i}(t) \delta(\mathbf{x} - \mathbf{x}_{s,i}(t)).
        \label{eq:micro_flux_density}
    \end{equation}
\end{itemize}

To recover the physical charge density of the ``extended body'' model, it is recognized that the delta function represents the \textit{kinematic skeleton} of the matter. The physical charge density $\rho_{\text{micro}}$ is obtained by convolving this centroid density with the particle's internal \textit{form factor} or \textit{shape function} $\sigma_s(\mathbf{r})$:
\begin{equation}
    \rho_{\text{micro}}(\mathbf{x},t) = q_s \int n_s(\mathbf{x}', t) \sigma_s(\mathbf{x} - \mathbf{x}') d^3x'.
\end{equation}
The exact form of the shape function is not of primary relevance here. The conceptual importance lies in modeling charge and mass not as singularities, but as extended bodies with a finite distribution. As argued previously, the transition to the macroscopic model inevitably involves averaging, rendering the precise microscopic geometry irrelevant. Thus, densities are employed from the outset to avoid the well-known singularities (infinities) associated with point charge models (see Wald \cite{Wald_AdvancedEM}).

These fields are rigidly linked by particle conservation. As the entities move, the local number density evolves according to the microscopic \textit{continuity equation}:
\begin{equation}
    \frac{\partial n_s}{\partial t} + \nabla \cdot \boldsymbol{\Gamma}_s = 0.
    \label{eq:micro_particle_continuity}
\end{equation}
This represents the conservation of particles, from which the conservation of mass and charge naturally follows.

\subsection{The Common Root of Mass and Charge}
\label{sec:CommonRoot}

The utility of kinematic field variables lies in the rigorous unification of mechanics and electromagnetism. Separate fluids are not postulated for charge and mass; rather, they are derived by projecting the \textit{same} kinematic fields ($n_s, \boldsymbol{\Gamma}_s$) onto different physical properties. The velocity field is not a free variable to be postulated separately for charge and mass. It is uniquely determined by the ratio of the flux to the number density. This enforces the first fundamental constraint of the theory:

\begin{principle}[The principle of matched velocity]
\label{principle:MatchedVelocity}
For a fundamental stable particle, the center of charge and the center of mass are coincident. Therefore, the kinematic velocity defining the electric current is identically the velocity carrying the inertial mass:
\begin{equation}
    \mathbf{v}_{\text{charge}} \equiv \mathbf{v}_{\text{mass}} \equiv \frac{\boldsymbol{\Gamma}_s}{n_s}.
\end{equation}
\end{principle}

The \textit{electromagnetic source terms} are obtained by weighting these centroid fields by the species charge $q_s$. The total microscopic charge density $\rho_{\text{micro}}$ and electric current density $\mathbf{j}_{\text{micro}}$ are:
\begin{equation}
    \rho_{\text{micro}} = \sum_s q_s n_s, \quad \quad \mathbf{j}_{\text{micro}} = \sum_s q_s \boldsymbol{\Gamma}_s.
    \label{eq:Micro_Electric_Def}
\end{equation}

Simultaneously, the mechanical state variables are obtained by weighting the \textit{exact same fields} by the species mass $m_s$. The total mass density $\rho_{m}$ and mechanical momentum density $\mathbf{g}_{\text{mech}}$ are:
\begin{equation}
    \rho_{m} = \sum_s m_s n_s, \quad \quad \mathbf{g}_{\text{mech}} = \sum_s m_s \boldsymbol{\Gamma}_s.
    \label{eq:Micro_Mech_Def}
\end{equation}
\footnote{$m_s$ is defined as the empirical, physical mass of the extended particle (the renormalized mass).}

\section{Quantum Boundary Conditions: Constraint Forces}
\label{sec:First_Simplification}

As justified in the preceding chapter, electrons and ions are modeled as continuous small distributions. To prevent them from dispersing due to self-repulsion, a stabilizing mechanism is required. This is operationalized by introducing the \textit{quantum constraint force} ($\mathbf{f}_{Q}$). This force acts as a rigorous effective boundary condition, summarizing the non-electromagnetic interactions (Pauli Exclusion, Strong Force) required to maintain the structural integrity of matter.
\begin{equation}
    \frac{d\mathbf{p}_{\text{mech}}}{dt} = \mathbf{f}_{\text{total}} = \mathbf{f}_{\text{em}} + \mathbf{f}_{Q}.
    \label{eq:Micro_Newton_Total}
\end{equation}

\subsection{The Condition of Energy Isolation}
\label{sec:IdealConstraints}

To treat the charge carrier as a stable mechanical entity, the energy restrictions on the stabilizing forces $\mathbf{f}_{Q}$ must be rigorously defined. They are classified strictly as \textit{forces of ideal constraint}.

For the particle to serve as a stable input, the internal binding forces must counteract the local electromagnetic stress without performing net work on the center of mass. Therefore, the standard conditions of \textit{global energy isolation} are imposed:
\begin{equation}
    \int_{\Omega_{\text{particle}}} \mathbf{f}_{Q}(\mathbf{x}) \cdot \mathbf{v}(\mathbf{x}) \, dV \equiv 0.
    \label{eq:WorklessConstraint}
\end{equation}

This condition ensures the \textit{orthogonality of stability and dynamics}. While $\mathbf{f}_{Q}$ is essential for the \textit{existence} of the particle, it vanishes identically from the \textit{energy balance} of the trajectory. This formally isolates the electromagnetic field as the \textit{sole source of external work} on the matter.

\section{Microscopic Electrodynamics}
\label{sec:MicroscopicElectrodynamics}

The conservation laws governing the interaction between microscopic charge carriers and the continuum field are now derived. The analysis proceeds from the fundamental equations of evolution and the constraints they impose on the sources.

\subsection{The Microscopic Maxwell Equations and Compatibility}
\label{sec:Micro_Maxwell_Continuity}

The dynamics of the unaveraged microscopic fields ($\mathbf{e}, \mathbf{b}$) driven by discrete sources ($\rho_{\text{micro}}, \mathbf{j}_{\text{micro}}$) are governed by the microscopic Maxwell equations:

\begin{subequations}
\label{eq:MicroMaxwell}
\begin{align}
    \nabla \cdot \mathbf{e} &= \frac{\rho_{\text{micro}}}{\varepsilon_0} \label{eq:Maxwell_GaussE} \\
    \nabla \times \mathbf{e} + \frac{\partial \mathbf{b}}{\partial t} &= \mathbf{0} \label{eq:Maxwell_Faraday} \\
    \nabla \cdot \mathbf{b} &= 0 \label{eq:Maxwell_GaussB} \\
    \nabla \times \mathbf{b} - \mu_0\varepsilon_0\frac{\partial \mathbf{e}}{\partial t} &= \mu_0\mathbf{j}_{\text{micro}} \label{eq:Maxwell_Ampere}
\end{align}
\end{subequations}

Charge conservation is intrinsic to this structure. Since the divergence of the curl vanishes identically ($\nabla \cdot (\nabla \times \mathbf{A}) \equiv 0$), taking the divergence of Ampère's law (Eq.~\ref{eq:Maxwell_Ampere}) and combining it with the time derivative of Gauss's law (Eq.~\ref{eq:Maxwell_GaussE}) enforces the continuity equation:
\begin{equation}
    \frac{\partial \rho_{\text{micro}}}{\partial t} + \nabla \cdot \mathbf{j}_{\text{micro}} = 0.
    \label{eq:Micro_Continuity_Derived}
\end{equation}
Given the Principle of Matched Velocity, particle conservation implies charge conservation.

\subsection{The Electromagnetic Momentum Identity}

Conservation of momentum is inherent in the algebraic structure of the field equations. By forming a specific linear combination—multiplying Eq.~\eqref{eq:Maxwell_GaussE} by $\varepsilon_0 \mathbf{e}$ and Eq.~\eqref{eq:Maxwell_GaussB} by $\mathbf{b}/\mu_0$, while taking the cross product of Faraday's Law (Eq.~\ref{eq:Maxwell_Faraday}) with $\varepsilon_0 \mathbf{e}$ and Ampère's Law (Eq.~\ref{eq:Maxwell_Ampere}) with $\mathbf{b}/\mu_0$—one obtains the exact balance law for the vacuum field.

Summing these terms yields the intermediate balance:
\begin{multline}
    \varepsilon_0 \left[ \mathbf{e} (\nabla \cdot \mathbf{e}) + \mathbf{e} \times (\nabla \times \mathbf{e}) \right]
    + \frac{1}{\mu_0} \left[ \mathbf{b} (\nabla \cdot \mathbf{b}) + \mathbf{b} \times (\nabla \times \mathbf{b}) \right] \\
    + \frac{\partial}{\partial t} \left( \varepsilon_0 \mathbf{e} \times \mathbf{b} \right) = - (\rho_{\text{micro}}\mathbf{e} + \mathbf{j}_{\text{micro}}\times\mathbf{b}).
\end{multline}

Applying standard vector calculus identities\footnote{Specifically $\mathbf{A} \times (\nabla \times \mathbf{A}) = \frac{1}{2}\nabla(A^2) - (\mathbf{A} \cdot \nabla)\mathbf{A}$.} transforms the bracketed terms into the divergence of a tensor, yielding the pivotal identity:

\begin{equation}
    - \frac{\partial}{\partial t}\left(\varepsilon_0\mathbf{e}\times\mathbf{b}\right) - \nabla\cdot \mathbf{t}_{\text{em}} = \rho_{\text{micro}}\mathbf{e} + \mathbf{j}_{\text{micro}}\times\mathbf{b}.
    \label{eq:Micro_Momentum_Identity}
\end{equation}

The terms on the left-hand side define the dynamics of the field:
\begin{itemize}
    \item \textbf{Field Momentum Density:} $\mathbf{g}_{\text{em}} = \varepsilon_0(\mathbf{e}\times\mathbf{b})$.
    \item \textbf{Maxwell Stress Tensor:} $\mathbf{t}_{\text{em}}$, defined component-wise as:\footnote{\textit{Note on Sign Convention:} We adopt the sign convention of Penfield and Haus \cite{PenfieldHaus1967}.}
    \begin{equation}
        (t_{\text{em}})_{ij} = \varepsilon_0\left(e_i e_j - \frac{1}{2}\delta_{ij}e^2 \right) + \frac{1}{\mu_0}\left(b_i b_j - \frac{1}{2}\delta_{ij}b^2 \right).
        \label{eq:MaxwellStressTensor_rev}
    \end{equation}
\end{itemize}

\paragraph{Epistemological Status: The Open System.}

Equation \eqref{eq:Micro_Momentum_Identity} contains the term $\rho_{\text{micro}}\mathbf{e} + \mathbf{j}_{\text{micro}}\times\mathbf{b}$, recognized as the Lorentz force density. This term couples to the mechanical domain via Newton's Law, as demonstrated in the subsequent section.

However, it is crucial to explicitly define the logical status of Eq.~\eqref{eq:Micro_Momentum_Identity}. This expression does \textit{not} yet represent a closed physical law of interaction; it is an algebraic identity derived entirely within the electromagnetic domain.

It describes an \textit{Open System}. The Maxwell equations are mathematically open; they are inhomogeneous differential equations. While they describe how sources generate fields, they do not, by themselves, describe how fields back-react upon the sources. The terms $\rho_{\text{micro}}$ and $\mathbf{j}_{\text{micro}}$ function as external drivers. To close the system and determine the time evolution of charge and current, an additional physical law is required: Newton's Second Law.

\subsection{The Physical Closure: Newton's Law}
\label{sec:Micro_Coupling}

To convert the electromagnetic identity (Eq.~\ref{eq:Micro_Momentum_Identity}) into a predictive physical theory, the system must be closed. The external input functions ($\rho, \mathbf{j}$) must be replaced with dynamic variables governed by their own evolution equations.

This is achieved by introducing the \textit{mechanical domain} and asserting the \textit{Axiom of Interaction}: the inertial evolution of particle mass is driven by the Lorentz force (while the internal structure is stabilized by the quantum constraint forces $\mathbf{f}_{Q}$).

\begin{equation}
    \frac{d \mathbf{p}_{\text{mech}}}{dt} \equiv \mathbf{f}_{\text{em}} + \mathbf{f}_{Q}.
\end{equation}

This axiom couples the two domains. A single particle is defined by its position and velocity, and is bound to the mechanical domain by its mass (via Newton's Law). However, Newton's Law alone is not closed without a specified force. The particle also possesses charge, which binds it to the electromagnetic domain. The Axiom of Interaction states that the connection between these domains is the Lorentz force.

For the system of extended particles defined in Sec.~\ref{sec:Micro_Architecture}, the macroscopic mechanical state is constructed directly from the centroid kinematic fields ($n_s, \boldsymbol{\Gamma}_s$). Summing over all species $s$:

\begin{itemize}
    \item \textbf{Mechanical Momentum Density:}
    \begin{equation}
        \mathbf{g}_{\text{mech}} = \sum_s m_s \boldsymbol{\Gamma}_s = \sum_s \rho_{m,s} \mathbf{v}_s.
    \end{equation}
    \item \textbf{Kinetic Stress Tensor (Convective Flux):}

\end{itemize}

Newton's Second Law for this continuum takes the form of a rigorous balance equation:
\begin{equation}
    \frac{\partial \mathbf{g}_{\text{mech}}}{\partial t} + \nabla\cdot\mathbf{t}_{\text{kin}} = \mathbf{f}_{\text{total}} = \mathbf{f}_{\text{em}} + \mathbf{f}_{Q}.
\end{equation}

This equation is rearranged to isolate the electromagnetic interaction term, acknowledging that the constraint forces $\mathbf{f}_Q$ are required for stability but do not couple to the field:
\begin{equation}
    \mathbf{f}_{\text{em}} = \left( \frac{\partial \mathbf{g}_{\text{mech}}}{\partial t} + \nabla\cdot\mathbf{t}_{\text{kin}} \right) - \mathbf{f}_{Q}.
\end{equation}

\paragraph{The Synthesized Balance Law}
Substituting the electromagnetic identity (Eq.~\ref{eq:Micro_Momentum_Identity}) for $\mathbf{f}_{\text{em}}$ yields the complete, synthesized balance law for the closed system:

\begin{equation}
\begin{aligned}
    \underbrace{ \left(\frac{\partial \mathbf{g}_{\text{mech}}}{\partial t} + \nabla\cdot\mathbf{t}_{\text{kin}}\right) - \mathbf{f}_{Q} }_{\substack{\textbf{The Mechanical Reservoir} \\ \text{(Net Momentum Gain)}}}
    \quad &\overset{\text{Physical Coupling}}{\textbf{=}} \quad
    \underbrace{ \rho_{\text{micro}}\mathbf{e} + \mathbf{j}_{\text{micro}}\times\mathbf{b} }_{\substack{\textbf{The Single Gateway} \\ \text{(Lorentz Force)}}} \\
    &\overset{\text{Math. Identity}}{=} \quad
    \underbrace{ \left(\frac{\partial \mathbf{g}_{\text{em}}}{\partial t} + \nabla\cdot \mathbf{t}_{\text{em}}\right) }_{\substack{\textbf{The Electromagnetic Reservoir} \\ \text{(Field Momentum Supply)}}}
\end{aligned}
\label{eq:Micro_Momentum_Architecture_Synthesized}
\end{equation}

This structure rigorously defines the system architecture as \textit{two parallel reservoirs connected by a single valve}:
\begin{enumerate}
    \item \textbf{The Gateway:} The exchange of momentum between the Electromagnetic and Mechanical domains occurs \textit{exclusively} via the Lorentz force.
    \item \textbf{The Response:} The change in observable mechanical momentum is the result of this influx of electromagnetic momentum plus the internal constraint forces ($\mathbf{f}_{Q}$).
\end{enumerate}

Physically, this interprets the force density $\mathbf{f}_{\text{em}}$ as the local sink of electromagnetic momentum. Whenever the divergence of the tensor is not balanced by the time variation of the field momentum density ($-\nabla \cdot \mathbf{t}_{\text{em}} \neq \partial \mathbf{g}_{\text{em}} / \partial t$), there is a coupling to the mechanical domain. The loss of electromagnetic momentum manifests as a gain in local mechanical momentum.

The system is now mathematically closed and unambiguously defined. The particles have a clear position and velocity; their motion is governed by Newton's Law and the Lorentz Force. 

\section{Energy Conservation and the Rate of Work}
\label{sec:Micro_Energy}

The power law corresponding to the momentum balance established in Section \ref{sec:Micro_Coupling} is now derived. This derivation identifies the mechanism of energy exchange by strictly applying the mechanical definition of work to the microscopic state.

\subsection{The Power Equation}

The rate of change of the kinetic energy of the charge carriers is determined by the power density $P_{\text{mech}}$ delivered by the total force. Mathematically, this is the scalar projection of the equation of motion (Eq.~\ref{eq:Micro_Newton_Total}) onto the kinematic velocity field $\mathbf{v}$:
\begin{equation}
    P_{\text{mech}} \equiv \mathbf{f}_{\text{total}} \cdot \mathbf{v} = (\mathbf{f}_{\text{em}} + \mathbf{f}_{Q}) \cdot \mathbf{v}.
\end{equation}

Crucially, the velocity $\mathbf{v}$ appearing here is the transport velocity of the mass, which, by the \textit{Principle of Matched Velocity} (Principle~\ref{principle:MatchedVelocity}), is identical to the transport velocity of the charge.

Applying the Condition of Global Energy Isolation (Eq.~\ref{eq:WorklessConstraint}), the contribution of the stabilizing constraint forces vanishes identically:
\begin{equation}
    P_{Q} = \mathbf{f}_{Q} \cdot \mathbf{v} \equiv 0.
\end{equation}

Consequently, the power driving the mechanical evolution of the system is determined \textit{exclusively} by the electromagnetic term:
\begin{equation}
    P_{\text{mech}} = \mathbf{f}_{\text{em}} \cdot \mathbf{v}.
    \label{eq:ExclusiveGateway}
\end{equation}

This result establishes the \textit{energetic orthogonality} of the constraints. While the quantum constraint forces are structurally necessary to define the particle (preventing collapse), they are mathematically invisible to the energy balance of the trajectory. The electromagnetic field is the sole source of external work.

\subsection{Magnetic Orthogonality and the Electric Work Term}

The Lorentz force density is now projected onto the kinematic velocity field to determine the electromagnetic power density $P_{\text{em}}$.

Substituting the source definition $\mathbf{j}_{\text{micro}} = \rho_{\text{micro}}\mathbf{v}$ (as mandated by the \textit{Principle of Matched Velocity}) into the force equation yields:
\begin{equation}
    P_{\text{em}} = \mathbf{f}_{\text{em}} \cdot \mathbf{v} = (\rho_{\text{micro}}\mathbf{e})\cdot\mathbf{v} + \left( (\rho_{\text{micro}}\mathbf{v}) \times \mathbf{b} \right) \cdot \mathbf{v}.
\end{equation}

The second term vanishes identically due to the vector identity for the scalar triple product ($\mathbf{A} \cdot (\mathbf{A} \times \mathbf{B}) \equiv 0$). This enforces the condition of \textit{magnetic orthogonality}:
\begin{equation}
    P_{\text{mag}} = (\mathbf{j}_{\text{micro}} \times \mathbf{b}) \cdot \mathbf{v} \equiv 0.
    \label{eq:MagneticWorkZero}
\end{equation}

It is important to emphasize that the vanishing of magnetic work is not merely a property of the B-field; it is a direct consequence of the \textit{charge-mass bridge}. The magnetic field acts solely as a deflecting agent (altering curvature) on the particle. It can change the trajectory of the mechanical moment but not its magnitude. Kinetic energy transfer is mediated \textit{exclusively} by the electric field:
\begin{equation}
    \boxed{ P_{\text{total}} = P_{\text{em}} = \mathbf{j}_{\text{micro}} \cdot \mathbf{e} }
    \label{eq:Micro_Energy_Gateway}
\end{equation}
The particle's speed (and thus kinetic energy) cannot be changed by the magnetic field.

The power equation thus follows directly from projecting the momentum equation onto the velocity field of the particles. It is not an independent law but a projection of the momentum equation:
\begin{equation}
    \underbrace{ \mathbf{f}_{\text{mech}} \cdot \mathbf{v} }_{\text{Power}} = \underbrace{ \mathbf{f}_{\text{em}} \cdot \mathbf{v} }_{\text{Work}} = \mathbf{j}_{\text{micro}} \cdot \mathbf{e}.
\end{equation}

\subsection{The Electromagnetic Identity (The Exchange Rate)}
Maxwell's equations inherit a definition of the commonly known electromagnetic field energy. By taking the scalar product of Ampère's Law (Eq.~\ref{eq:Maxwell_Ampere}) with $\mathbf{e}$ and subtracting the scalar product of Faraday's Law (Eq.~\ref{eq:Maxwell_Faraday}) with $\mathbf{b}/\mu_0$, we obtain the intermediate balance:

\begin{equation}
    \frac{1}{\mu_0} \left[ \mathbf{e} \cdot (\nabla \times \mathbf{b}) - \mathbf{b} \cdot (\nabla \times \mathbf{e}) \right] - \left[ \varepsilon_0 \mathbf{e} \cdot \frac{\partial \mathbf{e}}{\partial t} + \frac{1}{\mu_0} \mathbf{b} \cdot \frac{\partial \mathbf{b}}{\partial t} \right] = \mathbf{j}_{\text{micro}} \cdot \mathbf{e}.
\end{equation}

Applying standard vector calculus identities transforms the bracketed terms, yielding Poynting's Theorem:

\begin{equation}
    - \left( \frac{\partial u_{\text{em}}}{\partial t} + \nabla\cdot \mathbf{s}_{\text{em}} \right) = \mathbf{j}_{\text{micro}} \cdot \mathbf{e},
    \label{eq:Micro_Poynting}
\end{equation}
where $u_{\text{em}} = \frac{1}{2}(\varepsilon_0 e^2 + \frac{1}{\mu_0} b^2)$ and $\mathbf{s}_{\text{em}} = \frac{1}{\mu_0}(\mathbf{e} \times \mathbf{b})$.

This is known as Poynting's Theorem. Again, it is important to emphasize that this equation (Eq.~\ref{eq:Micro_Poynting}) alone has no physical meaning; it is entirely within the electromagnetic domain—an identity in the electromagnetic domain. Just as the Lorentz force was able to be reformulated using Maxwell's equations (Eq. \ref{eq:Micro_Momentum_Identity}), the same can be done for the energy equation. However, the physical meaning of the equation is only closed when also considering the mechanical domain.

\subsection{The Energy Chain of Transfer}
Combining the algebraic identity with this mechanical projection reveals the complete architecture:

\begin{equation}
\boxed{
\begin{aligned}
   \underbrace{ \frac{\partial u_{\text{kin}}}{\partial t} + \nabla \cdot \mathbf{s}_{\text{kin}}  }_{\substack{\textbf{The Mechanical Reservoir} \\ \text{(Kinetic Evolution)}}}
    \quad &\overset{\text{FECC}}{\textbf{=}} \quad
    \underbrace{ \mathbf{j}_{\text{micro}} \cdot \mathbf{e} }_{\substack{\textbf{The Single Gateway} \\ \text{(Bidirectional Valve)}}} \\
    \quad &\overset{\text{Math. Identity}}{=} \quad
    \underbrace{ -\left(\frac{\partial u_{\text{em}}}{\partial t} + \nabla \cdot \mathbf{s}_{\text{em}}\right) }_{\substack{\textbf{The Electromagnetic Reservoir} \\ \text{(Field Evolution)}}}
   \label{eq:Micro_Energy_Architecture}
\end{aligned}
}
\end{equation}

This structure defines the \textit{topology of interaction}. In regions where $\mathbf{j}_{\text{micro}}=0$, the gateway is closed ($P_{\text{em}} = 0$). The electromagnetic field energy is conserved independently:
\begin{equation}
    \frac{\partial u_{\text{em}}}{\partial t} + \nabla \cdot \mathbf{s}_{\text{em}} = 0.
\end{equation}
There is no sink or source of field energy; it is merely redistributed locally.

Now, consider the case where $\mathbf{j}_{\text{micro}} \cdot \mathbf{e}$ is non-zero. This implies the presence of a local particle moving with velocity $\mathbf{v}$ and experiencing a Coulomb force along its direction of motion. There is a local change in the magnitude of the particle's velocity; the particle's kinetic energy is changing. The electric field performs work on the moving particle. This signifies a local exchange of electromagnetic field energy and mechanical energy of the particle.

When $\mathbf{j}_{\text{micro}} \cdot \mathbf{e}$ is non-zero, there is a sink or source of electromagnetic field energy:
\begin{equation}
    \frac{\partial u_{\text{em}}}{\partial t} + \nabla \cdot \mathbf{s}_{\text{em}} \neq 0.
\end{equation}
If there is a sink, the energy is transferred towards the kinetic energy of the particle upon which the Lorentz force acts at that precise location of interaction. If there is a source of electromagnetic field energy, the energy is transferred away from the kinetic energy of the particle and into the electromagnetic field energy.

The equation represents a completely well-defined interaction: forces, velocities, sinks, and sources of the field energy are all well-defined and dependent. This reveals a fundamental insight: \textit{Momentum and Energy are two projections of the same physical reality.} The energy equation is simply the momentum equation viewed along the trajectory of the mass.

Therefore, the relationship $P = \mathbf{f} \cdot \mathbf{v}$ is not merely a definition; it is the \textit{constraint of consistency}. It ties the energy budget $P$ irrevocably to the force $\mathbf{f}$ and the specific mass carrier $\mathbf{v}$. This is the origin of the \textit{Force-Energy Consistency Criterion (FECC)}, utilized here to resolve the macroscopic paradoxes.

\section{The Force-Energy Consistency Criterion (FECC)}
\label{sec:FECC_Definition}

The derivation in the preceding section yielded a specific expression for energy transfer: $P = \mathbf{j}_{\text{micro}} \cdot \mathbf{e}$. However, this result is not merely a formula derived from Maxwell's equations; it is the electromagnetic manifestation of a universal mechanical constraint.

This constraint is now formalized into the central diagnostic tool of this manuscript: the \textit{force-energy consistency criterion} (FECC).

\subsection{The Mechanical Axiom: Projection, Not Postulate}

In many standard treatments of macroscopic electrodynamics, the balance laws for momentum (force density $\mathbf{f}$) and energy (power density $P$) are treated as independent postulates, often derived from separate divergences of the Energy-Momentum tensor.

However, it is demonstrated here that Momentum and Energy are not distinct physical entities; they are \textit{indivisible facets of the same particle dynamics}.
\begin{itemize}
    \item \textbf{Momentum} is the vector measure of the interaction.
    \item \textbf{Energy} is the scalar measure of the interaction, projected along the trajectory of the mass.
\end{itemize}

These two descriptions are strictly coupled by the \textbf{kinematic constraint}: the velocity $\mathbf{v}$ of the specific mass target upon which the force acts. The power density $P$ is, by axiomatic definition, the rate at which the force $\mathbf{f}$ performs work on that mass:
\begin{equation}
    P \equiv \mathbf{f} \cdot \mathbf{v}.
    \label{eq:mech_power_def}
\end{equation}

This identity is fundamental. A physical theory cannot define $\mathbf{f}$ and $P$ in isolation. If a theory postulates a power density, it has implicitly defined the velocity field to which that power couples. If that velocity does not correspond to the motion of the massive constituents, the theory is kinematically ill-defined.

\subsection{The FECC for Electrodynamics}

The derivation proves that the microscopic Maxwell-Lorentz system rigorously satisfies this axiom. The term $\mathbf{j}_{\text{micro}} \cdot \mathbf{e}$ is not an arbitrary choice for the "Poynting Sink"; it is the unique interaction energy demanded by the force structure and the kinematic constraints of the Charge-Mass Bridge.

The \textit{force-energy consistency criterion} (FECC) is therefore established as the primary benchmark for physical validity. For any proposed electromagnetic theory (microscopic or macroscopic), the defined force density $\mathbf{f}_{\text{em}}$ and the defined energy transfer $P_{\text{em}}$ must satisfy:

\begin{equation}
    \boxed{ \mathbf{f}_{\text{em}} \cdot \mathbf{v}_{\text{mass}} \equiv P_{\text{em}} }
    \label{eq:FECC}
\end{equation}

\subsection{The Diagnostic Power of the FECC}

This criterion provides a rigorous method for auditing complex macroscopic theories. Real materials are often mixtures of distinct components (e.g., electrons, ions, neutral lattice) moving with distinct velocities.

In such cases, the FECC must be applied to each massive component $i$ individually. If the field exerts a specific force density $\mathbf{f}_i$ on component $i$, the FECC demands that the power delivered to that component is strictly $P_i = \mathbf{f}_i \cdot \mathbf{v}_i$.

The \textit{total} interaction power density $P_{\text{int}}$ delivered to the mechanical system must be the sum over all massive components:
\begin{equation}
    P_{\text{int}} = \sum_i P_i = \sum_i (\mathbf{f}_i \cdot \mathbf{v}_i).
    \label{eq:fecc_generalized}
\end{equation}

This establishes the primary diagnostic tool for the parts that follow. It follows that any field-matter theory (such as Minkowski's) that proposes an interaction force and power which diverge from this criterion—by postulating a force that performs no work while simultaneously claiming a non-zero energy transfer—is fundamentally disconnected from the axiomatic principles of mechanics.

\section{The Unified System in Covariant Formulation}
\label{sec:FreeCharges_CovariantSummary}

This chapter concludes by demonstrating that the derived microscopic forms can be presented in a covariant formulation. While the concise structure of the covariant notation often obscures distinct physical mechanisms (masking the separate roles of force and energy), the inherent physics remains unchanged. This covariant formulation serves not merely as a summary, but as the rigorous validation of the internal constraints established in this analysis.

The complete physical system is defined by the interaction of two distinct Energy-Momentum tensors:

\begin{itemize}
    \item \textbf{The Electromagnetic Field ($T_{\text{em}}^{\mu\nu}$):} 
    Defined from the field strength tensor $F^{\mu\nu}$ via the standard vacuum relation:
    \begin{equation}
        T_{\text{em}}^{\mu\nu} = \frac{1}{\mu_0}\left(F^{\mu\alpha}F^{\nu}_{\;\;\alpha} - \frac{1}{4}g^{\mu\nu}F_{\alpha\beta}F^{\alpha\beta}\right).
    \end{equation}
    Its components map directly to the vector quantities derived previously (where $-\mathbf{t}_{\text{em}}$ accounts for the metric signature):
    \begin{equation}
        T_{\text{em}}^{\mu\nu} = 
        \begin{pmatrix}
        u_{\text{em}} & c\mathbf{g}_{\text{em}}^T \\
        c\mathbf{g}_{\text{em}} & -\mathbf{t}_{\text{em}}
        \end{pmatrix}.
    \end{equation}

    \item \textbf{The Massive Particles ($T_{\text{mech}}^{\mu\nu}$):}
    This tensor describes the mechanical state. For a continuum of extended particles (Kinematic Dust) with rest-mass density $\rho_0$ and 4-velocity $u^\mu$, it is strictly defined by the dyadic product:
    \begin{equation}
        T_{\text{mech}}^{\mu\nu} = \rho_0 u^\mu u^\nu = 
        \begin{pmatrix}
        \gamma^2 \rho_0 c^2 & \gamma^2 \rho_0 c v^i \\
        \gamma^2 \rho_0 c v^j & \gamma^2 \rho_0 v^i v^j
        \end{pmatrix}.
    \end{equation}
    Expressed in terms of the lab-frame densities:
    \begin{equation}
        T_{\text{mech}}^{\mu\nu} = 
        \begin{pmatrix}
        u_{\text{kin}} & c\mathbf{g}_{\text{mech}}^T \\
        c\mathbf{g}_{\text{mech}} & \mathbf{t}_{\text{kin}}
        \end{pmatrix}.
    \end{equation}
\end{itemize}

\subsection{The Structural Constraint (The Covariant FECC)}
The explicit form of $T_{\text{mech}}^{\mu\nu}$ reveals a critical property often overlooked in macroscopic postulations: \textbf{The tensor components are kinematically dependent.}

Because the tensor is constructed from the outer product of the 4-velocity, the energy density ($T^{00}$) and momentum density ($T^{0i}$) are intrinsically coupled. Their ratio yields the transport velocity:
\begin{equation}
    \frac{c\mathbf{g}_{\text{mech}}}{u_{\text{kin}}} = \frac{\gamma^2 \rho_0 c \mathbf{v}}{\gamma^2 \rho_0 c^2} = \frac{\mathbf{v}}{c}.
\end{equation}
This is the covariant manifestation of the \textit{force-energy consistency criterion}. One cannot alter the momentum density of the matter without simultaneously altering its energy density and its velocity.

\subsection{The Covariant Balance Laws}
The interaction is described by the 4-divergence of these tensors, linked by the Lorentz 4-force density $f^\mu_{\text{em}} = F^{\mu\nu}J_\nu$. This relationship divides the problem into two valid perspectives:

\paragraph{1. The Material Subsystem (Exchange).}
Viewing the matter as an open system driven by the field, the divergence of the mechanical tensor is non-zero:
\begin{equation}
    \partial_\nu T_{\text{mech}}^{\mu\nu} = f^{\mu}_{\text{em}} + f^{\mu}_{Q}.
    \label{eq:TensorDivergence_Mech}
\end{equation}
The temporal component ($\nu=0$) recovers the "Work-Energy" relation ($P=\mathbf{f}\cdot\mathbf{v}$), representing the rate at which the subsystem gains energy from the gateway.

\paragraph{2. The Total System (Conservation).}
Viewing the universe as a whole, the system is closed. Since the Lorentz force is also the negative divergence of the vacuum field tensor ($\partial_\nu T_{\text{em}}^{\mu\nu} = -f^{\mu}_{\text{em}}$), the sum satisfies strict local conservation:
\begin{equation}
    \partial_\nu \left( T_{\text{em}}^{\mu\nu} + T_{\text{mech}}^{\mu\nu} \right) = f^{\mu}_{Q}.
    \label{eq:TotalTensorDivergence}
\end{equation}
In the absence of external work done by constraints (the Energy Isolation condition, $f^{\mu}_{Q} u_\mu = 0$), the total energy-momentum flux is conserved.

\subsection{The Covariant Synthesis}

The derivation is now condensed into a single, Lorentz-invariant statement. By combining the interaction law (Eq.~\ref{eq:TensorDivergence_Mech}) with the field identity, we arrive at the unified covariant architecture.

This equation formally encapsulates the Microscopic Baseline. It decomposes the universe into two tensor reservoirs connected by a single 4-vector gateway:

\begin{equation}
\boxed{
\begin{aligned}
    \underbrace{ \partial_\nu T_{\text{mech}}^{\mu\nu} - f^{\mu}_{Q} }_{\substack{\textbf{The Mechanical Reservoir} \\ \text{(4-Momentum Gain)}}}
    \quad &\overset{\text{Physical Coupling}}{\textbf{=}} \quad
    \underbrace{ f^{\mu}_{\text{em}} }_{\substack{\textbf{The Covariant Gateway} \\ \text{(Lorentz 4-Force)}}} \\
    &\overset{\text{Field Identity}}{=} \quad
    \underbrace{ -\partial_\nu T_{\text{em}}^{\mu\nu} }_{\substack{\textbf{The electromagnetic reservoir} \\ \text{(4-Momentum Supply)}}}
\end{aligned}
}
    \label{eq:Covariant_Architecture}
\end{equation}

This single expression encapsulates every physical principle derived in this chapter:

\begin{enumerate}
    \item \textbf{The Temporal Component ($\mu=0$):} Represents the Energy Balance.
    \begin{itemize}
        \item The mechanical term becomes $\frac{\partial u_{\text{kin}}}{\partial t} + \dots$
        \item The gateway becomes the work rate $\mathbf{f} \cdot \mathbf{v}$.
        \item The constraint force $f^0_Q$ vanishes (due to orthogonality), isolating the field as the sole energy source.
    \end{itemize}
    
    \item \textbf{The Spatial Components ($\mu=i$):} Represent the Momentum Balance.
    \begin{itemize}
        \item The mechanical term becomes Newton's Second Law.
        \item The gateway becomes the Lorentz force density.
        \item The constraint force $f^i_Q$ remains active, enforcing stability.
    \end{itemize}
\end{enumerate}

\section[Epistemology of Observables]{Epistemology of Observables}
\label{sec:FreeCharges_Epistemology}

The derivations in this chapter lead to a critical epistemological constraint regarding the verification of electromagnetic theories. This constraint defines the precise boundary between physical measurement and mathematical definition.

\subsection{The Limit of Observability}

Empirical physics is governed by a fundamental asymmetry: Classical measurement is mediated exclusively by mechanical exchange. Therefore, electromagnetic fields, potentials, or momentum fluxes are not directly perceived; only their \textit{mechanical consequences} are detected. All measurements---whether by a galvanometer, a piezoelectric sensor, or a photon detector---fundamentally measure the \textbf{mechanical response} of matter. This establishes the \textbf{limit of observability}. The equation of motion allows us to decompose the measurable force into field-centric components:
\begin{equation}
    \underbrace{ \mathbf{f}_{\text{mech}} }_{\substack{\textbf{Observable} \\ \text{(Mechanical Metric)}}} 
    \quad \equiv \quad 
    \underbrace{ -\frac{\partial \mathbf{g}_{\text{em}}}{\partial t} - \nabla\cdot \mathbf{t}_{\text{em}} }_{\substack{\textbf{Inferred} \\ \text{(Theoretical Definition)}}}.
    \label{eq:Momentum_Decomposition_Epist}
\end{equation}

The experimentalist measures the sum on the left. In principle, one may "reshuffle" the terms on the right-hand side—modifying the definition of the tensor components or the momentum density—as long as the total divergence remains unchanged. If the total mechanical force is invariant, the alternative definitions are empirically indistinguishable, creating a space for theoretical debate that experiment cannot resolve.

\section{The Path Forward: From Exactness to Emergence}

This template provides the ingredients; the remaining task is construction. In the parts that follow, the derivation will demonstrate how to derive the complex macroscopic world from this classical starting point.

Before proceeding to the construction of the macroscopic framework in Part II, the analysis will remain within the microscopic domain for two additional chapters. First, the principle of the "Arbitrary Split" will be investigated on simplified microscopic grounds, exploring the consequences of dividing the total system into subsystems. Following this, the Abraham-Minkowski controversy will be briefly examined within this rigorous microscopic baseline. Only after establishing these foundational conclusions will the text proceed to the construction of the macroscopic theory.

\chapter{The Arbitrary Split: A Microscopic Analysis}
\label{chap:Critique_ArbitrarySplit}

Chapter \ref{chap:FreeCharges} established the \textit{microscopic baseline}: a system uniquely constrained by the ``two reservoirs, one valve'' architecture. The derivation confirmed that the coupling between field and matter is not a free variable but is rigidly fixed by the kinematic lock ($\mathbf{f} \cdot \mathbf{v} = P$).

This chapter examines the ``Hypothesis of Equivalence''—formalized by Penfield and Haus \cite{PenfieldHaus1967} and Pfeifer et al. \cite{Pfeifer2007}—which posits that the division of the total energy-momentum tensor is a matter of convention. It is crucial to note that this principle was originally proposed within the macroscopic regime, where its application is significantly more subtle. However, the microscopic reduction employed here is designed to clarify the conceptual underpinnings of this principle. The goal of this chapter is to demonstrate that distinct algebraic arrangements imply distinct physical realities. A similar forensic argumentation will then be applied to the macroscopic regime in Part \ref{part:NewPart6}.

\section{The Principle of Physical Exclusivity}
\label{sec:Singular_Reality}

The prevailing consensus in macroscopic electrodynamics—the ``arbitrary split'' paradigm—derives from the rigorous proof of Penfield and Haus \cite{PenfieldHaus1967}. These authors demonstrated that no electromagnetic energy-momentum tensor is complete in isolation; any choice of field tensor necessitates a specific, accompanying material tensor to close the system's conservation laws.

Penfield and Haus interpreted this mathematical freedom as a license for convention, asserting that the division between field and matter is arbitrary so long as the total remains invariant. This view is re-examined here.

\subsection{The Causal Bridge: Distinguishing Algebra from Physics}
\label{sec:Causal_Bridge}

It is argued that the core difficulty in the ``Arbitrary Split'' argument lies in the conflation of algebraic equivalence with physical identity. It treats physics as \textit{mathematical bookkeeping}, wherein the validity of a theory is determined solely by whether the ledger balances (algebraic conservation), while essentially omitting the provenance of the terms (physical source).

In pure mathematics, the equality sign ($=$) denotes numerical equivalence; terms can be moved freely between the left and right sides of an equation without affecting its validity. In physics, however, the equation of motion represents a \textit{causal bridge}. It connects two ontologically distinct domains:
\begin{itemize}
    \item \textit{The Left Side (Inertia):} Represents the intrinsic state of the \textit{matter} (The Effect).
    \item \textit{The Right Side (Force):} Represents the extrinsic agency of the \textit{field} (The Cause).
\end{itemize}
This equation exhibits \textit{semantic rigidity}. Reassigning terms from the "Force" side to the "Inertia" side alters the physical definitions of the entities involved. This distinction requires the careful separation of mathematical equivalence from physical identity.

\subsection{The Newtonian Limit: Testing the Hypothesis}
\label{sec:Artifactual_Vector}

To isolate the structural mechanics of the ``Arbitrary Split,'' the analysis momentarily strips away the complexity of the macroscopic continuum and applies the logic to a simplified \textit{Newtonian baseline}. This serves to illustrate the mechanical implications of the hypothesis when algebraic freedom is applied to a causal equation.

Consider a single electron governed by the standard Lorentz force. The microscopic baseline is given by Newton's Second Law:
\begin{equation}
    \underbrace{ m \mathbf{a} }_{\text{Kinematic Response}} = \underbrace{ q(\mathbf{e} + \mathbf{v} \times \mathbf{b}) }_{\text{Lorentz Interaction}}
    \label{eq:GroundTruth_Base}
\end{equation}
Mathematically, the ``Arbitrary Split'' asserts that one is free to add a term to the field side, provided the mechanical side is adjusted to compensate. Let us add an arbitrary ``anomalous vector'' field $\mathbf{\Phi}(\mathbf{x}, t)$ to both sides. The algebraic equality remains perfectly valid:
\begin{equation}
    m \mathbf{a} + \mathbf{\Phi} = q(\mathbf{e} + \mathbf{v} \times \mathbf{b}) + \mathbf{\Phi}
\end{equation}
The terms can now be grouped to define a ``New'' theory of electrodynamics:
\begin{itemize}
    \item A \textit{new force} is defined: $\mathbf{F}_{\text{new}} \equiv q(\mathbf{e} + \mathbf{v} \times \mathbf{b}) + \mathbf{\Phi}$.
    \item A \textit{new inertia} is defined: $\mathbf{P}_{\text{new}}' \equiv m \mathbf{a} + \mathbf{\Phi}$.
\end{itemize}
According to the logic of the hypothesis, this new formulation ($\mathbf{P}_{\text{new}}' = \mathbf{F}_{\text{new}}$) is a ``valid alternative'' because it conserves the total momentum of the system/field complex.

However, this operation fundamentally alters the terms on each side. A non-zero $\mathbf{\Phi}$ introduces a new electromagnetic force and, consequently, a new mass term. The new formulations might predict forces acting on vacuum regions where no mass exists ($\rho_m=0$). We identify this artifact here as ``Vacuum Inertia.''

\subsection{The Principle of Mutual Exclusivity}
\label{sec:Mutual_Exclusivity}

This thought experiment leads to the proposition contrary to the Arbitrary Split: the \textit{principle of physical exclusivity}. When a term $\mathbf{\Phi}$ is added to a differential equation, the physics is changed.

Different tensor formulations are not equivalent descriptions of the same mechanical reality. They describe distinct physical topologies:
\begin{itemize}
    \item If \textit{Reality A (Lorentz)} is true, then $\mathbf{\Phi} = 0$. The alternative formulations are physically inconsistent because they postulate forces that do not exist.
    \item If \textit{Reality B (Alternative)} is true, then $\mathbf{\Phi} \neq 0$. The Lorentz formulation is physically incomplete because it fails to account for the vacuum inertia.
\end{itemize}
There is no middle ground. The introduction of a non-zero $\mathbf{\Phi}$ definitively changes the force and the mechanical physical descriptions of the system.

\subsection{Invariance by Adding a Zero Scalar Field}
A nuance must be added to this classification. It is possible to add a term $\Phi$ to the equation which sums identically to zero.

As defined in the previous chapter, the quantity that is eventually measured—or that has a physical consequence—is the total force density:
\begin{equation}
    \underbrace{ \mathbf{f}_{\text{mech}} }_{\substack{\textbf{Observable} \\ \text{(Mechanical Metric)}}} 
    \quad \equiv \quad 
    \underbrace{ -\frac{\partial \mathbf{g}_{\text{em}}}{\partial t} - \nabla\cdot \mathbf{t}_{\text{em}} }_{\substack{\textbf{Inferred} \\ \text{(Theoretical Definition)}}}.
    \label{eq:Momentum_Decomposition_Epist_Ch2}
\end{equation}
This is the total force density that requires a mass target on the LHS. Consequently, it is the \textit{total} of the RHS that defines the physics, not the individual terms.

Consider the addition of a scalar field (or divergence-free tensor term) $\Phi$ which sums to zero:
\begin{equation}
    \Phi = -\frac{\partial \mathbf{g}_{\text{new}}}{\partial t} - \nabla\cdot \mathbf{t}_{\text{new}} \equiv \mathbf{0}.
\end{equation}

One may always add this null term to the total physical equation without changing the observable. Thus, this would \textit{not} constitute a mutually exclusive theory, as it mathematically amounts to adding zero to the equation. The mechanical response of the system ($\mathbf{f}_{\text{mech}}$) remains invariant.

This is the origin of the ambiguity regarding the exact definitions of momentum flux and momentum density (specifically the mathematical non-uniqueness) in the context of the Poynting theorem. One may add variables that sum to zero, thereby changing the definitions of local momentum and flux, without altering the observable effect on the system.

However, when this text refers to ``mutually exclusive theories'' (Lorentz vs. Minkowski), it does not refer to adding a zero. It refers to the case where $\Phi \neq \mathbf{0}$ holds, and the mechanical response of the system is fundamentally changed.

\paragraph{The Macroscopic Nuance.}
It must be stated clearly that in the macroscopic regime—where our hypothesis of mutually exclusive realities will ultimately be tested—the divergence is significantly more subtle. In continuum mechanics, the term $\mathbf{\Phi}$ typically manifests as the divergence of a stress tensor that integrates to zero over a closed volume. Consequently, it changes the local \textit{force density} while leaving the \textit{total force} on an isolated body invariant. This complex mechanism will be analyzed in detail in \textbf{Part VI}.

However, the microscopic reduction employed here is designed to clarify this distinction. It demonstrates that distinct algebraic arrangements imply distinct physical realities. To accept $\mathbf{F}_{\text{new}}$ as the true force, one must strictly accept that the particle possesses a ``hidden inertia'' corresponding to $\mathbf{\Phi}$—momentum that exists even when the kinematic velocity is constant. The ``New Theory'' is not merely a redescription; it is the definition of a \textit{distinct physical model} containing anomalous forces and hidden reservoirs that the standard Lorentz formulation excludes.

\section{The Mathematical Reformulation of Microscopic Sources}
\label{sec:Source_Reformulation}

The consequences of the \textit{principle of physical exclusivity} are now rigorously applied to the microscopic baseline established in the previous chapter. The objective is to identify precisely where the ``Arbitrary Split'' diverges from physical reality. To achieve this, a distinction must first be made between valid algebraic manipulation and invalid physical redefinition.

The first step is a \textit{purely mathematical reformulation}. This demonstrates that it is possible to rewrite the microscopic Maxwell equations in a form that \textit{resembles} the macroscopic equations, without changing the physical meaning of a single term. This ``safe mode'' operation proves that the variables $\mathbf{D}$ and $\mathbf{H}$ are not initially born from new physical laws, but are generated by a simple algebraic regrouping of the microscopic continuity equation.

\subsection{The Representation Theorem: Decomposing the Flow}
\label{sec:Math_Potentials}

The ability to introduce auxiliary fields is a guaranteed property of any vector field. Rather than starting with static charge, the analysis begins with the fundamental dynamic quantity: the microscopic current density $\mathbf{j}_{\text{micro}}$.

According to the Helmholtz Decomposition Theorem, any sufficiently smooth vector field can be resolved into two orthogonal components: an \textit{irrotational} (longitudinal) part and a \textit{solenoidal} (transverse) part.

\paragraph{1. The Decomposition of Current.}
The microscopic current $\mathbf{j}_{\text{micro}}$ is therefore mathematically expressed as the sum of a changing longitudinal vector field and a circulating vector field. The potentials $\mathbf{p}$ (polarization) and $\mathbf{m}$ (magnetization) are \textit{defined} such that:

\begin{equation}
    \mathbf{j}_{\text{micro}} \equiv \frac{\partial \mathbf{p}}{\partial t} + \nabla \times \mathbf{m}.
    \label{eq:helmholtz_j}
\end{equation}

This definition assigns clear kinematic roles to the potentials:
\begin{itemize}
    \item $\frac{\partial \mathbf{p}}{\partial t}$ represents the \textit{linear displacement current}: the flow of charge accumulating at a boundary.
    \item $\nabla \times \mathbf{m}$ represents the \textit{circulating current}: the closed-loop flow that contributes no net accumulation.
\end{itemize}

\paragraph{2. The Derivation of Charge Density.}
To find the corresponding charge representation, this decomposition is substituted directly into the continuity equation ($\frac{\partial \rho}{\partial t} + \nabla \cdot \mathbf{j} = 0$). Since the divergence of a curl is identically zero, the circulating term vanishes, leaving:
\begin{equation}
    \frac{\partial}{\partial t} \left( \rho_{\text{micro}} + \nabla \cdot \mathbf{p} \right) = 0.
\end{equation}
Integration from a neutral state yields the necessary definition of the charge density:
\begin{equation}
    \rho_{\text{micro}} = - \nabla \cdot \mathbf{p}.
    \label{eq:rho_from_p}
\end{equation}

\paragraph{Ontological Status (The Kinematic Alias).}
This ordering reveals that, within the context of the microscopic Lorentz theory, $\mathbf{p}$ and $\mathbf{m}$ are not new physical entities. They are \textit{kinematic aliases} for charge in motion. Therefore, $\mathbf{p}$ and $\mathbf{m}$ possess no intrinsic dynamics separate from the mass and charge they describe. Crucially, physical forces do not couple to the potential $\mathbf{p}$ itself (which may be non-zero even in regions where the net charge is zero), but strictly to the \textit{source densities} defined by its derivatives: the charge density ($-\nabla \cdot \mathbf{p}$) and the current density ($\partial \mathbf{p}/\partial t$). The same holds for the magnetization field $\mathbf{m}$.

\subsection{Partitioning and Regrouping}

The algebraic procedure that generates the standard macroscopic formalism is now executed also microscopically. This procedure is not physical; it is a choice of bookkeeping. The total microscopic sources are partitioned into two distinct groups, labeled ``1'' (Target Sources) and ``2'' (Background Sources):
\begin{align*}
    \rho_{\text{micro}} &= \rho_{\text{micro,1}} + \rho_{\text{micro,2}} \\
    \mathbf{j}_{\text{micro}} &= \mathbf{j}_{\text{micro,1}} + \mathbf{j}_{\text{micro,2}}
\end{align*}

Crucially, \textit{Group 1 and Group 2 are physically indistinguishable.} They are both composed of massive point charges obeying Newton's Laws. The distinction is purely arbitrary—Group 1 is tracked explicitly, while Group 2 is ``hidden'' inside auxiliary potentials ($\mathbf{p}_2, \mathbf{m}_2$).

Substituting this partition into the microscopic Maxwell equations, the terms associated with Group 2 are mathematically migrated from the source side (RHS) to the field side (LHS):

\begin{align}
    \underbrace{\varepsilon_0\nabla \cdot \mathbf{e}}_{\text{Vacuum field}} + \underbrace{\nabla \cdot \mathbf{p}_2}_{\text{Migrated source}} &= \rho_{\text{micro,1}} \\
    \underbrace{\frac{1}{\mu_0}\nabla \times \mathbf{b} - \varepsilon_0\frac{\partial \mathbf{e}}{\partial t}}_{\text{Vacuum field}} - \underbrace{\left( \frac{\partial \mathbf{p}_2}{\partial t} + \nabla \times \mathbf{m}_2 \right)}_{\text{Migrated source}} &= \mathbf{j}_{\text{micro,1}}
\end{align}

By collecting the terms on the left-hand side, the new \textit{composite field variables} are defined:
\begin{equation}
    \mathbf{d} \equiv \varepsilon_0\mathbf{e} + \mathbf{p}_2, \quad \quad \mathbf{h} \equiv \frac{1}{\mu_0}\mathbf{b} - \mathbf{m}_2.
    \label{eq:HybridDefinitions}
\end{equation}

This derivation illustrates that we can mathematically reform the macroscopic Maxwell equations already microscopically. Here, $\mathbf{d}$ and $\mathbf{h}$ are \textit{composite constructs}. They represent the sum of the true vacuum field ($\mathbf{e}, \mathbf{b}$) and the re-labeled source currents ($\mathbf{p}_2, \mathbf{m}_2$).

Most importantly, the absorption of Group 2 into the field definition does not change any physics. We purely reformulated here. The bound currents (encoded in $\partial \mathbf{p}_2/\partial t$ and $\nabla \times \mathbf{m}_2$) are still flows of massive particles.

\section{The Minkowski Algorithm Applied to Microscopic Reality}
\label{sec:Micro_Minkowski_Algorithm}

Having reconstituted the macroscopic Maxwell equations at the microscopic level, it is now possible to demonstrate the arbitrary split hypothesis directly within the microscopic domain using the exact tensor definitions of Minkowski or Abraham. We can derive the microscopic equivalent of their macroscopic formulations. The mathematical protocol established by Hermann Minkowski \cite{Minkowski1908}—originally designed for macroscopic media—is applied directly to the unambiguous microscopic ground truth.

It is crucial to emphasize that this serves as a diagnostic procedure. Minkowski himself did not apply his theory to the microscopic scale; he postulated it for the macroscopic continuum. However, by applying his logic to the discrete level where the baseline is defined, one can observe exactly how the formalism handles (or mishandles) the momentum of the source carriers microscopically.

\subsection{Derivation of the Minkowski Tensor}
The derivation herein adheres to the procedure outlined by M{\o}ller \cite{moller1952}. The process commences with the ``Safe Mode'' equations derived in Section \ref{sec:Source_Reformulation}, which mimic the macroscopic form but strictly govern the ``Group 1'' (Target) sources:
\begin{align}
    \nabla \cdot \mathbf{d} &= \rho_{\text{micro},1} \label{eq:Micro_Mink_Maxwell1} \\
    \nabla \times \mathbf{h} - \frac{\partial \mathbf{d}}{\partial t} &= \mathbf{j}_{\text{micro},1} \label{eq:Micro_Mink_Maxwell2}
\end{align}
(With $\nabla \times \mathbf{e} + \dot{\mathbf{b}} = 0$ and $\nabla \cdot \mathbf{b} = 0$ remaining unchanged).
Recall that here, $\mathbf{d} \equiv \varepsilon_0\mathbf{e} + \mathbf{p}_2$ and $\mathbf{h} \equiv \frac{1}{\mu_0}\mathbf{b} - \mathbf{m}_2$.

Minkowski's procedure initiates with a purely algebraic manipulation. He substitutes the hybrid variables ($\mathbf{d}, \mathbf{h}$) into the Lorentz force expression for the ``Free'' sources ($f^\mu_{1}$).

By utilizing Eq.~\eqref{eq:Micro_Mink_Maxwell2} to eliminate the source current, an identity is derived that holds for \textit{any} arbitrary pair of field tensors ($F^{\mu\nu}, G^{\mu\nu}$). This step contains no new physics; it is a reformulation within the electromagnetic domain.

\begin{equation}
    f^\mu_{1} = \underbrace{ -\partial_\nu \left(F^{\mu\lambda}G_{\lambda}^{\;\;\nu} - \frac{1}{4} g^{\mu\nu} F_{\kappa\beta}G^{\kappa\beta}\right) }_{\text{Tensor Term}} + \underbrace{ \frac{1}{4} \left( F_{\kappa\beta} \frac{\partial G^{\kappa\beta}}{\partial x_\mu} - \frac{\partial F_{\kappa\beta}}{\partial x_\mu} G^{\kappa\beta} \right) }_{\text{Remainder Term}}.
    \label{eq:Micro_Minkowski_Identity}
\end{equation}

This constitutes the pivotal moment. The equation above is merely within the electromagnetic domain. To turn it into a physical theory, one must assign agency to the terms.
Minkowski made the \textbf{theoretical choice} to define the ``Remainder Term'' as the force density exerted by the field on the ``Bound'' matter (Group 2), $f^\mu_{2, M}$(\cite{moller1952}).

\begin{equation}
    f^\mu_{2, M} \equiv \frac{1}{4} \left( F_{\kappa\beta} \frac{\partial G^{\kappa\beta}}{\partial x_\mu} - \frac{\partial F_{\kappa\beta}}{\partial x_\mu} G^{\kappa\beta} \right).
    \label{eq:Micro_Minkowski_ForceDef}
\end{equation}

Having defined the force term, the first term in Eq.~\eqref{eq:Micro_Minkowski_Identity} is automatically identified as the negative divergence of the Energy-Momentum Tensor. This construction yields the \textbf{Minkowski Tensor} ($S^{\mu\nu}_M$):

\begin{equation}
    S^{\mu\nu}_M = F^{\mu\lambda}G_{\lambda}^{\;\;\nu} - \frac{1}{4} g^{\mu\nu} F_{\kappa\beta}G^{\kappa\beta}.
\end{equation}
Its momentum density component corresponds to the characteristic asymmetry $\mathbf{g}_{\text{Minkowski}} = \mathbf{d} \times \mathbf{b}$.

The global architecture of the Minkowski formulation is now exhibited. By rearranging the identity, the following structure is obtained:

\begin{equation}
\boxed{
\begin{aligned}
    \underbrace{ \partial_\nu T_{\text{mech, eff}}^{\mu\nu} }_{\substack{\textbf{The implied mechanical reservoir} \\ \text{(Must match Physical Reality)}}}
    \quad &\overset{\text{?}}{\textbf{=}} \quad
    \underbrace{ f^\mu_{1} + f^\mu_{2, M} }_{\substack{\textbf{The Minkowski gateway} \\ \text{(Postulated Force)}}} \\
    &\overset{\text{Definitional Identity}}{=} \quad
    \underbrace{ -\partial_\nu S^{\mu\nu}_M }_{\substack{\textbf{The electromagnetic reservoir} \\ \text{(Minkowski Field Supply)}}}
\end{aligned}
}
    \label{eq:Covariant_Minkowski_Architecture}
\end{equation}

This completes the algorithm. A mathematically self-consistent framework has been constructed (the RHS balances the LHS). However, it is not yet a physical theory. The mechanical counterpart must be checked for consistency with experimental observation, which is the subject of the following analysis.

\subsection{The Divergence of Lorentz and Minkowski Frameworks}
\label{sec:Divergence_LorentzMinkowski}

The microscopic Minkowski formulation is herein compared against the Microscopic Baseline of the Lorentz framework. To fully check the internal consistency of these competing worldviews, one must step back from the covariant 4-tensor notation. While elegant, the tensor formalism can obscure the causal mechanism of the interaction. To see the mechanics, the system must be analyzed in the 3-vector language of Newton and Lorentz.

In Section \ref{sec:Singular_Reality}, it was argued abstractly that adding an ``anomalous vector'' $\mathbf{\Phi}$ to the force definition necessarily creates a disparate physical universe. The analysis now quantifies exactly what this vector is for the Minkowski formulation. The exact gap between the Lorentz ``Ground Truth'' and the Minkowski ``Postulate'' is calculated, thereby identifying the precise physical anomalies introduced by the ``arbitrary split.''

\subsection{The Divergence of Force (Momentum Exchange)}
\label{sec:Momentum_Exchange_Div}

In the rigorous baseline, ``Force'' is defined exclusively as the Lorentz interaction acting on the discrete sources. The momentum balance is structurally clear:

\begin{equation}
\boxed{
\resizebox{0.95\textwidth}{!}{
$\begin{aligned}
    \underbrace{ (\rho_1 - \nabla \cdot \mathbf{p})\mathbf{e} + \left(\mathbf{j}_1 + \frac{\partial \mathbf{p}}{\partial t} + \nabla \times \mathbf{m}\right)\times\mathbf{b} }_{\substack{\textbf{The Lorentz definition} \\ \text{(Force on All Matter)}}}
    \quad &\overset{\text{Math. Identity}}{\textbf{=}} \quad
    \underbrace{ -\left(\frac{\partial \mathbf{g}_{\text{em}}}{\partial t} + \nabla\cdot \mathbf{t}_{\text{em}}\right) }_{\substack{\textbf{The vacuum definition} \\ \text{(Momentum Supply)}}}
\end{aligned}$
}
}
    \label{eq:Micro_Momentum_Architecture_Split}
\end{equation}

Minkowski's formulation proposes a topological redefinition. By algebraically absorbing the ``Bound'' source terms ($\mathbf{p}, \mathbf{m}$) into the field tensor, it implicitly \textbf{redefines} the concept of field momentum. To maintain mathematical equality, the definition of ``Mechanical Force'' must theoretically shift to compensate.

\begin{equation}
\boxed{
\begin{aligned}
    \underbrace{ (\rho_1 \mathbf{e} + \mathbf{j}_1 \times \mathbf{b}) + \mathbf{f}_{\text{matter}, M} }_{\substack{\textbf{The Minkowski postulate} \\ \text{(Free Force + Proposed Matter Force)}}}
    \quad &\overset{\text{Definitional Identity}}{\textbf{=}} \quad
    \underbrace{ -\left( \frac{\partial \mathbf{g}_M}{\partial t} + \nabla \cdot \mathbf{t}_M \right) }_{\substack{\textbf{The Minkowski definition} \\ \text{(Field + Conflated Momentum)}}}
\end{aligned}
}
    \label{eq:Micro_Minkowski_Momentum_Architecture}
\end{equation}

Here, the variables are defined by the microscopic limit of the Minkowski postulates:
\begin{align}
    \mathbf{g}_M &\equiv \mathbf{d} \times \mathbf{b} \\
    \mathbf{t}_M &\equiv \mathbf{e} \otimes \mathbf{d} + \mathbf{h} \otimes \mathbf{b} - \frac{1}{2}(\mathbf{e} \cdot \mathbf{d} + \mathbf{b} \cdot \mathbf{h})\mathbf{I}
\end{align}

The Minkowski force density ($\mathbf{f}_{\text{matter}, M}$) is a new mathematical object. Unlike the Lorentz force, which has a clear vector origin ($\mathbf{j}\times\mathbf{b}$), this force is defined by the components of the tensor divergence. Explicitly, the $k$-th component is:

\begin{equation}
    (\mathbf{f}_{\text{matter}, M})_k \equiv \frac{1}{2} \sum_{j=1}^{3} \left[ p_j (\partial_k e_j) - e_j (\partial_k p_j) + m_j (\partial_k b_j) - b_j (\partial_k m_j) \right].
    \label{eq:MinkowskiForce_Explicit}
\end{equation}

\paragraph{Identification of the Anomalous Vector.}
The \textbf{anomalous force} introduced by this formulation can now be isolated. It is simply the difference between the postulated force and the ground truth:

\begin{equation}
    \mathbf{\Phi}= \Delta \mathbf{f} = \mathbf{f}_{\text{matter}, M} - \mathbf{f}_{\text{Lorentz}} = \frac{\partial (\mathbf{p} \times \mathbf{b})}{\partial t} + \nabla \cdot \mathbf{t}_{\Delta}.
\end{equation}

Here, $\mathbf{t}_{\Delta}$ denotes the stress tensor difference, defined as:
\begin{equation}
    \mathbf{t}_{\Delta} \equiv \mathbf{t}_M - \mathbf{t}_{\text{em}} = \mathbf{e} \otimes \mathbf{p} + \mathbf{m} \otimes \mathbf{b} - \frac{1}{2}(\mathbf{e} \cdot \mathbf{p} + \mathbf{m} \cdot \mathbf{b})\mathbf{I}.
\end{equation}

Crucially, \textbf{this constitutes the anomalous vector $\mathbf{\Phi}$} predicted in the principle of physical exclusivity (Section \ref{sec:Singular_Reality}). The total terms represent the difference between the Lorentz formulation (vacuum tensor) and Minkowski's formulation microscopically. 

\subsection{The Divergence of Power (Energy Exchange)}
\label{sec:Energy_Divergence}

A continuous divergence is observed in the energy domain. The Ground Truth asserts that energy transfer occurs strictly via electric work on the currents:

\begin{equation}
\boxed{
\begin{aligned}
    \underbrace{ \left( \mathbf{j}_1 + \frac{\partial \mathbf{p}}{\partial t} + \nabla \times \mathbf{m} \right) \cdot \mathbf{e} }_{\substack{\textbf{The Lorentz definition} \\ \text{(Work on All Matter)}}}
    \quad &\overset{\text{Math. Identity}}{\textbf{=}} \quad
    \underbrace{ -\left(\frac{\partial u_{\text{em}}}{\partial t} + \nabla \cdot \mathbf{s}_{\text{em}}\right) }_{\substack{\textbf{The vacuum definition} \\ \text{(Field Energy Supply)}}}
\end{aligned}
}
    \label{eq:Micro_Energy_Architecture_Split}
\end{equation}

Minkowski, however, \textbf{postulates} a new interaction term $P_{\text{matter}, M}$, implying a completely new definition of energy transfer:

\begin{equation}
\boxed{
\begin{aligned}
    \underbrace{ \mathbf{j}_{1} \cdot \mathbf{e} + P_{\text{matter}, M} }_{\substack{\textbf{The Minkowski postulate} \\ \text{(Free Work + Proposed Matter Work)}}}
    \quad &\overset{\text{Definitional Identity}}{\textbf{=}} \quad
    \underbrace{ -\left( \frac{\partial u_M}{\partial t} + \nabla \cdot \mathbf{s}_M \right) }_{\substack{\textbf{The Minkowski definition} \\ \text{(Field + Conflated Energy)}}}
\end{aligned}
}
    \label{eq:Micro_Minkowski_Energy_Architecture}
\end{equation}

Here, the variables are defined as:
\begin{align}
    u_M &\equiv \frac{1}{2}(\mathbf{e} \cdot \mathbf{d} + \mathbf{b} \cdot \mathbf{h}) \\
    \mathbf{s}_M &\equiv \mathbf{e} \times \mathbf{h}
\end{align}

Explicitly, the proposed Minkowski power density corresponds to:
\begin{equation}
    P_{\text{matter}, M} = \frac{1}{2} \left( \frac{\partial \mathbf{e}}{\partial t} \cdot \mathbf{d} - \mathbf{e} \cdot \frac{\partial \mathbf{d}}{\partial t} + \frac{\partial \mathbf{b}}{\partial t} \cdot \mathbf{h} - \mathbf{b} \cdot \frac{\partial \mathbf{h}}{\partial t} \right).
    \label{eq:Micro_Minkowski_Power_Explicit_General}
\end{equation}

\paragraph{Identification of the Hidden Reservoir.}
This leads to a measurable discrepancy in the proposed energy budget:
\begin{equation}
    \Delta P = P_{\text{matter}, M} - P_{\text{Lorentz}} = \frac{1}{2} \frac{\partial}{\partial t}(\mathbf{e} \cdot \mathbf{p} - \mathbf{m} \cdot \mathbf{b}) - \nabla \cdot (\mathbf{e} \times \mathbf{m}).
\end{equation}

By applying the logic of the \textbf{principle of physical exclusivity} to the energy domain, this term is identified as the ``hidden energy reservoir.'' It represents energy that the Minkowski formalism calculates as transferred to the matter, but which is not accounted for by the standard electric work done on the moving charges.

\section{The Analysis of Mechanical Consistency}
\label{sec:Anatomy_Conflation}

The microscopic Minkowski formulation is herein evaluated for mechanical consistency. We consider the illustrative example of an expanding electric dipole, consisting of an electron and an ion, to investigate its behavior in the limit of discrete separation.

\subsection{The Diagnostic Task: Two Distinct Critiques}
\label{sec:Diagnostic_Task}

The exact ``Anomalous'' terms required by the Minkowski postulate have now been successfully formulated. The presence of these terms ($\Delta \mathbf{f}$ and $\Delta P$) forces a confrontation with two separate and equally structural inconsistencies.

\paragraph{Critique 1: The Topological Divergence (Vacuum Inertia).}
The first failure is found in the Force equation itself. By predicting an anomalous force $\Delta \mathbf{f}$, the formulation risks assigning force to regions where no matter exists. If $\Delta \mathbf{f} \neq 0$ in a vacuum, the theory implies the existence of \textbf{Vacuum Inertia}—a massive carrier that accelerates without substance.

\paragraph{Critique 2: The Energetic Divergence (Internal Mismatch).}
The second divergence is a lack of internal coordination. Even if the new force is accepted, a consistent mechanical theory requires that the power $P$ delivered to the system equals the mechanical work done by the force ($\mathbf{f} \cdot \mathbf{v}$). The \textbf{Force-Energy Consistency Criterion (FECC)} tests this link. If $\Delta P \neq \Delta \mathbf{f} \cdot \mathbf{v}$, the internal logic of the theory is inconsistent; the energy side of the ledger describes a different physical process than the momentum side.

In the following sections, the analysis demonstrates that the Minkowski formulation fails on \textbf{both} counts microscopically. Use of the Expanding Dipole (Section \ref{sec:Micro_Dipole_Case}) proves the existence of Vacuum Inertia, while the Static Dipole (Section \ref{sec:FECC_Violation_Example}) proves the violation of the FECC.

\subsection{Case Study 1: The Expanding Microscopic Dipole (Topological Divergence)}
\label{sec:Micro_Dipole_Case}

The analysis first addresses \textbf{Critique 1: The Topological Divergence}. The Minkowski postulate implies that the force distribution $\mathbf{f_M}$ differs from the Lorentz baseline $\mathbf{f}_{L}$ by the anomalous vector $\mathbf{\Phi}_{\text{force}}$. We now search for this discrepancy in a physical system.

To examine the local consistency of the Minkowski formulation, it is applied to a limiting case where the distinction between "matter" and "vacuum" is unambiguous: the dissociation of a neutral atom into free charges. This scenario serves as a stress test, isolating the specific regions where the mathematical definitions of the macroscopic theory may diverge from the physical distribution of mass.

\subsubsection{The Physical Setup: Separation of Charge}

Consider a simplified model of a neutral atom composed of a massive positive ion cloud and a negative electron cloud. Initially, they are superimposed and static at the origin. At time $t=0$, an external non-electromagnetic force causes the clouds to separate with velocity $\pm \mathbf{v}$.

\paragraph{The Physical Reality (Ground Truth).}
From the perspective of the Microscopic Baseline established in Chapter \ref{chap:FreeCharges}, the topology of the system is strictly defined:
\begin{itemize}
    \item \textbf{Mass Distribution:} Mass exists \textit{exclusively} at the coordinates of the ion cloud ($x_+$) and the electron cloud ($x_-$). The growing volume \textit{between} them is physical vacuum ($n=0, \rho_m=0$).
    \item \textbf{Current Distribution:} The motion of the charged clouds constitutes a real physical current $\mathbf{j}_{\text{micro}}$ localized to the particles.
    \item \textbf{Force Distribution:} The Lorentz force acts \textit{only} on the charge carriers. No electromagnetic force acts on the vacuum gap between them, as $\rho=0$ and $\mathbf{j}=0$ in that region.
\end{itemize}

The microscopic current density is mathematically encoded as the time derivative of a polarization field:
\begin{equation}
    \mathbf{j}_{\text{micro}} = \frac{\partial \mathbf{p}}{\partial t}.
\end{equation}
Because the field $\mathbf{p}$ is defined by the time integral of the current flux, a non-zero polarization vector is mathematically generated throughout the entire volume \textit{traversed} by the separating charges. As the clouds move apart, a uniform polarization field $\mathbf{p}$ effectively "fills" the growing vacuum gap between them.

It is essential to recognize that this field $\mathbf{p}$ acts strictly as a \textbf{kinematic alias} for the charge displacement. In the vacuum gap, it is merely a record of past transport. The field itself assumes no intrinsic physical meaning; it possesses no mass and no physical substance. It is a mathematical bookkeeping device, not a material entity. Only the divergence ($\nabla \cdot \mathbf{p}$) and the time rate of change ($\partial \mathbf{p}/\partial t$) have physical meaning, corresponding to the current charges they represent.

\subsubsection{The Baseline Prediction: The Lorentz Interaction}

The mechanical prediction of the microscopic theory is established first. The force density is determined by the Lorentz law, which respects the topological distribution of the sources:
\begin{equation}
    \mathbf{f}_{\text{Lorentz}} = (-\nabla \cdot \mathbf{p})\mathbf{e} + \frac{\partial \mathbf{p}}{\partial t} \times \mathbf{b}.
\end{equation}
Applying this to the expanding dipole: inside the vacuum gap, the polarization field is spatially uniform ($\nabla \cdot \mathbf{p} = 0$). Once the separation stops, the field is static ($\partial \mathbf{p} / \partial t = 0$). Consequently, the Lorentz force density vanishes identically in the vacuum region. The force acts \textit{exclusively} at the boundaries (the particle clouds), where the divergence of $\mathbf{p}$ is non-zero.

The mechanical evolution of this system is governed by the exact momentum balance derived in Chapter 1:

\begin{equation}
\boxed{
\resizebox{0.95\textwidth}{!}{
$\begin{aligned}
    \underbrace{ \left(\frac{\partial \mathbf{g}_{\text{mech}}}{\partial t} + \nabla\cdot\mathbf{t}_{\text{kin}}\right) - \mathbf{f}_{\text{Lorentz}} }_{\substack{\textbf{The mechanical reservoir} \\ \text{(Net Momentum Gain)}}}
    \quad &\overset{\text{Physical Coupling}}{\textbf{=}} \quad
    \underbrace{ (\rho_1 - \nabla \cdot \mathbf{p})\mathbf{e} + \left(\mathbf{j}_1 + \frac{\partial \mathbf{p}}{\partial t} + \nabla \times \mathbf{m}\right)\times\mathbf{b} }_{\substack{\textbf{The true gateway} \\ \text{(Lorentz Force on Discrete Mass)}}} \\
    &\overset{\text{Math. Identity}}{=} \quad
    \underbrace{ -\left(\frac{\partial \mathbf{g}_{\text{em}}}{\partial t} + \nabla\cdot \mathbf{t}_{\text{em}}\right) }_{\substack{\textbf{The electromagnetic reservoir} \\ \text{(Momentum Supply)}}}
\end{aligned}$
}
}
    \label{eq:Micro_Momentum_Reference}
\end{equation}

This equation correctly maps the transfer of momentum. The electromagnetic force on the right couples precisely to the change in mechanical momentum on the left, respecting the localized nature of the mass target. The Lorentz theory consistently describes the behavior of the expanding electric dipole, consisting of ion and electron.

\subsubsection{The Prediction of Topological Mismatch (Vacuum Force)}
\label{sec:Vacuum_Force}

The mechanical consistency of the Minkowski formulation is now tested by applying it to the expanding dipole. The goal is to identify the mechanical response $\mathbf{g}_{\text{mech}, M}$ required to balance the proposed Minkowski force density.

\begin{equation}
\boxed{
\begin{aligned}
    \underbrace{ \frac{\partial \mathbf{g}_{\text{mech}, M}}{\partial t} + \nabla\cdot\mathbf{t}_{\text{kin}, M} }_{\substack{\textbf{The implied mechanical reservoir} \\ \text{(Required Response)}}}
    \quad &\overset{\text{Forced Balance}}{\textbf{=}} \quad
    \underbrace{ \mathbf{f}_{\text{matter}, M} }_{\substack{\textbf{The Minkowski force} \\ \text{(Derived from Definition)}}}
\end{aligned}
}
    \label{eq:Micro_Momentum_Architecture_Minkowski_Hybrid}
\end{equation}
(For clarity, the focus is on the force exerted on a single dipole where $\rho_1=0$).

The general definition of the Minkowski force density on a polarized medium is given by the divergence of the stress tensor. Neglecting magnetization for simplicity, the $k$-th component is:
\begin{equation}
    (\mathbf{f}_{\text{matter}, M})_k = \frac{1}{2} \sum_{j=1}^{3} \left[ p_j (\partial_k e_j) - e_j (\partial_k p_j) \right].
    \label{eq:MinkowskiForce_PureP}
\end{equation}

\paragraph{The Vacuum Force Divergence.}
The mechanical implications become apparent when examining the spatial dependency of this force. The first term in the bracket depends directly on the polarization vector $\mathbf{p}$, rather than its divergence:
\begin{equation}
    \mathbf{f}_{\text{Kelvin-like}} \propto (\mathbf{p} \cdot \nabla) \mathbf{e}.
\end{equation}
In the expanding dipole model, the polarization field $\mathbf{p}$ is non-zero and uniform throughout the entire volume \textit{between} the separating charges (the vacuum gap).

Consequently, if a non-uniform external electric field ($\nabla \mathbf{e} \neq 0$) is applied to this system, the Minkowski formulation predicts a non-zero force density \textbf{inside the vacuum gap}. This prediction creates an immediate conflict with the local laws of mechanics. Comparing the predicted force to the available mass:
\begin{itemize}
    \item \textbf{The Force Prediction:} $\mathbf{f}_{\text{matter}, M} \neq 0$ in the gap.
    \item \textbf{The Mechanical Requirement:} To satisfy Newton's Second Law ($\mathbf{f} = \dot{\mathbf{g}}$), there must be a time rate of change of mechanical momentum density in the gap.
    \item \textbf{The Physical Reality:} The gap is a vacuum. The mass density is strictly zero ($\rho_m = 0$).
\end{itemize}

This confirms the prediction of the \textbf{principle of physical exclusivity}. Because Minkowski's force vector $\mathbf{f}_M$ targets the kinematic alias $\mathbf{p}$ rather than the physical charge $\rho$, it inevitably distributes the interaction into the void.

To maintain the validity of the Minkowski tensor, the theory implies the definition of \textbf{effective vacuum inertia}—a hypothetical mechanical effect that fills the void between the charges. The theory implies that the force acts on this field of "defined momentum" rather than on the discrete particles themselves.

This comparison reveals that the  microscopic Minkowski formulation achieves algebraic conservation only by "distributing" the mechanical interaction into regions where no matter exists.

\subsection{Case Study 2: The Static Dipole (Energetic Divergence)}
\label{sec:FECC_Violation_Example}

The analysis now extends the structural critique to \textbf{Critique 2: The Energetic Divergence.} Having shown that the Minkowski force definition violates the topological definition of mass (Vacuum Inertia), the analysis tests whether the formalism is at least \textbf{internally} consistent. We verify if the energy budget matches the mechanical work predicted by the force, effectively running a diagnostic on the "gears" of the theory.

\subsubsection{The Setup: A Static Dipole in a Time-Varying Field}
Consider an illustrative test case: a rigid, stationary electric dipole with constant moment $\mathbf{p}$, subjected to a time-varying external electric field $\mathbf{e}(t)$.
\begin{itemize}
    \item \textbf{Kinematics:} Because the dipole is held fixed in a rigid constraint, the velocity of all mass carriers is strictly zero ($\mathbf{v}=0$).
    \item \textbf{Mechanics:} Basic mechanics dictates that the power delivered to a stationary system must be zero: $P_{\text{mech}} = \mathbf{f} \cdot \mathbf{v} \equiv 0$.
\end{itemize}

\paragraph{1. The Baseline Consistency (Lorentz).}
First, the Microscopic Baseline is verified. The Lorentz formulation tracks energy transfer strictly via work done on currents. Since the dipole moment is constant, the polarization current is zero ($\mathbf{j} = \partial \mathbf{p} / \partial t = 0$). Consequently, the interaction term vanishes identically:
\begin{equation}
    P_{\text{Lorentz}} = \mathbf{j} \cdot \mathbf{e} = 0.
\end{equation}
The electromagnetic energy balance is perfectly closed. Any local fluctuation in vacuum energy density $u_{\text{em}}$ is exactly balanced by the divergence of the vacuum flux $\mathbf{s}_{\text{em}}$, with zero energy lost to the matter:

\begin{equation}
\boxed{
\begin{aligned}
   \underbrace{ -\left(\frac{\partial u_{\text{em}}}{\partial t} + \nabla \cdot \mathbf{s}_{\text{em}}\right) }_{\substack{\textbf{The electromagnetic reservoir} \\ \text{(Field Supply)}}}
    \quad \overset{\text{FECC}}{\textbf{=}} \quad
    \mathbf{j} \cdot \mathbf{e} = 0
\end{aligned}
}
\end{equation}
The theory is consistent: Mechanical Work ($0$) equals Field Energy Loss ($0$).

\paragraph{2. The Minkowski Prediction.}
The Minkowski formulation is now applied to this same system. Minkowski postulates a modified energy density $u_M$ and flux $\mathbf{S}_M$, which leads to a distinct energy balance equation. The difference between this postulated balance and the Lorentz baseline is:
\begin{equation}
    \Delta P = P_{\text{Minkowski}} - P_{\text{Lorentz}} = \frac{1}{2} \frac{\partial}{\partial t}(\mathbf{e} \cdot \mathbf{p} - \mathbf{m} \cdot \mathbf{b}) - \nabla \cdot (\mathbf{e} \times \mathbf{m}).
\end{equation}
For the specific case of a static electric dipole ($\mathbf{p}=\text{const}, \mathbf{m}=0$), this expression simplifies to a single term:
\begin{equation}
    \Delta P = \frac{1}{2} \mathbf{p} \cdot \frac{\partial \mathbf{e}}{\partial t}.
\end{equation}
Because the external field is varying in time ($\dot{\mathbf{e}} \neq 0$), this term is strictly \textbf{non-zero}.

This leads to a surprising prediction: the Minkowski energy balance implies a local "sink" of energy. Energy is calculated to be leaving the electromagnetic field budget, despite the absence of any moving mass to receive it:

\begin{equation}
\boxed{
\begin{aligned}
    \underbrace{ -\left( \frac{\partial u_M}{\partial t} + \nabla \cdot \mathbf{S}_M \right) }_{\substack{\textbf{The Minkowski reservoir} \\ \text{(Field Supply)}}}
    \quad \textbf{=} \quad
    \underbrace{ \frac{1}{2} \mathbf{p} \cdot \frac{\partial \mathbf{e}}{\partial t} }_{\substack{\textbf{Predicted energy transfer} \\ \text{(Non-Zero)}}} \quad \neq 0
\end{aligned}
}
\end{equation}

\paragraph{Confirmation of Critique 2 (Internal Inconsistency).}
To assess the validity of this prediction, one must compare the Energy side of the theory with its Force side:
\begin{itemize}
    \item \textbf{Force Prediction:} The mass is stationary ($\mathbf{v}=0$). Consequently, the mechanical power delivered by the Minkowski force is rigorously zero:
    \begin{equation}
        P_{\text{force}} = \mathbf{f}_M \cdot \mathbf{v} = 0.
    \end{equation}
    \item \textbf{Energy Prediction:} The Minkowski Poynting theorem simultaneously calculates a non-zero rate of energy transfer:
    \begin{equation}
        P_{\text{energy}} = \frac{1}{2} \mathbf{p} \cdot \dot{\mathbf{e}} \neq 0.
    \end{equation}
\end{itemize}

The Minkowski formulation predicts that the electromagnetic field loses energy ("anomalous work") that does not appear as kinetic energy of the matter. This internal inconsistency—where $P \neq \mathbf{f} \cdot \mathbf{v}$—indicates that the microscopic Minkowski formalism is not a connected mechanical system.

\section{The Role of Integration}
\label{sec:Total_Force_Filter}

Despite the microscopic inconsistencies identified above, the Minkowski formulation has served as a successful predictive tool for over a century. This success is explained by the mathematics of integration.

In most experimental contexts, the internal stress distribution of a material is secondary to its net trajectory. The primary interest is the \textbf{total force} $\mathbf{F}_{\text{total}} = \int \mathbf{f} \, dV$. It is demonstrated here that for isolated bodies, the integration of the force density over the entire volume eliminates the local divergence-free terms, yielding the correct total force vector for the center of mass. This explains why the Minkowski formulation, while microscopically distinct, remains physically predictive for calculating total forces on isolated bodies.

\subsection{The Equivalence of the Integral.}
Comparing the total force predicted by the Lorentz baseline against the Minkowski formulation for a static, localized object (such as the dipole). The difference in local force density is defined by the divergence of the difference tensor:
\begin{equation}
    \Delta \mathbf{f} = \mathbf{f}_{\text{Minkowski}} - \mathbf{f}_{\text{Lorentz}} = \nabla \cdot \mathbf{T}_\Delta
\end{equation}
where the tensor $\mathbf{T}_\Delta$ contains the terms involving $\mathbf{p}$ and $\mathbf{m}$.To calculate the discrepancy in the total force, this difference is integrated over a volume $V$ that entirely encloses the object:
\begin{equation}
    \Delta \mathbf{F}_{\text{total}} = \int_V (\nabla \cdot \mathbf{T}_\Delta) \, dV.
\end{equation}
Applying the Divergence Theorem, this volume integral transforms into a surface integral over the boundary $S$:
\begin{equation}
    \Delta \mathbf{F}_{\text{total}} = \oint_S \mathbf{T}_\Delta \cdot d\mathbf{a}.
\end{equation}
Because the object is localized, the polarization and magnetization fields ($\mathbf{p}, \mathbf{m}$) vanish at the integration boundary $S$, which lies in the vacuum outside the object. Consequently, $\mathbf{T}_\Delta = 0$ on the surface, and the total force difference vanishes identically:
\begin{equation}
    \Delta \mathbf{F}_{\text{total}} \equiv 0.
\end{equation}

It proves that in static cases, \textbf{Lorentz and Minkowski predict exactly the same Total Force.}

\subsection{The Point Dipole Approximation.}
A similar masking effect is formalized in the standard derivation of the Kelvin force. The total force on a distributed charge distribution $\rho = -\nabla \cdot \mathbf{p}$ is rigorously defined by the volume integral:
\begin{equation}
    \mathbf{F}_{dp} = \int_V (-\nabla \cdot \mathbf{p})\mathbf{e}_{\text{ext}} \, dV.
\end{equation}
Using the vector identity $(\nabla \cdot \mathbf{A})\mathbf{B} = \nabla \cdot (\mathbf{A} \otimes \mathbf{B}) - (\mathbf{A} \cdot \nabla)\mathbf{B}$, we expand the integrand:
\begin{equation}
    \mathbf{F}_{\text{dp}} = \int_V \left[ (\mathbf{p} \cdot \nabla) \mathbf{e}_{\text{ext}} - \nabla \cdot (\mathbf{p} \otimes \mathbf{e}_{\text{ext}}) \right] dV.
\end{equation}
The second term is a total divergence. By Gauss's theorem, it transforms into a surface integral over the boundary of the dipole volume. Since the polarization $\mathbf{p}$ vanishes outside the dipole, this term is zero. Assuming the external field gradient $\nabla \mathbf{e}_{\text{ext}}$ is approximately constant over the small volume, it can be pulled out of the integral:
\begin{equation}
    \mathbf{F}_{\text{dp}} = \left( \int_V \mathbf{p} \, dV \right) \cdot \nabla \mathbf{e}_{\text{ext}}.
\end{equation}
By assuming the external field gradient $\nabla \mathbf{e}_{\text{ext}}$ is constant over the microscopic volume, this integral simplifies to the familiar form:
\begin{equation}
    \mathbf{F}_{dp} = (\mathbf{p}_{dp} \cdot \nabla)\mathbf{e}_{\text{ext}}, \quad \text{where } \mathbf{p}_{dp} = \int_V \mathbf{p} \, dV.
\end{equation}

It is critical to identify the precise step where the physical model diverges. This occurs in the transition where the divergence term $-\nabla \cdot (\mathbf{p} \otimes \mathbf{e}_{\text{ext}})$ is discarded because it integrates to zero over the boundary.

As established in Part I regarding the ``Newtonian Limit'' (Section~\ref{sec:Artifactual_Vector}), this discarded term corresponds exactly to the \textbf{Artifactual Vector} $\mathbf{\Phi}$ of the dipole system. By removing it, the derivation rigorously preserves the \textbf{Total Force} (the global ledger) but silently rewrites the \textbf{Force Density} (the local physics), redistributing the mechanical interaction from the boundaries (where charge exists) to the bulk volume (where it does not). This is the subtle mechanism of the ``Arbitrary Split'' in action: the total force remains invariant, but the internal topology is destroyed.

While mathematically valid for calculating the net translation, the neglect of the stress tensor divergence (which integrates to zero) and the point-dipole reduction effectively destroy the internal mechanical topology. The model preserves the global vector but erases the local physics.

\subsection{Conclusion.}
The agreement of Total Forces is a necessary but insufficient condition for physical validity. While the Minkowski and Lorentz formulations agree on the \textit{sum}, they disagree on the \textit{location}. Since the laws of mechanics ($\mathbf{f} = \rho_m \mathbf{a}$) are local, a consistent theory must identify the correct location of interaction. Only the Lorentz formulation satisfies this condition by respecting the vacuum gap.

\section{The Universality of the Critique: Beyond Minkowski}
\label{sec:Universality_of_Critique}

The microscopic analysis has focused on the Minkowski formulation, as it is the canonical example for the "Canonical" approach. However, the principle identified in Section \ref{sec:Singular_Reality} is strictly universal: \textbf{Different formulations are mutually exclusive.}

Accepting the Lorentz force microscopically implies the logical exclusion of \textit{any} other formulation. Conversely, accepting the Minkowski or Abraham tensor microscopically implies excluding the Lorentz force. There is no middle ground where both descriptions are simultaneously "correct" but "different."

To demonstrate this, the vacuum diagnostic is briefly applied to the primary historical competitor: the Abraham tensor microscopically.

\subsection{The Abraham Formulation}
\label{subsec:Abraham_Critique}

The Abraham formulation is frequently cited as the mechanically consistent alternative to Minkowski because it resolves the asymmetry of the energy-momentum tensor ($\mathbf{g} = \mathbf{s}/c^2$). To analyze its microscopic consistency, its proposed field components are explicitly defined:
\begin{itemize}
    \item \textbf{Momentum Density:} $\mathbf{g}_{Abr} = \frac{1}{c^2}(\mathbf{e} \times \mathbf{h})$
    \item \textbf{Stress Tensor:} $\mathbf{t}_{Abr} = \frac{1}{2} \left( \mathbf{e} \otimes \mathbf{d} + \mathbf{d} \otimes \mathbf{e} + \mathbf{h} \otimes \mathbf{b} + \mathbf{b} \otimes \mathbf{h} \right) - u_{Abr} \mathbf{I}$
\end{itemize}
The Abraham force density is then defined by the standard divergence relation:
\begin{equation}
    \mathbf{f}_{Abr} \equiv -\nabla \cdot \mathbf{t}_{Abr} - \frac{\partial \mathbf{g}_{Abr}}{\partial t}.
\end{equation}

\paragraph{The Abraham Force Gap.}
To reveal the physical implications of this definition, the "new" force it predicts is rigorously isolated by defining the \textbf{Abraham Force Gap} ($\mathbf{f}_{\Delta, Abr}$). This is the difference between the proposed Abraham force density and the Lorentz baseline:
\begin{equation}
    \mathbf{f}_{\Delta, Abr} \equiv \mathbf{f}_{Abr} - \mathbf{f}_{vac} = -\nabla \cdot \mathbf{T}_{\Delta} - \frac{\partial \mathbf{g}_{\Delta}}{\partial t}.
\end{equation}

For a non-magnetic medium ($\mathbf{m}=0$), the momentum densities are identical ($\mathbf{g}_{Abr} = \mathbf{g}_{vac} = \varepsilon_0 \mathbf{e} \times \mathbf{b}$), so the discrepancy arises entirely from the symmetrization of the stress tensor. The difference tensor $\mathbf{T}_{\Delta}$ is given by:
\begin{equation}
    \mathbf{T}_{\Delta} = \mathbf{t}_{Abr} - \mathbf{t}_{vac} = \frac{1}{2} \left( \mathbf{e} \otimes \mathbf{p} + \mathbf{p} \otimes \mathbf{e} \right) - \frac{1}{2} (\mathbf{e} \cdot \mathbf{p}) \mathbf{I}.
\end{equation}

The divergence of this difference tensor defines the "Anomalous Force" predicted by the Abraham formulation. Expanding the term $-\nabla \cdot \mathbf{T}_{\Delta}$ reveals:
\begin{equation}
    \mathbf{f}_{\Delta, Abr} = - \frac{1}{2} \left[ (\nabla \cdot \mathbf{e})\mathbf{p} + (\mathbf{p} \cdot \nabla)\mathbf{e} + (\mathbf{e} \cdot \nabla)\mathbf{p} + \mathbf{e}(\nabla \cdot \mathbf{p}) - \nabla(\mathbf{e} \cdot \mathbf{p}) \right].
\end{equation}
Applying this result to the vacuum gap of the expanding dipole yields the same result as in the Minkowski case. Because the force expression depends explicitly on $\mathbf{p}$ and its spatial derivatives, $\mathbf{f}_{\Delta, Abr}$ is non-zero in the void. Like Minkowski, the microscopic Abraham formulation necessitates the creation of Vacuum Inertia to satisfy local conservation. Thus, the Abraham formulation is equally \textbf{subject to Critique 1}.

\section{Conclusion: The Principle of Physical Exclusivity}
\label{sec:Chapter2_Conclusion}

The consequences of the \textbf{principle of physical exclusivity} (Section \ref{sec:Singular_Reality}) have now been demonstrated at the microscopic level. By conducting a rigorous topological verification of the force definitions, the analysis has arrived at a position distinct from the modern consensus of the ``arbitrary split.''

Microscopically, it has been demonstrated that the debate over the correct energy-momentum tensor is not a question of convention; it is a binary choice between two distinct physical realities at the fundamental level:
\begin{itemize}
    \item \textbf{Option A (The Baseline):} The Lorentz force is the correct description of microscopic reality. In this universe, force acts exclusively on charge. Consequently, Minkowski, Abraham, and all other conflated tensors are \textit{structurally divergent} as microscopic descriptions.
    \item \textbf{Option B (The Alternative Models):} The Minkowski (or Abraham) force is correct \textit{microscopically}. In this universe, the Lorentz force is incorrect. The axioms of classical electrodynamics must be rewritten to allow fields to exert force on vacuum potentials ($\Phi \neq 0$), creating ``Vacuum Inertia'' and ``Hidden Reservoirs.''
\end{itemize}

Empirical evidence strongly supports Option A. The vacuum Maxwell equations and the Lorentz framework remain valid interaction laws down to the resolution of the Compton wavelength of the electron ($\lambda_c = h/mc \approx 2.43 \times 10^{-12}$ m) \cite{Zangwill2013}. Consequently, we reject any mutually exclusive theory—such as the Abraham or Minkowski formulation—as a physically consistent microscopic description of reality.

\subsection{The Path Forward}

The transition to the macroscopic domain is an act of \textbf{high-frequency field filtering}. It is a method of compressing the bandwidth of reality to make it computable.

Crucially, it is argued that \textbf{the principle of physical exclusivity is scale-invariant.} If a formulation is topologically inconsistent at the microscopic level (requiring vacuum inertia), averaging does not fix the physics. A force that acts on the void at the microscopic scale does not acquire a valid mass target simply through lower-resolution viewing. The ``vacuum inertia'' implies a topological divergence that is not resolved by scale.

The insights of this chapter invite a re-evaluation of the historical context. In the following chapter, the historical experiments of the Abraham-Minkowski debate will be reviewed through the perspective of the Microscopic Baseline.

Following this review, Part II of this book will commence. The objective will be to derive a consistent macroscopic theory by strictly filtering the microscopic Lorentz ground truth, ensuring that the macroscopic variables remain faithful to the underlying mass topology.

\chapter{Experiment and the Interpretation of Evidence}
\label{chap:Experiment}

For over a century, the physics community has engaged in an extensive program to determine "which tensor is correct." The bibliography of this debate, meticulously compiled by McDonald \cite{McDonald_Biblio}, spans over three hundred publications—from the foundational work of Poynting to modern laser-tweezer experiments. This reflects a significant and sustained intellectual effort to resolve what appeared to be a binary choice between two competing models. However, before evaluating the experimental evidence, a rigorous hierarchy of information resolution must be established.

\section{The Hierarchy of Information Compression}
\label{sec:Info_Compression}

To interpret the validity of the experiments, the semantic ambiguity that has sustained the confusion—the precise definitions of the "Microscopic" and "Macroscopic" domains—must first be resolved.

This work rigorously separates the analysis into layers of information resolution:
\begin{enumerate}
    \item \textbf{Stage 1 (Quantum Baseline):} Quantum Electrodynamics (QED). The domain of probabilistic wavefunctions and uncertainty. Here, the boundary between particle and field dissolves (allowing hybrid objects like \textit{polaritons}).
    \item \textbf{Stage 2 (Classical Microscopic Baseline):} The Lorentz theory. The domain of deterministic trajectories, continuous fields, and discrete charges. Here, the separation is absolute.
    \item \textbf{Stage 3 (Macroscopic Model):} The continuum theory. The domain of averaged fields and smooth media.
\end{enumerate}

In this first part of the book, the analysis has operated strictly within the second layer of theoretical abstraction (Stage 2). It is crucial to recognize that this layer is, by definition, an \textbf{intentional simplification} of Stage 1. If one demands an explanation for an experimental output that retains the full fidelity of the quantum state, one must employ QED.

We cannot describe photons, polaritons, or any quantum object with any classical continuous theory at Stage 2. This limitation applies to all proposed energy-momentum tensors under investigation here: Abraham, Minkowski, or the Lorentz Vacuum tensor. These are deterministic theories that cannot describe the probabilistic QED behavior that constitutes the fundamental truth of Stage 1. None of them are capable of explaining experimental outputs in the category of the double-slit experiment of a single electron; QED is required for such phenomena.

Consequently, it is argued that one cannot judge a classical theory—which is by intent a simplification of QED—by purely quantum events. This may be viewed as a \textbf{category error}. The aim is not to replace QED with classical theories. Rather, the goal is to establish a simplified, deterministic model that can serve as an input for the derivation of the macroscopic theory. The utility of the classical model lies precisely in its ability to \textbf{compress} the complexity of the quantum state into a manageable, deterministic signal.

\paragraph{Analogy: The Emergence of Temperature.}
The conceptual step from Microscopic to Macroscopic is identical to the step from QED to Classical dynamics. Detailed descriptions are inextricably filtered to gain computable variables.
Consider the macroscopic variable of \textbf{Temperature}. It is stated: "This block has Temperature $T$."
A critic might object: "But relying on Temperature is false! I have looked closer, and the matter is made of particles with individual velocities. Temperature is an illusion; if your theory cannot predict the trajectory of every particle, it must be wrong!"
This objection represents a methodological divergence. Temperature is valid \textit{precisely because} it ignores the individual trajectories. It is an intentional step of information compression.

\subsection{The Rule of Judgment.}

Significant confusion arises if one is not aware of the resolution scale being considered when interpreting the outcome of an experiment. When judging an experiment, we propose that one must therefore strictly enforce the scale of resolution. It is argued that one cannot judge a theory based on phenomena that are \textbf{intentionally filtered out} of its bandwidth.

\subsection{The Question of this Chapter (Stage 2 Selection)}
\label{subsec:TwoQuestions}

The structural analysis of the preceding chapters established a fundamental distinction: the microscopic Lorentz framework is the only physically consistent description of the microscopic reality. This analysis therefore does not accept any mutually exclusive theory—such as the Abraham or Minkowski formulation—as a physically consistent microscopic description of reality. By assigning force to the vacuum or energy to a hidden reservoir, these alternative models imply physical properties that diverge from the fundamental definitions of mass and work.

In this chapter, the available experimental outcomes are re-evaluated through this lens, with a constant focus on the hierarchy of information resolution.

\textbf{Note:} For the experimental validation in this chapter, we do not judge the macroscopic tensor formulations (Stage 3). Macroscopic formalisms will be the subject of the remainder of this book (Parts II–VII).

\section{Remark: The Scope of the Historical Controversy}
\label{sec:Historical_Scope}

It is important to acknowledge the scale of the intellectual effort that has been dedicated to this problem. As the extensive bibliography compiled by McDonald \cite{McDonald_Biblio} attests, hundreds of papers have been written debating the merits of the Abraham versus Minkowski formulations. This debate has been fought by some of the most capable minds in physics, and the arguments for both sides are often compelling within their own frameworks.

It is intuitively natural to search for a "correct" tensor that predicts the \textit{total force} on a body, as this is the observable outcome of an experiment. The assumption that one of these two famous tensors (Abraham or Minkowski) must be the answer is a reasonable starting point. The critique presented here suggests, however, that the question itself is ill-posed, as both options fundamentally smear the microscopic topology.

Historically, the question has often been framed as a dual choice: "Minkowski or Abraham?" The total force on an object (e.g., an isolated dipole) is calculated and compared with experiment. The outcome is then taken to validate one or the other.

This is a totally reasonable approach if one is trying to distinguish between Abraham and Minkowski. The outcome of the experiment will indeed match the prediction of the correct effective theory. As shown in the previous chapter, the total force on an isolated object can be reconciled with either the Minkowski or the Abraham tensor (by ignoring the internal stress distribution).

The central observation of this history is that the participants were not arguing against the microscopic baseline; they were, in almost every case, trying to defend it. Distinguished physicists did not set out to challenge the microscopic foundations of electrodynamics. Often, the starting point to derive the total force on an isolated object was the Lorentz force on charges and currents. The integrated total force was then taken as an argument for or against Abraham or Minkowski.

As demonstrated in the previous chapter, this analysis suggests that the question itself represented a \textbf{category error}. Assuming the Lorentz framework is correct implicitly rejects all mutually exclusive theories, such as those of Abraham or Minkowski, at the microscopic level. The experiments have validated the algebra of the limiting cases (integration and total force behavior), but they are often interpreted as validating the \textit{local physics} of tensors that describe a distinct structural reality. This work therefore critiques the \textit{conclusion} of these historical works while validating their \textit{method}. By showing that their successful predictions derive directly from the Lorentz force, the need for the mutually exclusive tensors they sought to justify is removed.

\section{Challenges in Phenomenological Derivation}
\label{sec:Phenomenological_Challenges}

Having established the microscopic baseline, the theoretical arguments historically used to validate the conflated tensors must be addressed. Prior to examining the experimental record in detail, two specific logical challenges that have historically impeded the resolution of this debate are identified.

\subsection{Challenge 1: Field Kinematics}
\label{subsec:FieldKinematics}

The first challenge arises from a persistent intuition regarding the refractive index $n$. A common argument for a modified electromagnetic momentum—such as Minkowski's $n^2$ scaling or Abraham's $1/n$ scaling—relies on the observation that the phase velocity of light inside a medium slows to $v_{phase} = c/n(\omega)$.

The mechanism of the refractive index is, however, entirely kinematic. It is not a property of a "slow photon," but a property of wave interference. This process is rigorously described by the \textbf{Ewald-Oseen extinction theorem}, which establishes that the fundamental propagation speed of the field never changes.

The macroscopic slowing is an emergent phenomenon derived from the superposition of vacuum-speed waves:
\begin{enumerate}
    \item \textbf{Vacuum Propagation:} The incident electromagnetic wave travels fundamentally at the vacuum speed $c$, even inside the material. There is no such thing as "slow light" at the microscopic level.
    \item \textbf{Induced Response:} This incident field drives the bound charges (electrons/ions), creating a collective oscillation $\mathbf{P}(t)$.
    \item \textbf{Secondary Radiation:} These oscillating dipoles act as microscopic antennas, generating secondary spherical waves that also propagate strictly at $c$.
    \item \textbf{Macroscopic Superposition:} The total field is the linear superposition of the incident vacuum wave and the secondary scattered waves. The complex interference between them creates a new macroscopic wave pattern whose \textit{phase front} shifts in a way that mathematically mimics a slower velocity $v_{phase} = c/n$.
\end{enumerate}

From this continuous perspective, the refractive index $n$ is an \textbf{emergent kinematic parameter} describing an interference pattern, not a dynamic parameter scaling the momentum carrier itself.

Moreover, as stated in the previous section, it is a category error to judge a continuous theory of Stage 2 using the argument of a single photon (Stage 1). A photon is not predicted by any tensor formulation—neither Minkowski, Abraham, nor Lorentz. The concept of "a photon with momentum $p = \hbar k$" belongs strictly to \textbf{Stage 1}. Stage 2 (the domain of classical stress tensors) contains only continuous fields and momentum densities.

\subsection{Challenge 2: The Canonical Distinction (Specific Application)}
\label{subsec:CanonicalDistinction}

The most sophisticated version of the "Quantum Rescue" is the defense of the Minkowski momentum advanced by Barnett and Loudon \cite{BarnettLoudon2010}. They concede that the Abraham momentum corresponds to the mechanical (kinetic) momentum ($p_{kin} = mv$), but argue that the Minkowski momentum represents the \textit{canonical momentum} ($p_{can}$), the generator of spatial translations.

Based on the principle established above, this specific argument falls into the class of a \textbf{category error}:

The classical stress tensor is structurally defined by its exclusion of quantum operators. It is a model for calculating mechanical force $\mathbf{f}$, purely and intentionally defined to be a classical deterministic model. One should therefore avoid assigning properties of a non-deterministic theory of Stage 1 (QED) to a Stage 2 theory.

If a tensor fails to predict the correct mechanical damage (kinetic work) in the classical domain, labeling it "Canonical" does not restore its mechanical validity. A theory should be judged on the ground of what it is able to predict mechanically. It is argued here that one should not judge a deterministic theory of Stage 2 with quantum arguments from Stage 1.

\section{Revisiting Key Experiments: The Verification of the Survivor}
\label{sec:Exp_Revisit}

To conclude this analysis, the historical experiments often cited in favor of the Abraham or Minkowski tensors are revisited. In the existing literature, these experiments are typically presented as deciding factors between one of these two tensors.
On the contrary, the following analysis demonstrates that the experimental outcome is in every case perfectly predicted by the \textbf{microscopic ground truth} (the Lorentz force) and thus by the \textbf{vacuum stress tensor}.
It is proposed that the "Minkowski" or "Abraham" labels attached to these results in the literature are rather taxonomic aliases for specific mechanical limits of the Lorentz force, not signatures of new fundamental physics.

\subsection{Analysis of the Jones-Leslie Radiation Pressure Experiment}
\label{subsec:ExpJonesLeslie}

The high-precision radiation pressure experiments of Jones and Richards \cite{Jones1954} and later Jones and Leslie \cite{Jones1978} are foundational to this debate. In these experiments, the recoil force on a mirror was measured, first in air and then suspended in various dielectric liquids.

\textbf{The Observation:} The experimenters observed that the total force on the mirror increases in direct proportion to the refractive index $n$ of the surrounding liquid ($F \propto n P / c$).

\textbf{The Interpretation in Literature:} Historically, this $n$-scaling was interpreted as experimental evidence that the photon momentum in a dielectric is the Minkowski momentum $p = n \hbar k$.

\textbf{The Lorentz Force Viewpoint:}
As shown by Loudon \cite{Loudon2004}, the force on the mirror is only one part of the total interaction volume. The Vacuum Tensor formulation decomposes the interaction into two spatially distinct mechanical effects:
\begin{enumerate}
    \item \textbf{Force on the Mirror (The Measured Quantity):} The vacuum field drives surface currents $\mathbf{j}$ on the reflective coating. The Lorentz force $\mathbf{f} = \mathbf{j} \times \mathbf{B}$ on these currents scales with $n$. This is not because the "photon" is heavier, but because the magnetic field amplitude $B$ inside a dielectric is enhanced by the refractive index ($B = n E / c$) due to impedance matching.
    \item \textbf{Force on the Liquid (The Ignored Quantity):} Simultaneously, the field exerts a gradient force $\mathbf{f} \approx (\mathbf{P} \cdot \nabla)\mathbf{E}$ on the liquid dipoles immediately adjacent to the mirror face, pushing the liquid \textbf{away} from the surface.
\end{enumerate}

The experiment measured only the force on the mirror. When viewed through the lens of the Lorentz force, the result shows that the vacuum field, when acting on a conductor submerged in a dielectric, generates a surface pressure proportional to $n$. The Lorentz force accurately describes the outcome of the experiment.

\subsection{Analysis of the Ashkin-Dziedzic Liquid Surface Experiment}
\label{subsec:ExpAshkinDziedzic}

Another pivotal experiment often cited in this debate is that of Ashkin and Dziedzic \cite{Ashkin1973}, who observed the deformation of a free liquid surface under illumination from a focused laser beam.

\textbf{The Observation:} The liquid surface formed an \textbf{outward bulge} at the point of illumination.

\textbf{The Historical Interpretation:} This result was initially framed as a decisive binary test. The Minkowski formulation predicts an outward surface tension, while the Abraham formulation predicts an inward tension. Because the observed bulge was outward, it was widely claimed as experimental evidence for the validity of the Minkowski momentum.

\textbf{The Lorentz Force Viewpoint (Hydrodynamic Amplification):}
The resolution, derived by Gordon \cite{Gordon1973} and confirmed by subsequent analyses by Brevik \cite{Brevik1979}, reveals that the bulge is driven by transverse forces acting deep within the bulk liquid:

\begin{enumerate}
    \item \textbf{Radial Gradient Force:} The experiment utilized a focused Gaussian beam, not a plane wave. The strong radial intensity gradient ($\nabla |E|^2$) exerts a powerful inward \textbf{Lorentz force density} $\mathbf{f} \approx (\mathbf{P} \cdot \nabla)\mathbf{E}$ on the dipoles throughout the volume of the liquid.
    \item \textbf{Hydrodynamic Amplification:} This radial constriction creates a hydrodynamic pressure gradient analogous to extrusion. This squeezes the fluid radially towards the beam axis, dramatically increasing the internal hydrostatic pressure.
    \item \textbf{Surface Deformation:} The liquid relieves this internal volumetric pressure by bulging outwards against surface tension.
\end{enumerate}

Detailed calculations demonstrate that this radial hydrodynamic force is approximately \textbf{$10^3$ times stronger} than the direct longitudinal radiation pressure. The Lorentz force predicts the radial squeeze and the subsequent bulge perfectly.

\subsection{Analysis of the Gibson et al. Photon Drag Experiment}
\label{subsec:ExpGibson}

The photon drag experiment, notably performed in semiconductors by Gibson et al. (1980) \cite{Gibson1980}, provides a crucial test case involving free charge carriers.

\textbf{The Observation:} An infrared light pulse propagates through a semiconductor rod. The momentum from the light "drags" the free charge carriers (electrons/holes), generating a measurable voltage along the rod. The result was unambiguous: the momentum transferred to these charge carriers corresponds to the Minkowski value, $n \hbar k$ per photon.

\textbf{The Historical Interpretation:} As with previous examples, this result was interpreted as direct evidence for the fundamental correctness of the Minkowski tensor.

\textbf{The Lorentz Force Viewpoint:}
The experimental outcome is explained in detail by Loudon, Barnett, and Baxter \cite{Loudon2005}. It is found by applying the Lorentz force to the \textit{complete} physical system, which consists of two distinct components:

\begin{enumerate}
    \item \textbf{Force on Free Carriers (Measured):} The Lorentz force exerted on the mobile charge carriers pushes them down the rod. This charge accumulation creates the electric field measured by the voltmeter. Detailed calculations confirm that the integrated impulse on these carriers scales with $n$.
    \item \textbf{Force on the Host Lattice (Unmeasured):} Simultaneously, the electromagnetic field interacts with the bound charges of the stationary crystal lattice. The same Lorentz calculations show this results in a counter-momentum transfer to the lattice. Because the lattice is mechanically fixed and electrically neutral, this force generates no voltage.
\end{enumerate}

The analysis showed that the Lorentz force framework predicted the experimental outcome correctly.

\subsection{Analysis of the Walker et al. "Abraham Force" Experiment}
\label{subsec:ExpWalker}

The series of experiments by G. B. Walker et al. (1975) \cite{Walker1975} were designed specifically to isolate the elusive "Abraham Force" term [$\frac{\varepsilon_r - 1}{c^2} \frac{\partial}{\partial t}(\mathbf{E} \times \mathbf{H})$], a force density predicted by the Abraham tensor but absent in Minkowski's.

\textbf{The Observation:} The experiment measured a torque on a dielectric disk suspended in orthogonal, time-varying electric and magnetic fields. The magnitude of the torque was consistent with the prediction of the Abraham force term. This result was initially cited as definitive experimental evidence for the Abraham formulation.

\textbf{The Lorentz Force Viewpoint:}
The mechanism is a classic application of the Vacuum/Lorentz framework:
\begin{enumerate}
    \item \textbf{Induction:} The time-varying magnetic field ($\dot{\mathbf{B}}$) induces a rotational electric field ($\mathbf{E}_{\text{induced}}$) throughout the apparatus, according to Faraday's Law ($\nabla \times \mathbf{E} = -\dot{\mathbf{B}}$).
    \item \textbf{Interaction:} This induced field exerts a standard Lorentz force ($\mathbf{f} = \rho_{\text{free}} \mathbf{E}_{\text{induced}}$) on the free surface charges residing on the metal electrodes.
    \item \textbf{Measurement:} The apparatus measured the torque resulting from this electrode force, which, due to the specific geometry of the setup, mathematically mimicked the form of the "Abraham force."
\end{enumerate}

The Lorentz force describes the outcome of the experiment accurately.

\subsection{Analysis of the Campbell et al. Atomic Recoil Experiment}
\label{subsec:ExpCampbell}

Another experimental evidence cited in favor of the Minkowski momentum comes from the atom interferometry experiment of Campbell et al. (2005) \cite{Campbell2005}. Using a Bose-Einstein condensate (BEC) as a dispersive medium, the team measured the recoil momentum transferred to a single atom absorbing a photon.

\textbf{The Observation:} The experiment unambiguously demonstrated that the atom recoils with a momentum of $n\hbar k$, where $n$ is the refractive index of the condensate.

\textbf{The Interpretation:} Because the atom absorbed one photon and gained a momentum of $p = n\hbar k$, proponents argued that the photon "inside" the medium must inherently possess the Minkowski momentum.

\textbf{The Lorentz Force Prediction (The Classical Limit):}
The debate rests on a \textbf{category error}. The experiment measures an interaction at the single-atom level—the domain of \textbf{Stage 1 (Quantum Reality)}. However, as explicitly noted by Campbell et al. in their analysis, the \textbf{standard Lorentz force} (Stage 2) correctly predicts this recoil momentum ($p = n \hbar k$) in the classical limit.

Citing Haugan and Kowalski \cite{Haugan1982}, they explain the mechanism: because the electric field is shielded in the medium ($E \propto 1/n$), the atomic transition (Rabi cycle) takes longer to complete by a factor of $n$. Consequently, the Lorentz force acts for a longer duration, imparting a total mechanical impulse exactly equal to $n \hbar k$.

\section{The Limits of Validity: The Standard Model Boundary}
\label{sec:DefensiveBoundary}

It is essential to clear away a potential misunderstanding. The "Microscopic Baseline" defended in this book—the Lorentz Force coupled to the Vacuum Stress Tensor—is \textbf{not a novel theory}. It is the \textbf{Standard Model} of classical electrodynamics found in every foundational textbook, from Jackson \cite{Jackson1999} to Landau \cite{LandauLifshitzVol8}.

No standard reference is known that challenges the validity of the vacuum Energy-Momentum Tensor at the microscopic level. Thus, no new physics is presented; the Standard Model is simply defended against the "arbitrary split." The contribution is the rigorous demonstration that this Standard Model describes a physical reality that is \textbf{mutually exclusive} to the conflated tensors.

Furthermore, there is no attempt to replace Quantum Electrodynamics (QED). The defense of the Lorentz force is a defense of the \textbf{classical limit}. As noted by Jackson, the classical description of the electron is an effective model valid down to the Compton wavelength. Denying the validity of the vacuum tensor at this scale is inconsistent with the principles of the foundational texts of the field. It is proposed that where the classical framework is valid, it must be \textbf{internally consistent}. The conflated tensors fail this consistency test.

\section{Conclusion: Microscopic Electrodynamics}
\label{sec:Exp_Conclusion}

The microscopic analysis of Part I is now complete. It has been demonstrated that the microscopic Lorentz force is capable of explaining the outcomes of the key experimental observations often cited in the "Abraham-Minkowski" debate.

The contribution of this first part of the book is the rigorous demonstration that the Standard Lorentz Force Model describes a physical reality that is \textbf{mutually exclusive} to other tensors like Abraham and Minkowski within the microscopic domain. It has been shown that the Abraham and Minkowski tensors are not physically valid descriptions of the microscopic topology of matter, as they require the existence of vacuum inertia and hidden energy reservoirs.

\subsection{The Path Forward}

In this first part of the book, the analysis has exclusively considered the microscopic descriptions of reality in the classical limit (Stage 2).

The remainder of this book (Parts II–VII) is dedicated to deriving a consistent macroscopic description of matter (Stage 3). A single cup of coffee consists of $10^{23}$ chaotic particles. To explain the behavior of systems of many particles, the microscopic description, while topologically accurate, is mathematically intractable. To describe the macroscopic world, one needs to find a new language—a macroscopic language—to describe the system's behavior.

The transition from microscopic to macroscopic is an act of averaging, an act of information compression, and an act of high-frequency filtering. To derive a mechanically consistent macroscopic description of matter—one that preserves the topological truth of the vacuum while enabling the computational power of the continuum—is the primary objective of the following chapters.

This concludes Part I. The construction of the \textbf{First Pillar} begins in Part II.

\part{Part II: The Macroscopic Filter}
\label{part:MacroscopicFilter}

\chapter{Introduction: The Bridge of Scale}
\label{chap:Part2_Intro}

\section{Introduction: The Filtering Perspective}

Part I of this manuscript established the unambiguous "microscopic baseline" of classical physics: a microscopic reality governed by the deterministic interplay of discrete charged particles and vacuum fields. It was demonstrated that at this fundamental level, the Lorentz force is the sole gateway for interaction, and the system is mechanically consistent by definition.

To describe the world at the human scale, however, a fundamental barrier arises. It is mathematically impossible to predict the deformation of a solid, the turbulence of a fluid, or the temperature of a gas by tracking the individual microscopic trajectories of $10^{23}$ particles.

The transition from the high-frequency complexity of the microscopic world (the ``high-bandwidth signal'') to the predictable laws of macroscopic physics is commonly achieved through coarse-graining—a convolution with a smoothing kernel. Statistical mechanics provides the rigorous mathematical framework for this transition via the concept of the \textbf{ensemble}~\cite{kardar_statistical_2007, Jackson1999}.

We propose a complementary conceptual view on the same mathematical process. This transition is framed here using the intuitive and physically descriptive language of signal processing.

The core methodological insight of this framework is that \textbf{any averaging operation---whether spatial, temporal, or ensemble-based---is mathematically equivalent to a low-pass filter}. The microscopic state is modeled as a \textbf{signal} containing information across all spatial and temporal frequencies. The smooth, continuous macroscopic world emerges naturally as the low-frequency component of that signal. This perspective clarifies why distinct physical laws emerge: they are descriptions of different bands of the system's total frequency spectrum, handled in fundamentally different ways by the filtering process.

\paragraph{The Physical Nature of the Filter.}
The core physical insight is that this filtering process is not merely a mathematical tool; it is a physical reality inherent to the definition of macroscopic properties. The measurement process itself defines the cutoff frequency of the filter. When temperature or elasticity is measured, the instrument itself physically filters out the high-frequency responses of the underlying particles. The act of measurement effectively defines the averaging volume.

\paragraph{The Bandwidth-Limited Observer.}
Fundamentally, this filtering is inherent to the act of macroscopic observation. The languages of thermodynamics, continuum mechanics, and electrodynamics are emergent descriptions conditioned by the resolution of the observer. Our senses—and our instruments—act as low-pass filters: a thermometer does not perceive individual particles colliding with its surface, but rather the collective intensity of the thermal bath; a microphone does not perceive air molecules striking a membrane, but rather the collective wave of sound. Classical physics, in this view, is the emergent language of a bandwidth-limited observer.

All of classical physics relies on variables that have no meaning at the single-particle level. Concepts such as temperature, pressure, viscosity, and resistance are not fundamental properties of the microscopic constituents. They are \textbf{emergent manifestations} of the collective, becoming visible only when the high-frequency chaos of the microscopic signal is integrated out.

This part of the book presents macroscopic physics not as a separate set of fundamental laws, but as an \textbf{emergent physics derived via information compression.} We demonstrate that the high-frequency fluctuations—which are filtered out of the macroscopic kinematic description—re-enter the theory as emergent variables (such as temperature and polarization), describing a \textit{filtered reality}.

The aim of this part is not to re-derive the laws of classical physics, but to derive a conceptual understanding of the \textit{transformation itself}. The following analysis traces how the rigorous, two-domain architecture of the microscopic world (Fields and Particles) transforms into the complex, multi-domain architecture of the macroscopic laboratory.

\section{The Microscopic Signal: The "Ground Truth"}
\label{sec:micro_foundation}

The analysis begins by revisiting the starting point of this derivation, established in Part I: the underlying microscopic reality from which all macroscopic phenomena must emerge. This microscopic system represents the deterministic "microscopic baseline"---defined herein as the complete, \textbf{high-bandwidth signal} of reality.

\subsection{Conceptual Boundaries and Assumptions}
\label{sec:conceptual_boundaries}

The analysis is confined to a closed, classical system governed solely by the Newton-Lorentz laws of motion and the microscopic Maxwell equations. This idealized model relies on two foundational assumptions to set its boundaries and maintain internal consistency:

\begin{enumerate}
    \item \textbf{Dynamic Exclusions (The Filtering of Gravity):} Nuclear and gravitational forces are excluded. Nuclear forces operate at scales far below the atomic interactions of interest. Gravity, conversely, presents a different case: it is typically treated as an inherently smooth, low-frequency field at terrestrial scales. Consequently, gravity passes through the conceptual ``filter'' essentially unchanged. It therefore offers no insight into the \emph{emergence} of complex macroscopic domains like thermodynamics or continuum mechanics. The focus is exclusively on electromagnetic interactions, which are the source of the high-frequency complexity (fluctuations) that the filter must transform.
    
    \item \textbf{The "Static Stage" (Quantum Stability):} A purely classical model of charged particles is famously unstable (e.g., electrons spiraling into nuclei). To resolve this, a standard semi-classical approach is adopted. The quantum-mechanical stability of matter is treated as a \textbf{pre-existing boundary condition}. The laws that give matter its structure (e.g., the Pauli exclusion principle) act as the static ``stage'' upon which the classical dynamics play out. Crucially, these quantum constraints are treated as non-dissipative mechanical supports. The energy and momentum exchanged \emph{by} these stable entities via classical fields are analyzed, not the internal quantum dynamics that grant them stability.
\end{enumerate}

\subsection{The Anatomy of the Signal}
\label{sec:micro_anatomy}

Within these boundaries, the microscopic state is defined physically by the distinct massive charge carriers---the electrons and ions ($s$). As established in Chapter~\ref{chap:FreeCharges}, the kinematic state of this system is rigorously described by the singular density fields:
\begin{itemize}
    \item \textbf{Number Density:} $n_s(\mathbf{x}, t) = \sum \delta(\mathbf{x} - \mathbf{x}_i)$
    \item \textbf{Particle Flux:} $\boldsymbol{\Gamma}_s(\mathbf{x}, t) = \sum \mathbf{v}_i \delta(\mathbf{x} - \mathbf{x}_i)$
\end{itemize}

These singular sources drive the unaveraged vacuum fields ($\mathbf{e}, \mathbf{b}$) via the microscopic Maxwell equations (Eqs.~\ref{eq:MicroMaxwell}). The evolution of this signal is driven by the \textbf{microscopic Newton-Lorentz} system. The field exerts force exclusively on the massive carriers via the Lorentz law, and the carriers respond with exact Newtonian inertia.

This mathematical description presents the central conceptual challenge of the micro-to-macro transition. It is a "forest of highly localized densities" fluctuating rapidly in both space and time. From a signal processing perspective, this state represents a signal with immense bandwidth.

While this microscopic description is highly accurate on a fundamental scale, it is conceptually discontinuous with macroscopic observation. Emergent concepts such as \textbf{temperature, pressure, and elasticity} are entirely absent at this level. These are not fundamental properties, but \emph{manifestations} of the collective, becoming visible only when the high-frequency chaos of the signal is filtered out.

\section{The Filtering Formalism and Conditions}
\label{sec:FilteringFormalism}

The objective is to decompose any microscopic field into a smooth macroscopic component and a rapidly varying fluctuation. The fundamental decomposition of a generic microscopic field $\phi(\mathbf{x}, t)$ is a generalized Reynolds decomposition:
\begin{equation}
    \phi(\mathbf{x}, t) = \underbrace{\Phi(\mathbf{x}, t)}_{\substack{\text{Macroscopic} \\ \text{(Signal)}}} + \underbrace{\delta\phi(\mathbf{x}, t)}_{\substack{\text{Fluctuation} \\ \text{(High-Frequency)}}}.
    \label{eq:DecompositionGeneral}
\end{equation}
The macroscopic field $\Phi(\mathbf{x}, t)$ is defined as the local spatio-temporal average of $\phi(\mathbf{x}, t)$, denoted by the operator $\langle \cdot \rangle$. Mathematically, this is expressed as a convolution with a smoothing kernel, $h(\mathbf{x}, t)$:
\begin{equation}
    \Phi(\mathbf{x}, t) \equiv \langle \phi(\mathbf{x}, t) \rangle \equiv (\phi * h)(\mathbf{x}, t).
    \label{eq:ConvolutionDefinition}
\end{equation}
The kernel $h(\mathbf{x}, t)$ defines the spatial volume and temporal duration over which the average is taken. In the frequency domain, this operation acts as a \textbf{low-pass filter}, isolating the large-scale, slowly varying features. The fluctuation $\delta\phi(\mathbf{x}, t)$ contains only the filtered-out high-frequency information, such that, by definition, the average of the fluctuation vanishes: $\langle \delta\phi(\mathbf{x}, t) \rangle = 0$.

As stated previously, it is crucial to recognize that the kernel $h(\mathbf{x}, t)$ is not an arbitrary mathematical construct. It represents the \textbf{point spread function} (spatial resolution) and the \textbf{response time} (temporal resolution) defined by the measuring instrument. The measurement process of any macroscopic material property implicitly defines this kernel.

\subsection{The Condition of Scale Separation: The Spectral Gap}
\label{sec:ScaleSeparation}

The mechanical validity of the macroscopic map rests on a single physical guarantor: the Condition of Scale Separation. For the filtering process to yield a meaningful theory, there must exist a clear \textbf{spectral gap} or \textbf{valley of silence}.

The microscopic reality is characterized by inherent scales of fluctuation, most notably the interatomic spacing $a$ and the collision time $\tau_{\text{micro}}$. The macroscopic phenomena occupying the distinct, low-frequency band (characteristic length $L_{\text{macro}} \gg a$).

Consider the energy spectrum of a crystalline solid in spatial frequency (wavenumber $k$) space:
\begin{itemize}
    \item \textbf{The macroscopic peak ($k_{\text{macro}} \approx 2\pi/L_{\text{macro}} \approx 0$):} Energy is concentrated at long wavelengths corresponding to the macroscopic scale (the shape of the object, the smooth external field).
    \item \textbf{The microscopic peak ($k_{\text{micro}} \approx 2\pi/a$):} Energy spikes again at the reciprocal lattice vector, corresponding to the rapid variation of fields between nuclei and electrons.
    \item \textbf{The valley of silence:} Between these two peaks lies a vast, empty region where the spectrum is effectively zero.
\end{itemize}

This \textbf{valley of silence} is the guarantor of macroscopic physics. A valid macroscopic theory effectively truncates the spatial Fourier transform within this gap. The stability of the theory relies on the filter cutoff ($k_c$) falling safely within this region:
\begin{equation}
    \frac{1}{L_{\text{macro}}} \ll k_c \ll \frac{1}{a}.
    \label{eq:ScaleSeparation}
\end{equation}

To appreciate the physical robustness of this condition, one must quantify the sheer magnitude of the separation. The distinction is not merely mathematical; it is enforced by the very nature of experimental observation and material definition.

Consider the spatial contrast between the microscopic reality and the macroscopic assignment:
\begin{itemize}
    \item \textbf{The Atomic Scale ($a \approx 10^{-10}$ m):} The microscopic fields fluctuate wildly over the angstrom scale of the lattice. This is the domain of the fundamental discrete sources.
    \item \textbf{The Engineering Scale ($L_{\text{assign}} \approx 10^{-3} \dots 10^{-2}$ m):} When an experimentalist assigns a property like polarization $P(\mathbf{x})$ to a dielectric block, it is invariably an integral measure. Whether defined by the size of the sample (e.g., $1 \text{ cm}^3$) or the manufacturing tolerance of a graded interface, the "resolution" of the property assignment is orders of magnitude larger than the atomic spacing.
\end{itemize}

The ratio between these scales is typically on the order of $10^8$. To the macroscopic observer, the atomic fluctuations are not merely "noise"; they occupy a spatial frequency band a hundred million times higher than the resolution of the physical model.

The experimental process—often involving the measurement of external far-field responses—serves to physically integrate out the atomic scale. Consequently, material properties are not intrinsic constants of the vacuum; they are emergent effective parameters conditioned by the scale of observation. When "macroscopic electrodynamics" is discussed, it implicitly defines a theory valid only for wavelengths larger than the structural heterogeneity of the matter.

The existence of the spectral gap ensures that this hypothesis is safe. Because the "valley of silence" is so vast, the precise shape of the filter (or the exact integration volume) becomes irrelevant. The macroscopic signal ($k_{\text{macro}} \approx 0$) and the microscopic noise ($k_{\text{micro}} \sim 10^{10} \text{ m}^{-1}$) are mathematically orthogonal using any reasonable definition of the macroscopic scale.

This spectral isolation is the only reason "Material Properties" can be treated as objective data. It provides the guarantee of \textbf{device independence}. If this gap were not present, the very definition of a macroscopic material would be impossible. The "measured" properties would depend inextricably on the resolution of the probe; "permittivity" would not be an intrinsic attribute of the object but a transient artifact of the instrument. The vast spectral gap ensures that all valid macroscopic instruments—whether they resolve 1 mm or 0.1 mm—integrate over the same stable "silence," yielding a consistent, objective assignment of material state.

Only when this gap closes—for instance, in metamaterials or nanostructures where $a \sim \lambda$—does the distinction between "signal" and "fluctuation" collapse, causing the standard macroscopic description to fail.

\subsubsection{Cascading Scales: The Example of Ferromagnetism}
The relativity of the macroscopic limit is best illustrated by ferromagnetism, which presents not one, but two distinct spectral gaps.
\begin{enumerate}
    \item \textbf{Gap 1 (Micro $\to$ Meso):} The fundamental atomic spins fluctuate wildly. Averaging over the exchange correlation length suppresses this microscopic chaos, yielding the smooth, saturated magnetization of a single \textbf{magnetic domain}. To a nanoscopic probe, this domain is the "macroscopic" signal.
    \item \textbf{Gap 2 (Meso $\to$ Macro):} Across a bulk material, these domains themselves vary in direction (fluctuate). A standard engineering probe averages over thousands of domains, interpreting their collective orientation as the bulk magnetization vector $\mathbf{M}$.
\end{enumerate}
This cascading structure demonstrates that "macroscopic" is a relative term defined by the specific spectral gap in which the observer (the filter) resides.

\section[The Kinematic Decomposition]{The Kinematic Decomposition: \\ Unordered, Ordered, and Relative Motion}
\label{sec:KinematicDecomposition}

We propose a conceptual decomposition of the microscopic motion into three distinct kinematic modes: \textbf{Unordered}, \textbf{Ordered}, and \textbf{Relative} motion. We posit that these modes correspond directly to the three pillars of classical physics:
\begin{itemize}
    \item \textbf{Unordered Motion:} The domain of \textbf{Thermodynamics}.
    \item \textbf{Ordered Motion:} The domain of \textbf{Continuum Mechanics}.
    \item \textbf{Relative Motion:} The domain of \textbf{Macroscopic Electrodynamics}.
\end{itemize}

This framework is elaborated below.

\subsection{Defining the Spectral Modes}

\subsubsection{1. The Macroscopic Flow (Ordered Motion)}
First, the smooth, macroscopic velocity field, $\mathbf{V}_s(\mathbf{x}, t)$, is defined. This represents the ordered collective motion of species $s$. It is defined as the ratio of the macroscopic particle flux density $\langle \boldsymbol{\Gamma}_s \rangle$ to the macroscopic number density $N_s = \langle n_s \rangle$:
\begin{equation}
    \mathbf{V}_s(\mathbf{x}, t) \equiv \frac{\langle \boldsymbol{\Gamma}_s(\mathbf{x}, t) \rangle}{\langle n_s(\mathbf{x}, t) \rangle} = \frac{\langle \boldsymbol{\Gamma}_s(\mathbf{x}, t) \rangle}{N_s(\mathbf{x}, t)}.
    \label{eq:MacroVelocityDefinition}
\end{equation}
This defines the averaged center-of-mass velocity of species $s$ within the averaging volume.

\subsubsection{2. The Fundamental Decomposition (Unordered Motion)}
With the macroscopic flow established, the total velocity $\mathbf{v}_{s,i}$ of each individual particle $i$ is decomposed:
\begin{equation}
    \mathbf{v}_{s,i}(t) = \mathbf{V}_s(\mathbf{x}_{s,i}(t), t) + \mathbf{v}_{s,i, \text{unord}}(t).
\end{equation}
The \textbf{unordered velocity} $\mathbf{v}_{s,i, \text{unord}}$ represents the particle's random thermal motion relative to the mean flow. By construction, its average is zero. This component forms the basis of thermodynamics.

\subsubsection{3. From Species to System (Relative Motion)}
In a system composed of multiple species (positive ions and negative electrons), the ordered velocities of each species may differ:
\begin{itemize}
    \item \textbf{Ordered Flow of Ions ($\mathbf{V}_+$):} Represents the bulk motion of the material lattice.
    \item \textbf{Relative Flow of Electrons ($\mathbf{V}_-$ vs. $\mathbf{V}_+$):} The differential ordered motion of the electron fluid relative to the ion lattice constitutes the macroscopic \textbf{electric current}, the source of macroscopic electrodynamics.
\end{itemize}

\section{The Architecture of Emergence: The Hypothesis}
\label{sec:ArchitectureEmergence}

We propose that the architecture of macroscopic physics can be organized into a $2 \times 3 \times 2$ Matrix:
\begin{enumerate}
    \item The microscopic reality is divided into two source domains: \textbf{Mechanical} (Mass) and \textbf{Electromagnetic} (Charge).
    \item The dynamics are subdivided into the three kinematic modes: \textbf{Unordered}, \textbf{Ordered}, and \textbf{Relative}.
    \item Each cell is further decomposed into a \textbf{Signal} (Low-Frequency) and a \textbf{Compressed Fluctuation} (High-Frequency).
\end{enumerate}

We hypothesize that the high-frequency components, which are filtered out of the explicit kinematic description, re-emerge as new macroscopic low-frequency material properties—specifically \textbf{temperature}, \textbf{elasticity}, and \textbf{binding energy}.

\subsection{The Unified Matrix}

The following \textbf{unified matrix} (Table~\ref{tab:UnifiedMatrix_Part2}) serves as the comprehensive map for this investigation.

\begin{table}[h]
\centering
\footnotesize
\caption{\textbf{The Unified Matrix of Classical Physics.} Macroscopic physics acts as information compression: retaining the low-frequency signal while compressing the high-frequency signal.}
\label{tab:UnifiedMatrix_Part2}
\begin{tabular}{|c|c||c|c|c|}
\hline
\multicolumn{2}{|c||}{} & \textbf{\shortstack{Thermo-\\dynamics}} & \textbf{\shortstack{Continuum\\Mechanics}} & \textbf{\shortstack{Electro-\\dynamics}} \\
\multicolumn{2}{|c||}{} & \textbf{1. Unordered} & \textbf{2. Ordered} & \textbf{3. Relative} \\
\multicolumn{2}{|c||}{} & (Fluctuation) & (Common Motion) & (Diff. Motion) \\
\hline \hline
\multirow{4}{*}{\textbf{\shortstack{Low\\Frequency\\Signal}}} & \textbf{Mech.} & \textbf{---} & \textbf{\shortstack{Bulk\\Momentum}} & \textbf{\shortstack{Carrier\\Momentum}} \\
 & & & $\mathbf{p} = m \mathbf{v}_{\text{bulk}}$ & $\mathbf{p}_e = m_e \mathbf{v}_e$ \\
\cline{2-5}
 & \textbf{EM} & \textbf{---} & \textbf{---} & \textbf{\shortstack{Electro-\\dynamics}} \\
 & & & & $\mathbf{J}, \rho$ \\
\hline \hline
\multirow{4}{*}{\textbf{\shortstack{Compressed\\High\\Frequency}}} & $\langle \textbf{Mech.} \rangle$ & \textbf{Temperature} & \textbf{---} & \textbf{---} \\
 & & $T$ & & \\
\cline{2-5}
 & $\langle \textbf{EM} \rangle$ & \textbf{\shortstack{Thermal\\Radiation}} & \textbf{\shortstack{Elastic\\Energy}} & \textbf{\shortstack{Binding\\Energy}} \\
 & & & & \\
\hline
\end{tabular}
\end{table}

The goal of the rest of the book is to explicate how the terms in this matrix emerge from the underlying microscopic reality. We will explore the symmetries between these domains and, most importantly, identify the interactions that link them. For instance, we will identify \textbf{mechanical work} as the interaction transferring energy to the continuum domain, and \textbf{dissipation} as the interaction transferring energy to the thermal domain.

\subsection{Plan of Analysis for Part II}
With this architectural blueprint established, the analysis proceeds as follows:

\begin{itemize}
    \item \textbf{Chapter \ref{chap:ThermoAndContinuum}: The Emergence of Continuum Mechanics and Thermodynamics.} The macroscopic laws for these mechanical domains are derived conceptually from the kinematic decomposition.
    \item \textbf{Chapter \ref{chap:HostInterface}: Macroscopic Electrodynamics.} The relative motion is established as the origin of macroscopic electromagnetism. The macroscopic Maxwell equations and force/energy balances are derived via filtering. The analysis demonstrates that filtering high-frequency fields necessitates the emergence of a new coupling term towards the continuum and thermodynamic domains: the \textbf{Host Interface}.
\end{itemize}

\chapter{Thermodynamics and Continuum Mechanics}
\label{chap:ThermoAndContinuum}

The purpose of this chapter is to establish a conceptual framework rather than to perform a rigorous derivation of established physical laws. We do not claim to re-derive Thermodynamics or Continuum Mechanics from first principles; these disciplines stand on their own foundation. Instead, this analysis offers a \textit{perspective of synthesis}.

We aim to demonstrate how these distinct macroscopic domains emerge naturally from the microscopic reality when viewed through the \textit{filtering lens} defined in the previous chapter. By explicitly treating macroscopic physics as an exercise in signal processing, we clarify the architectural relationships between the domains, revealing them not as separate worlds, but as different partitions of the same underlying information.

\section{Domain 1: Unordered Motion and the Emergence of Thermodynamics}
\label{sec:ThermoDomain}

The analysis examines the first emergent domain, corresponding to the \textbf{Unordered Column} of the unified matrix (Column 1). This domain accounts for the energy contained within the high-frequency components of the microscopic reality—the chaotic motion of particles ($\mathbf{v}_{s,i, \text{unord}}$) and the rapid fluctuations in their positions and fields.

As established, the average momentum of the unordered motion vanishes. Consequently, this domain carries no net macroscopic momentum and generates no net macroscopic current. Yet, this domain is crucial. While its average is zero, its \textit{mean-squared amplitude} is not. It constitutes a vast reservoir of energy. This is the domain of microscopic chaos, and the language used to describe its collective energy balance is \textbf{Thermodynamics}.

\subsection{Temperature and Conduction: The Mechanical Shadow}
\label{sec:temperature}

The primary variable of this domain emerges from the need to quantify the energy contained within the \textbf{Mechanical Unordered} motion (Mechanical Row, Unordered Column).

While the average velocity of this domain is zero ($\langle \mathbf{v}_{\text{unord}} \rangle = 0$), the average \textit{squared} velocity is not. This chaotic motion constitutes a massive reservoir of \textbf{internal kinetic energy} ($U_K$):
\begin{equation}
    U_K(\mathbf{x}, t) = \left\langle \sum_{s,i} \frac{1}{2} m_s |\mathbf{v}_{s,i, \text{unord}}(t)|^2 \delta(\mathbf{x} - \mathbf{x}_{s,i}(t)) \right\rangle.
\end{equation}
 This quantity represents the energy density of the "thermal fluctuation" of the particles.

\textbf{Temperature} ($T$) is the emergent macroscopic field that acts as the direct measure of this kinetic energy. For systems in local thermodynamic equilibrium, the equipartition theorem provides the rigorous link between the statistical microscopic motion and the macroscopic variable:
\begin{equation}
    U_{K}(\mathbf{x}, t) = \frac{d_f}{2} N(\mathbf{x}, t) k_B T(\mathbf{x}, t),
    \label{eq:TemperatureDefinition}
\end{equation}
where $N$ is the total macroscopic number density, $k_B$ is the Boltzmann constant, and $d_f$ is the relevant number of degrees of freedom.

From a signal processing perspective, temperature is an averaged measure of the \textbf{mean-squared amplitude} (the variance or "power") of the microscopic velocity noise. It quantifies the intensity—the "volume"—of the chaotic motion filtered out of the macroscopic velocity field. The chaotic high-frequency mechanical motion is compressed into a low-frequency scalar field $T(\mathbf{x}, t)$; it is an act of information compression.

The transport of this mechanical energy through the material is known as \textbf{thermal conduction}. It is the process where high-frequency kinetic energy diffuses through the system via random collisions and short-range interactions. Macroscopically, this diffusive process is typically modeled by Fourier's Law ($\mathbf{Q}_{\text{cond}} = -\kappa \nabla T$), where the thermal conductivity $\kappa$ describes the efficiency of this mechanical energy transfer.

\subsection[Thermal Radiation]{Thermal Radiation: \\ The Electromagnetic Shadow}
\label{sec:thermal_radiation}

The second component of the unordered domain corresponds to the \textbf{Electromagnetic Unordered} motion (Electromagnetic Row, Unordered Column).

Microscopically, the chaotic acceleration of charged particles ($\mathbf{a}_{\text{unord}}$) inevitably sources high-frequency electromagnetic fields. While the macroscopic average of these fields may vary locally ($\mathbf{E} = \langle \mathbf{e} \rangle \approx 0$), their mean-squared amplitude is non-zero. This constitutes a reservoir of \textbf{high-frequency electromagnetic energy}:
\begin{equation}
    U_{EM, \text{unord}} = \frac{\varepsilon_0}{2} \langle |\delta \mathbf{e}|^2 \rangle + \frac{1}{2\mu_0} \langle |\delta \mathbf{b}|^2 \rangle.
\end{equation}

This domain allows energy to be broadcast across space, even in vacuum. The transport mechanism is \textbf{thermal radiation}. Macroscopically, individual wavelets are not perceived. Instruments, acting as low-pass filters, measure the time-averaged energy flow as a smooth radiative heat flux, $\mathbf{Q}_{\text{rad}}$ (e.g., the Stefan-Boltzmann law). This again is an act of information compression: from high-frequency fields to low-frequency thermal radiation.

\subsection{Entropy: The Measure of Filtered Information}
\label{sec:entropy}

The specific energy forms of the unordered domain have now been defined: Mechanical ($U_K$, measured by Temperature) and Electromagnetic ($U_{EM}$, measured by Field Variance). The unification of these two disparate physical forms lies in the nature of the discarded information.

\textbf{Entropy} ($S$) is not a variable \textit{within} the material matrix; it is the statistical variable that characterizes the \textbf{Unordered Column} itself.

While temperature and field variance quantify the \textit{amplitude} (the "volume") of the high-frequency signal, entropy quantifies its \textbf{bandwidth} (the "complexity"). It serves as a measure of the \textbf{information loss} inherent in the macroscopic filtering process.

In the language of signal processing, the complexity of the microscopic signal is reflected in its \textbf{frequency spectrum}. A simple, ordered signal (like a coherent wave) has a narrow bandwidth—it is predictable and incompressible (low entropy). A chaotic, unordered signal (like thermal noise) has an infinite bandwidth—it is unpredictable and incompressible (high entropy).

Entropy, therefore, applies equally to the mechanical "jiggle" of particles and the electromagnetic "jiggle" of fields. It is the universal metric of the \textbf{spectral complexity} of the high-frequency microscopic system state.

\subsubsection{The Second Law: Spectral Spreading}
This perspective offers an intuitive explanation for the \textbf{Second Law of Thermodynamics}. The Second Law may be interpreted as the statistical tendency of energy to migrate from narrow-band modes to broad-band modes.

Consider the "landscape of spectral possibility." The region of phase space corresponding to "white noise" (broadband chaos) is vastly larger than the region corresponding to "pure tones" (ordered motion). When a system evolves, it performs a random walk through this landscape. It is overwhelmingly likely to wander into the vast regions of high spectral complexity simply because there are more ways for energy to be distributed across a dense, broadband spectrum than across a sparse one.

This is \textbf{spectral spreading}. Ordered energy (work) is locally concentrated in the frequency domain. Unordered energy (heat) is spread across the entire spectrum. The degradation of work into heat is simply the natural diffusion of the system's state into the high-bandwidth "valley of silence" that the macroscopic filter cannot resolve.

\subsection{Summary: The Emergence of Thermodynamics}

Thermodynamics is conceptually identified as the emergent language for the \textbf{unordered shadow} of the material matrix. It replaces the impossibly complex microscopic dynamics with a set of powerful macroscopic low-frequency variables:
\begin{itemize}
    \item \textbf{The Mechanical Shadow:} \textbf{Temperature} ($T$) measures the intensity of the chaotic particle motion ($U_K$).
    \item \textbf{The Electromagnetic Shadow:} \textbf{Thermal Radiation} measures the intensity of the chaotic field motion ($U_{EM}$).
    \item \textbf{The Statistical Governor:} \textbf{Entropy} ($S$) measures the complexity (bandwidth) of the entire unordered domain.
\end{itemize}

\section{Domain 2: Ordered, Neutral Motion and the Emergence of Continuum Mechanics}
\label{sec:OrderedDomain}

The focus now shifts to the world of collective, directed behavior, corresponding to the \textbf{Ordered Column} of the unified matrix. This domain describes the continuous flow of matter. We propose that the defining feature of this domain is the emergence of \textbf{Continuum Mechanics}, the effective theory governing the \textbf{momentum balance} of the ordered motion.

This section focuses specifically on the physics of macroscopically \textbf{neutral matter}, where the positive and negative charges move in lockstep ($\mathbf{V}_{+} \approx \mathbf{V}_{-} \equiv \mathbf{V}_{\text{matter}}$). This assumption implies that macroscopic charge and current densities vanish ($\rho \approx 0, \mathbf{J} \approx \mathbf{0}$), eliminating large-scale electromagnetic forces locally. This is the realm of fluid dynamics, elasticity, and acoustics.

\subsection{The Continuum: Mass and Momentum Density}

The fundamental variables of this domain emerge by applying the low-pass filter to the microscopic state. This is an example of \textbf{conceptual preservation}; the concepts remain the same, merely smoothed in scale.

The \textbf{macroscopic mass density}, $\rho_m(\mathbf{x}, t)$, is the filtered representation of the microscopic mass distribution. Formally, this is a convolution of the microscopic density with a smoothing kernel $w(\mathbf{x})$:
\begin{equation}
    \rho_m(\mathbf{x}, t) = \int \rho_{m, \text{micro}}(\mathbf{x}', t) w(\mathbf{x} - \mathbf{x}') \, d^3x'.
\end{equation}
This transforms the "forest of delta functions" of the atomic positions into a smooth, continuous field.

The \textbf{macroscopic mechanical momentum density}, $\mathbf{G}_{\text{mech}}(\mathbf{x}, t)$, represents the momentum of the ordered flow:
\begin{equation}
    \mathbf{G}_{\text{mech}}(\mathbf{x}, t) = \rho_m(\mathbf{x}, t) \mathbf{V}_{\text{matter}}(\mathbf{x}, t).
\end{equation}
This definition is a crucial act of partitioning. $\mathbf{G}_{\text{mech}}$ exclusively represents the momentum of the \textbf{ordered domain}. The momentum of the unordered domain is filtered out, but its \emph{flux} reappears as a macroscopic force.

These fields obey the fundamental law of \textbf{mass conservation} (the macroscopic continuity equation), derived by filtering the microscopic conservation law (Eq.~\ref{eq:micro_particle_continuity}):
\begin{equation}
    \frac{\partial \rho_m}{\partial t} + \nabla \cdot \mathbf{G}_{\text{mech}} = 0.
    \label{eq:MacroContinuity}
\end{equation}

\subsection{The Governing Law: Filtering the Momentum Balance}

The dynamics of the continuum are governed by the momentum balance. This macroscopic law emerges by applying the low-pass filter to the microscopic law of motion (Newton's second law, Eq.~\ref{eq:Micro_Newton_Total}).

This filtering process is mathematically complex due to the non-linearity of the momentum flux and the nature of the interparticle forces. The filtering operation acts on the microscopic convective flux term (the kinetic stress), separating it into ordered and unordered components. The conceptual result is presented here: the filtering operation transforms the microscopic momentum balance into the \textbf{Cauchy Momentum Equation}:
\begin{equation}
    \frac{\partial \mathbf{G}_{\text{mech}}}{\partial t} + \nabla \cdot (\rho_m \mathbf{V}_{\text{matter}}\mathbf{V}_{\text{matter}}) = \nabla \cdot \mathbf{T} + \mathbf{f}_{\text{ext}}.
    \label{eq:CauchyMomentum}
\end{equation}
The terms on the left represent the rate of change of ordered momentum and the flux of ordered momentum (macroscopic convection). The terms on the right represent the forces acting on the continuum. $\mathbf{f}_{\text{ext}}$ represents external body forces (such as gravity, which passes through the filter unchanged, or macroscopic EM forces, discussed in Sec.~\ref{sec:bridge_em_mech}).

The crucial emergent term is $\mathbf{T}$, the \textbf{Macroscopic Stress Tensor}.

\subsection{The Emergence of the Stress Tensor: Three Energy Gateways}

The stress tensor is the central concept of continuum mechanics. It is a macroscopic field describing the short-range internal forces within the material. Its emergence is a profound example of \textbf{conceptual transformation} (information compression).

We separate the total stress tensor conceptually into three components, related to the \textbf{Three Interaction Mechanisms} which we define in the following:
\begin{equation}
    \mathbf{T} = \underbrace{\mathbf{T}_{\text{storage}}}_{\text{Mechanism 3}} + \underbrace{\mathbf{T}_{\text{work}} + \mathbf{T}_{\text{dissipation}}}_{\text{Mechanisms 1 \& 2}}.
\end{equation}

\subsubsection{1. Storage Stress: The Elastic Mechanism}
\label{sec:potential_stress}

The first component, $\mathbf{T}_{\text{storage}}$, arises from the average of the \textbf{interatomic forces}. This mechanism dominates in solids and corresponds to the \textbf{Electromagnetic Shadow Row} (Ordered Column) of the unified matrix.

In a neutral material, the macroscopic electric field $\mathbf{E} = \langle \mathbf{e} \rangle$ may be zero. But what holds a solid together? The answer lies in the microscopic structure of the fields that are filtered out. The structured arrangement of positive ions and negative electrons (e.g., in a crystal lattice) creates an intense, rapidly varying, but highly structured microscopic electric field ($\mathbf{e}_{\text{ord}}$).

\textbf{Elasticity} is the macroscopic manifestation of this field. When the material is deformed (Strain), work is done against $\mathbf{e}_{\text{ord}}$, displacing atoms from equilibrium. This energy is stored reversibly in the variance of the binding fields. Thus, \textbf{Storage Stress} is the mechanical realization of the \textbf{Storage Mechanism}. It is a conservative, "spring-like" force.

\subsubsection{2. Interaction Stress: Work and Dissipation}
\label{sec:interaction_stress}

The remaining terms, $\mathbf{T}_{\text{work}}$ and $\mathbf{T}_{\text{dissipation}}$, do not represent energy stored \textit{within} the ordered domain's structure. Instead, they represent the \textbf{interaction} between the ordered continuum and the unordered thermodynamic domain. They arise from the \textbf{flux of unordered momentum} (Kinetic Stress).

\paragraph{Pressure (The Work Mechanism).}
The component $\mathbf{T}_{\text{work}}$ corresponds to the isotropic flux of momentum. In gases and liquids, particles in constant thermal motion ($\mathbf{v}_{unord}$) continuously bombard any varying surface.
Macroscopically, this high-frequency bombardment is compressed into a single scalar field: \textbf{Pressure} ($P$).

Pressure acts as a bridge for doing \textbf{macroscopic work}. It allows the high-frequency energy of the unordered domain (heat) to drive the low-frequency motion of the ordered domain (work). A primary example is the \textbf{steam engine}: the chaotic, high-frequency oscillation of water molecules creates a macroscopic pressure that drives a piston, successfully converting disordered thermal energy into ordered mechanical work.

\paragraph{Viscosity (The Dissipation Mechanism).}
The component $\mathbf{T}_{\text{dissipation}}$ corresponds to the anisotropic flux of momentum (shear stress). In the presence of a velocity gradient, momentum diffuses between adjacent layers of the flow.
Macroscopically, this is \textbf{Viscosity}. Unlike pressure, which bridges energy to do useful work, viscosity acts as a sink. It transfers energy from the ordered macroscopic flow back into the unordered microscopic chaos. It represents the irreversible \textbf{dissipation} of ordered mechanical energy into heat.

\section[Macroscopic Electrodynamics]{Macroscopic Electrodynamics \\ as the Third Domain}
\label{sec:UnifiedHost}

In this chapter, we have conceptually explained the domains of Continuum Mechanics and Thermodynamics, delineating how they arise from the microscopic world and how they interact.

In the remainder of the book, we will consider the third domain in detail: the \textbf{Macroscopic Electrodynamic Domain}. We propose that macroscopic electrodynamic effects arise as soon as \textbf{ordered motion of electrons relative to the ion lattice} occurs. The goal will be to define that domain internally, as well as to explain how it couples towards the Continuum Mechanics and Thermodynamic domains.

\chapter{The Host Interface: The Macroscopic Transmission}
\label{chap:HostInterface}

\section{Introduction: Domain 3 and the Relative Motion}

In the previous chapters, we delineated the origins of the Continuum Mechanics (Ordered Motion) and Thermodynamic (Unordered Motion) domains. In this chapter, we address the third column of the Unified Matrix: \textbf{Domain 3}, defined by the \textbf{Relative Motion} of charged fluids.

This domain is the birthplace of \textbf{Macroscopic Electrodynamics}. Our goal is to derive how this domain couples to the established mechanical domains. We do not aim to re-derive the internal laws of thermodynamics or continuum mechanics; rather, we seek to rigorously define the \textbf{Interface}—the gateway through which electromagnetic energy and momentum are transferred into the material host. In Parts \ref{part:ElectricResponse}, \ref{part:MagneticResponse}, and \ref{part:MovingMatter}, this interface will be examined in detail regarding the interaction with specific forms of matter. Here, the mechanism itself is derived.

\section{Filtering the Linear Laws: Conceptual Preservation}
\label{sec:filtering_fields}

The filtering framework is first applied to the microscopic Maxwell equations. The crucial feature of these laws is their \textbf{linearity} in both the fields and the sources. Linearity ensures that the differential operators ($\nabla, \partial/\partial t$) commute with the averaging operator $\langle \cdot \rangle$. This leads to \textbf{conceptual preservation}: the structure of the physical law survives the micro-to-macro transition intact.

\subsection{The Macroscopic (Signal) Equations}
Applying the filter $\langle \cdot \rangle$ to the microscopic equations yields the macroscopic Maxwell equations in their exact vacuum form:

\begin{subequations}
\label{eq:Macro_Maxwell_VacuumForm}
\begin{align}
    \nabla \cdot \mathbf{E} &= \frac{\rho_{\text{total}}}{\varepsilon_0} \\
    \nabla \times \mathbf{E} + \frac{\partial \mathbf{B}}{\partial t} &= \mathbf{0} \\
    \nabla \cdot \mathbf{B} &= 0 \\
    \nabla \times \mathbf{B} - \mu_0\varepsilon_0\frac{\partial \mathbf{E}}{\partial t} &= \mu_0 \mathbf{J}_{\text{total}}
\end{align}
\end{subequations}
where $\mathbf{E} = \langle \mathbf{e} \rangle$, $\mathbf{B} = \langle \mathbf{b} \rangle$, and the sources $\rho_{\text{total}}, \mathbf{J}_{\text{total}}$ are the averaged, low-frequency components of the total microscopic sources.

This derivation confirms that the "vacuum form" is the rigorous macroscopic description of the signal. The averaging process itself does not necessitate the introduction of auxiliary fields ($\mathbf{D}, \mathbf{H}$).

\subsection{The Kinematic Status of Polarization and Magnetization}

It is crucial to define the status of Polarization ($\mathbf{P}$) and Magnetization ($\mathbf{M}$) within this framework. Microscopically, the sources can always be represented by polarization and magnetization potentials ($\mathbf{p}, \mathbf{m}$) via a Helmholtz decomposition, as detailed in Chapter \ref{chap:Critique_ArbitrarySplit} (Critique of the Arbitrary Split).

This decomposition is general; one may apply it to any subset of microscopic sources. Conventionally, currents and charges that are not "free" to move—such as those bound within atoms or the lattice—are renamed via these potentials. It is vital to emphasize that these fields have no independent physical existence; they are \textbf{kinematic descriptors} of the microscopic sources.

Let us conceptually separate all microscopic sources into free charges ($\rho_{\text{free}}$) and currents ($\mathbf{j}_{\text{free}}$), and bound charges ($\rho_{\text{bound}}$) and currents ($\mathbf{j}_{\text{bound}}$). One must remain aware that microscopically, there is no "matter" in the sense of a continuous material host; there are only charges, currents, and fields. The averaging process generates the macroscopic concept of "material hosts."

Because the averaging operator commutes with the linear differential operators, the macroscopic sources rigorously inherit this exact structure:
\begin{align}
    \mathbf{J}_{\text{total}} &= \mathbf{J}_{\text{free}} + \mathbf{J}_{\text{bound}} = \mathbf{J}_{\text{free}} + \frac{\partial \mathbf{P}}{\partial t} + \nabla \times \mathbf{M}, \\
    \rho_{\text{total}} &= \rho_{\text{free}} - \nabla \cdot \mathbf{P},
    \label{eq:Macro_PM_definition}
\end{align}
where $\mathbf{P} = \langle \mathbf{p} \rangle$ and $\mathbf{M} = \langle \mathbf{m} \rangle$.

Just like their microscopic counterparts, $\mathbf{P}$ and $\mathbf{M}$ are strictly \textbf{kinematic descriptors} of the averaged sources. In the filter framework, they represent the low-frequency components of the sources inside matter: the \textbf{induced charge separation} and \textbf{induced current circulation}, respectively. Consequently, the material parameters that scale them (Susceptibility, Permittivity, Permeability) are not energetic properties of the vacuum, but \textbf{geometric induction coefficients} describing the magnitude of these induced sources. They mathematically quantify \textit{how much} source ($\rho, \mathbf{J}$) generates in response to a field.

\subsection{The Fluctuation Equations: Linear Decoupling}
Due to the linearity of Maxwell's equations, one may separate the fields and sources cleanly into low- and high-frequency parts. Subtracting the macroscopic equations from the microscopic ones ($\delta\phi = \phi - \Phi$) isolates the dynamics of the high-frequency fluctuations. These fluctuations obey a parallel set of Maxwell-like equations, driven exclusively by the source fluctuations ($\delta\rho, \delta\mathbf{j}$).

In the linear domain of field generation, the separation is absolute; the Signal and the Fluctuation are mathematically decoupled. However, they re-couple inextricably when analyzing the \textbf{non-linear} processes of force and energy exchange.

It is interesting to note that the "neutral" material host—the domain of Continuum Mechanics and Thermodynamics—is constituted by these high-frequency parts. A macroscopic current flows \textbf{in} a conductor, but the conductor itself is made of the high-frequency components of the microscopic particles. A "neutral conductor" in macroscopic terms simply means there are no low-frequency net charges or currents present.

\subsection{Macroscopic Equations in terms of D and H}
We have presented the macroscopic Maxwell equations in their fundamental vacuum form. Inserting the separation of bound and free charges, one may present them in the traditional form including the auxiliary fields $\mathbf{D}$ and $\mathbf{H}$:
\begin{subequations}
\begin{align}
    \nabla \cdot \mathbf{D} &= \rho_{\text{free}} \\
    \nabla \times \mathbf{E} + \frac{\partial \mathbf{B}}{\partial t} &= \mathbf{0} \\
    \nabla \cdot \mathbf{B} &= 0 \\
    \nabla \times \mathbf{H} - \frac{\partial \mathbf{D}}{\partial t} &= \mathbf{J}_{\text{free}}
\end{align}
\end{subequations}
However, in the following analysis, we strictly adhere to the formulation which clearly separates sources (charges and currents) from fields ($\mathbf{E}$ and $\mathbf{B}$), as this is required to maintain mechanical transparency.

\section{The Macroscopic Balances: The Emergence of the Host Interface}
\label{sec:bridge_em_mech}

We now propose an approach to derive the macroscopic force balance. The conceptual validity of this approach lies in its ability to synthesize the domains; its application will be verified in Parts \ref{part:ElectricResponse}--\ref{part:MovingMatter}.

\subsection{Strategy: The Extraction of the Electrodynamic Signal}

The starting point of our derivation is the microscopic momentum balance derived in Chapter \ref{chap:FreeCharges}. It is essential to recognize that this single equation—the combination of Newton's laws and the microscopic Lorentz force—is effectively the "source code" for the entire universe of classical physics. Hidden within its singular densities and fluctuating fields are the seeds of virtually every macroscopic mechanical phenomenon—from fluid dynamics and acoustics to thermodynamics, viscosity, plasticity, and elasticity.

It is impossible to re-derive all of these macroscopic laws at once. Our goal is more specific: we aim to perform a \textbf{mathematical surgery} on this equation.

\begin{enumerate}
    \item \textbf{Extraction:} We will rewrite the microscopic equation to explicitly separate the terms corresponding to \textbf{Macroscopic Electrodynamics} (the Ordered Lorentz Force) from all other terms.
    \item \textbf{The Host Interface:} All remaining terms—representing viscosity, pressure, thermal fluctuations, and quantum binding forces—will be grouped into a single aggregate term: the \textbf{Host Force} ($\mathbf{F}_{\text{host}}$).
    \item \textbf{Averaging:} We will then apply the filtering operator to transition this system to the macroscopic scale.
\end{enumerate}

This process allows us to isolate the electromagnetic signal while rigorously defining the "interface" through which it couples to the rest of the physical world.

\subsection{The Microscopic Baseline}
In Chapter \ref{chap:FreeCharges}, we derived the microscopic momentum balance for the fundamental particles (ions and electrons). This equation is the "Microscopic Ground Truth":
\begin{equation}
\begin{aligned}
    \underbrace{ \left(\frac{\partial \mathbf{g}_{\text{mech}}}{\partial t} + \nabla\cdot\mathbf{t}_{\text{kin}}\right) - \mathbf{f}_{Q} }_{\substack{\textbf{The Mechanical Reservoir} \\ \text{(Net Momentum Gain)}}}
    \quad &\overset{\text{Physical Coupling}}{\textbf{=}} \quad
    \underbrace{ \rho_{\pm , \text{micro}} \mathbf{e} + \mathbf{j}_{\text{micro}}\times\mathbf{b} }_{\substack{\textbf{The Single Gateway} \\ \text{(Lorentz Force)}}} \\
    &\overset{\text{Math. Identity}}{=} \quad
    \underbrace{ -\left(\frac{\partial \mathbf{g}_{\text{em}}}{\partial t} + \nabla\cdot \mathbf{t}_{\text{em}}\right) }_{\substack{\textbf{The Electromagnetic Reservoir} \\ \text{(Field Momentum Supply)}}}
\end{aligned}
\label{eq:Micro_Momentum_Architecture}
\end{equation} 
where:
\begin{align}
        \mathbf{g}_{\text{mech}} &= \rho_{m,\pm} \mathbf{v}_\pm, \\
        \mathbf{t}_{\text{kin}} &= \rho_{m,\pm} (\mathbf{v}_\pm \otimes \mathbf{v}_\pm), \\
        \mathbf{j}_{\text{micro}} &= \rho_{\pm, \text{micro}} \mathbf{v}_\pm.
\end{align}

It is essential to understand that this equation inherently contains \textit{all} of classical physics on a microscopic scale. It contains the seeds of thermodynamics, continuum mechanics, and electrodynamics. The goal is not to re-derive all of physics, but to extract the \textbf{Macroscopic Electrodynamic} component and identify the bridge to the other domains.

\subsection{Decomposition: Ordered vs. Unordered}

To achieve this, we first subdivide the microscopic equation. As shown before, the microscopic velocity $\mathbf{v}$ may be decomposed into a macroscopic ordered velocity $\mathbf{V}$ and a fluctuating unordered velocity $\mathbf{c}$:
\begin{equation}
    \mathbf{v}_\pm(\mathbf{x},t) = \mathbf{V}_\pm(\mathbf{x},t) + \mathbf{c}_\pm(\mathbf{x},t).
\end{equation}
Here, it is crucial to differentiate between the ions ($+$) and the electrons ($-$). In the macroscopic EM domain, we adopt a \textbf{Two-Fluid Model}. The ordered velocities of the two species are not necessarily the same; their relative motion is the source of current. Thus, the differentiation into ordered and unordered motion happens separately for each species.

Similarly, we decompose the microscopic fields $\mathbf{e}$ and $\mathbf{b}$ into their low and high-frequency parts:
\begin{equation}
    \mathbf{e} = \mathbf{E} + \delta\mathbf{e}, \quad \quad \mathbf{b} = \mathbf{B} + \delta\mathbf{b}.
\end{equation}

Our goal is to extract the \textbf{ordered mechanical momentum response} of the positive and negative fluids and the corresponding \textbf{ordered Lorentz force}, which is part of the Lorentz force generated by the low frequency fields $\mathbf{E}$ and $\mathbf{B}$.

Mathematically, we perform this by inserting the Reynolds decompositions ($\mathbf{v}=\mathbf{V}+\mathbf{c}$ and $\mathbf{e}=\mathbf{E}+\delta\mathbf{e}$) into the microscopic momentum balance. We then \textbf{rearrange} the terms, moving everything that is NOT the ordered kinematic response or the macroscopic Lorentz force to the left-hand side.

This grouping yields the following "intermediate" microscopic form:
\begin{equation}
\begin{aligned}
    &\underbrace{ \left(\frac{\partial \mathbf{g}_{\pm,ord}}{\partial t} + \nabla\cdot\mathbf{t}_{\pm,ord}\right) - \mathbf{f}_{\text{host},\pm,micro} }_{\substack{\textbf{The Mechanical Reservoir} \\ \text{(Net Momentum Gain)}}} \\
    &\quad \overset{\text{Physical Coupling}}{\textbf{=}}
    \underbrace{ \rho_{\pm , \text{micro}} \mathbf{E} + \mathbf{j}_{\text{micro,ord}}\times\mathbf{B} }_{\substack{\textbf{The Macroscopic Gateway} \\ \text{(Ordered Lorentz Force)}}}
\end{aligned}
\end{equation} 
where the ordered kinematic terms are defined structurally identical to the microscopic ones:
\begin{align}
        \mathbf{g}_{\pm,ord} &= \rho_{m,\pm} \mathbf{V}_\pm, \\
        \mathbf{t}_{\pm,ord} &= \rho_{m,\pm} (\mathbf{V}_\pm \otimes \mathbf{V}_\pm),\\
        \mathbf{j}_{micro,ord} &= \rho_{\pm, \text{micro}} \mathbf{V}_\pm.
\end{align}

\subsection{The Host Force: The Definition of the Aggregate Term}

The term $\mathbf{f}_{\text{host},\pm,micro}$ acts as a \textbf{container} for all remaining terms. Mathematically, it defines the discrepancy between the full microscopic reality and our idealized macroscopic electrodynamics. It collects the residue of the filtration process.

Ideally, one would derive every term from first principles, but that would require deriving all of thermodynamics and fluid mechanics from scratch. Instead, we define $\mathbf{f}_{\text{host}}$ as the aggregate sum of all non-macroscopic-EM interactions:

\begin{align}
    \mathbf{f}_{\text{host},\pm,micro} \equiv \quad & \underbrace{(\rho_{\pm, \text{micro}} \mathbf{c})\times\mathbf{b}}_{\text{Unordered Magnetic Force}} \\
    &+ \underbrace{\rho_{\pm , \text{micro}} \delta \mathbf{e} + \mathbf{j}_{\text{micro,ord}}\times \delta\mathbf{b}}_{\text{Microscopic Field Fluctuations of ordered Lorentz Force}} \\
    &- \underbrace{\nabla \cdot (\rho_{m,\pm} \mathbf{c}_\pm \otimes \mathbf{c}_\pm)}_{\text{Kinetic Stress (Pressure/Viscosity)}} \\
    &- \underbrace{\nabla \cdot (\rho_{m,\pm} (\mathbf{V}_\pm \otimes \mathbf{c}_\pm + \mathbf{c}_\pm \otimes \mathbf{V}_\pm))}_{\text{Mixed Stress}} \\
    &+ \mathbf{f}_{Q} \quad (\text{Quantum Constraints}).
\end{align}

We have strictly extracted the low-frequency electromagnetic field response and the ordered mechanical momentum response. All other parts are inherent in this \textbf{Phenomenological Interface}. It is the \textbf{Universal Coupling Term} where the macroscopic electromagnetic response couples to the continuum mechanics and thermodynamic domains.

\subsection{The Micro-Macro Transition: Averaging}

So far, we have only rearranged the microscopic terms. However, we have grouped them such that the separation of scales is already structurally prepared. 

Now that we have separated the signals microscopically, the transition to the macroscopic world is simply a matter of applying the filter $\langle \cdot \rangle$.

This mathematical operation represents the definitive step from \textbf{Stage 2 (Microscopic Baseline)} to \textbf{Stage 3 (Macroscopic Model)}, as defined in Chapter \ref{chap:part1_intro}. It fundamentally changes the system under consideration: we are no longer tracking discrete particles, but continuous fluid fields.

The ordered velocity $\mathbf{V}_{\pm}$ is by definition a low-frequency variable, so the averaging kernel simply spreads the mass and charge density of the species, generating a \textbf{Two-Fluid Continuous Model}.

The macroscopic mechanical momentum of the positive and negative fluids is given by:
\begin{equation}
   \frac{\partial \mathbf{G}_{\pm}}{\partial t} + \nabla\cdot\mathbf{T}_{\pm}
\end{equation}
Likwise, the Macroscopic Host Force is simply the average of the microscopic aggregate:
\begin{equation}
    \mathbf{F}_{\text{host},\pm} \equiv \langle \mathbf{f}_{\text{host},\pm,micro} \rangle.
\end{equation}
A similar effect applies to the force density. Because the external fields $\mathbf{E}, \mathbf{B}$ are smooth, they factor out of the average:
\begin{equation}
   \langle \rho_{\pm , \text{micro}} \mathbf{E} + \mathbf{j}_{\text{micro,ord}}\times\mathbf{B}  \rangle = \rho_{\text{total},\pm} \mathbf{E} + \mathbf{J}_{\text{total},\pm}\times\mathbf{B}.
\end{equation}

We generalize this interface to the macroscopic scale without yet specifying its internal mechanics. We write the macroscopic momentum balance as:
\begin{equation}
\boxed{
\begin{aligned}
    \underbrace{ \left(\frac{\partial \mathbf{G}_{\pm}}{\partial t} + \nabla\cdot\mathbf{T}_{\pm}\right) - \mathbf{F}_{\text{host},\pm} }_{\substack{\textbf{The Mechanical Reservoir} \\ \text{(Net Momentum Gain)}}}
    \quad &\overset{\text{Physical Coupling}}{\textbf{=}} \quad
    \underbrace{ \rho_{\text{total},\pm} \mathbf{E} + \mathbf{J}_{\text{total},\pm}\times\mathbf{B} }_{\substack{\textbf{The Macroscopic Gateway} \\ \text{(Macroscopic Lorentz Force)}}} \\
\end{aligned}
}
\label{eq:Macro_Momentum_Architecture}
\end{equation}

\subsection{Interpretation of the Host Force}

Let us interpret this derived equation. We have a 2-fluid model where positive and negative fluids move independently. They are driven by a macroscopic Lorentz force. 
$\mathbf{F}_{\text{host}}$ inherits all other parts of classical physics. It acts as the \textbf{Universal Interface}—the source or sink of momentum entering or leaving the macroscopic EM domain from the Continuum or Thermodynamic domains.

We do not claim to derive the internal laws of these domains (e.g., the Navier-Stokes equations or the precise constitutive laws of elasticity) from this single coupling term. However, the completeness of this term is easily demonstrated by a simple thought experiment:

Consider the limit where the macroscopic electromagnetic fields vanish ($\mathbf{E} = \mathbf{0}, \mathbf{B} = \mathbf{0}$). In this limit, the "Gateway" closes, and the equation reduces to a purely mechanical balance driven solely by $\mathbf{F}_{\text{host}}$. Since we know that non-electromagnetic matter is governed by Continuum Mechanics and Thermodynamics, it follows that $\mathbf{F}_{\text{host}}$ \textbf{must inherit the entirety of these physical laws}. It contains the pressure gradients, the viscous stresses, and the thermal diffusion that drive the universe when no external fields are present.

We can conceptually identify the physical nature of the terms contained within it, essentially "peering into the black box":

\paragraph{1. Microscopic Field Interactions (Binding Forces).}
The term $\langle \rho \delta \mathbf{e} + \dots \rangle$ represents the microscopic field interactions. These are the \textbf{binding forces}.
\begin{itemize}
    \item In conductors, these are the forces that keep free currents inside the material (work function/surface barriers).
    \item These are the drag forces acting on moving fluids (origin of \textbf{resistance}).
    \item In dielectrics/magnets, these are the restoring forces acting on bound charges (origin of \textbf{permittivity/permeability}).
\end{itemize}

\paragraph{2. Unordered Momentum Transfer (Pressure).}
The term $\langle \mathbf{t}_{\pm,unord} \rangle$ represents the high-frequency dynamic momentum transfer. This corresponds to \textbf{pressure} in the continuum mechanics domain.

\paragraph{3. Quantum Stabilization.}
The term $\langle \mathbf{f}_Q \rangle$ is responsible for the stability of matter itself. While we treat the material host as a given, this term reminds us that the very existence of stable atoms and lattices relies on non-classical constraints (such as the Pauli exclusion principle) that balance the attractive electromagnetic forces.

\paragraph{4. Thermal Momentum.}
The flux term $\langle \mathbf{g}_{\pm,unord} \rangle$ inherently contains the \textbf{temperature} of the thermodynamic domain.

\subsection{Summary}
We have derived a macroscopic momentum equation for properties electrodynamics. We propose a \textbf{Two-Fluid Model}, considering the ions and electrons as separate fluids with independent velocity fields.
We defined the \textbf{Host Force} ($\mathbf{F}_{\text{host}}$) as the gateway for transformation. It captures the interaction with the Thermodynamic and Continuum Mechanics domains. The detailed application of this framework will be the subject of the following parts of the book.

\section{The Energy Equation}

Our proposed energy equation follows directly by projecting the macroscopic momentum equation onto the velocity fields of the two fluids ($\mathbf{V}_{\pm}$). We multiply the momentum equation scalarly by the corresponding velocities. The derivation parallels the microscopic derivation in Chapter \ref{chap:FreeCharges}.

It follows:
\begin{equation}
\boxed{
\begin{aligned}
    \underbrace{ \left( \frac{\partial U_{\pm}}{\partial t} + \nabla\cdot\mathbf{S}_{\pm} \right) - P_{\text{host}, \pm} }_{\substack{\text{Total mechanical response}}}
    \quad &\overset{\text{Physical Coupling}}{\textbf{=}} \quad
    \underbrace{ \mathbf{J}_{\pm}\cdot\mathbf{E} }_{\substack{\textbf{The Energy Gateway} \\ (P_{\text{Lorentz}})}} \\
\end{aligned}
}
    \label{eq:Macro_Energy_Architecture_Def}
\end{equation}

The \textbf{Host Power} is equally defined by multiplying the Host Force by the respective velocities of the two fluids:
\begin{equation}
    P_{\text{host},\pm} = \mathbf{F}_{\text{host},\pm} \cdot \mathbf{V}_{\pm}.
\end{equation}

The Host Power is the gateway for any energy transfer into or out of the macroscopic electrodynamic domain. As we strictly derived the energy balance by projecting the momentum balance onto the kinematic velocities, the \textbf{Force-Energy Consistency Criterion (FECC)} is satisfied by construction.

\paragraph{The Orthogonality of Energy (Spectral Consistency)}

It is interesting to note that the energy density, while a quadratic (non-linear) quantity ($u \propto e^2$), separates clearly into two parts. The mutual terms of the low and high-frequency parts are orthogonal.

The total microscopic electromagnetic energy density $u_{\text{em}}$ is averaged:
\begin{equation}
    \langle u_{\text{em}} \rangle = \left\langle \frac{\varepsilon_0}{2} e^2 + \frac{1}{2\mu_0} b^2 \right\rangle.
\end{equation}
Focusing on the electric term and applying the Reynolds decomposition ($\mathbf{e} = \mathbf{E} + \delta\mathbf{e}$):
\begin{equation}
    \langle e^2 \rangle = \langle (\mathbf{E} + \delta\mathbf{e})^2 \rangle = \langle E^2 + 2\mathbf{E}\cdot\delta\mathbf{e} + (\delta\mathbf{e})^2 \rangle.
\end{equation}
Applying the averaging operator:
\begin{equation}
    \langle e^2 \rangle = E^2 + \underbrace{2\langle \mathbf{E}\cdot\delta\mathbf{e} \rangle}_{=0} + \langle (\delta\mathbf{e})^2 \rangle.
\end{equation}
The cross-term vanishes due to the \textbf{Orthogonality of Scales}. Physically, this occurs because the macroscopic field $\mathbf{E}$ is effectively constant over the integration volume, while the fluctuation $\delta\mathbf{e}$ oscillates rapidly with zero mean. The total averaged electromagnetic energy therefore separates cleanly:
\begin{equation}
    \langle u_{\text{em}} \rangle = \underbrace{ \left( \frac{\varepsilon_0}{2}E^2 + \frac{1}{2\mu_0}B^2 \right) }_{U_{\text{EM}} \text{ (Signal)}} + \underbrace{ \left\langle \frac{\varepsilon_0}{2}(\delta\mathbf{e})^2 + \frac{1}{2\mu_0}(\delta\mathbf{b})^2 \right\rangle }_{\langle \delta u \rangle \text{ (Fluctuation - Heat/Elasticity)}}.
    \label{eq:Energy_Orthogonality}
\end{equation}
This result is pivotal. It demonstrates that the averaging process effectively filters all high-frequency electromagnetic energy out of the electromagnetic domain. This energy does not vanish; it is reassigned.

In the Macroscopic (Stage 3) perspective:
\begin{itemize}
    \item The \textbf{Signal Energy} ($U_{\text{EM}}$) corresponds to the macroscopic fields $\mathbf{E}$ and $\mathbf{B}$. This is the only energy recognized as "electromagnetic."
    \item The \textbf{Fluctuation Energy} ($\langle \delta u \rangle$) is no longer inherent to the electromagnetic description. Instead, it is subsumed into the material host.
\end{itemize}
Thus, all "messy" microscopic energy terms are inherent in $\mathbf{F}_{\text{host}}$ and $P_{\text{host}}$, leaving the macroscopic electromagnetic definition purely as the energy of the signal.

\section{On the Separation of Scales: The Three Stages of Abstraction}

Before applying this framework in the following chapters, we must clarify a fundamental epistemological point. The averaging process—the transition from the microscopic to the macroscopic—fundamentally changes the system under consideration.

As we have shown, the averaging process acts as a \textbf{low-pass filter}, discarding all high-frequency information. This is not merely a mathematical convenience; it generates a mathematically distinct macroscopic system.

This Part of the book has illustrated conceptually that the process of averaging is inherently a process of \textbf{information compression}. The high-frequency field fluctuations and the microscopic inertia are filtered out, only to re-emerge as new, macroscopic variables (such as Temperature or Elasticity) in the macroscopic description.

Thus, when we consider the macroscopic field description, we have lost the explicit information of the underlying microscopic reality. For example, when we describe a material by its temperature, we possess no information about the exact positions and velocities of the individual particles. Even though we know the underlying reality consists of moving particles, the macroscopic description is distinct. It is a \textbf{self-contained system}.

\subsection{The Three Stages of Reality}

We can now formalize the \textbf{Three Layers of Abstraction} used in this text (as first introduced in Section \ref{sec:Info_Compression}):

\begin{itemize}
    \item \textbf{Stage 1: Quantum Electrodynamics (QED).} The probabilistic, infinite-bandwidth description of the fundamental quantum reality.
    \item \textbf{Stage 2: Microscopic Classical Physics.} The deterministic "Ground Truth" established in Part I. It serves as the input signal for our filter.
    \item \textbf{Stage 3: Macroscopic Classical Physics.} The filtered, low-frequency output. This is the domain of engineering, thermodynamics, and continuum mechanics.
\end{itemize}

We have now moved from the Microscopic (Stage 2) towards the Macroscopic (Stage 3) domain. The Stage 3 description is, by definition, unable to explicate the internal phenomena of Stage 2. It is a filtered, simplified description of the same reality.

Crucially, these three stages are \textbf{mutually exclusive} and \textbf{self-contained} descriptions. They have their own laws, their own variables, and their own internal consistency. While one might switch between the stages to gain deeper insight, we propose that a theory must be judged on its own terms within its own stage.

\subsection{The Perspective of the Macroscopic Observer}

In the following chapters, we will switch "gears" between the domains to show how macroscopic variables emerge from the Stage 2 reality. However, from the view of the \textbf{Macroscopic System (Stage 3)}, the only apparent electromagnetic fields are the low-frequency signals ($\mathbf{E}, \mathbf{B}$).

When we speak in the language of macroscopic classical physics, "electromagnetic phenomena" refers strictly to these low-frequency interactions. The underlying microscopic fields have been filtered out. They may reappear as emergent mechanical properties—such as \textbf{elasticity} or \textbf{binding energy}—but in Stage 3, these are \textit{mechanical} properties, not electromagnetic ones. Even though they are electromagnetic in origin (Stage 2), they are mechanical in function (Stage 3).

We will explain this in detail with concrete examples in the rest of the book. In summary, the reader must always remain aware of the \textbf{Stage of Abstraction}. QED, Microscopic electrodynamics, and Macroscopic classical physics are three different languages describing the same underlying territory.

\section{The Unified Macroscopic Architecture}
\label{sec:Validation}

We can now synthesize the results into the complete architecture of Macroscopic Electrodynamics. This architecture rigorously mirrors the structure of the microscopic baseline.

\begin{equation}
\boxed{
\begin{aligned}
    \underbrace{ \left( \frac{\partial \mathbf{G}_\pm}{\partial t} + \nabla\cdot\mathbf{T}_{\pm} \right) - \mathbf{F}_{\text{host},\pm} }_{\substack{\text{Total Mechanical Response} \\ \text{(Ordered + Host)}}}
    \quad &\overset{\text{Physical Coupling}}{\textbf{=}} \quad
    \underbrace{\rho_{\text{total},\pm} \mathbf{E} + \mathbf{J}_{\text{total},\pm}\times\mathbf{B} }_{\substack{\textbf{The Momentum Gateway} \\ (\mathbf{F}_{\text{Lorentz}})}} \\
    &\overset{\text{Math. Identity}}{=} \quad
    \underbrace{ -\left(\frac{\partial \mathbf{G}_{\text{EM}}}{\partial t} + \nabla\cdot \mathbf{T}_{\text{EM}}\right) }_{\substack{\text{Macroscopic EM Response} \\ \text{(Vacuum Tensor)}}}
\end{aligned}
}
\end{equation}

This structure clearly delineates the domains:
\begin{itemize}
    \item \textbf{LHS (Mechanical Domain):} Contains the ordered inertia of the fluids and the total material response ($\mathbf{F}_{\text{host}}$).
    \item \textbf{RHS (EM Domain):} Described strictly by the Vacuum-Form Maxwell Stress Tensor.
    \item \textbf{Gateway:} The macroscopic Lorentz force connects them.
\end{itemize}

Similarly, the energy architecture is:

\begin{equation}
\boxed{
\begin{aligned}
    \underbrace{ \left( \frac{\partial U_{\pm}}{\partial t} + \nabla\cdot\mathbf{S}_{\pm} \right) - P_{\text{host},\pm} }_{\substack{\text{Total Mechanical Response}}}
    \quad &\overset{\text{Physical Coupling}}{\textbf{=}} \quad
    \underbrace{ \mathbf{J}_{\pm}\cdot\mathbf{E} }_{\substack{\textbf{The Energy Gateway} \\ (P_{\text{Lorentz}})}} \\
    &\overset{\text{Math. Identity}}{=} \quad
    \underbrace{ -\left(\frac{\partial U_{\text{EM}}}{\partial t} + \nabla\cdot \mathbf{S}_{\text{EM}}\right) }_{\substack{\text{Macroscopic EM Response}}}
\end{aligned}
}
\end{equation}

Both equations together define the \textbf{Macroscopic Electromagnetic Stress-Energy Tensor} as the low-frequency projection of the microscopic tensor.
\begin{itemize}
    \item Energy Density: $U_{\text{EM}} = \frac{1}{2}\left(\varepsilon_0 E^2 + \frac{1}{\mu_0} B^2\right)$
    \item Energy Flux: $\mathbf{S}_{\text{EM}} = \frac{1}{\mu_0} (\mathbf{E} \times \mathbf{B})$
    \item Momentum Density: $\mathbf{G}_{\text{EM}} = \varepsilon_0 (\mathbf{E} \times \mathbf{B})$
    \item Stress Tensor: $\mathbf{T}_{\text{EM}} = \varepsilon_0 \mathbf{E} \otimes \mathbf{E} +  \frac{1}{\mu_0} \mathbf{B} \otimes \mathbf{B} - U_{\text{EM}} \mathbf{I}$
\end{itemize}

\section{Conclusion and Outlook}
\label{sec:HostInterpretation}

We proposed the macroscopic electrodynamics framework as a low-frequency projection of the underlying microscopic reality. All high-frequency components have been conceptually separated into the \textbf{Host Interface}. We propose that this is the gateway that couples the macroscopic electrodynamic domain towards the Continuum Mechanics and Thermodynamic domains.

The proposed macroscopic framework utilizes a \textbf{Two-Fluid Model}, where the velocities of the ion fluid and the electron fluid are considered separately. This directed framework is \textbf{mechanically consistent} by definition: all forces have a clear mass target, all mass targets have well-defined velocities, and the power equation describes the power output corresponding to the Lorentz force.

The framework is now ready to be verified. In the following three parts of the book (Parts \ref{part:ElectricResponse}--\ref{part:MovingMatter}), we will consider all macroscopic electrodynamic interactions, separated into an electric domain (Part \ref{part:ElectricResponse}), a magnetic domain (Part \ref{part:MagneticResponse}), and a domain including moving matter (Part \ref{part:MovingMatter}). We will apply the derived framework in detail to macroscopic phenomena. We propose that the vacuum framework---the low-frequency part of the microscopic reality---is able to describe the \textit{electromagnetic input} to all macroscopic phenomena, including dielectric, magnetic, and moving media, in a mechanically consistent manner.

Finally, in Part \ref{part:Synthesis} of the book, we will close the circle and return to the Unified Matrix proposed at the beginning of Part \ref{part:MacroscopicFilter}.

\part{Part III: The Electric Response -- Dynamics of Charge Separation and Accumulation}
\label{part:ElectricResponse}

\chapter{Introduction to Part III}
\label{chap:IntroPart3}

Following the derivation of the macroscopic equations via the filtering process, the analysis now proceeds to their application in the macroscopic domain.

\section{The Physics of Divergent Currents}

Standard pedagogy distinguishes "Conductors" and "Dielectrics" as separate classes of matter, governed by distinct laws (Ohm's Law vs. constitutive relations) and described by disparate parameters (Conductivity vs. Permittivity).

From the perspective of the vacuum (the "microscopic baseline"), no fundamental difference exists between "conducting matter" and "dielectric matter." There is only \textbf{charge density} ($\rho$) and \textbf{current density} ($\mathbf{J}$). The electron moving in a copper wire and the electron shifting in a glass insulator represent fundamentally the same process. They obey the same Lorentz force and are coupled to the same vacuum fields. The only physical distinction is the \textbf{topology of their constraint}.

This Part is dedicated to the study of the \textbf{longitudinal current} ($\mathbf{J}_{\parallel}$). This is the component of the current density that possesses divergence ($\nabla \cdot \mathbf{J} \neq 0$). By the continuity equation ($\partial \rho / \partial t = - \nabla \cdot \mathbf{J}$), this kinematic property leads inevitably to the accumulation of charge density.
The subject, therefore, is the physics of \textbf{charge separation}.

\section{The Spectrum of Constraint}

Material behavior in this regime is unified by a single parameter: \textbf{The scale of the constraint.}

Fundamentally, essentially all currents are "bound." True "free current" exists only in the macroscopic domain. In the domain of macroscopic electrodynamics, every charge carrier is constrained to a massive lattice by the host interface. The classification of "free" or "bound" currents is defined simply by the length scale of this constraint.

\paragraph{1. The Macroscopic Constraint (The Conductor).}
In a metal, the electron is "free" only in the sense that its constraint is the size of the macroscopic object. It functions within the volume but is rigorously bound to it. Charge accumulates at the macroscopic boundaries of a conductor and is confined by microscopic field interactions at the surface.

\paragraph{2. The Microscopic Constraint (The Dielectric).}
In a dielectric, the electron's constraint is the atomic scale. Displacement occurs, but is limited to angstroms. However, the conceptual physics is identical. The charge moves until it encounters the boundary of its constraint. Charge accumulates at the atomic boundary defined by the microscopic restoring forces within a dipole.

\subsection{The Synthesis}
This reveals a fundamental symmetry: \textbf{A dielectric is effectively a conductor with a microscopic mean-free-path.} Conversely, a conductor is a dielectric with infinite polarizability.
The analysis tracks standard currents ($\mathbf{J}$) operating under a defined geometric constraint.
The \textbf{permittivity} ($\varepsilon$) or \textbf{susceptibility} ($\chi_e$) serves strictly as the geometric measure of the macroscopic charge separation induced in a material by a given field ($\mathbf{P} \propto \chi_e \mathbf{E}$).

\section*{The Governing Equations}

In this Part III, macroscopic electrodynamics is consistently explained from the perspective of the \textbf{momentum balance} derived in the preceding chapter:
\begin{equation}
\boxed{
\begin{aligned}
    \underbrace{ \left( \frac{\partial \mathbf{G}_\pm}{\partial t} + \nabla\cdot\mathbf{T}_{\pm} \right) - \mathbf{F}_{\text{host}} }_{\substack{\text{Total Mechanical Response} \\ \text{(Ordered + Host)}}}
    \quad &\overset{\text{Physical Coupling}}{\textbf{=}} \quad
    \underbrace{\rho_{\text{total}} \mathbf{E} + \mathbf{J}_{\text{total}}\times\mathbf{B} }_{\substack{\textbf{The Momentum Gateway} \\ (\mathbf{F}_{\text{Lorentz}})}} \\
    &\overset{\text{Math. Identity}}{=} \quad
    \underbrace{ -\left(\frac{\partial \mathbf{G}_{\text{EM}}}{\partial t} + \nabla\cdot \mathbf{T}_{\text{EM}}\right) }_{\substack{\text{Macroscopic EM Response} \\ \text{(Vacuum Tensor)}}}
\end{aligned}
}
\end{equation}

and the corresponding energy balance:
\begin{equation}
\boxed{
\begin{aligned}
    \underbrace{ \left( \frac{\partial U_{\text{mech}}}{\partial t} + \nabla\cdot\mathbf{S}_{\text{mech}} \right) - P_{\text{host}} }_{\substack{\text{Total Mechanical Response}}}
    \quad &\overset{\text{Physical Coupling}}{\textbf{=}} \quad
    \underbrace{ \mathbf{J}\cdot\mathbf{E} }_{\substack{\textbf{The Energy Gateway} \\ (P_{\text{Lorentz}})}} \\
    &\overset{\text{Math. Identity}}{=} \quad
    \underbrace{ -\left(\frac{\partial U_{\text{EM}}}{\partial t} + \nabla\cdot \mathbf{S}_{\text{EM}}\right) }_{\substack{\text{Macroscopic EM Response}}}
\end{aligned}
}
\end{equation}

\subsection{The Interaction Density: \texorpdfstring{$\mathcal{D}$}{D}}

This text frequently utilizes the power density scalar field, $\mathcal{D}$, defined as:
\begin{equation}
    \mathcal{D} \equiv \mathbf{J} \cdot \mathbf{E} = -\left(\frac{\partial U_{\text{EM}}}{\partial t} + \nabla\cdot \mathbf{S}_{\text{EM}}\right).
\end{equation}
This scalar represents the fundamental density of mechanical-electromagnetic energy exchange. It acts as the \textbf{Universal Gateway} through which energy leaves or enters the electromagnetic domain. Importantly, at this stage of the definition, $\mathcal{D}$ encompasses \textbf{all three} possible destinations:
\begin{enumerate}
    \item \textbf{Dissipation:} Irreversible transfer to heat (Thermodynamics).
    \item \textbf{Work:} Ordered transfer to macroscopic kinetic energy (Continuum Mechanics).
    \item \textbf{Storage:} Reversible transfer to microscopic potential energy (Binding).
\end{enumerate}
The distinction between these forms is not a property of the field, but of the material host structure that receives the energy.

In regions where $\mathcal{D} = 0$, the energy gateway is closed, and the electromagnetic field energy is conserved independently:
\begin{equation}
    \frac{\partial U_{\text{EM}}}{\partial t} + \nabla \cdot \mathbf{S}_{\text{EM}} = 0.
\end{equation}
No sink or source of field energy exists; it is merely redistributed locally.

When $\mathcal{D} \equiv \mathbf{J} \cdot \mathbf{E}$ is non-zero, a macroscopic fluid (ion or electron fluid) moves with velocity $\mathbf{v}$ under the influence of the Coulomb force. Work is performed on this fluid. This work either increases the mechanical momentum of the fluid or is transferred to the host medium through the host interface. 

Considering the acceleration of the electron fluid, inertia is essentially negligible. As the mass tends to zero, the fluid accelerates quasi-statically, and the energy is effectively transferred directly to the host interface.

This work done on the fluid signifies a local exchange of electromagnetic field energy. When $\mathcal{D} \neq 0$, a sink or source of electromagnetic field energy is present:
\begin{equation}
    \mathcal{D} \neq 0 \implies \frac{\partial U_{\text{EM}}}{\partial t} + \nabla \cdot \mathbf{S}_{\text{EM}} \neq 0.
\end{equation}
A sink implies energy transfer to the kinetic energy of the fluid or transmission to the host. A source implies energy transfer from the host into the electromagnetic field.

This equation represents a rigorously defined interaction: forces, velocities, sinks, and sources of the field energy are interconnected. The \textbf{Force-Energy Consistency Criterion (FECC)} is upheld at all times, tying the energy budget $P$ irrevocably to the force $\mathbf{f}$ and the specific mass carrier $\mathbf{v}$.

\section{The Mechanics of Interaction: Three Interaction Mechanisms}

The "host force" is not a single interaction; it represents the aggregate of all non-electromagnetic forces acting on the charge carriers. "Non-electromagnetic" here refers to the \textbf{macroscopic} classification. These forces (such as friction or binding) are mechanical in nature within Stage 3, even if their ultimate origin lies in Stage 2 electromagnetic fields.

In the regime of divergent currents, we propose that the interaction of the host force can be categorized into three distinct forms, representing the three mechanical pathways of electromagnetic energy:

\subsection{1. Dissipation: The Bridge to Thermodynamics}
When the host interface acts as friction, it opposes the motion of the charges relative to the lattice.
\begin{itemize}
    \item \textbf{Action:} The field accelerates the fluids, and the relative motion generates heat, exciting high-frequency random mechanical motion.
    \item \textbf{Result:} Energy leaves the electromagnetic domain ($\mathcal{D} > 0 \to P_{\text{host}} > 0$) and dissipates into the \textbf{thermodynamic domain} (heat).
    \item \textbf{Universality:} This occurs in both imperfect conductors (Resistance) and imperfect dielectrics (Dielectric Loss).
\end{itemize}

\subsection{2. Work: The Bridge to Continuum Mechanics}
When the host interface acts as a binding agent, it transmits the force from the charge carriers to the bulk object.
\begin{itemize}
    \item \textbf{Action:} The field pulls the charge; the charge pulls the lattice. If the object moves or is deformed, macroscopic work is performed.
    \item \textbf{Result:} Energy is exchanged between the electromagnetic domain and the \textbf{ordered kinetic domain} of the bulk mass.
    \item \textbf{Universality:} This is the mechanism of electrostatic actuation (motors, dielectric actuators).
\end{itemize}

\subsection{3. Storage: The Bridge to Microscopic Degrees of Freedom}
This interaction fundamentally distinguishes dielectrics from conductors. The host interface acts as an \textbf{elastic restoring force}, permitting reversible energy storage.
\begin{itemize}
    \item \textbf{Action:} The field displaces bound charges against a restoring force.
    \item \textbf{Result:} The dielectric acts as a local sink ($\mathcal{D} > 0$) or source ($\mathcal{D} < 0$) of macroscopic field energy. The energy shifts vertically within the spectrum: from the macroscopic signal to the \textbf{microscopic degrees of freedom} (potential energy in the binding fields).
\end{itemize}

\section{Summary of Contributions}

The analysis of these interactions presents two structural clarifications to resolve ambiguities in the field:

\begin{enumerate}
    \item \textbf{The Identification of Electro-Mechanical Storage:} The physical location of "material potential energy" is identified. It is demonstrated that the energy stored in a dielectric is not a property of the macroscopic field (as implied by $\mathbf{D} \cdot \mathbf{E}$), but resides in the \textbf{variance of the microscopic binding fields}.
    
    \item \textbf{The Theorem of Macroscopic Indeterminacy:} A thought experiment demonstrates that the local deformation force density is \textbf{mathematically indeterminate} from macroscopic field variables alone. It is shown that the partition of the total Lorentz force into "Deformation" (Work) and "Internal Stress" (Structure) depends on the microscopic topology of the material.
\end{enumerate}

\section{Roadmap of Part III}

The logical progression of the analysis is as follows:

\begin{enumerate}
    \item \textbf{Chapter \ref{chap:Conductors}: Free Currents (Conductors).} Analysis of free charges where the constraint is macroscopic, focusing on \textbf{Dissipation} (Resistance) and \textbf{Work} (Actuation).
    
    \item \textbf{Chapter \ref{chap:Dielectrics}: Bound Currents (Dielectrics).} Analysis of bound charges where the constraint is microscopic, introducing the \textbf{Storage} interaction and defining "binding energy" as the energy stored in the high-frequency fields of the host.
    
    \item \textbf{Chapter \ref{chap:ForceDensity}: The Physics of Coupling.} Detailed investigation of the \textbf{Work} interaction, proposing the "Theorem of Indeterminacy" regarding force density coupling and microscopic topology.
\end{enumerate}

\chapter{Free Charges on Conductors}
\label{chap:Conductors}

This chapter analyzes the behavior of free charges in conductors by applying the framework derived in the preceding parts. To address conductors, the general two-fluid model is adapted. For the examples considered herein, the following assumptions are established:

\begin{itemize}
    \item \textbf{The massive lattice ($\mathbf{V}_+$):} The positive ion fluid carries effectively all of the system's mass ($m_+ \gg m_-$). Its velocity defines the bulk mechanical velocity of the material itself ($\mathbf{V}_{\text{mech}} \approx \mathbf{V}_+$). This is the material constituting the \textbf{Continuum Mechanics} domain.
    \item \textbf{The electron gas ($\mathbf{V}_-$):} The mobile electron fluid carries charge but comparatively negligible mass. Its change in mechanical momentum is therefore negligible. The electromagnetic force is practically instantaneously counterbalanced by the host force.
\end{itemize}

\section{Case Study 1: The Rigid Ideal Conductor}
\label{sec:IdealConductor}

The analysis begins with the fundamental baseline: a rigid, ideal conductor placed in a static external electric field $\mathbf{E}_{\text{ext}}$.
\begin{itemize}
    \item \textbf{Ideal ($R=0$):} The electron fluid experiences no resistive drag ($\mathbf{f}_{\text{drag}} = 0$).
    \item \textbf{Rigid ($\mathbf{V}_+ = \mathbf{0}$):} The ion lattice is held fixed in the laboratory frame.
\end{itemize}
This scenario demonstrates a system where the Electromagnetic Domain is dynamically active (charges move) but energetically isolated from the Mechanical/Thermal domains.

\subsection{Dynamics of the Bulk: The Null-Field Condition}

Consider the interior of the conductor where the electron fluid is free to move. The momentum balance for the electron fluid is applied. Given the extreme smallness of the electron mass ($m_- \ll m_+$) and the absence of drag forces, the inertial term is negligible. The force balance is instantaneous:
\begin{equation}
    \underbrace{ \rho_- \frac{d \mathbf{v}_{-}}{dt} }_{\approx 0} \approx \underbrace{ \rho_- \mathbf{E} }_{\text{Lorentz Force}} + \underbrace{ \mathbf{F}_{\text{host}} }_{0} \quad \implies \quad \mathbf{E}_{\text{internal}} \approx \mathbf{0}
\end{equation}
This imposes a strict constraint: the electron fluid redistributes \textit{instantaneously} to ensure the macroscopic electric field is cancelled everywhere within the bulk.

\subsubsection{The Closed Energy Gateway}
The energy balance determines the flow. Although a transient current $\mathbf{J}$ flows to redistribute the charge, it flows through a region where the electric field is effectively zero. Therefore, the power delivered to the matter vanishes, and the \textbf{Interaction Density} is zero:
\begin{equation}
    \mathcal{D} = \mathbf{J} \cdot \mathbf{E} \approx 0
\end{equation}
This confirms that the "energy gateway" is closed. No energy is transferred from the field to the kinetic energy of electrons (since $m \to 0$) or to heat (since $R=0$).

The \textbf{macroscopic energy balance} simplifies to the conservation of field energy alone:
\begin{equation}
    0 = \underbrace{ \mathbf{J} \cdot \mathbf{E} }_{0} = \underbrace{ - \left( \frac{\partial u_{\text{EM}}}{\partial t} + \nabla \cdot \mathbf{S}_{\text{EM}} \right) }_{\text{Field Dynamics}}
\end{equation}
\textbf{Interpretation:} The ideal conductor does not participate in the energy budget. The electromagnetic energy is simply redistributed in space ($\nabla \cdot \mathbf{S}_{\text{EM}}$) to accommodate the new field configuration, but it remains entirely within the electromagnetic domain.

\subsection{Statics of The Boundary: Force Transmission}

The redistribution drives charge to the physical boundaries of the object. Here, the constraint $\mathbf{E}=0$ breaks down just outside the surface, leading to the transmission of force.

Consider the surface layer with accumulated charge density $\sigma$. The external field exerts a massive outward pull on this electron layer:
\begin{equation}
    \mathbf{f}_{\text{Lorentz}} = \rho_{\text{surf}} \mathbf{E}_{\text{local}} \neq \mathbf{0}
\end{equation}
The stability of the surface charge is maintained by the \textbf{host force} ($\mathbf{f}_{\text{host}}$). 

Crucially, this force is not merely a quantum mechanical abstraction. It represents the aggregate of the intense, short-range \textbf{microscopic electromagnetic forces} exerted by the positive ion lattice. These high-frequency binding fields are "filtered out" of the smooth macroscopic field $\mathbf{E}$, and thus reappear in the macroscopic momentum balance as the emergent material force density $\mathbf{f}_{\text{host}}$.

The momentum balance at the surface resolves the transmission mechanism:
\begin{equation}
    \underbrace{ \mathbf{f}_{\text{Lorentz}} }_{\text{Outward Pull on Electrons}} + \underbrace{ \mathbf{f}_{\text{host}} }_{\text{Inward Pull by Lattice}} = \mathbf{0} \quad \implies \quad \mathbf{f}_{\text{host}} = - \rho \mathbf{E}
\end{equation}
This is the physical "coupling": The field pulls the electrons, and the electrons pull the lattice (via the confining host force). Thus, the macroscopic electromagnetic force is perfectly transmitted to the bulk mechanical object without performing internal work. In the present framework: the \textbf{momentum gateway} is wide open (force is transmitted), but the \textbf{energy gateway} is closed (velocity is zero).

\section{Case Study 2: Resistance (The Dissipation Gateway)}
\label{sec:ResistiveConductor}

Material imperfection is subsequently introduced. Consider the same rigid conductor ($\mathbf{V}_+ = \mathbf{0}$), but with a host interface that includes a frictional component.
\begin{itemize}
    \item \textbf{Resistive ($R>0$):} The electron fluid experiences a drag force proportional to its velocity ($\mathbf{f}_{\text{drag}} \propto - \mathbf{V}_-$).
\end{itemize}
This scenario activates the \textbf{dissipation gateway}. Unlike the ideal case, the rearrangement of charge now incurs an energetic cost.

\subsection{Phase I: The Transient Response (The Energy Sink)}

As the external field is applied, the electrons begin to drift. However, they are no longer freely accelerated; they collide with the thermal vibrations of the lattice.

\subsubsection{Force Balance: The Necessity of an Internal Field}
Due to the negligible inertia of the electrons, the driving Lorentz force must be instantaneously balanced by the drag force from the host:
\begin{equation}
    \underbrace{ \rho_- \mathbf{E} }_{\text{Driving Force}} + \underbrace{ \mathbf{f}_{\text{drag}} }_{\text{Resistive Host Force}} \approx \mathbf{0}
\end{equation}

This balance has a crucial implication: to sustain the motion required to shield the field, a non-zero electric field $\mathbf{E}$ must persist \textit{inside} the conductor to overcome the drag. The shielding is not instantaneous; it is a process governed by the conductivity.

\subsubsection{Power Balance: The Joule Heating Mechanism}
This phase is defined by a continuous transfer of energy. The power transfer across the host interface is examined:
\begin{equation}
    P_{\text{host}} = \mathbf{f}_{\text{drag}} \cdot \mathbf{V}_-
\end{equation}
Since the drag force opposes the velocity, this term is strictly negative ($P_{\text{host}} < 0$). In this framework, a negative host power signifies a \textbf{sink action}: energy flows \textit{out} of the electro-mechanical system and \textit{into} the thermodynamic domain.

The \textbf{macroscopic energy balance} reveals the complete process. The divergence of the field energy is identified as the \textbf{Interaction Density} ($\mathcal{D}$), mapping the event of energy conversion:
\begin{equation}
    \underbrace{ \mathcal{D}(\mathbf{x}, t) }_{\text{Interaction Density}} \equiv \underbrace{ - \left( \frac{\partial u_{\text{EM}}}{\partial t} + \nabla \cdot \mathbf{S}_{\text{EM}} \right) }_{\text{Field Energy Loss}} = \underbrace{ \mathbf{J} \cdot \mathbf{E} }_{\text{Gateway}} = \underbrace{ - P_{\text{host}} }_{\text{Heat Generation}}
\end{equation}
This confirms the physical pathway of \textbf{Joule heating}: Electromagnetic forces accelerate the electron fluid against the resistive lattice drag. The electromagnetic field performs work on the fluid ($\mathbf{J} \cdot \mathbf{E} > 0$). There is a local \textbf{sink} in the electromagnetic field energy ($\frac{\partial u_{\text{EM}}}{\partial t} + \nabla \cdot \mathbf{S}_{\text{EM}} < 0$) that is converted instantly into heat.

\subsection{Phase II: Convergence to Static Equilibrium}

Crucially, the presence of resistance does not change the final state, only the time required to reach it. As charge accumulates at the boundaries, the internal counter-field grows until it cancels the external field ($\mathbf{E}_{\text{total}} \to 0$).

When the field vanishes, the driving force stops, and the current ceases ($\mathbf{J} \to 0$). Consequently, the drag force and dissipation both vanish ($P_{\text{host}} \to 0$).

The system settles into the exact same state as the ideal conductor: the surface charges are held in place by the non-dissipative "wall" component of the host force.

\paragraph{Conclusion.}
This comparison illustrates the versatility of the host force $\mathbf{F}_{\text{host}}$. In the static limit, it acts as a \textbf{constraint} (storing potential energy via stress). In the dynamic limit, it acts as a \textbf{drain} (dissipating energy via friction). The unified framework captures both behaviors without modification.

\section{Case Study 3: Actuation (The Work Gateway)}
\label{sec:MechanicalWork}

The conversion of electromagnetic energy into macroscopic mechanical work is now examined.

Consider a rigid conductor carrying a surface charge density $\sigma$, placed in an external electric field. Unlike the previous cases, the entire object (lattice + electrons) is allowed to move with a macroscopic velocity $\mathbf{V}_{\text{mech}}$.

\subsection{The Mechanism of Transmission: The Coupling}

The mechanism of force transmission to a neutral bulk is examined. An electric field cannot directly engage the neutral interior; it couples to the object exclusively due to the charged surface layer.

\subsubsection{Force Balance at the Surface}
As derived in Case 1, the external field exerts a massive outward pull on the surface electrons:
\begin{equation}
    \mathbf{f}_{\text{Lorentz}} = \rho_{\text{surf}} \mathbf{E}_{\text{local}}
\end{equation}
The electrons transmit this pull to the ion lattice via the \textbf{host force} ($\mathbf{f}_{\text{host}}$), which acts as a binding agent. The lattice then transmits this surface stress to the rest of the bulk material via internal mechanical macroscopic stress.

Therefore, the net force density accelerating the object's center of mass is effectively the host force:
\begin{equation}
    \mathbf{f}_{\text{total}} = \mathbf{f}_{\text{host}} \approx \rho_{\text{surf}} \mathbf{E}_{\text{local}}
\end{equation}
This confirms that the host interface is the physical mechanism of force transmission.

\subsection{Dynamics of Rigid Motion}

The energy exchange when this force causes the conductor to move with velocity $\mathbf{V}_{\text{mech}}$ is analyzed.

\subsubsection{Power Balance: Motor vs. Generator Action}
The rate at which the field performs work on the object is determined by the host power density:
\begin{equation}
    P_{\text{host}} = \mathbf{f}_{\text{host}} \cdot \mathbf{V}_{\text{mech}}
\end{equation}
The sign of this term dictates the mode of operation and the direction of energy flow:

\paragraph{1. Motor Mode (The Mechanical Sink):}
If the conductor moves \textit{with} the electric force ($\mathbf{V}_{\text{mech}} \cdot \mathbf{f}_{\text{host}} > 0$), the field performs positive work.
\begin{equation}
    P_{\text{host}} > 0 \quad \implies \quad \text{Energy flows } \textbf{field} \to \textbf{mechanics}
\end{equation}
The host interface acts as a \textbf{local energy sink}. Energy is drawn from the electromagnetic field and converted into the kinetic energy of the bulk matter.

\paragraph{2. Generator Mode (The Mechanical Source):}
If an external mechanical agent pushes the conductor \textit{against} the electric force ($\mathbf{V}_{\text{mech}} \cdot \mathbf{f}_{\text{host}} < 0$), the mechanical system performs work on the field.
\begin{equation}
    P_{\text{host}} < 0 \quad \implies \quad \text{Energy flows } \textbf{mechanics} \to \textbf{field}
\end{equation}
The host interface acts as a \textbf{local energy source}. Kinetic energy is consumed to pump energy back into the electrostatic field (increasing potential).

\subsubsection{The Unified Balance}
In both cases, the macroscopic energy balance holds rigorously using only field and current densities:
\begin{equation}
    \underbrace{ - P_{\text{host}} }_{\text{Mechanical Power}} = \underbrace{ \mathbf{J}_{\text{conv}} \cdot \mathbf{E} }_{\text{Gateway}} = \underbrace{ - \left( \frac{\partial u_{\text{EM}}}{\partial t} + \nabla \cdot \mathbf{S}_{\text{EM}} \right) }_{\text{Field Budget}}
\end{equation}
Here, the ``current'' is the convection current of the moving surface charge ($\mathbf{J}_{\text{conv}} = \rho_{\text{surf}} \mathbf{V}_{\text{mech}}$). This confirms that mechanical work is a distinct manifestation of current-field interaction, mediated by the host interface. The energy exchange is strictly local: it occurs exclusively within the charged surface layer, entering the mechanical domain as a boundary power source.

\section{Case Study 4: Elastic Deformation (The Elastic Boundary)}
\label{sec:ElasticDeformation}

Finally, consider a conductor that is held stationary but is mechanically \textbf{elastic}. The lattice is no longer rigid; it acts as a macroscopic spring that can stretch or compress.
This scenario introduces a new energy destination: \textbf{Macroscopic Elastic Potential Energy} (Strain), which resides within the Continuum Mechanics domain.

\subsection{The Mechanism of Deformation}

When the external field is applied, the surface charges accumulate as before. The electric field exerts an outward pull $\mathbf{f}_{\text{Lorentz}}$ on the surface layer.

\subsubsection{Force Transmission: From Surface to Bulk}
The electrons transmit this pull to the surface of the lattice via the host force. However, since the lattice is elastic, it does not move rigidly. Instead, the surface layer is pulled outward, stretching the atomic bonds in the bulk material.

This creates an internal \textbf{mechanical stress} ($\mathbf{T}_{\text{mech}}$) within the conductor, representing the elastic restoring forces of the stretched atomic bonds. Equilibrium is reached when the internal elastic tension balances the surface pull:
\begin{equation}
    \underbrace{ \mathbf{n} \cdot \mathbf{T}_{\text{mech}} }_{\text{Elastic Stress at Boundary}} + \underbrace{ \mathbf{f}_{\text{host}} }_{\text{Surface Pull}} = \mathbf{0}
\end{equation}

The object physically expands because the surface is being pulled by the host force, placing the entire volume under mechanical tension.

\subsection{Energy Balance: Reversible Storage}

During the transient phase where the material stretches, the surface moves with a deformation velocity $\mathbf{v}_{\text{def}}$.

\subsubsection{The Unified Balance}
The energy exchange is rigorously described by the macroscopic energy balance:
\begin{equation}
    \underbrace{ - P_{\text{host}} }_{\text{Mechanical Power}} = \underbrace{ \mathbf{J}_{\text{conv}} \cdot \mathbf{E} }_{\text{Gateway}} = \underbrace{ - \left( \frac{\partial u_{\text{EM}}}{\partial t} + \nabla \cdot \mathbf{S}_{\text{EM}} \right) }_{\text{Field Budget}}
\end{equation}
Here, the host power density is $P_{\text{host}} = \mathbf{f}_{\text{host}} \cdot \mathbf{v}_{\text{def}}$. The direction of energy flow is clear:

\paragraph{The Reversible Mechanical Sink.}
As the material stretches, it moves \textit{with} the force ($\mathbf{f}_{\text{host}} \cdot \mathbf{v}_{\text{def}} > 0$).
\begin{equation}
    P_{\text{host}} > 0 \quad \implies \quad \text{Energy flows } \textbf{field} \to \textbf{elasticity}
\end{equation}
The host interface acts as a \textbf{local energy sink}. Energy is drawn from the electromagnetic field and stored as potential energy in the stretched atomic bonds of the lattice ($\Delta U_{\text{elastic}} > 0$). There is a local sink of field energy:
\begin{equation}
    -\left(\frac{\partial U_{\text{EM}}}{\partial t} + \nabla\cdot \mathbf{S}_{\text{EM}}\right) > 0.
\end{equation}

\paragraph{Reversibility.}
Unlike the resistive case, this energy transfer is fully reversible. If the external field is reduced, the elastic tension pulls the surface back ($\mathbf{v}_{\text{def}}$ reverses). The host force now does negative work ($P_{\text{host}} < 0$), acting as a \textbf{local energy source} that pumps the stored elastic energy back into the electromagnetic field.

\section*{Synthesis: The Universal Gateway for Free Charges}

This classification of conductor interactions serves as the proof-of-concept for the unified framework. It has been demonstrated that a single electromagnetic law---the macroscopic Lorentz force on the conduction current---is sufficient to explain phenomena as diverse as static shielding, resistive heating, motor action, and elastic deformation.

The strength of this explanation lies in the rigorous definition of the \textbf{host interface} ($\mathbf{F}_{\text{host}}$). This force is not merely a friction term; it is the universal bridge that connects the electrodynamic domain to the rest of physics:
\begin{enumerate}
    \item \textbf{The Bridge to Thermodynamics:} In the \textbf{resistive} case, $\mathbf{F}_{\text{host}}$ acts as a dissipative drag, converting ordered field energy into the random thermal motion of the lattice (Heat).
    \item \textbf{The Bridge to Continuum Mechanics:} In the \textbf{movable} case, $\mathbf{F}_{\text{host}}$ acts as an actuator, converting field energy into the ordered kinetic energy of the bulk mass. In the \textbf{deformable} case, $\mathbf{F}_{\text{host}}$ acts as a stress source, converting field energy into the potential energy of the lattice bonds.
\end{enumerate}

\paragraph{Validation of Consistency.}
In every scenario analyzed, the physical picture remains rigorously defined because the \textbf{Force-Energy Consistency Criterion (FECC)} is strictly preserved.
\begin{itemize}
    \item \textbf{Clear Mass Target:} The electromagnetic force $\mathbf{f}_{\text{Lorentz}}$ always acts on a specific, massive entity (the electron fluid).
    \item \textbf{Defined Velocity:} The power transfer is always defined by the velocity of that specific mass ($\mathbf{J} \cdot \mathbf{E} = \mathbf{f}_- \cdot \mathbf{v}_-$). Note that $\mathbf{v}_+ \approx 0$ in the laboratory frame, so the work is primarily done on the mobile electrons, who then transfer it to the lattice via the host force.
    \item \textbf{Separated Domains:} Energy never "disappears" into ambiguous macroscopic terms. It flows through the single gateway ($\mathbf{J} \cdot \mathbf{E}$) and is immediately accounted for by the host interface as either Heat (Thermodynamic Domain) or Work (Continuum Mechanics Domain).
\end{itemize}

Having validated the framework for free charges, where the "host" acts as a barrier (constraint), drain (dissipation), or coupling (work), the analysis turns to the challenge of \textbf{dielectrics}. Here, it is proposed that next to Work and Dissipation, a third pathway for the energy is introduced: a \textbf{Storage Gateway} for energy stored reversibly in the \textbf{microscopic} high-frequency fields of the dielectric.

\chapter{Dielectrics: Binding Energy}
\label{chap:Dielectrics}

\section{Introduction: The Problem of Reversible Storage}

The investigation now shifts to the physics of bound charges. The preceding analysis of conductors established that the interaction is mediated exclusively by the \textbf{Host Interface} ($\mathbf{F}_{\text{host}}$). This interface acts as a physical gateway, facilitating the transfer of energy \textit{out} of the electromagnetic domain—either dissipating it as heat (Resistance) or converting it into macroscopic mechanical work (Actuation).

In dielectrics, a fundamentally different phenomenon is encountered: \textbf{reversible storage}. When a dielectric material is polarized, energy appears to be stored within the material without being converted into heat or bulk motion. Upon removal of the external stimulus, this energy is returned to the electromagnetic field.

Conventional formulations address this by defining a ``material energy density'' term ($\mathbf{P} \cdot \mathbf{E}$). However, from a rigorous mechanical perspective, this raises a profound ontological question: \textit{Where does this energy physically reside?} 

To resolve this, the definition of the interface is expanded. The \textbf{Host Interface} is recognized to operate in three distinct modes. In addition to the dissipative drag (resistance) and mechanical work (actuation), it provides a conservative restoring force, herein termed \textbf{Binding}. It is argued that this ``Binding Mode'' serves as the macroscopic accounting mechanism for energy that is stored reversibly in the ``hidden'' microscopic degrees of freedom.

\section{The Conventional Definition of Dielectric Energy}
\label{sec:Conventional_Conflation}

Prior to introducing the proposed framework, the standard definition of dielectric energy—corresponding to the Minkowski or Abraham formulations—is briefly reviewed. In standard texts, the energy density of a linear dielectric is given by $u_{conv} = \frac{1}{2}\mathbf{D}\cdot\mathbf{E}$. This macroscopic energy density is traditionally decomposed into vacuum and material components:
\begin{equation}
    u_{conv} = \frac{1}{2}(\varepsilon_0 \mathbf{E} + \mathbf{P}) \cdot \mathbf{E} = \underbrace{\frac{1}{2}\varepsilon_0 E^2}_{\text{Vacuum Field Energy}} + \underbrace{\frac{1}{2}\mathbf{P}\cdot\mathbf{E}}_{\text{Material Energy}}.
\end{equation}
The first term represents the genuine energy of the electromagnetic field (the vacuum stress capability). The physical nature of the second term is the primary subject of this investigation. A deeper comparative analysis of different energy formulations is reserved for Part VI.

\section{The Conceptual Testbed: The Lattice of Isolated Conductors}
\label{sec:Conductor_Lattice_Model}

Dielectric matter is comprised of atoms where electrons are bound to nuclei by microscopic electrodynamic forces, stabilized by quantum mechanical boundary conditions. However, this quantum complexity often obscures the classical mechanism of energy storage. To isolate the electromagnetic mechanism, a \textbf{conceptual testbed} is employed: the \textbf{Lattice of Isolated Conductors}.

The dielectric is modeled microscopically as a regular array of isolated, rigid, ideal conductor regions (representing atoms) embedded in a vacuum. By stripping away inertia, friction, and thermal motion, this ``proxy model'' permits a surgical comparison between the microscopic baseline and the filtered macroscopic description.

\subsection{The Goal: Isolating Electromagnetic Binding}
The specific objective is to isolate the phenomenon of \textbf{electromagnetic binding}. In standard treatments, the macroscopic field energy is absorbed into macroscopic parameters (such as $\varepsilon_r$). The present analysis aims to look one level deeper to identify where \emph{exactly} the energy is stored, and how the averaging process from micro to macro occludes the internal microscopic field energy.

Binding is identified not as a constituent parameter, but as a distinct macroscopic force density ($\mathbf{F}_{\text{bind}}$) of the Host Interface. This is not a new fundamental interaction, but a \textbf{spectral reclassification}. This approach allows the material response to be treated not as a passive modification of the vacuum, but as an active mechanical participant that internally stores energy.

\subsection{The Microscopic Baseline: Passive Quantum Constraints}
\label{sec:PEC_Physics}

The system is defined as a regular 2D lattice of isolated Perfect Electric Conductor (PEC) cylinders embedded in a vacuum. This geometry provides a mathematically clean definition of discrete matter, stripping away the complexity of orbital mechanics while preserving the essential feature of charge confinement.

\begin{figure}[htbp]
    \centering
    \includegraphics[width=0.5\textwidth]{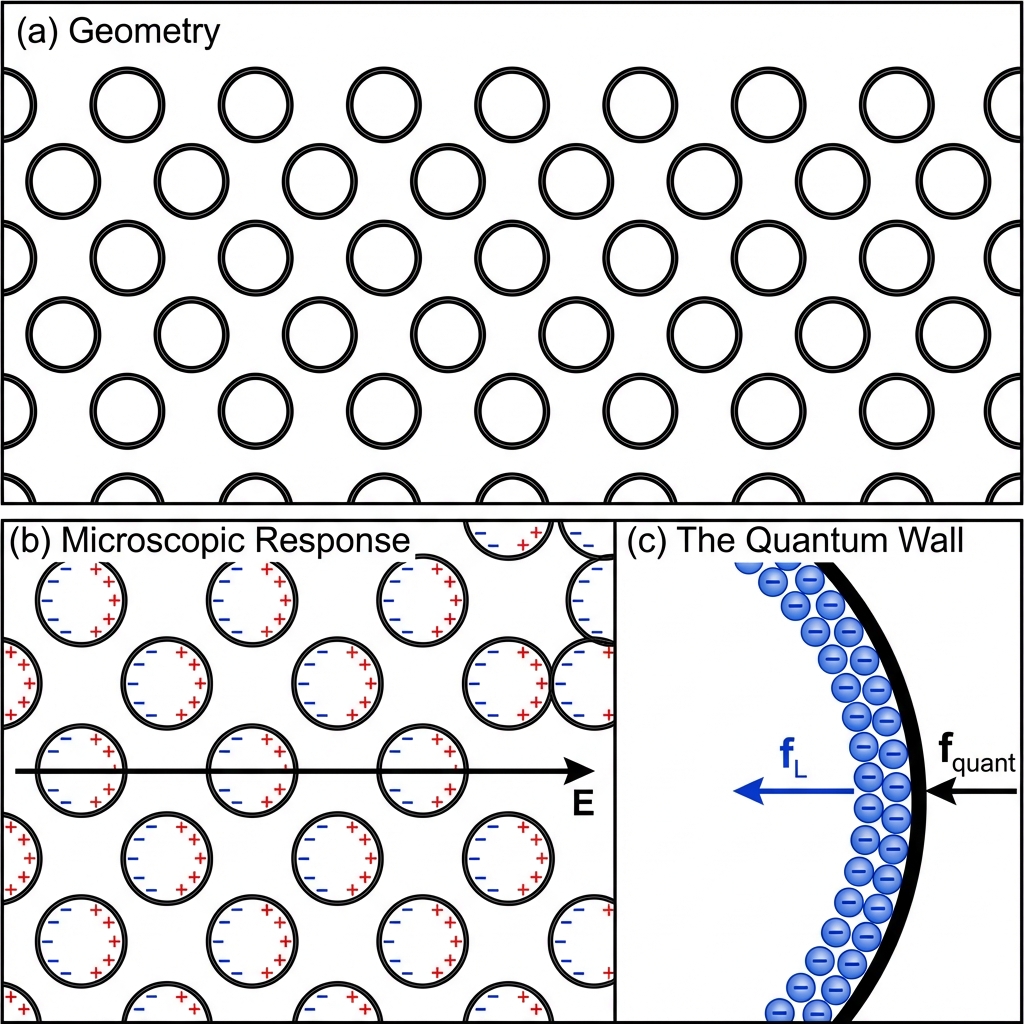}
    \caption{\textbf{The Lattice of Isolated Conductors.}
    \textit{Description:}
    \textbf{(a) Geometry:} A 2D array of isolated circular PEC cylinders separated by vacuum.
    \textbf{(b) Microscopic Response:} An external field is applied. The mobile electron fluid shifts to the surface, creating exact dipoles. The internal field is cancelled ($\mathbf{e}=\mathbf{0}$ inside).
    \textbf{(c) The Quantum Wall (Zoom-in):} A detailed view of the surface. The Lorentz force ($\mathbf{f}_{L}$) pushes the electron fluid outward, balanced exactly by the rigid ``hard wall'' of the conductor boundary ($\mathbf{f}_{quant}$).}
    \label{fig:Conductor_Lattice_Micro}
\end{figure}

Each cylinder is modeled as a fixed ionic lattice containing a mobile electron fluid, behaving as a Perfect Conductor as detailed in Chapter \ref{chap:Conductors}. 
Crucially, the physical origin of the forces holding these surface charges within the conductor boundaries must be identified. Why do the electrons not escape into the vacuum under the influence of the external field?
In a full physical description, the stability of dipoles and matter in general is governed by quantum mechanics. In this classical framework, this quantum stability is treated not as a dynamic force, but as a pre-existing \textbf{quantum mechanical boundary condition}.
The quantum binding potential is modeled as a rigid geometric constraint. It exerts a normal force ($\mathbf{f}_{quant}$) sufficient to confine the charge, but it permits zero displacement perpendicular to the boundary ($v_{\perp} = 0$).

\textbf{Methodological Note:} This analysis does not attempt to simulate the complex orbital dynamics of real atoms as described by QED. This ``hard-wall'' proxy is used to isolate a specific theoretical mechanism: to demonstrate how a \textbf{passive geometric constraint} (which performs no work) transforms into an \textbf{active macroscopic binding force} through the process of averaging.

\subsection{Microscopic Analysis: The Conservative Baseline}
\label{sec:Microscopic_Analysis}

The momentum balance of the isolated conductors determines the microscopic baseline. This behavior parallels that of the perfect conductor described in Chapter \ref{chap:Conductors}. 
This establishes the microscopic view of the system prior to the application of averaging.
The rigorous microscopic momentum balance for the mobile electron fluid within each cylinder is given by:
\begin{equation}
    \underbrace{ \frac{d \mathbf{g}_{mech}}{dt} }_{\to 0 \text{ (Zero-Inertia)}} = \underbrace{\mathbf{f}_{Lorentz}}_{\text{EM Driver}} + \underbrace{\mathbf{f}_{quant}}_{\text{Constraint}}.
    \label{eq:Micro_Momentum_Balance}
\end{equation}
The mass of the charge carriers is assumed to be sufficiently small that the inertial response term is negligible.
It is crucial to distinguish between \textit{negligible} mass and \textit{absent} mass. The charge-mass bridge is strictly adhered to: carriers must possess finite mass to serve as the physical target for the Lorentz force, satisfying the Force-Energy Consistency Criterion (FECC: $P = \mathbf{f} \cdot \mathbf{v}$). However, in this limit, the transient energy transferred into the kinetic reservoir ($\frac{1}{2}mv^2$) during redistribution of internal charges is negligible compared to the energy stored in the field configuration. This allows the mechanical system to be treated as having no kinetic energy storage capacity.

When exposed to an external field, the electrons redistribute instantaneously until the net field inside the conductor is zero ($\mathbf{e}_{\text{micro}} = 0$). This results in a surface charge accumulation (an induced dipole). The Lorentz force pushing the electrons outward is perfectly counterbalanced by the quantum boundary force keeping them confined.

\subsection{Microscopic Energy Balance}
The classification of $\mathbf{f}_{\text{quant}}$ as a passive constraint is energetically decisive. Because the boundary is rigid, the charge carriers cannot move in the direction of the constraint force. Therefore, the quantum boundary condition performs \textbf{zero mechanical work}:
\begin{equation}
    P_{\text{quant}} = \mathbf{f}_{\text{quant}} \cdot \mathbf{v} \equiv 0.
\end{equation}
Furthermore, since the internal field is perfectly screened in the bulk ($\mathbf{e}=0$), the Lorentz force also performs no work at equilibrium ($P_{\text{Lorentz}} = \mathbf{j}_{\text{micro}} \cdot \mathbf{0} = 0$).
Just as we explained in Chapter \ref{chap:Conductors}, the system is in a state of static equilibrium involving intense forces, but the total energy transfer is strictly zero ($P_{\text{micro}} = 0$). When an external field is applied, the electromagnetic field energy is merely redistributed within the electrodynamics domain obeying the conservation law:
\begin{equation}
    0 = \underbrace{ \mathbf{j}_{\text{micro}} \cdot \mathbf{e} }_{0} = \underbrace{ - \left( \frac{\partial u_{\text{em}}}{\partial t} + \nabla \cdot \mathbf{S}_{\text{em}} \right) }_{\text{Field Dynamics}}
\end{equation}

In this microscopic reality, energy is spatially redistributed by the Poynting vector, but it never leaves the electromagnetic domain. There is no ``material potential energy''; there is only the configuration of the electromagnetic field itself.

This raises the central question of this analysis:
\begin{center}
\textit{If the microscopic system inherits only electromagnetic field energy, where is the additional macroscopic energy term originating from? The ``Material Energy'' we identified as $\frac{1}{2}\mathbf{P}\cdot\mathbf{E}$?}
\end{center}

We propose that the answer lies in the process of \textbf{averaging}. It will be demonstrated that the energy is hidden in the variance of the field---a reservoir that the averaging process mathematically converts into the emergent concept of what we call \textbf{``Binding Energy''}.

\section{The Micro-to-Macro Momentum Transition: Emergence of Macroscopic Binding}
\label{sec:Macro_Transition}

The filtering framework derived in Part II is now applied to the Lattice of Isolated Conductors. It will be demonstrated how the averaging process generates an emergent macroscopic force, herein defined as the \textbf{Binding Force}.

The analysis begins with the microscopic momentum balance of the system in the static case:
\begin{equation}
    \mathbf{0} = \underbrace{\rho_{\text{micro}}\mathbf{e}}_{\mathbf{f}_{electric}} + \underbrace{\mathbf{f}_{quant}}_{\text{Constraint}}.
\end{equation}
Using the Reynolds decomposition, the fields and charges are partially separated into high and low frequency components:
\begin{align}
\rho_{\text{micro}} &= \rho + \delta\rho\\
\mathbf{e} &= \mathbf{E} + \delta\mathbf{e}
\end{align}

Inserting this separation into the momentum equation and isolating the low-frequency field components ($\rho \mathbf{E}$) yields:
\begin{equation}
    \rho \mathbf{E} = -\delta\rho \mathbf{E}- \delta\rho \delta\mathbf{e} - \rho \delta\mathbf{e} - \mathbf{f}_{quant}.
\end{equation}
As detailed in Part II, the residual terms are assigned to the microscopic host force:
\begin{equation}
    \mathbf{f}_{host} = -\delta\rho \mathbf{E}- \delta\rho \delta\mathbf{e} - \rho \delta\mathbf{e} - \mathbf{f}_{quant}.
\end{equation}

Next, the momentum balance is averaged to obtain the macroscopic model. This step represents a conceptual transition from microscopic resolution to macroscopic description. The averaging process fundamentally alters the description of the medium; it effectively deletes the high-frequency field components from the explicit system. Moreover, the averaging of these high-frequency parts causes them to emerge as ``new'' properties within the low-frequency spectrum of the macroscopic system---an act of information compression.

Applying the averaging operation to the momentum equation for the low-frequency fields is robust:
\begin{equation}
   \langle \rho \mathbf{E} \rangle = \rho \mathbf{E} 
\end{equation}
The low-frequency fields are, by definition, invariant under the averaging process. 

The high-frequency fields, however, are deeply affected. They are compressed into new emergent low-frequency macroscopic field variables:
\begin{equation}
   \langle -\delta\rho \mathbf{E}- \delta\rho \delta\mathbf{e} - \rho \delta\mathbf{e} - \mathbf{f}_{quant}\rangle = \langle - \delta\rho \delta\mathbf{e} - \mathbf{f}_{quant}\rangle 
\end{equation}
(Note: Cross terms like $\langle \delta\rho \mathbf{E} \rangle$ vanish by orthogonality).

This defines the new \textbf{Macroscopic Host Force}:
\begin{equation}
    \mathbf{F}_{host} = \langle \delta\rho \delta\mathbf{e} - \mathbf{f}_{quant}\rangle 
\end{equation}
This force inherits the interactions of the microscopic fields. It is a new emergent force within the macroscopic domain that did not exist as a unified entity in the microscopic description. It compresses the high-frequency interactions into a low-frequency macroscopic force. This force, representing the high-frequency interactions of the microscopic fields, is termed the \textbf{Binding Force}. It is a constituent of the general Host Force, distinct from resistive or work-related forces.

\subsection{Comparison of Momentum Balance: Micro vs. Macro}

A comparison of the two descriptions of the same system follows.

\textbf{Microscopic View:}
It has been established that all charges are counterbalanced by the boundary quantum forces of the system. The charges accumulate at the boundary of each individual conductor of the lattice. Inside the perfect conductors, there are no electric fields and no forces. The forces are highly local, existing solely at the boundaries.

\textbf{Macroscopic View:}
In the macroscopic system (the dielectric), the charges are distributed over the entire volume. Even though the net charge might be zero inside the volume, there is still a superposition of positive and negative charge fluids everywhere. Furthermore, there is a non-zero electric field inside the dielectric. The charges (positive and negative) experience a constant force which pulls them in opposite directions. The counterbalancing force is the binding force, which is part of the Host Force.

\paragraph{Illustrative Example.}
Consider a square lattice of isolated conductors. The external E-field is assumed constant over the lattice.

Microscopically, currents are only present inside the perfect conductors. The corresponding charges accumulate at the boundary. The microscopic current and corresponding charges can be rewritten using the microscopic polarization field formulation:
\begin{equation}
   \rho_\text{micro} = - \nabla \cdot \mathbf{p} 
\end{equation}
Inside the conductors, there is a superposition of $+$ and $-$ charges, but no net charge. As the E-field is zero, no acceleration occurs inside. As soon as the external field increases, the positive and negative charges inside the perfect conductor redistribute at the boundaries to counterbalance the external field.

The macroscopic system presents a different picture. The averaged charge density is given by:
\begin{equation}
   \rho_{\text{bound}} = \langle \rho_\text{micro}\rangle = \langle -\nabla \cdot \mathbf{p} \rangle  = -\nabla \cdot \mathbf{P} 
\end{equation}
Assuming the macroscopic polarization $\mathbf{P}$ is constant over the square lattice, the averaging has filtered out all high-frequency parts. The macroscopic bound charge density thus appears only at the edges of the macroscopic medium.
However, inside the dielectric, the fluid superposition model applies. The charges experience a force in the direction of the external field, but they are held in position by the \textbf{Binding Force}. This binding force may be envisioned as a spring between the charge fluids, representing the microscopic field interactions. One may imagine the two fluids, $+$ and $-$, shearing against each other, with the binding force resisting until the charges completely counterbalance the external field.

\begin{figure}[htbp]
    \centering
    \includegraphics[width=0.5\textwidth]{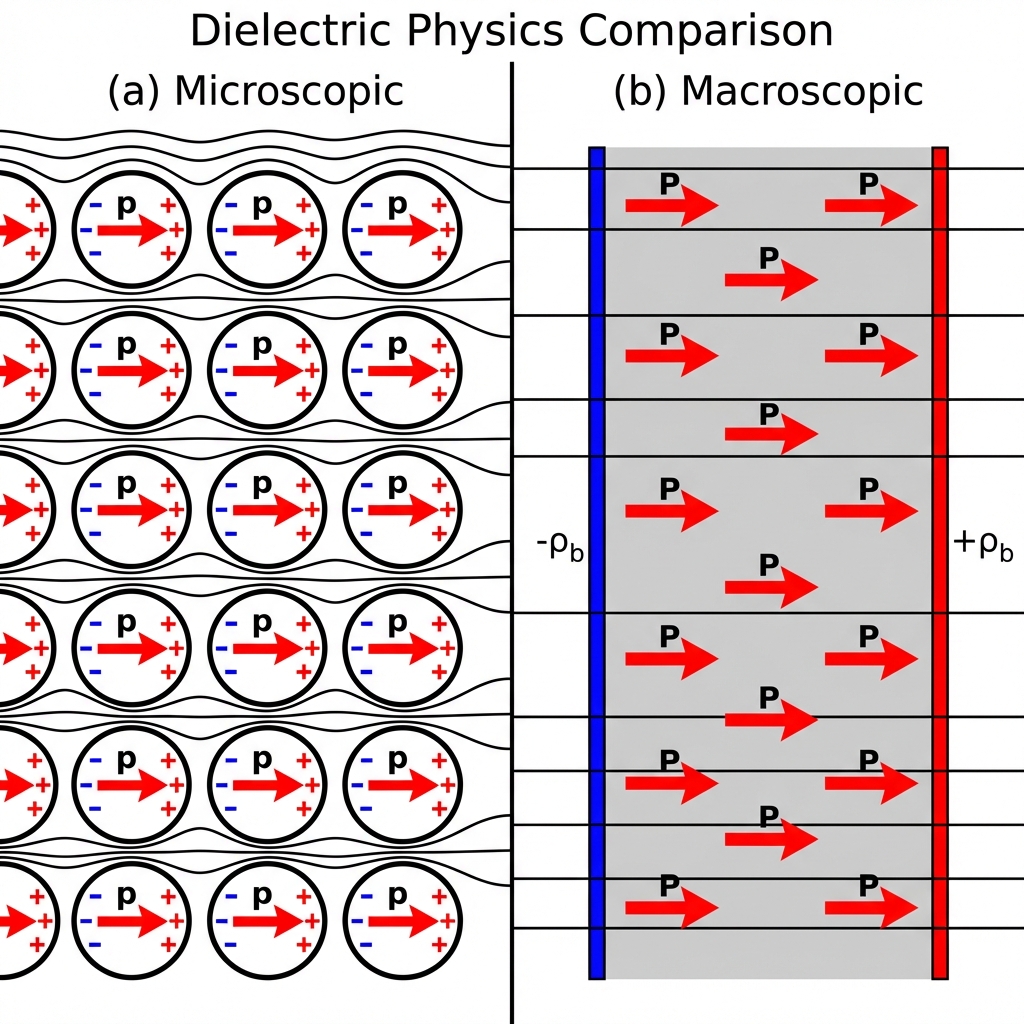}
    \caption{\textbf{The Ontological Shift: Fields, Polarization, and Charges.}
    \textit{Description:}
    \textbf{(a) Microscopic Reality (Input):} Discrete lattice of PEC cylinders. 
    \textit{Fields:} High-frequency E-field lines weave around the obstacles.
    \textit{Polarization:} Localized microscopic dipole vectors $\mathbf{p}$ (red arrows) exist inside each conductor.
    \textit{Charges:} Surface charges accumulate on the boundaries ($+$ red, $-$ blue) to screen the internal field.
    \textbf{(b) Macroscopic Reality (Output):} Homogenized continuous medium.
    \textit{Fields:} Smooth, uniform macroscopic E-field lines.
    \textit{Polarization:} A constant macroscopic polarization field $\mathbf{P}$ (red arrows) permeates the volume.
    \textit{Charges:} The local surface charges are filtered out, reappearing as continuous bound charge layers ($\pm \rho_b$) accumulation only at the macroscopic boundaries (blue and red strips).}
    \label{fig:Ontological_Shift}
\end{figure}

\section{Energy Balance: Microscopic vs. Macroscopic}

A comparative analysis of the energy balances of the microscopic and macroscopic systems is now conducted.

\subsection{Microscopic Energy Balance}
Microscopically, the energy balance is straightforward. The energy is redistributed, but remains completely within the electromagnetic fields. As the E-field inside the perfect conductor is zero, the instantaneous work is zero:
\begin{equation}
    0 = \underbrace{ \mathbf{j}_{\text{micro}} \cdot \mathbf{e} }_{0} = \underbrace{ - \left( \frac{\partial u_{\text{em}}}{\partial t} + \nabla \cdot \mathbf{S}_{\text{em}} \right) }_{\text{Field Dynamics}}
\end{equation}

\subsection{Macroscopic Energy Balance}
Macroscopically, the energy equation follows from multiplying the momentum equation with the corresponding velocities of each fluid, as derived in Part II.
The macroscopic current density follows directly from averaging the microscopic current density:
\begin{equation}
   \mathbf{J} = \langle \mathbf{j}_\text{micro}\rangle = \langle \frac{\partial \mathbf{p}}{\partial t} \rangle  = \frac{\partial \mathbf{P}}{\partial t} 
\end{equation}
As the external field increases, a current is induced in the macroscopic dielectric. This is the low-frequency component of the microscopic currents inside the perfect conductors. The positive and negative charges inside the dielectric, which are counterbalanced by the binding force, move in opposite directions and perform work against this counterbalancing binding force.

Since the macroscopic current density is non-zero and the E-field is non-zero, the energy equation of the macroscopic system is:
\begin{equation}
\boxed{
\begin{aligned}
    \underbrace{ \left( \frac{\partial U_{\text{mech}}}{\partial t} + \nabla\cdot\mathbf{S}_{\text{mech}} \right) - P_{\text{host}} }_{\substack{\text{Total Mechanical Response}}}
    \quad &\overset{\text{Physical Coupling}}{\textbf{=}} \quad
    \underbrace{ \mathbf{J}\cdot\mathbf{E} }_{\substack{\textbf{The Energy Gateway} \\ (P_{\text{Lorentz}})}} \\
    &\overset{\text{Math. Identity}}{=} \quad
    \underbrace{ -\left(\frac{\partial U_{\text{EM}}}{\partial t} + \nabla\cdot \mathbf{S}_{\text{EM}}\right) }_{\substack{\text{Macroscopic EM Response}}}
\end{aligned}
}
\end{equation}

The charges experience a force in the direction of the external field. As the external field increases, the charges start to move, a current is created, and the moving charges perform work against the binding force. As the external field increases, electromagnetic work is done and there is a local sink inside the dielectric. Energy leaves the separate macroscopic electromagnetic domain and is transferred into the binding energy ($U_{\text{binding}}$), corresponding to the work done by the binding force.
\begin{equation}
    \underbrace{ \mathcal{D}(\mathbf{x}, t) }_{\text{Interaction Density}} \equiv \underbrace{ - \left( \frac{\partial u_{\text{EM}}}{\partial t} + \nabla \cdot \mathbf{S}_{\text{EM}} \right) }_{\text{Field Energy Loss}} < 0
\end{equation}

If the external field decreases, the current flows in the opposite direction. Work is done \textit{by} the binding forces on the charges, creating a source of electromagnetic field energy.
The energy leaves the binding reservoir and returns to the electromagnetic domain:
\begin{equation}
\boxed{
\begin{aligned}
    \underbrace{     \frac{\partial U_{\text{binding}}}{\partial t}  = - P_{\text{host}} }_{\substack{\text{Binding Energy Rate}}}
    \quad &\overset{\text{Physical Coupling}}{\textbf{=}} \quad
    \underbrace{ \mathbf{J}\cdot\mathbf{E} }_{\substack{\textbf{The Energy Gateway} \\ (P_{\text{Lorentz}})}} \\
    &\overset{\text{Math. Identity}}{=} \quad
    \underbrace{ -\left(\frac{\partial U_{\text{EM}}}{\partial t} + \nabla\cdot \mathbf{S}_{\text{EM}}\right) }_{\substack{\text{Macroscopic EM Response}}}
\end{aligned}
}
\end{equation}

\subsection{Conclusion on the Conceptual Testbed}

The microscopic and macroscopic systems are distinct descriptions of the same reality at different resolutions.

An artificial system has been constructed to illustrate the emergence of a new macroscopic force—defined as the \textbf{Binding Force}—to represent microscopic field interactions. The binding force is part of the Host Force and thus appears as \textbf{non-electromagnetic} in the macroscopic system. Within the macroscopic system, the ``electromagnetic domain'' is exhaustively defined by the signal energy $U_{EM}(\mathbf{E}, \mathbf{B})$. The high-frequency fields are not part of the system anymore; they have been filtered out.

As shown in Part II, microscopically, the field energy clearly separates into high and low frequency parts:
\begin{equation}
    \langle u_{total} \rangle = \underbrace{ U_{EM}(\mathbf{E}, \mathbf{B}) }_{\text{Macroscopic Field Energy}} + \underbrace{ \langle \delta u(\delta\mathbf{e}, \delta\mathbf{b}) \rangle }_{\text{High-Frequency Field Variance}}.
\end{equation}

The fluctuation energy $\langle \delta u \rangle$ is not part of the macroscopic system anymore. In the macroscopic equations, this energy appears as the accumulated work done against the \textbf{Host Force}. Since $\mathbf{F}_{host}$ is classified as a non-electromagnetic constitutive force in the macro-frame, $U_{\text{binding}}$ acts as a \textbf{non-electromagnetic potential energy reservoir}. It is a new energy bucket, distinct from the field energy $U_{EM}$, that exists solely to balance the macroscopic budget.

\section{Generalization to Real Matter: The Validity of the Isomorph}
\label{sec:RealMatter}

The analysis has utilized a specific model: a lattice of ideal macroscopic conductors. One may question whether results derived from this "artificial" texture apply to real dielectric matter (atoms and molecules).

Crucially, the validity of this generalization relies on the \textbf{Constraint Topology} established in Part I (Chapter \ref{chap:FreeCharges}).

\subsection{The Universality of Workless Constraints}
In Section \ref{sec:IdealConstraints}, it was established that the stability of real microscopic matter (atoms) is maintained by fundamental non-electromagnetic forces (quantum constraints, $\mathbf{f}_Q$). A key axiom of the microscopic baseline is that these constraints are \textbf{workless}:
\begin{equation}
    P_{Q} = \mathbf{f}_Q \cdot \mathbf{v} \equiv 0.
\end{equation}
They maintain the structure of the atom (preventing collapse) but do not participate in the energy exchange. Consequently, the energy transfer to the charge carriers is \textbf{exclusively electromagnetic}:
\begin{equation}
    P_{\text{total}} = \mathbf{j} \cdot \mathbf{e}.
\end{equation}

Now, consider the "Lattice of Conductors" analyzed in this chapter. The charges on the conducting islands are confined by the boundary of the metal. These boundaries act as rigid, workless constraints. They maintain the structure of the artificial dielectric but do not perform work on the charges.

\subsection{The Isomorphism}
Therefore, the "Lattice of Conductors" is a rigorous \textbf{Classical Isomorph} of real matter.
\begin{itemize}
    \item \textbf{Real Matter:} Charges confined by a specific potential well (Quantum Laws), acting as a workless constraint.
    \item \textbf{Isomorph:} Charges confined by a specific potential well (Conducting Boundaries), acting as a workless constraint.
\end{itemize}

Since the macroscopic energy laws depend \textit{only} on the topology of the energy exchange (Exclusively Electromagnetic + Reversible), the detailed shape of the underlying potential well "filters out" in the averaging process. Whether the restraining force is a "hard wall" (conductor) or a "soft spring" (atom) is irrelevant to the macroscopic structure of the energy balance.

Both systems exhibit the same fundamental phenomenology:
\begin{enumerate}
    \item \textbf{No Dissipation:} The constraints are ideal ($R=0$).
    \item \textbf{Reversible Storage:} Work done displacement against the constraint is stored as potential energy in the field configuration.
\end{enumerate}

Thus, the "Storage Gateway" mechanism derived here is not merely an analogy; it is the structurally correct macroscopic description for any system of bound charges stabilized by workless constraints.

\subsection{The Rotational Sibling: Orientational Polarization}
\label{sec:Orientational_Polarization}

The discussion thus far has focused on ``induced polarization''—the stretching of neutral atoms (shear mode). However, a second fundamental mechanism exists: \textbf{orientational polarization}.

Polar molecules (like water) possess a \textbf{permanent electric dipole moment} ($\mathbf{p}_0$) due to their fixed chemical asymmetry. These are the electric analogs of the permanent magnetic spins encountered in Chapter \ref{part:MagneticResponse}.
In the absence of a field, these dipoles are randomly oriented, summing to zero. When a field is applied, microscopically, the dipoles rotate into the direction of the external field. Again, microscopically, next to the quantum boundary conditions, there are only electrodynamic fields at play. Whether the dipoles stretch or rotate, microscopically this establishes a change in the high-dynamic electromagnetic fields.

\paragraph{Macroscopic Output (The Filtering of Rotation).}
Crucially, the averaging process acts as a low-pass filter, stripping away the high-frequency geometric detail of the ``pivot.'' Macroscopically, the \textbf{rotation disappears}. The observer sees only the net transport of charge—a simple \textbf{longitudinal current} ($\mathbf{J} \parallel \mathbf{E}$). The angular origin of the motion is information that has been discarded.

This rotation of \textbf{constant microscopic dipoles} generates a macroscopic current density that is physically indistinguishable from the ``stretching'' current:
\begin{equation}
    \mathbf{J}_{reorient} = \frac{\partial \mathbf{P}}{\partial t}.
\end{equation}

The work done by the macroscopic binding force represents the microscopic change of high-dynamic fields.

\section{Case Study 1: Ideal Polarization (The Storage Gateway)}
\label{sec:IdealDielectric}
Consider now two cases of macroscopic polarization from the viewpoint of the derived framework. This extends the analysis from Chapter \ref{chap:FreeCharges}, considering dielectrics in parallel to conductors.

The analysis begins with the ideal baseline: a rigid, lossless dielectric. The bound charges are confined to their microscopic unit cells. This confinement creates a restorative force, but no friction.

\subsection{Phase I: The Transient Response (Charging)}

When the external field is applied, the bound electron clouds shift relative to the nuclei. This is not a free drift; it is a displacement against the internal restoring binding force. The driving Lorentz force is balanced exclusively by the \textbf{Binding Force} ($\mathbf{f}_{\text{bind}}$).

The force balance for the inertialess bound charge fluid is:
\begin{equation}
    \underbrace{ \rho_- \mathbf{E} }_{\text{Driving Force}} + \underbrace{ \mathbf{f}_{\text{bind}} }_{\text{Restoring Force}} \approx \mathbf{0}
\end{equation}
This balance dictates the state of \textbf{Polarization}. The charge fluid shift is counterbalanced by the restoring force ($\mathbf{f}_{\text{bind}}$). 

\subsubsection{Power Balance: Reversible Storage}
In the dynamic process, this displacement requires work. The power transfer is:
\begin{equation}
      \frac{\partial U_{\text{binding}}}{\partial t} = -P_{\text{bind}} = -\mathbf{f}_{\text{bind}} \cdot \mathbf{V}_-
\end{equation}
Crucially, this force is conservative. The energy is not lost to heat; it is stored.

The energy balance reveals the process:
\begin{equation}
    \underbrace{ \frac{\partial U_{\text{binding}}}{\partial t} }_{\text{Potential Energy Storage}} = \underbrace{ \mathbf{J}_P \cdot \mathbf{E} }_{\text{Power from Field}} = \underbrace{ - \left( \frac{\partial u_{\text{EM}}}{\partial t} + \nabla \cdot \mathbf{S}_{\text{EM}} \right) }_{\text{Loss of Macroscopic Field Energy}}
\end{equation}
This confirms the physical pathway of \textbf{Dielectric Storage}: Energy is drawn from the macroscopic field budget ($u_{\text{EM}}$) and is stored reversibly in the binding energy. 

The FECC is upheld: All forces, masses, and velocities of all interactions are clearly defined. 

\section{Case Study 2: Lossy Polarization}
\label{sec:LossyDielectric}

Material imperfection is now introduced. In a real dielectric, the shifting electron clouds interact with the thermal vibrations of the lattice. The response is no longer purely elastic; it includes friction. The process is conceptually identical to the resistive heating explained in Chapter \ref{chap:FreeCharges}.
This scenario activates \textbf{both} non-electromagnetic interfaces simultaneously: the storage gateway and the dissipation gateway.

\subsection{Phase I: The Transient Response (The Split Stream)}

As the polarization current flows ($\mathbf{J}_P = \partial \mathbf{P}/\partial t$), the charges experience two opposing forces: the binding force ($\mathbf{f}_{\text{bind}}$) and the drag ($\mathbf{f}_{\text{drag}}$).

The driving Lorentz force must overcome both the conservative restoring force and the dissipative drag force:
\begin{equation}
    \underbrace{ \rho_- \mathbf{E} }_{\text{Driving Force}} + \underbrace{ \mathbf{f}_{\text{bind}} }_{\text{Binding}} + \underbrace{ \mathbf{f}_{\text{drag}} }_{\text{Friction}} \approx \mathbf{0}
\end{equation}

\subsubsection{Power Balance: Partitioning the Energy}
The total power delivered by the field ($\mathbf{J}_P \cdot \mathbf{E}$) is now partitioned into two distinct physical channels:

\begin{equation}
    \mathbf{J}_P \cdot \mathbf{E} = \underbrace{ -P_{\text{stiff}} }_{\text{Stored}} + \underbrace{ -P_{\text{drag}} }_{\text{Dissipated}}
\end{equation}

\begin{itemize}
    \item \textbf{The Binding Channel (Storage):} The work done against binding ($P_{\text{bind}} = \mathbf{f}_{\text{bind}} \cdot \mathbf{V}_-$) is reversibly stored as polarization energy.
    \item \textbf{The Drag Channel (Heat):} The work done against friction ($P_{\text{drag}} = \mathbf{f}_{\text{drag}} \cdot \mathbf{V}_-$) is irreversibly dissipated into the lattice as heat.
\end{itemize}

This precisely mirrors the physics of the resistive conductor, but with the addition of the storage term. The \textbf{dielectric loss} is simply the host power component of the polarization process.

\subsection{Phase II: The Static Equilibrium}

When the field stabilizes, the polarization stops changing ($\mathbf{V}_- \to 0, \mathbf{J}_P \to 0$).
\begin{itemize}
    \item \textbf{Friction Vanishes:} Since velocity is zero, the drag force and host power vanish ($P_{\text{host}} \to 0$). The heating stops.
    \item \textbf{Tension Remains:} The displacement remains. The binding force ($\mathbf{f}_{\text{bind}}$) persists to balance the static electric field, maintaining the stored energy.
\end{itemize}

\section{Conclusion}

The unified framework successfully elucidates the physics of dielectric energy. For free charges and currents on conductors, two energy gateways were identified: work and dissipation. This chapter introduced the third: \textbf{Storage}. The Binding Force, as part of the Host Interface, counterbalances the microscopic forces and stores energy in a separate Binding Energy reservoir. It physically stores the electrodynamic field energy that has effectively left the macroscopic domain.

Force density in dielectric materials is the subject of the subsequent chapter. Having identified the dissipation and storage energy gateways, the analysis will extend to the gateway of \textbf{Work}—the physical coupling of the macroscopic electromagnetic domain to the Continuum Mechanics domain.

\chapter{Force Density: The Hierarchy of Deformation}
\label{chap:ForceDensity}

\section{Introduction: The Indeterminacy of Local Force}
\label{sec:ForceDensity_Intro}

Having addressed the mechanisms of energy storage in the previous chapter, the inquiry now turns to the complementary question of mechanical force: \textit{How does the electromagnetic field exert physical force on a dielectric medium?}

This question resides at the center of a historical schism in macroscopic electrodynamics. While the polarization field $\mathbf{P}$ is universally accepted as the source of bound charge ($\rho_b = -\nabla \cdot \mathbf{P}$), there is no consensus on the mechanical force acting on this charge. Standard texts present competing expressions---most notably the energy-based descriptions (Korteweg-Helmholtz) or point dipole approaches like the Einstein-Laub or Kelvin force density ($(\mathbf{P} \cdot \nabla)\mathbf{E}$). The simple Lorentz force density ($\rho_b \mathbf{E}$) is often dismissed as a naive candidate for deformation.

In this chapter, this ambiguity is addressed through a rigorous forensic reconstruction of the deformation mechanism. The investigation proceeds in four stages:

\begin{enumerate}
    \item \textbf{The Demonstration of Indeterminacy:} First, it is proven that the ``Force Density'' cannot be derived from macroscopic fields alone. Using the ``Laminated Conductor'' thought experiment, it is demonstrated that two materials with identical macroscopic fields experience radically different internal stresses. This result implies that energy-based methods are mechanically ill-posed if they attempt to derive force solely from a macroscopic energy functional without topological side-channel information.
    
    \item \textbf{Force Partitioning:} Second, it is proposed that the \textbf{Total Macroscopic Force} (Lorentz Force) partitions into two mechanically distinct components: the \textbf{Inner Dipole Force} (counter-balanced by internal quantum structures) and the \textbf{Continuum Deformation Force} (counter-balanced by neighboring lattice sites). It is proposed that only the latter is relevant for the mechanical deformation of macroscopic objects.
    
    \item \textbf{Validation of Point Dipole Methods:} Third, it is clarified that definition-based approaches (such as Einstein-Laub or Kelvin) remain valid. Unlike variational methods, they implicitly assume a specific microstructure (the point dipole), which qualifies as valid ``Side-Channel Information'' regarding the material topology.
    
    \item \textbf{The Lattice Adaptation Term:} Finally, the Point Dipole approach is refined. A \textbf{Hypothesis of Microstructure} is proposed to account for the finite size of real atoms and the exclusion of self-fields. This yields a \textbf{Lattice Adaptation Term} that corrects the standard formulas for the self-field exclusion effects inherent in solid matter.
\end{enumerate}

\section{Analysis of Macroscopic Indeterminacy: The Laminated Conductor Experiment}
\label{sec:proof_indeterminacy}

This section presents a demonstration that the macroscopic force density relevant for material deformation cannot, in principle, be calculated solely from macroscopic fields. It is proven that the deformation force is intrinsically dependent on the microscopic topology of the material---information that is mathematically eliminated during the spectral filtering process.

To demonstrate this, we construct a thought experiment using a composite material composed of simple, ideal components. By constructing the dielectric out of perfect conductors and vacuum, ambiguity regarding the microscopic physics is avoided, as the behavior of free charges on conductors is axiomatically established.

\subsection{The Conceptual Experiment}

Consider a simplified 2D composite material consisting of alternating layers of \textbf{Perfect Electric Conductor (PEC)} and an inert, non-conductive vacuum (or substrate). 
This material is subjected to a uniform external electric field $\mathbf{E}_{\text{ext}}$. We assume the conductors are ideal: free charges redistribute instantaneously to counteract the external field ($\mathbf{E}_{\text{metal}} = 0$).
The local charge distribution is described by a \textbf{microscopic polarization field} $\mathbf{p}_{\text{micro}}$. 

The internal physics is now analyzed for two distinct topological configurations of this composite, designated \textbf{Topology I} and \textbf{Topology II}.

\subsubsection{Topology I: The Continuous Laminate}
In the first configuration, the conducting slabs are arranged as long, continuous strips oriented \textbf{parallel} to the external field ($\mathbf{E}_{\text{ext}} \parallel \text{strips}$). The electric field lines run along the vacuum gaps like guides.

We assume the layers are thin compared to the macroscopic dimensions of the block, but long enough that edge effects at the ends of the sample can be neglected for the local analysis.

\paragraph{Analysis of Topology I: The Continuous Laminate (Stress Free).}
The local physical state of the continuous strips is governed by the induction condition for perfect conductors ($\mathbf{e}_{\text{metal}}=0$).

\begin{itemize}
    \item \textbf{Charge Accumulation:} In this parallel orientation, the local field has no component normal to the lateral interfaces of the strips. Consequently, the boundary condition is satisfied without any charge accumulation on the internal faces ($\sigma_{\text{interface}} = 0$). Free charges accumulate solely at the \textbf{physical boundaries of the macroscopic sample}.
    \item \textbf{Local Force Density:} Since the internal interfaces are electrically neutral, the local Lorentz force in the bulk is identically zero ($\mathbf{f} = \sigma \mathbf{e} = 0$).
    \item \textbf{Mechanical Result:} The interior of the block is mechanically inert. The electromagnetic force is concentrated entirely at the physical boundaries, putting the strips in tension but generating no interaction forces between them.
\end{itemize}

\subsubsection{Topology II: The Discontinuous Laminate (Checkerboard)}

The microstructure is now modified while maintaining the same material composition. Consider cutting the continuous strips into small, discrete squares of length $l \ll V^{1/3}$ (where $l$ is still macroscopic compared to atomic scales). The rows are then offset to create a ``Checkerboard'' pattern of isolated conducting squares embedded in the vacuum matrix.

This topology is conceptually identical to the ``Lattice of Isolated Conductors'' introduced in Chapter \ref{chap:Dielectrics}, but applied here to a deformation problem.

\begin{itemize}
    \item \textbf{Charge Accumulation:} Because each square is electrically isolated, charges cannot flow to the sample boundaries. Instead, positive and negative charges accumulate on the \textbf{upstream and downstream faces of every individual constituent}.
    \item \textbf{Local Force Density:} This creates a dense array of microscopic dipoles. The positive face of one square attracts the negative face of its neighbor across the vacuum gap. The force is no longer confined to the boundary; it is distributed homogeneously throughout the entire volume.
    \item \textbf{Mechanical Result:} The material experiences a substantial \textbf{Internal Stress} acting on every constituent element. The deformation force is a bulk density.
\end{itemize}

\begin{figure}[htbp]
    \centering
    \includegraphics[width=0.4\textwidth]{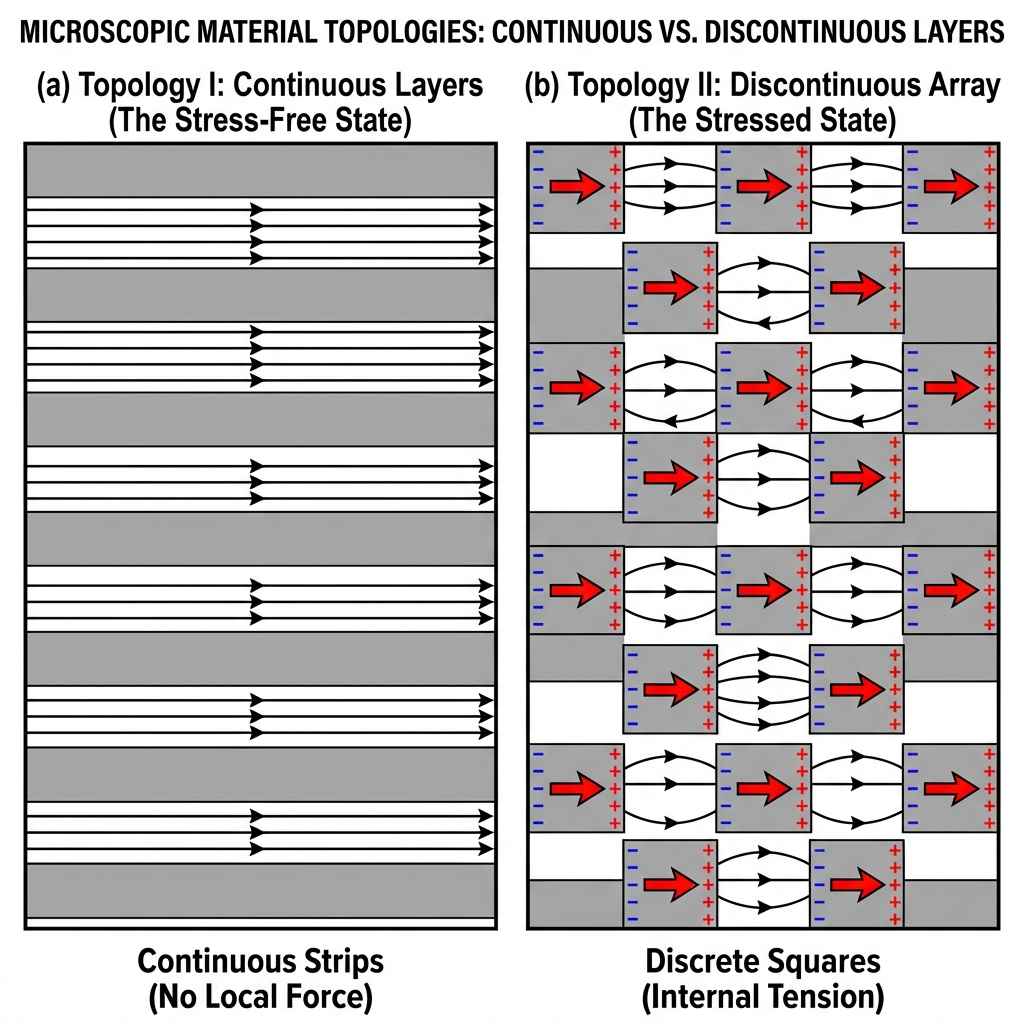}
    \caption{\textbf{Macroscopic Indeterminacy: Two Topologies, One Field.}
    \textit{Description:}
    \textbf{(a) Topology I: The Continuous Laminate (Stress Free).} Continuous metal strips guide the field. Locally, the interfaces are charge-free, and there is no microscopic polarization or force inside the metal.
    \textbf{(b) Topology II: The Discontinuous Laminate (Internal Tension).} Discrete metal squares. Charges accumulate on the vertical faces ($\pm$), creating internal dipoles (red arrows) and strong field bridging across the gaps (tension).
    \textit{Result:} Both systems appear identical macroscopically (same macroscopic $\mathbf{P}$ and $\mathbf{E}$), but their local stress states are opposite. This proves that stress cannot be derived from $\mathbf{P}$ and $\mathbf{E}$ alone.}
    \label{fig:Indeterminacy_Laminate}
\end{figure}
 
\subsection{The Macroscopic Equivalence}

Having established that the internal physical states are distinct, the macroscopic filter is now applied to demonstrate that they are indistinguishable to an external observer.

From the perspective of a macroscopic observer, the discrete layers are homogenized into a single dielectric block described by average fields. The macroscopic polarization $\mathbf{P}(\mathbf{x})$ is defined generally as the convolution of the microscopic distribution with a smoothing kernel $K(\mathbf{x})$:
\begin{equation}
    \mathbf{P}_{\text{macro}}(\mathbf{x}) = \int K(\mathbf{x} - \mathbf{r}) \, \mathbf{p}_{\text{micro}}(\mathbf{r}) \, d^3r.
\end{equation}

By construction, Topology II is derived from Topology I merely by a microscopic segmentation and spatial shift of the conducting matter ($\Delta \mathbf{r} \sim l$). Since the spectral filter $K$ is smooth over scales much larger than the unit cell ($L \gg l$), this high-frequency spatial variation is strictly filtered out:
\begin{equation}
    \mathbf{P}_{\text{macro}}^{\text{I}} \equiv \mathbf{P}_{\text{macro}}^{\text{II}}.
\end{equation}

Since the macroscopic polarization fields are identical---and the external boundary conditions are identical---the macroscopic Maxwell equations yield identical solutions for both systems.

To an observer measuring only the bulk response vectors $\mathbf{P}$ and $\mathbf{E}$, the two materials are \textbf{field-wise indistinguishable}. They both appear as a linear dielectric with effective susceptibility $\chi_{\text{eff}}$. Yet, as shown above, one is subject to a distributed bulk force (Topology II), while the other experiences zero internal stress (Topology I).

\subsection{Conclusion: The Theorem of Macroscopic Indeterminacy}

This thought experiment reveals a fundamental epistemological limit of continuum electrodynamics, which is formalized here as the \textbf{Theorem of Macroscopic Indeterminacy}.

Two physical systems have been constructed where the macroscopic input variables are identical, yet the mechanical output is radically different:
\begin{itemize}
    \item \textbf{Macroscopic Inputs (State Variables):} Identical polarization and field ($\mathbf{P}_I = \mathbf{P}_{II}, \mathbf{E}_I = \mathbf{E}_{II}$).
    \item \textbf{Microscopic Output (Stress):} The internal force density is non-zero in Topology II ($\mathbf{f} \neq 0$) and zero in Topology I ($\mathbf{f} = 0$).
\end{itemize}

\paragraph{The Consequences.}
This contradiction proves that the local deformation force is \textbf{not a single-valued function of the macroscopic fields}. Any proposed universal formula for the force density, $\mathbf{F}_{\text{def}}(\mathbf{P}, \mathbf{E})$, is mechanically ill-posed. It attempts to map a single input state to two mutually exclusive output states.
\begin{itemize}
    \item If the formula predicts internal stress (matching Topology II), it implies a ``Fictitious Force'' for Topology I.
    \item If it predicts zero stress (matching Topology I), it omits the physical force in Topology II.
\end{itemize}

Therefore, \textbf{macroscopic electrodynamics is mechanically indeterminate regarding deformation.} The calculation of local stress requires explicit ``Side-Channel Information'' regarding the microscopic topology to correctly partition the total electromagnetic force into deformation versus internal constraint.

\paragraph{Limitations of Energy-Based Methods (Korteweg-Helmholtz).}
This result specifically highlights the boundaries of the standard ``Energy Methods'' derived from thermodynamics (e.g., the Korteweg-Helmholtz variational principle). These methods operate on the premise that a force density can be derived by differentiating a macroscopic energy functional $W[\mathbf{P}, \mathbf{E}]$.
\begin{itemize}
    \item The Laminated Conductor experiment demonstrates that two systems can possess identical macroscopic energy states $W$ yet exhibit distinct mechanical stress distributions.
    \item Therefore, the mapping $W \to \mathbf{f}$ is not unique without additional topological information.
\end{itemize}
The Energy Method implicitly assumes that the mechanical force density is determined solely by the local variation of macroscopic parameters. A deeper analysis of the variational mechanism is conducted in \textbf{Part VI, Chapter \ref{chap:VariationalMethod}} (The Variational Connection).

\paragraph{The Validity of Point Dipole Approaches.}
It is crucial to distinguish this failure from the status of \textbf{Point Dipole Approaches} (such as the Einstein-Laub or Kelvin force densities). Unlike the energy method, these formulations are \textbf{not invalidated} by the theorem of indeterminacy.
\begin{itemize}
    \item These methods do not attempt to derive force from macroscopic fields alone.
    \item Instead, they start with an explicit model of the microstructure: the \textbf{Point Dipole}.
\end{itemize}
Because they satisfy the requirement of side-channel information by positing a specific microscopic model, they are valid candidates for the force density of materials that match that model. We will consider these approaches in detail in the following sections.

\section{The Tripartite Mechanical Partition}
\label{sec:ForcePartitioning}

Having established the theorem of macroscopic indeterminacy, we now consider the exact microstructure of common dielectrics. We proceed to investigate the point dipole force density approaches. Moreover, we maintain that the total macroscopic Lorentz force of the Vacuum Tensor remains valid as the total force density. However, this total force density partitions into different counterbalancing forces and different corresponding energy gateways.

The \textbf{Total Force Budget} ($\mathbf{f}_{\text{total}}$) partitions into three distinct physical roles, corresponding to the three energy gateways identified in the previous chapters:

\begin{enumerate}
    \item \textbf{Storage (Internal Constraint Forces):} Forces that stretch the individual dipoles against their internal quantum binding. These sum to zero over the bulk but create internal dipole stress. This corresponds to the Host Interface gateway that channels energy into the \textbf{microscopic degrees of freedom} (binding energy).
    \item \textbf{Dissipation (Friction Forces):} Forces that oppose the local motion of carriers relative to the lattice. These generate heat and do not deform the bulk structure. This corresponds to the Host Interface gateway that channels energy into the \textbf{thermodynamic domain}.
    \item \textbf{Work (Continuum Deformation Forces):} The net forces acting between separate dipoles. This is the component that enters the \textbf{Continuum Mechanics Domain} as a deformation force. This corresponds to the macroscopic force density proposed by Kelvin and Einstein-Laub.
\end{enumerate}

In the following sections, this decomposition is rigorously derived from a mesoscopic model. It is shown that the deformation force is not the standard Kelvin force, but requires a topological adaptation to account for the discreteness of matter.

\section{Modeling the Dipole}
\label{sec:Mesoscopic_Model}

In this section, a rigorous model is established for the fundamental constituent of the dielectric: the microscopic dipole. The goal is to derive a simplified representation that preserves the essential internal physics required for a consistent force density calculation, while acknowledging the limitations of standard approximations found in the literature.

\subsection{The Microscopic Baseline: The Distributed Dipole}

The analysis begins with the ``Microscopic Baseline.'' A single physical dipole is not a point singularity; it is a complex, distributed system consisting of a massive positive ion core and a diffuse negative electron cloud.

The dynamics of these constituents are governed by QED. However, we utilize the classical limit to find a consistent description using microscopic Lorentz theory. We describe the charge separation using a continuous microscopic polarization field $\mathbf{p}_{\text{micro}}(\mathbf{x}, t)$, defined such that it exactly reproduces the true microscopic source distributions:
\begin{align}
    \rho_{\text{micro}} &= - \nabla \cdot \mathbf{p}_{\text{micro}} \\
    \mathbf{j}_{\text{micro}} &= \frac{\partial \mathbf{p}_{\text{micro}}}{\partial t}
\end{align}
The fundamental fields ($\mathbf{e}, \mathbf{b}$) associated with this distribution are governed by:
\begin{align}
    \varepsilon_0 \nabla \cdot \mathbf{e} &= \rho_{\text{micro}} \\
    \nabla \times \mathbf{e} + \frac{\partial \mathbf{b}}{\partial t} &= \mathbf{0} \\
    \nabla \cdot \mathbf{b} &= 0 \\
    \frac{1}{\mu_0} \nabla \times \mathbf{b} - \varepsilon_0\frac{\partial \mathbf{e}}{\partial t} &= \mathbf{j}_{\text{micro}}
\end{align}

It is critical to recognize that at this level, $\mathbf{p}_{\text{micro}}$ represents the \textbf{physical, distributed reality}. It describes the exact positions of classical sub-particles. The internal fields have high gradients and rapid spatial fluctuations. The stability of this structure is maintained by non-electromagnetic quantum constraints that prevent the electron cloud from collapsing into the nucleus.

While exact knowledge of the electron cloud wavefunction is not required, it is essential to retain the understanding that the dipole consists of \textbf{separated, distributed entities}. It is not a single point.

To bridge the gap to macroscopic theory, the complex distributed model must be simplified. This simplification step is often performed implicitly in the literature, leading to ambiguity---particularly regarding the definition of force density. Standard treatments typically reduce the distributed dipole immediately to a point singularity.

The analysis first explicitly analyzes the standard \textbf{Point Dipole Approximation} to understand its utility and its limitations. An alternative \textbf{Mesoscopic Sphere Model} is then introduced that preserves the topological information necessary to define internal stress and mechanical work.

\subsection{The Point Dipole Approximation: A Standard Simplification}
\label{sec:Point_Dipole_Approx}

In standard literature, the complex distributed dipole is typically replaced by an \textbf{idealized point dipole approximation}. This simplification relies on a multipole expansion of the field, retaining only the lowest-order term.

\subsubsection{Idealization of the Source Distribution}
The total dipole moment $\mathbf{p}_{\text{dp}}$ of an arbitrary charge distribution $\rho(\mathbf{x})$ is defined by the volume integral:
\begin{equation}
    \mathbf{p}_{\text{dp}} = \int_V \mathbf{x}' \rho(\mathbf{x}') \, dV'.
\end{equation}
Using the identity $\rho = - \nabla \cdot \mathbf{p}_{\text{micro}}$ and integration by parts (assuming $\mathbf{p}_{\text{micro}}$ vanishes at the boundary), this confirms that the dipole moment is simply the volume integral of the continuous polarization field:
\begin{equation}
    \mathbf{p}_{\text{dp}} = \int_V \mathbf{p}_{\text{micro}} \, dV.
\end{equation}

The point dipole approximation replaces the physical distribution $\mathbf{p}_{\text{micro}}(\mathbf{x})$ with a mathematical singularity located at $\mathbf{x}_i$:
\begin{equation}
    \mathbf{p}_{\text{ideal}}(\mathbf{x}) = \mathbf{p}_{\text{dp}} \, \delta(\mathbf{x} - \mathbf{x}_i).
\end{equation}
Consequently, the charge density becomes the derivative of a delta function:
\begin{equation}
    \rho_{\text{ideal}} = - \nabla \cdot \mathbf{p}_{\text{ideal}}(\mathbf{x}) = - (\mathbf{p}_{\text{dp}} \cdot \nabla) \delta(\mathbf{x} - \mathbf{x}_i).
\end{equation}

This idealization generates a singular field profile. For example, the electric field of an ideal dipole contains a delta-function singularity at the origin (the ``contact term'') alongside the standard $1/r^3$ outer field~\cite{Zangwill2013}:
\begin{equation}
    \mathbf{E}_{\text{dipole}}(\mathbf{x}) = \underbrace{\frac{1}{4\pi\varepsilon_0} \frac{3(\mathbf{p}_{\text{dp}}\cdot\hat{\mathbf{n}})\hat{\mathbf{n}} - \mathbf{p}_{\text{dp}}}{|\mathbf{x}|^3}}_{\text{Outer Field}} - \underbrace{\frac{\mathbf{p}_{\text{dp}}}{3\varepsilon_0}\delta(\mathbf{x})}_{\text{Inner Singularity}}.
    \label{eq:point_dipole_field}
\end{equation}
The key assumption of this model is that while it destroys the local field structure near the source, it accurately preserves the far-field properties and the \textbf{net force} experienced by the object in an external field.

\subsubsection{Derivation of the Total Force}
The validity of this approximation for calculating net forces is confirmed by integrating the Lorentz force density over the dipole volume.

For the electric force, integrating $\mathbf{f}_{\text{em}} = \rho_{\text{micro}} \mathbf{E}_{\text{ext}}$ gives:
\begin{equation}
    \mathbf{F}_{\text{dp}} = \int_V \rho_{\text{micro}} \mathbf{E}_{\text{ext}} \, dV = \int_V (-\nabla \cdot \mathbf{p}_{\text{micro}}) \mathbf{E}_{\text{ext}} \, dV.
\end{equation}
Using the vector identity $(\nabla \cdot \mathbf{A})\mathbf{B} = \nabla \cdot (\mathbf{A} \otimes \mathbf{B}) - (\mathbf{A} \cdot \nabla)\mathbf{B}$, we expand the integrand:
\begin{equation}
    \mathbf{F}_{\text{dp}} = \int_V \left[ (\mathbf{p}_{\text{micro}} \cdot \nabla) \mathbf{E}_{\text{ext}} - \nabla \cdot (\mathbf{p}_{\text{micro}} \otimes \mathbf{E}_{\text{ext}}) \right] dV.
\end{equation}
The second term is a total divergence. By Gauss's theorem, it transforms into a surface integral over the boundary of the dipole volume. Since the polarization $\mathbf{p}_{\text{micro}}$ vanishes outside the dipole, this term is zero. Assuming the external field gradient $\nabla \mathbf{E}_{\text{ext}}$ is approximately constant over the small volume, it can be pulled out of the integral:
\begin{equation}
    \mathbf{F}_{\text{dp}} = \left( \int_V \mathbf{p}_{\text{micro}} \, dV \right) \cdot \nabla \mathbf{E}_{\text{ext}}.
\end{equation}
Substituting the definition of the total dipole moment $\mathbf{p}_{\text{dp}}$, we recover the standard Kelvin force expression:
\begin{equation}
    \mathbf{F}_{\text{dp}} = (\mathbf{p}_{\text{dp}} \cdot \nabla) \mathbf{E}_{\text{ext}}.
\end{equation}

\paragraph{Magnetic Force Component.}
A similar derivation applies to the magnetic force. The force on the microscopic current density is $\mathbf{f}_m = \mathbf{j}_{\text{micro}} \times \mathbf{B}_{\text{ext}}$. Using $\mathbf{j}_{\text{micro}} = \partial \mathbf{p}_{\text{micro}}/\partial t$:
\begin{equation}
    \mathbf{F}_{\text{dp, m}} = \int_V \left( \frac{\partial \mathbf{p}_{\text{micro}}}{\partial t} \times \mathbf{B}_{\text{ext}} \right) dV.
\end{equation}
Assuming the external magnetic field is uniform across the dipole, the time derivative and volume integral commute:
\begin{equation}
    \mathbf{F}_{\text{dp, m}} = \left( \frac{d}{dt} \int_V \mathbf{p}_{\text{micro}} \, dV \right) \times \mathbf{B}_{\text{ext}} = \frac{d\mathbf{p}_{\text{dp}}}{dt} \times \mathbf{B}_{\text{ext}}.
\end{equation}

\paragraph{The Total Dipole Force.}
Combining these results yields the complete Lorentz force on a dipole in terms of its integrated moments:
\begin{equation}
    \mathbf{F}_{\text{total}} = (\mathbf{p}_{\text{dp}} \cdot \nabla) \mathbf{E}_{\text{ext}} + \frac{d\mathbf{p}_{\text{dp}}}{dt} \times \mathbf{B}_{\text{ext}}.
\end{equation}
This confirms that the \textbf{Total Force} depends only on the integrated dipole moment $\mathbf{p}_{\text{dp}}$, validating the point dipole approximation for calculating net translation. However, as argued in the next section, this integration process destroys the critical internal dipole information required to calculate energy transfer between the domains.

\subsection{Critique: The ``Opaque System'' Singularity}
\label{sec:critique}

While the standard point dipole derivation correctly predicts the \textbf{Net Force} ($\mathbf{F}_{\text{total}}$), it achieves this by reducing the internal structure of the dipole into a mathematical singularity. This renders the model an ``Opaque System'': it correctly maps inputs (external fields) to outputs (net translation), but it destroys the internal physical mechanics of the dipole. We illustrate the consequences below.

\subsubsection{1. The Loss of Mechanical Topology}
It is critical to identify the precise step where the physical model diverges. This occurs in the transition where the divergence term $-\nabla \cdot (\mathbf{p}_{\text{micro}} \otimes \mathbf{E}_{\text{ext}})$ is discarded because it integrates to zero over the boundary.

As established in Part I regarding the ``Newtonian Limit'' (Section~\ref{sec:Artifactual_Vector}), this discarded term corresponds exactly to the \textbf{Artifactual Vector} $\mathbf{\Phi}$ of the dipole system. By removing it, the derivation rigorously preserves the \textbf{Total Force} (the global ledger) but silently rewrites the \textbf{Force Density} (the local physics), redistributing the mechanical interaction from the boundaries (where charge exists) to the bulk volume (where it does not). This is the mechanism of the ``Arbitrary Split'': the total force remains invariant, but the internal mechanical topology is destroyed. The change predicts a force density in regions where there is no physical mass to accelerate.

\subsubsection{2. The Loss of Stability Mechanisms}
A real dipole in an electric field experiences significant internal stresses. The positive and negative centers are pulled in opposite directions. The dipole maintains its structure only because of \textbf{internal restoring forces} (quantum binding) that balance the electromagnetic stress:
\begin{equation}
    \mathbf{f}_{\text{internal}} + \mathbf{f}_{\text{em}} = \mathbf{0} \quad (\text{locally})
\end{equation}
The point dipole model assumes this stability as an axiom but provides no mechanism for it. The singularity hides the internal interaction between the external field and the atomic binding forces. By omitting this internal interaction, the model fails to account for the energy stored in the distortion of the dipole structure (the binding energy identified in Chapter \ref{chap:Dielectrics}).

\subsubsection{3. The Loss of the Energy Gateway (Absence of Dissipation)}
Perhaps most critically, the point dipole model breaks the Force-Energy Consistency Criterion (FECC) for time-varying systems.
\begin{itemize}
    \item \textbf{The Physical Reality:} A change in polarization magnitude ($\partial \mathbf{P}/\partial t$) corresponds to the physical motion of charge carriers (ions and electron clouds moving relative to each other). This relative velocity $\mathbf{v}_{\text{internal}}$ is the handle for friction; it is the mechanism by which energy is dissipated as heat (dielectric loss).
    \item \textbf{The Model Artifact:} In the point dipole model, a change in $\mathbf{p}(t)$ is treated as a change in the magnitude of a static singularity. There are no internal moving parts, and thus no defined internal velocity $\mathbf{v}_{\text{internal}}$.
\end{itemize}
Without a defined velocity field for the constituents, it is impossible to rigorously derive the rate of mechanical work $P = \mathbf{f} \cdot \mathbf{v}$. This closes the door to a first-principles derivation of dissipation.

\subsection{Intermediate Step: The Mesoscopic ``Constant Sphere'' Model}
\label{sec:Meso_Sphere_Model}

To address the limitations of the point dipole approximation while maintaining mathematical tractability, an alternative mesoscopic model is proposed. This model retains the internal continuum description necessary for defining local stress and inner motion of the ions and electrons, yet it is constructed to yield the exact same total dipole moment and far-field behavior as the standard point model.

We propose the \textbf{uniformly polarized sphere} as the mesoscopic model.

\subsubsection{Model Definition}
We define the mesoscopic polarization field $\mathbf{p}_{\text{meso}}(\mathbf{x})$ as a sphere of radius $R$ with a constant polarization $\mathbf{p}_0$ inside, and zero outside:
\begin{equation}
    \mathbf{p}_{\text{meso}}(\mathbf{x}) = 
    \begin{cases} 
       \mathbf{p}_0 & \text{for } |\mathbf{x}| < R \\
       0 & \text{for } |\mathbf{x}| > R 
    \end{cases}
\end{equation}
Crucially, the magnitude $\mathbf{p}_0$ is chosen such that the volume integral matches the physical dipole moment of the atom:
\begin{equation}
    \int_V \mathbf{p}_{\text{meso}}(\mathbf{x}) \, dV = \frac{4}{3}\pi R^3 \mathbf{p}_0 = \mathbf{p}_{\text{dp}}.
\end{equation}

This distribution implies that the mesoscopic bound charge density $\rho_{\text{meso}} = -\nabla \cdot \mathbf{p}_{\text{meso}}$ vanishes everywhere inside the volume ($\nabla \cdot \mathbf{p}_0 = 0$). The physical charge is entirely concentrated at the boundary surface $\mathcal{S}$ as a bound surface charge density $\sigma_b$:
\begin{equation}
    \sigma_b = \mathbf{p}_{\text{meso}} \cdot \hat{\mathbf{n}} \quad \text{on } \mathcal{S}
\end{equation}
where $\hat{\mathbf{n}}$ is the outward normal vector.

\subsubsection{The Concrete Realization of $\mathbf{f}_Q$}

This model highlights the critical role of the non-electromagnetic constraints established in Part I (Section~\ref{sec:conceptual_boundaries}).

Electrostatically, the positive surface charges on one hemisphere and the negative surface charges on the other exert a strong attractive Coulomb force on one another. Left to electromagnetic forces alone, the sphere would instantaneously collapse inward (a manifestation of Earnshaw's Theorem).

The stability of this mesoscopic object therefore demands the explicit inclusion of the \textbf{Quantum Constraint Force} ($\mathbf{f}_Q$) localized at the surface. This force acts as an internal ``pressure'' or rigid constraint that perfectly counters the self-attractive electromagnetic stress at the boundary:
\begin{equation}
    \mathbf{f}_Q + \mathbf{f}_{\text{em, self}} = \mathbf{0} \quad \text{(at the boundary)}.
\end{equation}
In the language of our Effective Field Theory, $\mathbf{f}_Q$ here represents the \textbf{Poincaré Stress}---the necessary non-electromagnetic tension required to hold a distributed charge together against its own repulsion.

\paragraph{Energetic Role: The Passive Constraint.}
Consistent with the ``Passive Stage'' principle established in Eq.~\eqref{eq:WorklessConstraint}, $\mathbf{f}_Q$ acts here as a \textbf{workless constraint}. It maintains the structural integrity (the shape) of the sphere against the self-field, but it performs no work during the polarization process ($P = \mathbf{f}_Q \cdot \mathbf{v} = 0$), provided the ``shape'' of the sphere is assumed invariant.

This distinction allows us to cleanly separate the forces that maintain the \textit{existence} of the dipole ($\mathbf{f}_Q$) from the restoring forces that oppose its \textit{polarization} ($\mathbf{f}_{\text{restore}}$), a separation impossible in the point dipole limit.

\subsection{Conceptual Comparison: The Filtering Perspective}
\label{sec:Meso_Filtering_Perspective}

It is useful to contextualize these two models---the Point Dipole and the Mesoscopic Sphere---within the filtering framework established in Part II.

Assume the existence of the ``True'' complex, arbitrary dipole with charge distribution $\rho_{\text{micro}}(\mathbf{x})$. The detailed structure of the electron orbitals and the nucleus represents complete set of spatial frequencies in the charge density signal.

Both the Point Dipole approximation and our Mesoscopic Sphere model act as \textbf{intermediate filters}. They both replace the high-frequency reality with a simplified signal that preserves the exact same low-frequency content (the total moment).
\begin{itemize}
    \item The \textbf{Point Dipole} replaces the complex structure with a signal containing infinite high-frequency components (a delta function).
    \item The \textbf{Mesoscopic Sphere} replaces the complex structure with a signal containing manageable high-frequency components.
\end{itemize}

From the perspective of the final Macroscopic Theory, the choice between these two intermediate models is mathematically irrelevant. When we eventually apply the spectral filtering operator $\langle \cdot \rangle_{\text{macro}}$ (where the filtering volume $V \gg V_{\text{atom}}$), all high-frequency details---whether they are the singularities of point dipoles or the edges of spheres---are filtered out. The resulting macroscopic polarization $\mathbf{P}$ and field $\mathbf{E}$ will be identical in both cases.

However, in order to track all energy pathways of the dipole model, the choice is decisive. The Mesoscopic Sphere model preserves the \textbf{Distributed Domain}. It retains a defined boundary and an interior volume. This allows us to rigorously define the internal constraint forces ($\mathbf{f}_{\text{bound}}$) and the internal energy transfer, keeping the theory mechanically consistent during the transition.

The Mesoscopic Sphere is therefore selected not because it is ``more true'' than the point dipole (both are approximations), but because it possesses the necessary topological fidelity to support a consistent mechanical theory.

\subsection{The Ontological Status of Polarization}
\label{sec:Ontology_P}

Before proceeding, a common interpretation regarding the physical status of the polarization field $\mathbf{P}$ must be addressed.

In many texts, $\mathbf{P}$ is introduced as ``the density of dipole moments.'' While mathematically correct, this definition risks obscuring the fundamental nature of the field. It implies $\mathbf{P}$ is an independent physical entity---a new type of ``stuff'' that fills the medium---rather than what it actually is: a kinematic descriptor of charge displacement.

It is crucial to recognize that even in the extreme limit of the point dipole approximation, $\mathbf{p}_{\text{ideal}}$ does not acquire a new existence. The delta function in its definition is simply a high-frequency spatial representation of charge separation. The field $\mathbf{p}$ (whether microscopic or macroscopic) serves \textbf{one and only one physical function}: it acts as the vector potential for the bound charge and current densities:
\begin{equation}
    \rho_b = -\nabla \cdot \mathbf{P}, \quad \quad \mathbf{J}_b = \frac{\partial \mathbf{P}}{\partial t}.
\end{equation}
$\mathbf{P}$ has no intrinsic dynamics or energy separate from the mass and charge it describes.

\section{Mesoscopic Derivation of the Force Partition}
\label{sec:Meso_Force_Partitioning}

After defining the mesoscopic dipole model, we now analyze the active forces. We propose that the total force density acting on the charges that constitute the dipoles can be rigorously partitioned into distinct mechanical domains: internal binding forces (which do not cause bulk acceleration) and interaction forces (which drive material deformation).

The general force density is given by:
\begin{equation}
    \mathbf{f}_{\text{dp}} = \rho_{\text{meso}} \mathbf{e}_{\text{meso}} = (-\nabla \cdot \mathbf{p}_{\text{meso}}) \mathbf{e}.
\end{equation}
First, we separate the self-field contributions and external field contributions. 
The force density $\rho_{\text{meso}}\mathbf{e}_{\text{int}}$ represents the static repulsion of the charge distribution trying to expand outwards.

For the external field, we use the vector identity $(\nabla \cdot \mathbf{A})\mathbf{B} = \nabla \cdot (\mathbf{A} \otimes \mathbf{B}) - (\mathbf{A} \cdot \nabla)\mathbf{B}$ to decompose the force into two parts:
\begin{equation}
    \mathbf{f}_{\text{dp,ext}} = (\mathbf{p}_{\text{meso}} \cdot \nabla) \mathbf{e}_{\text{ext}} - \nabla \cdot (\mathbf{p}_{\text{meso}} \otimes \mathbf{e}_{\text{ext}}).
\end{equation}
The second term does not contribute to the total net force (as its volume integral vanishes), but it contributes locally to the stress.

We propose the following separation of all acting force densities:

\subsection{Category 1: The Internal Stress Budget (Zero Net Force)}
All electromagnetic forces that attempt to \textit{deform} the dipole structure but do not contribute to its net acceleration are isolated. These are grouped together as the \textbf{Total Deforming Stress}. To maintain the stability of the dipole, this entire electromagnetic budget must be perfectly counterbalanced by the dipole's internal physics:
\begin{equation}
    \underbrace{\left[ \rho_{\text{meso}}\mathbf{e}_{\text{int}} - \nabla \cdot (\mathbf{p}_{\text{meso}} \otimes \mathbf{e}_{\text{ext}}) \right]}_{\text{Total EM Deforming Stress}} + \underbrace{\mathbf{f}_{\text{Material}}}_{\text{Quantum \& Binding Forces}} = \mathbf{0}.
\end{equation}

Here, $\mathbf{f}_{\text{Material}}$ is the unified mechanical response of the atom. It includes both the static quantum boundary forces ($\mathbf{f}_Q$) that prevent collapse and the high-frequency electromagnetic fields present inside the dipole. 

Since these forces sum to zero and are handled entirely by the internal constitutive physics of the dipole, they are irrelevant for the mechanical transport of the bulk material. They are ``confined'' inside the boundary of the effective dipoles. 

\subsection{Category 2: Continuum Deformation (Net Coupling)}
The remaining terms constitute the force density that survives volume integration. These represent the net electromagnetic drive acting on the dipole as a single unit.

Assuming the external field gradients vary slowly across the small mesoscopic radius $R$, these terms yield a spatially uniform body force over the sphere:
\begin{equation}
    \mathbf{f}_{\text{cont}} = \underbrace{(\mathbf{p}_{\text{meso}} \cdot \nabla)\mathbf{e}_{\text{ext}}}_{\text{Kelvin Force Density}} + \underbrace{\frac{\partial \mathbf{p}_{\text{meso}}}{\partial t} \times \mathbf{b}_{\text{ext}}}_{\text{Magnetic Force Density}}.
\end{equation}

This is the component responsible for \textbf{Continuum Deformation}. Because this force pushes the dipole's center of mass, it must be balanced by reaction forces from the \textit{neighboring} lattice sites (macroscopic deformation force density transmission). Thus, this is the \textbf{only} component that enters the Continuum Mechanics domain as a force density source.

\subsection{The Universality of the Total Lorentz Force}
An interpretation of the macroscopic force density is now proposed that resolves the apparent conflict between the fundamental Lorentz force and the engineering need for a deformation calculation.

It is postulated that when transitioning from the microscopic to the macroscopic scale, the fundamental physics does not change. The \textbf{Total Macroscopic Force Density} acting on the material volume is still rigorously given by the averaged Lorentz force acting on the bound charges and currents.

Using the macroscopic bound charge density $\rho_b = -\nabla \cdot \mathbf{P}$ and bound current density $\mathbf{J}_b = \frac{\partial \mathbf{P}}{\partial t}$ (assuming non-magnetic media for simplicity), the total force density is:
\begin{equation}
    \mathbf{f}_{\text{total}} = \rho_b \mathbf{E} + \mathbf{J}_b \times \mathbf{B} = -(\nabla \cdot \mathbf{P}) \mathbf{E} + \frac{\partial \mathbf{P}}{\partial t} \times \mathbf{B}.
    \label{eq:Macro_Lorentz_Total}
\end{equation}

This expression, representing the source of the Vacuum Stress Tensor, represents the \textbf{Total Force Budget}. It captures the sum of \textit{all} macroscopic electromagnetic forces acting on the constituents of the material.

However, we propose that this total force budget is subdivided into three pathways, which we defined as storage, dissipation, and work:
\begin{enumerate}
    \item \textbf{Storage:} This pathway corresponds to the internal microscopic fields of the dipole itself. It is the inner-dipole counterbalancing force. This constitutes the energy pathway into the micro-field storage, presented in Chapter \ref{chap:Dielectrics}.
    \item \textbf{Dissipation:} This component is active only in a dynamic process. It is counterbalanced by friction on moving charges (electrons and ions) and dissipates energy into the thermodynamic domain.
    \item \textbf{Work:} This pathway enters the continuum mechanics domain. It performs macroscopic work on matter and is responsible for deformation and the internal stress field in matter.
\end{enumerate}

How these three pathways and forces effectively partition depends on the microstructure of matter. It generally cannot be deduced from the macroscopic fields alone, as illustrated by the indeterminacy derivation.

In the following, we consider the microstructure of common dielectric matter and aim to derive the force density that specifically drives deformation. We propose that this part is responsible for deformation, while the remainder of the total force budget is counterbalanced by internal microstructures.

\section{Derivation of the Deformation Force: The Lattice Adaptation}
\label{sec:Deformation_Force_Derivation}

For the practical engineering of materials, the \textbf{Deformation Force Density} is the quantity of primary interest. This is the force component that acts on the material lattice, driving elastic strain, acoustic waves, and bulk mechanical actuation. This is the force density that enters the Cauchy Momentum Equation describing the motion of the bulk material in continuum mechanics.

In this section, this force is derived from the mesoscopic model. It is shown that the standard literature result (the Einstein-Laub force) arises from a specific averaging assumption. It is proposed that while this standard result is a useful baseline, a rigorous treatment of real materials likely requires a \textbf{Lattice Adaptation Term} to account for the specific microscopic correlations of the lattice.

\subsection{The Mesoscopic Definition}
As established in Section~\ref{sec:Meso_Force_Partitioning}, the force responsible for \textbf{macroscopic elastic deformation} is exclusively the component of the total force that is balanced by the neighboring host lattice. 

In our mesoscopic sphere model, this is the force acting on the dipole's center of mass due to the fields generated by \textit{all other} dipoles:
\begin{equation}
    \mathbf{f}_{\text{deformation}} = (\mathbf{p}_{\text{meso}} \cdot \nabla)\mathbf{e}_{\text{ext}} + \frac{\partial \mathbf{p}_{\text{meso}}}{\partial t} \times \mathbf{b}_{\text{ext}}
    \label{eq:micro_def_force}
\end{equation}
where $\mathbf{e}_{\text{ext}}$ and $\mathbf{b}_{\text{ext}}$ represent the local microscopic fields at the position $\mathbf{x}$, excluding the self-field of the dipole itself.

\subsection{The Transition Problem: Naive vs. Correlated Averaging}
    
Our specific goal now is to establish the correct form of this force for standard matter. As established in Section~\ref{sec:proof_indeterminacy}, Point Dipole approaches (like Einstein-Laub) are fundamentally valid because they posit a specific, physically realistic microstructure. 
 
To obtain the macroscopic force density $\mathbf{F}_{\text{def}}$, we must average local force over a macroscopic volume:
\begin{equation}
    \mathbf{F}_{\text{def}} = \langle \mathbf{f}_{\text{deformation}} \rangle = \left\langle (\mathbf{p}_{\text{meso}} \cdot \nabla)\mathbf{e}_{\text{ext}} \right\rangle + \dots
\end{equation}
    
The mathematical challenge lies in the correlation between the dipole position and the local field. Dipoles are not placed randomly in space; they act as sources that modify the field around them, and they are located at specific lattice sites.

\subsubsection{The Standard Literature Assumption (Naive Averaging)}
The standard approach found in most textbooks assumes that the average of the local field experienced by a dipole is identical to the total averaged macroscopic field $\mathbf{E}$:
\begin{equation}
    \text{\textbf{Assumption:}} \quad \langle \mathbf{e}_{\text{ext}} \rangle \approx \mathbf{E}
\end{equation}

Under this assumption, the local field $\mathbf{e}_{\text{ext}}$ is replaced by $\mathbf{E}$ (which is constant over the averaging volume), and the average of the mesoscopic polarization $\langle \mathbf{p}_{\text{meso}} \rangle$ becomes the macroscopic polarization $\mathbf{P}$.

Applying this substitution to Eq.~\eqref{eq:micro_def_force} yields the well-known force density proposed historically by Einstein and Laub:
\begin{equation}
    \mathbf{F}_{\text{EL}} = (\mathbf{P} \cdot \nabla)\mathbf{E} + \frac{\partial \mathbf{P}}{\partial t} \times \mathbf{B}
\end{equation}

\subsection{The Correction: A Sampling Bias Argument}
\label{sec:Sampling_Bias}

It is proposed that this standard assumption is physically incomplete due to the \textbf{discrete topology} of matter.

\subsubsection{The ``Test Dipole'' Thought Experiment}
To illustrate our approach, we propose to consider a ``Test Dipole'' inserted at a random location within the polarized dielectric. The question arises: \textit{What is the average field this specific dipole experiences?}

The answer depends on the topology of the space available to the dipole.
\begin{itemize}
    \item \textbf{Mathematical Average (The Continuum Fallacy):} If the test dipole were a ghost that could exist anywhere---even overlapping with other atoms---it would sample the entire volume uniformly. In this case, the average field would indeed be the total macroscopic field $\mathbf{E}$.
    \item \textbf{Physical Average (Topological Exclusion):} However, matter is discrete and impenetrable. A physical dipole cannot reside inside the volume of another dipole. It is statistically excluded from the regions occupied by its neighbors.
\end{itemize}

The test dipole effectively resides in the interstitial vacuum regions between dipoles. It blindly misses the field singularities hidden within the ``bodies'' of its neighbors.
Consequently, it \textbf{does not encounter} the extremely high field singularities (the ``inner'' fields) that exist inside the electron clouds. The total macroscopic field $\mathbf{E}$ includes these singularities in its mathematical average; the physical dipole does not.

\subsubsection{The Effective Local Field}
Therefore, we propose an adaptation of the Einstein-Laub force density to account for this topological sampling bias. We propose that the correct forcing field $\mathbf{E}_{\text{eff}}$ is not the total average, but the total average \textit{minus} the contribution of the regions the dipole never visits:
\begin{equation}
    \mathbf{E}_{\text{eff}} = \mathbf{E}_{\text{total}} - \langle \mathbf{E}_{\text{inner}} \rangle_{\text{excluded}}
\end{equation}

\subsubsection{Quantifying the Correction: Singularity Subtraction}

This correction can be quantified by examining the field of an ideal point dipole. As shown in Eq.~\eqref{eq:point_dipole_field}, the field contains an explicit "Inner Singularity" term: $-\frac{\mathbf{p}}{3\varepsilon_0}\delta(\mathbf{x})$.

The total macroscopic field $\mathbf{E}$ contains the volume average of this singularity. However, our test dipole is physically excluded from the origin of every neighbor. Therefore, the component of the total field it systematically fails to sample is precisely the average of the delta-function singularities.

Using the fundamental definition of macroscopic polarization $\mathbf{P}(\mathbf{x}) = \langle \sum_i \mathbf{p}_i \delta(\mathbf{x} - \mathbf{x}_i) \rangle$, this excluded average is:
\begin{equation}
    \langle \mathbf{E}_{\text{inner}} \rangle = \left\langle \sum_i -\frac{\mathbf{p}_i}{3\varepsilon_0}\delta(\mathbf{x}-\mathbf{x}_i) \right\rangle = -\frac{1}{3\varepsilon_0} \underbrace{\left\langle \sum_i \mathbf{p}_i \delta(\mathbf{x}-\mathbf{x}_i) \right\rangle}_{\mathbf{P}} = -\frac{\mathbf{P}}{3\varepsilon_0}.
\end{equation}

\subsubsection{The Lattice Adaptation Term}
The effective local field (the field actually sampled by the dipole) is the total field minus this excluded contribution:
\begin{equation}
    \mathbf{E}_{\text{eff}} = \mathbf{E} - \langle \mathbf{E}_{\text{inner}} \rangle = \mathbf{E} - \left(-\frac{\mathbf{P}}{3\varepsilon_0}\right) = \mathbf{E} + \frac{\mathbf{P}}{3\varepsilon_0}.
    \label{eq:Lorentz_Local_Field}
\end{equation}

Consequently, we propose that the macroscopic force density responsible for deformation must be adapted. We replace $\mathbf{E}$ with $\mathbf{E}_{\text{eff}}$ in the electric term:

\begin{align}
    \mathbf{F}_{\text{deformation}} &\approx (\mathbf{P} \cdot \nabla)\mathbf{E}_{\text{eff}} + \frac{\partial \mathbf{P}}{\partial t} \times \mathbf{B} \nonumber \\
    &= \underbrace{\left[ (\mathbf{P} \cdot \nabla)\mathbf{E} + \frac{\partial \mathbf{P}}{\partial t} \times \mathbf{B} \right]}_{\text{Standard Einstein-Laub force}} + \underbrace{(\mathbf{P} \cdot \nabla)\left( \frac{\mathbf{P}}{3\varepsilon_0} \right)}_{\text{Lattice Adaptation Term}}.
\end{align}

This result suggests that the standard Einstein-Laub force density is the correct baseline, but it requires a specific topological correction term---a \textbf{Lattice Adaptation Term}---to account for the fact that discrete dipoles do not sample the macroscopic field uniformly.

\subsection{Limitations: The Hypothesis of Microstructure}
\label{sec:Adaptation_Critique}

The correction term derived above---$(\mathbf{P} \cdot \nabla)\left( \frac{\mathbf{P}}{3\varepsilon_0} \right)$---is explicitly framed as a \textbf{hypothesis of microstructure} rather than a universal constitutive law. It serves as a diagnostic tool to demonstrate the physical incompleteness of the standard Einstein-Laub formulation.

\paragraph{Structural Dependence.}
The derivation relied on the ``Test Dipole'' thought experiment, where a dipole was placed randomly within the vacuum space of the dielectric. This assumes a statistical isotropy characteristic of fluids or amorphous solids, where the spherical exclusion volume (leading to the factor $1/3$) is appropriate.
In real crystalline solids, dipoles are located at fixed lattice sites with specific symmetries. For anisotropic crystals (e.g., tetragonal or monoclinic lattices), the scalar factor $1/3$ might transform into a complex interaction tensor dependent on the specific lattice constants.

\paragraph{Model Dependence.}
Furthermore, the calculation of the excluded internal field $\langle \mathbf{E}_{\text{inner}} \rangle$ assumed the singularity of an ideal point dipole (equivalent to a spherical exclusion zone). If the microscopic subunits were modeled with more realistic shapes, the geometry of the subtracted self-field would change, altering the adaptation coefficient.
Moreover, it was assumed that the external field may be regarded as constant over the dipole volume. This simplification might not hold true in all cases; the high-gradient fields of neighboring dipoles might vary over the dipole's volume.

\subsection{Final Conclusion: The Invariance of the Total Lorentz Force}
Despite this complexity in partitioning, we maintain that the fundamental foundation remains secure. The \textbf{Total Macroscopic Force Density} is rigorously given by the Lorentz formulation:
\begin{equation}
    \mathbf{f}_{\text{total}} = \rho_b \mathbf{E} + \mathbf{J}_b \times \mathbf{B}.
\end{equation}
This expression acts as the invariant ``master equation.'' It correctly describes the total momentum budget of the material---the sum of the force driving the lattice (deformation), the force stretching the atoms (internal stress), and the force responsible for dissipation.

While the Vacuum Stress Tensor does not explicitly separate these mechanical roles, it remains an electromagnetically consistent description of the total system. The challenge of continuum mechanics is not to replace this tensor, but to couple it with a microstructurally informed material model that correctly interprets how this total force is distributed.

\chapter{Invariance of Total Momentum and Energy Budget}
\label{chap:BudgetInvariance}

In this chapter, we analyze the macroscopic Lorentz force, demonstrating that the total momentum budget of the electromagnetic field is rigorously partitioned into three identified domains. Moreover, we establish that the total momentum budget is invariant under the filtering process. However, we clarify a critical distinction: while the \textit{total} budget is fixed, the specific partition of that budget into internal channels cannot generally be discerned from the macroscopic fields alone. One must ``open the box'' and examine the microstructure to determine the precise allocation.

\section{The Invariance of Total Momentum: A Boundary Theorem}
\label{sec:Total_Force_Invariance}

We begin by establishing the principle of \textbf{Global Certainty}. This corresponds to \textbf{Level 2} of the uniqueness hierarchy (Chapter \ref{sec:Intro_Hierarchy}). This section demonstrates that the total force (the rate of change of total momentum) on an isolated body is \textit{invariant} through the filtering process. This provides a crucial consistency check: while the internal distribution of force is highly sensitive to microstructure and spectral filtering, the total momentum exchange with the vacuum is not.

Consider an isolated piece of matter occupying a volume $V$ bounded by a surface $S$, surrounded by vacuum. 
The microscopic momentum conservation equation is:

\begin{equation}
\begin{aligned}
    \underbrace{ \left(\frac{\partial \mathbf{g}_{\text{mech}}}{\partial t} + \nabla\cdot\mathbf{t}_{\text{kin}}\right) - \mathbf{f}_{Q} }_{\substack{\textbf{The Mechanical Reservoir} \\ \text{(Net Momentum Gain)}}}
    \quad &\overset{\text{Physical Coupling}}{\textbf{=}} \quad
    \underbrace{ \rho_{\text{micro}}\mathbf{e} + \mathbf{j}_{\text{micro}}\times\mathbf{b} }_{\substack{\textbf{The Single Gateway} \\ \text{(Lorentz Force)}}} \\
    &\overset{\text{Math. Identity}}{=} \quad
    \underbrace{ -\left(\frac{\partial \mathbf{g}_{\text{em}}}{\partial t} + \nabla\cdot \mathbf{t}_{\text{em}}\right) }_{\substack{\textbf{The Electromagnetic Reservoir} \\ \text{(Field Momentum Supply)}}}
\end{aligned}
\label{eq:Micro_Momentum_Architecture_v2}
\end{equation}
For simplicity, we summarize the mechanical reservoir into $\mathbf{g}_{\text{mech}}$: 
\begin{equation}
    \frac{\partial}{\partial t} (\mathbf{g}_{\text{mech}} + \mathbf{g}_{\text{em}}) + \nabla \cdot \mathbf{t}_{\text{em}} = 0
    \label{eq:Momentum_Decomposition_Micro_Review}
\end{equation}

What is preserved before and after averaging is the definition of the \textit{change of total momentum} inside an isolated piece of matter. Integrating over the volume, and assuming there is no mass flow in or out of the isolated volume, the change of the \textbf{Total System Momentum} $\mathbf{G}_{\text{sys}}$ is determined solely by the electromagnetic momentum flux through the boundary.

Using the divergence theorem:
\begin{equation}
    \frac{d}{dt} \mathbf{G}_{\text{sys}} = \frac{d}{dt} \int_V (\mathbf{g}_{\text{mech}} + \mathbf{g}_{\text{em}}) \, dV = - \oint_S \mathbf{t}_{\text{em}} \cdot d\mathbf{a}.
\end{equation}

\subsection{The Boundary Argument}
The key insight is topological. The fields $\mathbf{e}$ and $\mathbf{b}$ differ continuously between the microscopic and macroscopic descriptions only \textit{inside} the material volume, where high-frequency atomic variations exist.

However, at the boundary $S$ (located in the vacuum outside the object), the microscopic fields are smooth. The spectral filter $K(\mathbf{x})$ acts as a low-pass filter. Since the vacuum fields at a sufficient distance from the atoms effectively contain no high-frequency components relative to the macroscopic kernel, the filtering operation leaves them invariant:
\begin{equation}
    \mathbf{e}_{\text{vac}} = \mathbf{E}_{\text{vac}}, \quad \mathbf{b}_{\text{vac}} = \mathbf{B}_{\text{vac}}.
\end{equation}
Consequently, the Maxwell Stress Tensor at the boundary is identical in both descriptions:
\begin{equation}
    \oint_S \mathbf{t}_{\text{em}} \cdot d\mathbf{a} \equiv \oint_S \mathbf{T}_{\text{EM}} \cdot d\mathbf{A}.
\end{equation}

This proves that the \textbf{Rate of Change of Total Momentum} on an isolated body is not changed by the spectral filtering process. The system behaves identically in both descriptions to an external observer.

The filtering process changes the spatial distinction from microscopic ``field momentum'' vs. ``mechanical momentum'' \textit{inside} the bulk, into different macroscopic domains (mechanical momentum in the continuum mechanics domain, energy in the thermodynamic domain as heat, or as storage of high-frequency fields), but it cannot alter the \textbf{Sum} defined by the boundary flux.

\subsection{Total Change of Momentum of an Isolated Piece of Matter}
This insight also shows that the total force can always be calculated by the macroscopic Lorentz force and the vacuum tensor. The total change of momentum, representing the change of momentum of the center of mass of an isolated piece of matter, is defined by the vacuum stress tensor integrated over the volume's boundary. This is not affected by the averaging process.

Thus, while the \textit{distribution} of force (where exactly it acts inside the material) is dependent on the microstructure of an isolated piece of matter, the \textit{net change of momentum of the center of mass} on the object is well defined by purely macroscopic field variables. 

\subsection{Total Change of Angular Momentum}
The same argument applies to the torque on the total isolated piece of matter. The change of total angular momentum inside the volume is given by the moment of the stress tensor at the boundary:
\begin{equation}
    \frac{d}{dt} \mathbf{L}_{\text{total}} = - \oint_S \mathbf{r} \times (\mathbf{t}_{\text{em}} \cdot d\mathbf{a}).
\end{equation}
Since the boundary fields and the position vector $\mathbf{r}$ at the boundary are invariant under averaging, the flow of rotational electromagnetic momentum in or out of the total volume is not affected.
Thus, while the \textit{distribution} of torque (where exactly it acts inside the material) is dependent on the microstructure of an isolated piece of matter, the \textit{net torque} on the object is well defined by purely macroscopic field variables.

\section{The Invariance of Total Energy: A Boundary Theorem}
\label{sec:Energy_Invariance_Vacuum}

The same argument establishes consistency for the energy budget. Because the averaging kernel is normalized, and fields at the vacuum boundary are invariant under averaging ($\mathbf{e}_{\text{micro}}(S) = \mathbf{E}_{\text{macro}}(S)$), the total electromagnetic power flowing into the volume is identical in both descriptions:
\begin{equation}
    P_{\text{in}} = -\oint_S (\mathbf{e} \times \mathbf{b}) \cdot d\mathbf{a} \equiv -\oint_S (\mathbf{E} \times \mathbf{B}) \cdot d\mathbf{A}.
\end{equation}

Since the total energy supply ($P_{\text{in}}$) is identical, the sum of the internal sinks must be identical.
\begin{itemize}
    \item \textbf{Microscopic Sinks:} The energy flows only into the kinetic energy of particles and the fine-grained electromagnetic field energy.
    \item \textbf{Macroscopic Sinks:} The energy flows into Work, Heat, Potential Energy, and the coarse-grained electromagnetic field energy.
\end{itemize}
This establishes the rigorous conservation law of the Host Interface:
\begin{equation}
    \underbrace{ \frac{dK}{dt} + \frac{dU_{\text{em, micro}}}{dt} }_{\substack{\text{Microscopic Reality}}} \quad \equiv \quad \underbrace{ \frac{dK_{\text{carrier}}}{dt} + \frac{dU_{\text{em, macro}}}{dt} }_{\substack{\text{Direct Response}}} + \underbrace{ P_{\text{work}} + \frac{dU_{\text{potential}}}{dt} + P_{\text{heat}} }_{\substack{\text{Host Interface Output}}}.
\end{equation}

This theorem demonstrates that ``Heat'' and ``Binding'' are not auxiliary parameters added to the theory; they are real, physical partitions of the microscopic energy budget.

\section{The Indeterminacy of Internal Partition}
\label{sec:Internal_Indeterminacy}

We now juxtapose the \textbf{Global Certainty} established above with the principle of \textbf{Local Ambiguity} (corresponding to \textbf{Level 3} of the hierarchy).

We have previously shown that the force density transmitting into the continuum mechanics domain (Deformation) depends on the microstructure. The same is true for the force density transmitting into the thermodynamic domain (Heat), and the force density responsible for microscopic field interactions (Binding Energy). 

Thus, while the total bill is fixed, how the total electromagnetic momentum budget splits into these three domains cannot be uniquely determined from macroscopic fields alone. It is dependent on the microscopic topology of matter.

The energy equation couples towards the momentum equation by projection onto the velocities of the mass target. From the macroscopic electrodynamics system alone, one cannot inherently see into which path the energy flows. One observes that:
\begin{equation}
    \mathbf{f} \cdot \mathbf{v} = \mathbf{J}\cdot\mathbf{E} \neq 0
\end{equation}
is locally non-zero. This implies there is a velocity of a mass carrier and a sink or source of macroscopic electromagnetic field energy. However, the \textit{destination} of that energy remains ambiguous if one looks only at the electrodynamics macroscopic field variables.

This is the manifestation of the \textbf{Theorem of Macroscopic Indeterminacy} derived in Chapter \ref{chap:ForceDensity}. Macro-variables are trace-deficient; they describe the \textit{total} energy transfer but strictly fail to distinguish between its internal allocation to deformation, binding, or heat. Causal nuance has been filtered out.

This provides the rigorous basis for our conclusion regarding force density: Energy-based methods (like Korteweg-Helmholtz) are structurally constrained because they attempt to derive topology from a macroscopic functional. Conversely, microstructure-based methods (like Einstein-Laub) remain valid approximations precisely because they re-inject this missing causal nuance.

Let us consider definitive examples to illustrate this ``Macroscopic Indeterminacy'' in dynamic situations.

\subsection{Example 1: Heat vs. Energy Storage (The Dissipation Ambiguity)}

Consider a rigid body with changing polarization. From the perspective of macroscopic fields alone, one cannot differentiate whether the energy entering the system is being stored (reversible) or dissipated (irreversible) without further context.

When we characterize a polarizable material, we measure the macroscopic field response. From that, we can calculate the macroscopic Polarization field $\mathbf{P}$.
\textbf{Note:} It is vital to remember that $\mathbf{P}$ is merely a macroscopic proxy for the underlying microscopic bound currents ($\mathbf{J}_b = \partial_t \mathbf{P}$) and charges ($\rho_b = -\nabla \cdot \mathbf{P}$).

We can calculate the current and the power flow macroscopically. We can derive a hysteresis curve and determine how much energy has been dissipated (area of the loop) and how much has been stored (reactive part).

However, the exact \textit{microscopic event} at a specific instant cannot be discerned just by looking at the instantaneous macroscopic vectors $\mathbf{E}$ and $\mathbf{P}$. It is observed that power is transferred ($\mathbf{E} \cdot \mathbf{J} \neq 0$), but the macroscopic fields alone do not reveal the destination. Is the energy stretching a lossless spring (Storage) or resistive heating (Loss)? The fields are silent on this distinction.

The filtering process inherently deletes this information. When we characterize the macroscopic polarization field, we measure the far-field response outside the material. We cannot measure the high-frequency inner atomic gradient while characterizing it. This process is the inherent low-pass filtering that deletes the micro-structural information required to distinguish storage from loss.

\subsection{Example 2: Deformation vs. Storage (The Mechanical Ambiguity)}
This ambiguity extends to the mechanical domain. The total force is given by the Lorentz force, but it must be subdivided:
\begin{enumerate}
    \item \textbf{Internal Structural Forces:} Forces counterbalanced by microscopic constraints (e.g., dipole binding forces).
    \item \textbf{Continuum Forces:} Forces that translate into the continuum domain (e.g., bulk deformation).
\end{enumerate}

Consider a dielectric consisting of dipoles.
\begin{itemize}
    \item \textbf{Scenario A (Storage):} The dipoles expand individually---their charges separate further against internal quantum forces. Evolution is internal. Energy is transferred into macroscopic \textbf{storage} (binding energy).
    \item \textbf{Scenario B (Deformation):} The dipoles stay constant (rigid), but the distance \textit{between} them changes. Evolution is bulk. Energy is transferred into the macroscopic \textbf{deformation} of the material---the continuum mechanical domain.
\end{itemize}

Crucially, both scenarios can produce identical macroscopic polarization changes $\partial \mathbf{P}/\partial t$. To the macroscopic observer, charge is simply moving. Whether that motion corresponds to "stretching the atom" or "stretching the lattice" is topological information that has been filtered out. One must descend one resolution level (to the microscopic structure) to determine the exact mechanical partition.

\section{Case Study: Dielectric Mechanical Interaction (Volumetric Coupling)}
\label{sec:Dielectric_MechanicalWork}

Having established the decomposition of the total force and energy budgets, we proceed to analyze specific implementations.

Observation: We consider a rigid dielectric object moving with macroscopic velocity $\mathbf{V}_{\text{mech}}$ in a non-uniform external electric field (e.g., a dielectric slab being drawn into a capacitor). This parallels the conductor case analyzed in Part II, but the ``coupling'' is now volumetric rather than superficial.

\subsection{The Mechanism of Transmission: Volumetric Coupling}
In a dielectric, the coupling is distributed throughout the volume. The external field $\mathbf{E}$ pulls the positive bound charges ($\rho_+$) in one direction and the negative bound charges ($\rho_-$) in the other.
\begin{equation}
    \mathbf{f}_{\text{Lorentz}} = \rho_+ \mathbf{E} + \rho_- \mathbf{E}
\end{equation}
Unlike the conductor, these charges cannot flow to the surface. They are constrained to the lattice. The total Lorentz force is separated internally into the different budgets. In the static case, the total force partitions into the counterbalancing of internal microscopic structures (inner dipole binding forces) and the force density responsible for bulk deformation. 

While it has been proven that the local distribution of this deformation force is structurally dependent and cannot be defined by a universal field formula, the \textit{concept} remains clear: the field acts on the dipoles, and the dipole force is counterbalanced by the Host Interface that transmits that force to the bulk matter. This part of the host force is the gateway into the domain of continuum dynamics.

\subsection{Power Balance: Dielectric Actuation}
For now, we consider the simplified version of a rigid body, where the internal deformation is assumed to be zero. The power equation derived and used in the last chapters can be applied accordingly.
The motion of the polarized medium creates a convection current $\mathbf{J}_{\text{conv}} = \rho_{\text{bound}} \mathbf{V}_{\text{mech}}$. The interaction is governed by the standard gateway:
\begin{equation}
    \underbrace{ - P_{\text{host}} }_{\text{Mechanical Power}} = \underbrace{ \mathbf{J}_{\text{conv}} \cdot \mathbf{E} }_{\text{Gateway}} = \underbrace{ - \left( \frac{\partial u_{\text{EM}}}{\partial t} + \nabla \cdot \mathbf{S}_{\text{EM}} \right) }_{\text{Field Budget}}
\end{equation}
The direction of energy flow is dictated by the relative alignment of the velocity and the force.

\paragraph{The Mechanical Sink.}
If the dielectric is drawn into the field (motion aligned with the net pull), the Host Interface acts as a \textbf{Local Energy Sink}.
\begin{equation}
    P_{\text{host}} > 0 \quad \implies \quad \text{Energy flows } \textbf{Field} \to \textbf{Kinetic Energy}
\end{equation}
Energy leaves the electrodynamics domain, stripped from the electromagnetic field budget and converted into the kinetic energy of the block. The Host Interface facilitates this transfer without any intermediate thermal step.

\section{Case Study: Electrostriction and Elasticity}
\label{sec:Dielectric_Deformation}

Finally, we consider the case where the dielectric is \textbf{elastic}. The lattice is no longer rigid; it can stretch or compress under the stress transmitted by the bound charges. This is the phenomenon of \textbf{Electrostriction}.

\subsection{The Mechanism of Deformation: ``Inside-Out'' Stretching}

We now allow for the dielectric to be elastic. 
The Host Interface transmits the force density of the dipoles into the continuum mechanics domain. 
This force density acts as the input of external force density for the Cauchy equation in continuum mechanics.
The material expands or contracts until the internal \textbf{Mechanical Stress} ($\mathbf{T}_{\text{mech}}$) balances the electromagnetic pull.

\subsection{Energy Balance: The Multi-Reservoir Sink}

If we allow the dipoles themselves to stay constant and only reform the material, the energy is transferred into the continuum mechanics domain. The deformation of the material corresponds to partial redistribution and currents in the material. Macroscopically, this will emerge in a change of the macroscopic polarization field (representing a current density) for the deforming piece of matter. 

The power equation is:
\begin{equation}
    \underbrace{ - P_{\text{host}} }_{\text{Mechanical Power}} = \underbrace{ \mathbf{J}_{\text{conv}} \cdot \mathbf{E} }_{\text{Gateway}} = \underbrace{ - \left( \frac{\partial u_{\text{EM}}}{\partial t} + \nabla \cdot \mathbf{S}_{\text{EM}} \right) }_{\text{Field Budget}}
\end{equation}

When there is a local sink of energy, the energy leaves the electromagnetic domain and enters the continuum domain. There it is stored in the elastic deformation of the matter. The process is reversible.

\section{Summary of Part III: The Electric Response}
\label{sec:Summary_Part3}

In Part III of this book, we have rigorously analyzed all interactions of the \textbf{longitudinal current} ($\mathbf{J}_{\parallel}$). This is the component of the current density that possesses divergence ($\nabla \cdot \mathbf{J} \neq 0$), leading inevitably to the accumulation of charge density ($\rho$) and the generation of divergence-based fields ($\nabla \cdot \mathbf{E} \neq 0$).

We have demonstrated that a single unified framework---the \textbf{Vacuum Lorentz Framework}---rigorously describes the mechanics of both free charges (conductors) and bound charges (dielectrics).

\subsection{The Three Mechanical Gateways}
By enforcing the \textbf{Force-Energy Consistency Criterion (FECC)}, we identified that the Host Interface partitions the total electromagnetic work ($P = \mathbf{J} \cdot \mathbf{E}$) into three distinct physical channels:
\begin{enumerate}
    \item \textbf{Dissipation:} Irreversible transfer to heat (Resistance).
    \item \textbf{Work:} Ordered transfer to bulk motion (Actuation/Electrostriction).
    \item \textbf{Storage:} Reversible transfer to microscopic potential energy (Dielectric Binding).
\end{enumerate}

\subsection{The Universality of the Vacuum Tensor}
Finally, we addressed the ambiguity of force density. We proved the \textbf{Theorem of Macroscopic Indeterminacy}, showing that local stress cannot be derived from macroscopic fields alone without microstructural side-channel information.

However, we concluded that the \textbf{Total Momentum Budget} is structurally invariant. The vacuum energy-momentum tensor correctly describes the conservation of momentum for the entire system:
\begin{equation}
\boxed{
\begin{aligned}
    \underbrace{ \left( \frac{\partial \mathbf{G}_{\text{mech}}}{\partial t} + \nabla\cdot\mathbf{T}_{\text{mech}} \right) }_{\text{Total Material Response}}
    \quad &\overset{\text{Coupling}}{\textbf{=}} \quad
    \underbrace{\rho \mathbf{E} + \mathbf{J}\times\mathbf{B} }_{\text{Lorentz Force}} \\
    &\overset{\text{Identity}}{=} \quad
    \underbrace{ -\left(\frac{\partial \mathbf{G}_{\text{EM}}}{\partial t} + \nabla\cdot \mathbf{T}_{\text{EM}}\right) }_{\text{Vacuum Tensor Source}}
\end{aligned}
}
\end{equation}
This equation is the master key to the electric response. It ties the force $\mathbf{f}$ irrevocably to the energy flux $\mathbf{S}$ and the mass carrier $\mathbf{v}$.

With the longitudinal (electric) response complete, we now turn our attention to the transverse (magnetic) response: the physics of closed loops, solenoidal fields, and the physics of magnetization.

\part{Part IV: The Magnetic Response -- Dynamics of Rotational Currents}
\label{part:MagneticResponse}

\chapter{Introduction to Part IV}
\label{chap:IntroPart4}

The analysis now proceeds from the electrostatics of Charge Separation to the electrodynamics of \textbf{Charge Circulation}.

In Part III, the physics of \textbf{Divergent Currents} ($\nabla \cdot \mathbf{J} \neq 0$) was analyzed, where the kinematic accumulation of charge created the electrostatic restoring forces of the Dielectric. In this Part, the focus shifts to the physics of \textbf{Rotational Currents} ($\nabla \cdot \mathbf{J} = 0$). This kinematic property ensures that charge never accumulates; it acts as a continuous closed-loop flow.
This flow creates the \textbf{Magnetic Field} ($\mathbf{B}$).

This distinction defines the boundary between the Electric and Magnetic domains:
\begin{itemize}
    \item \textbf{Part III (Separation):} Currents are \textbf{Longitudinal} ($\nabla \cdot \mathbf{J} \neq 0$). Charge accumulates against constraints. The physics is defined by the \textbf{Divergent} Electric Field ($\nabla \cdot \mathbf{E} \neq 0$).
    \item \textbf{Part IV (Circulation):} Currents are \textbf{Transverse} ($\nabla \cdot \mathbf{J} = 0$). Charge flows in closed loops. The physics is defined by the \textbf{Solenoidal} Magnetic Field ($\nabla \cdot \mathbf{B} = 0$).
\end{itemize}

Standard pedagogical treatments often obscure this unity by treating "Current in a Wire" and "Magnetization in a Material" as distinct phenomena. 
In this Part, this distinction is re-evaluated and dissolved.
From the perspective of the Vacuum, there is only \textbf{Current Density} ($\mathbf{J}$). The current flowing in a macroscopic solenoid and the current spinning in an electron orbital are the same physical entity. They obey the same Maxwell equations. They couple to the same vacuum fields.
The only difference is the \textbf{Scale of the Circulation} and the nature of the internal constraint.

\section{The Governing Equations}

As in the electric case, the macroscopic theory must satisfy the rigorous mechanical constraint ($P = \mathbf{f} \cdot \mathbf{v}$). However, distinct from the electric case where momentum and energy are driven by the same force vector ($\mathbf{f} \parallel \mathbf{v}$), the magnetic domain reveals a fundamental decoupling.

The \textbf{Momentum Balance} is driven by the Lorentz Magnetic Force:
\begin{equation}
\boxed{
\begin{aligned}
    \underbrace{ \left( \frac{\partial \mathbf{G}_{\text{ord}}}{\partial t} + \nabla\cdot\mathbf{T}_{\text{ord}} \right) - \mathbf{F}_{\text{host}} }_{\substack{\text{Total Mechanical Response} \\ \text{(Ordered + Host)}}}
    \quad &\overset{\text{Physical Coupling}}{\textbf{=}} \quad
    \underbrace{\rho\mathbf{E} + \mathbf{J}\times\mathbf{B} }_{\substack{\textbf{The Momentum Gateway} \\ (\text{Dominant Term: } \mathbf{J} \times \mathbf{B})}} \\
    &\overset{\text{Math. Identity}}{=} \quad
    \underbrace{ -\left(\frac{\partial \mathbf{G}_{\text{EM}}}{\partial t} + \nabla\cdot \mathbf{T}_{\text{EM}}\right) }_{\substack{\text{Macroscopic EM Response} \\ \text{(Vacuum Tensor)}}}
\end{aligned}
}
    \label{eq:Macro_Momentum_Architecture_Part4_Intro}
\end{equation}

Applying the kinematic lock ($P = \mathbf{F} \cdot \mathbf{v}$) reveals the \textbf{Energy Balance}. Here the critical constraint is observed: even though the momentum is driven by $\mathbf{B}$, the energy transfer is strictly gated by $\mathbf{E}$:

\begin{equation}
\boxed{
\begin{aligned}
    \underbrace{ \left( \frac{\partial U_{\text{ord}}}{\partial t} + \nabla\cdot\mathbf{S}_{\text{ord}} \right) - P_{\text{host}} }_{\substack{\text{Total Mechanical Response}}}
    \quad &\overset{\text{Physical Coupling}}{\textbf{=}} \quad
    \underbrace{ \mathbf{J}\cdot\mathbf{E} }_{\substack{\textbf{The Energy Gateway} \\ (\text{Sole Mechanism})}} \\
    &\overset{\text{Math. Identity}}{=} \quad
    \underbrace{ -\left(\frac{\partial U_{\text{EM}}}{\partial t} + \nabla\cdot \mathbf{S}_{\text{EM}}\right) }_{\substack{\text{Macroscopic EM Response}}}
\end{aligned}
}
    \label{eq:Macro_Energy_Architecture_Part4_Intro}
\end{equation}

Unlike the electric force, the Lorentz magnetic force ($\mathbf{F}_m = q\mathbf{v} \times \mathbf{B}$) is strictly perpendicular to the velocity of the charge carrier. Consequently, \textbf{the magnetic field performs zero mechanical work} ($P = \mathbf{F}_m \cdot \mathbf{v} \equiv 0$).

This decoupling defines the central theoretical task of Part IV. The analysis must explain how the Host Interface reconciles a system where the force that steers the momentum is magnetic ($\mathbf{F}_m$) and does zero work, while the force that does the work is electric ($\mathbf{F}_e = q\mathbf{E}$).
It is demonstrated that the Magnetic Field acts not as an energy source, but as a \textbf{Steering Mechanism}. It redirects the momentum of the current, guiding the charge carriers to interact with the \textbf{Electric Field} (Induction). Thus, strictly speaking, all "Magnetic Energy" transfer is mediated by the Electric Gateway ($\mathbf{J} \cdot \mathbf{E}$). The Host Interface is the mechanism that reconciles this orthogonality.

\section{The Spectrum of Circulation}

The behavior of all magnetic systems is unified by viewing them through a single topological lens: \textbf{The Scale of the Loop.}

In a coil of wire, free electrons are constrained to the wire geometry by the host interface but are free to move along the wire path. The circulation radius is macroscopic.
In a magnetic material, the circulation radius of the current is atomic. The electrons orbit the nucleus or spin on their axis.
The fundamental physical distinction lies only in the scale at which the current is bound by the host interface. 

We view macroscopic magnetization not as a continuous abstract polarization field, but kinematically as an assembly of microscopic, superconducting solenoids of atomic radius. These represent the intrinsic angular momentum (spin) or orbital motion of the constituent electrons. 

Within this framework, Permeability and Susceptibility are reinterpreted not as opaque material constitutive properties, but as geometric coefficients describing the density and orientation of these elementary microscopic loops.

\section{The Mechanics of Interaction: Three Gateways}

As in Part III, the Host Interface ($\mathbf{F}_{\text{host}}$) mediates the interaction between the electromagnetic field and the mechanical world. For rotational currents, we identify the same three gateways, mirroring the electric domain:

\begin{enumerate}
    \item \textbf{Dissipation (The Thermal Sink):} Energy is irreversibly transferred to the random motion of the lattice (heat). This is the physics of Resistance and hysteretic losses, acting as a drain on the field energy.
    \item \textbf{Work (The Continuum Sink):} Energy is orderedly transferred to the bulk motion of the material (kinetic energy and deformation). This is the physics of Actuation and Magnetostriction.
    \item \textbf{Storage (The Potential Sink/Source):} Energy is reversibly transferred to the microscopic degrees of freedom (potential energy). Energy is stored in the variance of the microscopic Magnetic fields ($\langle \delta \mathbf{b}^2 \rangle$). This is the physics of Magnetic Binding Energy.
\end{enumerate}

\section{Summary of Contributions}

In analyzing these gateways, Part IV presents three key theoretical distinctions:

\begin{enumerate}
    \item \textbf{The Microscopic Model of Spin:} The energetics of the electron spin are rigorously addressed. The "Active Loop" model and the \textbf{Intrinsic Energy Reservoir} are proposed, demonstrating that for the Amperian model to hold, the electron must possess an internal energy source to maintain its constant current against induction.
    
    \item \textbf{The Definition of Magnetic Storage:} Magnetic Energy is defined not as a potential of the bulk material, but as the variance of the microscopic field. This provides a unified definition of "Field Energy" that covers both the vacuum and the material contribution, symmetric to the electric case.
    
    \item \textbf{The Ampere Topology of Force:} The derivation of the magnetic force density is refined by considering the topological exclusion of self-fields. A \textbf{Lattice Adaptation Term} is proposed to correct the standard Einstein-Laub formulation for the discrete nature of matter.
\end{enumerate}

\section{Roadmap of Part IV}

The logical progression of the analysis is as follows:

\begin{enumerate}
    \item \textbf{Chapter \ref{chap:RotationalCurrents}: Free Rotational Currents.} Analysis of the baseline of free circular currents in conductors, establishing the "Pivot" mechanism of the magnetic force.
    \item \textbf{Chapter \ref{chap:Spin}: Spin and The Reservoir.} Investigation of the energetics of the elementary current loop. Derivation of the \textbf{Intrinsic Reservoir} requirement and definition of the Microscopic Energy Source.
    \item \textbf{Chapter \ref{chap:Magnetic_Host}: The Host Interface.} Analysis of the material response (Bound Currents). Derivation of how Paramagnetism (Source) and Diamagnetism (Storage) emerge from the mechanics of the active and passive loops.
    \item \textbf{Chapter \ref{chap:MagneticForce}: Force Density in Magnetics: The Ampere Topology.} Derivation of the macroscopic force density on magnetized matter, parallel to the derivation of the electric force density in Part III, introducing the lattice adaptation term.
\end{enumerate}

\chapter{Free Rotational Currents}
\label{chap:RotationalCurrents}

\section{Case Study 1: The Rigid Ideal Superconductor (Flux Exclusion)}
\label{sec:IdealSuperconductor}

The analysis begins with the magnetic baseline: a rigid, ideal superconductor placed in an external magnetic field $\mathbf{B}_{\text{ext}}$.
\begin{itemize}
    \item \textbf{Ideal ($R=0$):} The electron fluid flows without friction.
    \item \textbf{Rigid ($\mathbf{V}_+ = \mathbf{0}$):} The lattice is fixed in the laboratory frame.
\end{itemize}

\subsection{Dynamics of the Bulk: The Quasistatic Balance}

Consider the interior of the superconductor. When the external magnetic field changes, it induces a rotational electric field ($\nabla \times \mathbf{E}_{\text{ext}} = -\dot{\mathbf{B}}_{\text{ext}}$).

However, just as in the electrostatic case, the electron fluid possesses negligible inertia ($m_- \to 0$); it cannot sustain a net force. The fluid accelerates instantaneously, generating a current $\mathbf{J}$ whose self-field $\mathbf{E}_{\text{self}}$ (arising from $\dot{\mathbf{B}}_{\text{self}}$) perfectly opposes the external driver.

The force balance is instantaneous and quasistatic:
\begin{equation}
    \underbrace{ \rho_- \frac{d \mathbf{v}_{-}}{dt} }_{\approx 0} \approx \rho_- (\mathbf{E}_{\text{ext}} + \mathbf{E}_{\text{self}}) \quad \implies \quad \mathbf{E}_{\text{total}} \approx \mathbf{0}
\end{equation}
This imposes the strict constraint of \textbf{Perfect Dynamic Screening}: the total electric field is cancelled everywhere within the bulk. Consequently, by Faraday's Law ($\dot{\mathbf{B}} = -\nabla \times \mathbf{E}$), the time rate of change of the internal magnetic flux is zero.

\subsection{The Closed Energy Gateway}
This balance has a direct energetic consequence. Although a massive persistent current $\mathbf{J}$ flows to maintain this shielding, it flows through a region where the total electric field is effectively zero.

The macroscopic energy balance simplifies to the conservation of field energy alone:
\begin{equation}
    0 = \underbrace{ \mathbf{J} \cdot \mathbf{E} }_{0} = \underbrace{ - \left( \frac{\partial u_{\text{EM}}}{\partial t} + \nabla \cdot \mathbf{S}_{\text{EM}} \right) }_{\text{Field Dynamics}}
\end{equation}
\textbf{Interpretation:} Just as in the electric conductor (Chapter \ref{chap:Conductors}), the energy gateway is closed. The superconductor does not absorb energy; it merely redistributes the magnetic field energy around its volume ($\nabla \cdot \mathbf{S}_{\text{EM}}$). The work done by the external source goes entirely into rearranging the vacuum fields, not into the kinetic energy of the electrons.

\subsection{Statics of the Boundary: Force Transmission}

The shielding currents are confined to a thin layer at the physical boundary. Here, the local magnetic field exerts a massive Lorentz force on the moving electron fluid:
\begin{equation}
    \mathbf{f}_{\text{Lorentz}} = \mathbf{J}_{\text{surf}} \times \mathbf{B}_{\text{local}}
\end{equation}

The electron fluid is constrained to the conductor by the \textbf{Host Force} ($\mathbf{f}_{\text{host}}$). Just as the lattice holds static surface charges in place against electrostatic repulsion, it holds the moving current loop in place against the magnetic Lorentz force.

The momentum balance at the surface resolves the mechanism of transmission:
\begin{equation}
    \underbrace{ \mathbf{J}_{\text{surf}} \times \mathbf{B}_{\text{local}} }_{\text{Lorentz Force on Electrons}} + \underbrace{ \mathbf{f}_{\text{host}} }_{\text{Constraint by Lattice}} = \mathbf{0}
\end{equation}
This confirms that whilst the magnetic force acts on the current, the mechanical force is delivered by the lattice. The Host Interface acts as a rigid \textbf{Constraint}, transmitting the stress without absorbing energy ($P = 0$).

\section{Case Study 2: The Resistive Conductor}
\label{sec:ResistiveMagnetic}

The reality of material imperfection is now introduced. The same rigid conductor ($\mathbf{V}_+ = \mathbf{0}$) is considered, but the Host Interface now includes a frictional component.
\begin{itemize}
    \item \textbf{Resistive ($R>0$):} The electron fluid experiences a drag force proportional to its velocity ($\mathbf{f}_{\text{drag}} \propto - \mathbf{V}_-$).
\end{itemize}
This scenario engages the \textbf{Dissipation Gateway}. Unlike the ideal case, the flow of magnetic current now comes with an energetic cost.

\subsection{Phase I: The Transient Response (The Energy Sink)}

When the magnetic environment changes ($\dot{\mathbf{B}} \neq 0$), a rotational electric field is induced. This field accelerates the electrons, which collide with the thermal vibrations of the lattice.

Due to the negligible inertia of the electrons, the total driving electric force (sum of external and self-induced fields) must be instantaneously balanced by the drag force from the Host:
\begin{equation}
    \underbrace{ \rho_- \mathbf{E}_{\text{total}} }_{\text{Net Driving Force}} + \underbrace{ \mathbf{f}_{\text{drag}} }_{\text{Resistive Host Force}} \approx \mathbf{0}
\end{equation}

This balance has a direct physical implication: the total electric field inside the conductor cannot be zero (as in the superconductor). Instead, a non-zero field must persist to overcome the drag. This is the microscopic origin of Ohm's Law ($\mathbf{E} = \eta \mathbf{J}$), where the field strength is dictated by the friction of the Host Interface.

This phase is defined by a continuous transfer of energy. We examine the power transfer across the Host Interface:
\begin{equation}
    P_{\text{host}} = \mathbf{f}_{\text{drag}} \cdot \mathbf{V}_-
\end{equation}
Since the drag force opposes the velocity, this term is strictly negative ($P_{\text{host}} < 0$). In this framework, a negative Host Power signifies a \textbf{Sink Action}: energy is flowing \textit{out} of the electromagnetic domain and \textit{into} the thermodynamic domain.

The Macroscopic Energy Balance reveals the complete description:
\begin{equation}
    \underbrace{ - P_{\text{host}} }_{\text{Heat Generation}} = \underbrace{ \mathbf{J} \cdot \mathbf{E}_{\text{total}} }_{\text{Power from Field}} = \underbrace{ - \left( \frac{\partial u_{\text{EM}}}{\partial t} + \nabla \cdot \mathbf{S}_{\text{EM}} \right) }_{\text{Loss of Field Energy}}
\end{equation}

This confirms the physical pathway of resistive heating.

\subsection{Phase II: Convergence and Equilibrium}

Crucially, the presence of resistance allows for two distinct outcomes depending on the nature of the external drive.

\subsubsection{Scenario A: Constant External Drive (Steady State)}
If the external source maintains a constant rate of flux change (constant voltage drive), the external induced field remains constant. The system settles into a dynamic equilibrium where the driving force is continuously counterbalanced by the resistive drag ($\rho_- \mathbf{E}_{\text{ext}} \approx -\mathbf{f}_{\text{drag}}$). A constant current flows, and the Host Interface acts as a continuous \textbf{Drain}, dissipating field energy into heat at a constant rate.

\subsubsection{Scenario B: Source Removal (Relaxation)}
If the external source is removed, the external drive vanishes. However, the current cannot stop instantly due to the system's geometric inductance.
The decaying current generates a self-induced electric field (Lenz's Law) that continues to push the electrons forward against the drag.
The system enters a relaxation phase where the stored magnetic energy drives the current. The current decays exponentially ($\mathbf{J} \to 0$) as the Host Interface drains the remaining magnetic energy into heat.

\paragraph{Conclusion.}
This comparison illustrates the critical role of the Host Interface. In the ideal case (Superconductor), the Host Force acts as a \textbf{Constraint}, supporting a persistent current and maintaining flux exclusion. In the resistive case, the Host Force acts as a \textbf{Drain}, dissipating energy until the system reaches thermal equilibrium or magnetic saturation. The unified framework captures both the transient heating and the relaxation dynamics without modification.

\section{Case Study 3: Mechanical Work (The Magnetic Motor)}
\label{sec:MagneticWork}

The case of a conductor moving with macroscopic velocity $\mathbf{V}_{\text{mech}}$ through a magnetic field (e.g., an electric motor) is now examined. This derivation addresses the apparent inconsistency of how a force proportional to $\mathbf{j}\times\mathbf{b}$ results in energy conversion, given that the magnetic force itself performs zero total work on any charge carrier.

\subsection{The Kinematic Decomposition}

Consider an ideal conductor ($R=0$) carrying a conduction current density $\mathbf{j}$ (tangential to the surface) and moving with a bulk mechanical velocity $\mathbf{V}_{\text{mech}}$ (normal to the surface).
The total velocity of the charge carrier fluid is the vector sum:
\begin{equation}
    \mathbf{v}_{\text{total}} = \underbrace{ \mathbf{v}_{\text{drift}} }_{\text{Internal Flow (tangential)}} + \underbrace{ \mathbf{V}_{\text{mech}} }_{\text{Bulk Motion (normal)}}
\end{equation}

The total Lorentz force acting on the carrier is:
\begin{equation}
    \mathbf{f}_{\text{Lorentz}} = q(\mathbf{v}_{\text{total}} \times \mathbf{B}) = q(\mathbf{v}_{\text{drift}} \times \mathbf{B}) + q(\mathbf{V}_{\text{mech}} \times \mathbf{B})
\end{equation}
Because the magnetic force is always perpendicular to the total velocity, the \textbf{Net Work done by the Magnetic Field is strictly zero}:
\begin{equation}
    P_{\text{mag}} = \mathbf{f}_{\text{Lorentz}} \cdot \mathbf{v}_{\text{total}} \equiv 0
\end{equation}
Energy is clearly being converted, but the magnetic field is not the source. To find the source, the force balance must be decomposed into components parallel and perpendicular to the conductor boundary.

\subsection{Step 1: The Tangential Balance (Electrical Energy Conversion)}

First, consider the force component tangent to the surface (along the current path inside the conductor). The bulk motion generates a motional Lorentz term $q (\mathbf{V}_{\text{mech}} \times \mathbf{B})$.
This force attempts to accelerate the electrons along the current path. To maintain equilibrium (Ohm's Law for $R=0$), a self-induced Electric Field ($\mathbf{E}_{\text{ind}}$) must instantaneously arise to perfect cancel it:
\begin{equation}
    q \mathbf{E}_{\text{ind}} + q (\mathbf{V}_{\text{mech}} \times \mathbf{B})_{\parallel} = 0
\end{equation}
Taking the dot product with the drift velocity $\mathbf{v}_{\text{drift}}$ (which is purely tangential):
\begin{equation}
    \mathbf{E}_{\text{ind}} \cdot \mathbf{v}_{\text{drift}} = - (\mathbf{V}_{\text{mech}} \times \mathbf{B}) \cdot \mathbf{v}_{\text{drift}}
\end{equation}

\textbf{The Work Mechanism:} This induced \textbf{Electric Field} performs real work on the charge carriers.
\begin{equation}
    P_{\text{elec}} = \mathbf{j} \cdot \mathbf{E}_{\text{ind}} \neq 0
\end{equation}
This represents the electrical power extracted from (or delivered to) the circuit (Back-EMF).

\subsection{Step 2: The Normal Balance (Mechanical Energy Output)}

Now consider the force component normal to the conductor boundary. The drift current generates the standard magnetic pressure term:
\begin{equation}
    \mathbf{f}_{\text{pressure}} = q (\mathbf{v}_{\text{drift}} \times \mathbf{B})_{\perp}
\end{equation}
This force pushes the electrons against the boundary of the conductor. As established in Case 1, this push is transmitted immediately to the lattice via the \textbf{Host Force}.
\begin{equation}
    \mathbf{f}_{\text{host}} = - \mathbf{f}_{\text{pressure}}
\end{equation}
Since the lattice is moving with velocity $\mathbf{V}_{\text{mech}}$, the Host Interface performs mechanical work:
\begin{equation}
    P_{\text{mech}} = \mathbf{f}_{\text{host}} \cdot \mathbf{V}_{\text{mech}} = - (\mathbf{j} \times \mathbf{B}) \cdot \mathbf{V}_{\text{mech}}
\end{equation}

\subsection{The Synthesis: The 90-Degree Pivot}

The two steps are now combined to reveal the mechanism of conservation, resolving the \textbf{Orthogonality Constraint}. Using the vector triple product identity $(\mathbf{A} \times \mathbf{B}) \cdot \mathbf{C} = -(\mathbf{C} \times \mathbf{B}) \cdot \mathbf{A}$, the mechanical output can be related to the electrical input:

\begin{equation}
    P_{\text{mech}} = - (\mathbf{j} \times \mathbf{B}) \cdot \mathbf{V}_{\text{mech}} = (\mathbf{V}_{\text{mech}} \times \mathbf{B}) \cdot \mathbf{j}
\end{equation}
Comparing this to the electrical work derived in Step 1 ($\mathbf{j} \cdot \mathbf{E}_{\text{ind}} = - (\mathbf{V}_{\text{mech}} \times \mathbf{B}) \cdot \mathbf{j}$), we arrive at the exact balance:

\begin{equation}
    \boxed{ P_{\text{elec}} + P_{\text{mech}} = 0 }
\end{equation}

\textbf{Physical Interpretation:}
The Lorentz force does not create energy. It acts as a \textbf{Pivot}—a localized kinematic transformer that perfectly transfers power between orthogonal domains.
\begin{enumerate}
    \item The \textbf{Induced Electric Field} inputs energy into the electron's momentum along the current path ($P_{\text{elec}}$).
    \item The \textbf{Magnetic Force} instantly redirects this momentum by 90 degrees, converting tangential velocity into normal pressure.
    \item The \textbf{Host Force} transmits this pressure to the bulk mass, extracting the energy as mechanical work ($P_{\text{mech}}$).
\end{enumerate}

The electron fluid acts as a lossless coupling medium. It receives energy electrically and delivers it mechanically, with the magnetic field serving as the kinematic pivot that turns the force vector. The ``Zero Work'' property of the magnetic field is strictly preserved: it does no work itself, but enables the transfer between two orthogonal work-performing agents (The Electric Field and The Lattice).

\subsection{Summary: The Unified Energy Balance}

This local kinematic mechanism can be mapped onto the global energy budget derived in Part II. The energy exchange is rigorously described by the Macroscopic Energy Balance equation:

\begin{equation}
    \underbrace{ - P_{\text{host}} }_{\text{Mechanical Power}} = \underbrace{ \mathbf{j} \cdot \mathbf{E}_{\text{ind}} }_{\text{Gateway}} = \underbrace{ - \left( \frac{\partial u_{\text{EM}}}{\partial t} + \nabla \cdot \mathbf{S}_{\text{EM}} \right) }_{\text{Field Budget}}
\end{equation}

The direction of energy flow is dictated entirely by the sign of the Host Power term $P_{\text{host}} = \mathbf{f}_{\text{host}} \cdot \mathbf{V}_{\text{mech}}$.

\paragraph{1. Motor Mode (The Mechanical Sink):}
If the conductor moves \textit{with} the magnetic force ($\mathbf{V}_{\text{mech}} \cdot \mathbf{f}_{\text{host}} > 0$), the system performs mechanical work.
\begin{equation}
    P_{\text{host}} > 0 \quad \implies \quad P_{\text{elec}} < 0 \quad \implies \quad \text{Energy flows } \textbf{Field} \to \textbf{Mechanics}
\end{equation}
The Host Interface acts as a \textbf{Local Energy Sink}. Energy is extracted from the electromagnetic circuit (Back-EMF opposes current) and converted into the kinetic energy of the bulk matter.

\paragraph{2. Generator Mode (The Mechanical Source):}
If an external mechanical agent pushes the conductor \textit{against} the magnetic force ($\mathbf{V}_{\text{mech}} \cdot \mathbf{f}_{\text{host}} < 0$), the mechanical system performs work on the field.
\begin{equation}
    P_{\text{host}} < 0 \quad \implies \quad P_{\text{elec}} > 0 \quad \implies \quad \text{Energy flows } \textbf{Mechanics} \to \textbf{Field}
\end{equation}
The Host Interface acts as a \textbf{Local Energy Source}. Kinetic energy is consumed to drive the electrons against the moving induced electric field, pumping energy back into the electromagnetic field.

\paragraph{Conclusion.}
This confirms that mechanical work in magnetic systems is just another form of current-field interaction mediated by the Host Interface. The magnetic field acts as the lossless pivot that allows the \textbf{Host Force} (Mechanics) and the \textbf{Induced Electric Field} (Electrodynamics) to exchange energy directly, preserving the conservation laws without requiring the magnetic field itself to perform work.

\section{Case Study 4: Elastic Deformation (Magnetic Pressure)}
\label{sec:MagneticDeformation}

Finally, consideration is given to a stationary but mechanically \textbf{elastic} conductor (e.g., a solenoid winding). The lattice acts as a spring that can deform under stress. This scenario introduces a new energy destination: \textbf{Elastic Potential Energy}.

\subsection{The Mechanism of Deformation}

When a current flows, the magnetic field exerts a Lorentz force on the charge carriers ($\mathbf{J} \times \mathbf{B}$). As established, this force is transmitted immediately to the lattice boundary via the Host Force.
If the lattice is elastic, it yields. The surface layer is pushed, stretching or compressing the atomic bonds in the bulk material.

Equilibrium is reached when the internal \textbf{Mechanical Stress} ($\mathbf{T}_{\text{mech}}$) balances the magnetic pressure transmitted by the Host:
\begin{equation}
    \underbrace{ \mathbf{n} \cdot \mathbf{T}_{\text{mech}} }_{\text{Elastic Stress}} + \underbrace{ \mathbf{f}_{\text{host}} }_{\text{Magnetic Pressure}} = \mathbf{0}
\end{equation}
The deformation is not a direct action of the field on the atoms, but a mechanical response to the Host Force pushing the current-carrying boundary.

\subsection{Energy Balance: Reversible Storage}

During the transient deformation phase, the conductor boundary moves with velocity $\mathbf{v}_{\text{def}}$.
The energy exchange follows the standard Macroscopic Energy Balance:
\begin{equation}
    \underbrace{ - P_{\text{host}} }_{\text{Mechanical Power}} = \underbrace{ \mathbf{j} \cdot \mathbf{E}_{\text{ind}} }_{\text{Gateway}} = \underbrace{ - \left( \frac{\partial u_{\text{EM}}}{\partial t} + \nabla \cdot \mathbf{S}_{\text{EM}} \right) }_{\text{Field Budget}}
\end{equation}
Here, the Host Power is $P_{\text{host}} = \mathbf{f}_{\text{host}} \cdot \mathbf{v}_{\text{def}}$.
As the material deforms with the force ($P_{\text{host}} > 0$), the Host Interface acts as a \textbf{Local Energy Sink}. Energy is drawn from the electromagnetic circuit (via the induced Back-EMF) and stored as potential energy in the lattice bonds ($\Delta U_{\text{elastic}} > 0$).

This process is fully reversible. If the current is reduced, the elastic tension restores the shape ($\mathbf{v}_{\text{def}}$ reverses). The Host Force performs negative work ($P_{\text{host}} < 0$), acting as a \textbf{Local Energy Source} that pumps the stored mechanical energy back into the electromagnetic domain (generating a Forward-EMF that sustains the current).

\section{Case Study 5: The Self-Contained Field (The LC Oscillator)}

Before considering magnetic materials, we present one further example of free charges on conductors: the idealized LC circuit. This system, consisting of two perfect conducting plates connected by a perfect conducting coil (the entire system is basically one single conductor brought into a certain form), serves as the illustrative case study for energy dynamics that remain confined \textit{within} the electromagnetic domain.

\subsection{The Closed Gateway}
Assume the capacitor is initially charged, storing energy entirely in the electric field $\mathbf{E}$ between the plates ($u_{\text{EM}} \approx \frac{1}{2}\varepsilon_0 E^2$). When the circuit is closed, charges flow through the coil, creating a current $\mathbf{J}_{\text{micro}}$ and a magnetic field $\mathbf{B}$ ($u_{\text{EM}} \approx \frac{1}{2\mu_0} B^2$).

Crucially, inside the perfect conductors (plates and coil wire), the net electric field must be effectively zero ($\mathbf{E} \approx \mathbf{0}$). This is a necessary self-consistency condition: in the massless carrier approximation, any non-zero net field would impart infinite acceleration to the current. The induced and capacitive fields must perfectly cancel inside the wire.

Consequently, the Energy Gateway within the material volume is effectively closed:
\begin{equation}
    P_{\text{gateway}} = \mathbf{J}_{\text{micro}} \cdot \mathbf{E} \approx 0
\end{equation}
Since the gateway term is zero, the framework confirms that \textbf{no energy is converted} to non-electromagnetic forms. There is no heating ($P_{\text{host}}=0$) and no mechanical work ($\Delta E_{\text{kin}}=0$). The "Host Interface" is effectively bypassed.

\subsection{Dynamics of the Self-Contained Field}
With the gateway closed, the energy evolution is governed entirely by the vacuum conservation law:
\begin{equation}
    0 = - \left( \frac{\partial u_{\text{EM}}}{\partial t} + \nabla \cdot \mathbf{S}_{\text{EM}} \right)
\end{equation}
This equation describes two distinct processes occurring simultaneously in the vacuum space surrounding the conductors:

\begin{enumerate}
    \item \textbf{Oscillation (Transformation):} The energy density $u_{\text{EM}}$ oscillates locally between electric potential energy (capacitor gap) and magnetic kinetic energy (coil interior).
    
    \item \textbf{Radiation (Redistribution):} The Poynting vector $\mathbf{S}_{\text{EM}}$ is non-zero and points outward toward infinity. This represents the phenomenon of \textbf{Radiation}.
\end{enumerate}

\subsection{The Entropy of the Field}
The radiation process offers a profound insight into the nature of field energy.
At $t=0$, the energy is highly ordered and concentrated within the localized volume of the capacitor. Over time, as the circuit oscillates, the Poynting flux carries this electromagnetic field energy outward, spreading it irreversibly into the infinite vacuum.

While electromagnetic field energy leaves the \textbf{localized circuit volume}, it never leaves the \textbf{electromagnetic domain}.
\begin{itemize}
    \item \textbf{Conversion vs. Spreading:} Unlike resistive heating (where EM energy becomes matter energy), radiation is simply EM energy moving to a new location.
    \item \textbf{Field Entropy:} This process can be viewed as an increase in the entropy of the field distribution. The total EM energy in this example is conserved, but it transitions from a concentrated, useful state (in the circuit) to a diffuse, disordered state (waves in deep space).
\end{itemize}

This case study reinforces the definition of the gateway. The term $\mathbf{J} \cdot \mathbf{E}$ is not merely a loss term; it is the specific \textit{coupler} to the non-electromagnetic world. When that coupler is zero, energy may spread within the electrodynamic domain via $\mathbf{S}_{\text{EM}}$, but it remains within the electromagnetic sector, never becoming Heat or Work.

\section[Conclusion]{Conclusion: \\ The Universal Routing Mechanism}

This classification serves as the structural validation of our unified framework. It reveals that the Host Interface acts as the \textbf{Universal Routing Mechanism} for electromagnetic energy.

Recall from Part III that the divergence of the field energy is not merely a loss term; it is the \textbf{interaction density} ($\mathcal{D}$), the scalar map of where the field physically "touches" the matter.
\begin{equation}
    \mathcal{D} = - (\nabla \cdot \mathbf{S}_{\text{EM}} + \dot{u}_{\text{EM}}) = \mathbf{J} \cdot \mathbf{E}
\end{equation}
The analysis of rotational currents confirms the \textbf{kinematic lock}: there is no sink without a moving fluid. The field can only transfer energy where there is a current $\mathbf{J}$ to accept the work.

Once admitted through this gateway, the Host Interface ($\mathbf{F}_{\text{host}}$) determines the mechanical destination of the energy. In the magnetic domain, the Host acts as a \textbf{Pivot}, pivoting the orthogonal magnetic force to route the energy into one of two distinct mechanical sinks:

\begin{enumerate}
    \item \textbf{The Bridge to Thermodynamics:} In the \textbf{Resistive} case, $\mathbf{F}_{\text{host}}$ acts as a dissipative drag, converting ordered field energy into the random thermal motion of the lattice (Heat).
    \item \textbf{The Bridge to Continuum Mechanics:} In the \textbf{Movable} or \textbf{Deformable} case, $\mathbf{F}_{\text{host}}$ acts as an actuator, converting field energy into the ordered kinetic or potential energy of the bulk mass (Work/Storage).
\end{enumerate}

\paragraph{Validation of Consistency.}
Crucially, this framework does not violate the \textbf{Force-Energy Consistency Criterion (FECC)}.
\begin{itemize}
    \item \textbf{Clear Mass Target:} The magnetic force $\mathbf{J} \times \mathbf{B}$ acts on the electron fluid, but performs no work ($P_{\text{mag}} = 0$).
    \item \textbf{Defined Gateway:} The energy transfer is strictly defined by the work done by the \textit{induced electric field} ($\mathbf{J} \cdot \mathbf{E}_{\text{ind}}$).
    \item \textbf{The Pivot Mechanism:} The magnetic force acts as a lossless pivot, instantaneously redirecting the momentum gained from the electric field (tangential) into pressure against the Host Interface (normal). 
\end{itemize}

Having validated the framework for free rotational currents, where the "Host" is the only partner, the analysis turns to the next phase: \textbf{Magnetic Materials}. Here, it is proposed that the same third pathway for energy is introduced as in Part III: a \textbf{Storage Gateway} for energy stored reversibly in the \textbf{microscopic} high-frequency fields of the magnetic material.

\chapter{Spin}
\label{chap:Spin}

This chapter proposes a model for the spin of the electron. Spin is inherently a quantum mechanical property. Two classical models have been historically proposed to model the electron spin and include its field and force contribution into microscopic electrodynamics:
The electric current model (Amperian) and the magnetic charge model (Gilbertian). One may either model the electron's magnetic dipole as a closed current loop or as dipole of separated magnetic charges. 

Experimental evidence analyzed by J.D. Jackson \cite{Jackson1977} from the \textbf{Hyperfine Structure} of atomic spectra confirms the current loop prediction. The electron behaves physically as a current loop, not as joined magnetic charges.
The microscopic current loop model of the electron spin is thus adopted by most modern literature (e.g., Jackson, Zangwill, Griffiths) as the standard classical approximation to include the spin's quantum effect within a classical electrodynamics theory. We proceed to use the same approach.

However, when the \textbf{energetics} of this model are analyzed using the rigorous work balance derived in the previous section, a discrepancy arises. The definition of a "constant current loop" is not inherently self-consistent without an energy source. The analysis below investigates this interaction in forensic detail.

\section{System Integration: The Energetics of Coupled Currents}
\label{sec:SystemIntegration}

Before proposing a microscopic model for the spin of the electron, we analyze the interaction of two current-carrying loops. 
Consider a system consisting of two distinct material domains, $\Omega_1$ and $\Omega_2$, both composed of ideal superconducting matter carrying parallel currents. The domains interact via their mutual magnetic field. The two pieces of material might have arbitrary shapes, but for simplicity, we treat them as two current-carrying loops.
\begin{itemize}
    \item \textbf{Domain 1 (The Stationary Source):} A rigid current loop fixed in the laboratory frame ($\mathbf{v}_{\text{mech}} = \mathbf{0}$).
    \item \textbf{Domain 2 (The Moving Sink):} A rigid current loop free to move with macroscopic velocity $\mathbf{v}_{\text{mech}} \neq \mathbf{0}$ under the influence of the attractive Lorentz force generated by the field of the other loop.
\end{itemize}

We briefly recapitulate the interaction of the two current loops, which was established in Chapter \ref{chap:RotationalCurrents}.

\subsubsection{Analysis of Stationary Loop 1: Pure Redistribution}

In the stationary loop $\Omega_1$, the lattice velocity is zero ($\mathbf{v}_{\text{mech}} = \mathbf{0}$). As established in Case Study 1, the Host Force acts solely as a static constraint to hold the current path in place. As the other loop moves towards the first loop, the magnetic field strength changes and an electric field is induced. This externally induced field is however counteracted by the self-induced field. Consequently, the mechanical work performed by the Host Interface is identically zero. The non-moving loop behaves as a conservative region of the continuum. Although the local energy density $u_{\text{EM}}$ changes with time (as the approaching second loop alters the total magnetic configuration), there is no energy conversion. The domain acts neither as a source nor a sink; it merely facilitates the \textbf{spatial redistribution} of field energy. The Poynting vector flows through it, reshaping the field, but no energy leaves the electromagnetic sector.

\subsubsection{Analysis of Moving Loop 2: Active Conversion}

In the moving domain $\Omega_2$, the lattice moves with velocity $\mathbf{v}_{\text{mech}}$ in the direction of the magnetic force. As established in Case Study 3, the Host Force acts as a dynamic actuator. The mechanical work performed is positive. Domain 2 acts as a \textbf{Local Energy Sink}. The term on the right-hand side represents a continuous extraction of energy from the field. In the language of the \textbf{Universal Routing Mechanism} (Chapter \ref{chap:IntroPart4}), the Host Interface routes this energy flux $\mathbf{S}_{\text{EM}}$ directly into the \textbf{Kinetic Sink} of the bulk matter ($P_{\text{mech}}$). The field energy is converted and instantaneously transduced into mass motion.

\subsubsection{Global Synthesis: Conservation via Depletion}

The global energy balance is obtained by integrating over the entire infinite volume $V_{\infty}$. The divergence theorem eliminates the flux term at infinity, leaving:
\begin{equation}
    \frac{d}{dt} \int_{V_{\infty}} u_{\text{EM}} \, dV = - \int_{\Omega_2} P_{\text{host}, 2} \, dV
\end{equation}

In the quasi-static limit characteristic of this interaction, the electric field energy is negligible compared to the magnetic storage. The total field energy is therefore defined strictly by the magnetic configuration:
\begin{equation}
    \int_{V_{\infty}} u_{\text{EM}} \, dV \approx \int_{V_{\infty}} \frac{1}{2\mu_0} \mathbf{B} \cdot \mathbf{B} \, dV.
\end{equation}

As the two current loops approach, the magnetic flux conservation constraint within the superconductors forces the current densities $\mathbf{J}_1$ and $\mathbf{J}_2$ to decrease. Consequently, the total integrated magnetic energy of the system drops.
The energy balance equation confirms the conservation mechanism: \textbf{The rate of decrease in the total magnetic field energy is exactly equal to the rate of mechanical work done on the moving domain.}

The interaction is thus resolved as a flow process:
\begin{enumerate}
    \item \textbf{Redistribution:} The stationary domain $\Omega_1$ passively shapes the field geometry, allowing energy to flow but absorbing none.
    \item \textbf{Conversion:} The moving domain $\Omega_2$ actively absorbs this energy flux, converting the system's magnetic potential energy into mechanical kinetic energy.
\end{enumerate}

\subsection{Driven Interaction: The Constant Current Regime}

We now extend the analysis to an open, driven system. We assume that each current loop is connected to an external energy supply (e.g., a voltage source) which actively adjusts the electromotive force to maintain the currents $\mathbf{J}_1$ and $\mathbf{J}_2$ at a \textbf{constant magnitude}, regardless of the system's evolution.

Power is injected into the system. In this framework, these sources are classified not as part of the internal field budget, but as an \textbf{External Power Input} ($P_{\text{ext}}$) entering the system via the Host Interface boundaries.

\subsubsection{Analysis of the Moving Loop: Driven Conversion}

As the loop $\Omega_2$ moves into the stronger field of $\Omega_1$, it experiences a motional electromotive force (Back-EMF) that opposes the current, as explained in Case Study 3 of Chapter \ref{chap:RotationalCurrents}. To maintain a constant current $\mathbf{J}_2$, the external source must now provide a counterbalancing force to cancel the induced electric force caused by the movement of the loop. As this external force already counterbalances the force in the direction of current flow, the self-induced field of the current itself is not needed to maintain equilibrium. The current in the moving loop stays constant.

The energy exchange is as follows: 
The external source injects energy into the electron fluid to sustain its velocity against the magnetic braking effect. The Lorentz force pivot instantaneously redirects this injected electrical energy into mechanical pressure. 
The external source pays directly for the mechanical kinetic energy of the bulk matter.
There is no sink or source of electromagnetic field energy within the moving loop; the externally injected energy is transferred directly into kinetic energy of the bulk matter.

\subsubsection{Analysis of the Stationary Loop: Driven Field Growth}

In the stationary loop $\Omega_1$, the approaching magnetic field of $\Omega_2$ induces a Back-EMF via mutual inductance. Although the loop is stationary and performs no mechanical work, the external source is now counterbalancing the externally induced electric field. 

However, the external source must bring in energy to counterbalance the externally induced electric field. This external source must perform work to maintain the constant current $\mathbf{J}_1$ against the externally induced voltage. This external source is paying for the energy of the electromagnetic field.

The energy exchange is as follows: 
The externally induced electric field is counterbalancing the external source. The external source is part of the Host Interface. Thus, the external source pumps in electromagnetic field energy into the system and acts as a local source of electromagnetic field energy.

\subsubsection{Global Synthesis: The Cost of Intervention}

This scenario inverts the energy dynamic of the isolated case where there is no external source to keep the current constant. In the isolated system, the field energy decreased to pay for the work. In this driven system, the total magnetic field energy \textit{increases} as the loops approach, because the currents are artificially prevented from dropping. The magnetic field intensity in the gap rises due to constructive interference.

The global energy balance reveals that the external sources must supply a dual requirement:
\begin{equation}
    P_{\text{total\_input}} = \underbrace{ P_{\text{mech}} }_{\text{Work Done}} + \underbrace{ \frac{d}{dt} \int_{V_{\infty}} \frac{1}{2\mu_0} \mathbf{B} \cdot \mathbf{B} \, dV }_{\text{Field Growth}}
\end{equation}

This rigorously demonstrates that in a constant-current system, the mechanical work is not drawn from the field energy; rather, both the work and the increased field energy are supplied simultaneously by the external driver.

\section{Spin: The Intrinsic Active Loop}
\label{sec:Spin}

The insights derived from the macroscopic "Constant Current" regime are now applied to the microscopic foundation of magnetism itself: the intrinsic spin of the electron.

In standard literature (e.g., Jackson, Zangwill, Griffiths), the magnetic moment of the electron ($\mathbf{m}$) is modeled as a microscopic Amperian current loop. This analogy is essential for applying the classical Lorentz force laws. However, when the \textbf{energetics} of this model are analyzed using the work balance derived in the previous section, a discrepancy emerges.

\subsection{The Energetic Problem}

Quantum mechanics dictates that the spin magnitude of an elementary particle is an intrinsic, invariant property. The electron's magnetic moment cannot decay or decrease due to interaction with external fields.
In classical terms, this imposes a rigid constraint: \textbf{The effective current loop representing the spin must operate at Constant Current ($I = \text{const}$), regardless of the external electromagnetic environment.}

This constraint creates an immediate work-energy deficit when the electron interacts with a time-varying magnetic field. As demonstrated in Section~\ref{sec:SystemIntegration}, maintaining a constant current against an induced Back-EMF requires work.
\begin{itemize}
    \item \textbf{Induction:} A changing magnetic flux through the spin-loop induces an electromotive force (EMF) via Faraday's Law.
    \item \textbf{Work Deficit:} To keep the current constant against this EMF, energy must be supplied to the loop to drive the charges against the electric field.
    \item \textbf{The Missing Source:} In our macroscopic current loop example, this energy was provided by an external voltage source. For an isolated elementary particle, there are no wires and no external battery.
\end{itemize}

\subsection{Literature Context and the Wald Solution}

This energy deficit is often omitted in standard textbooks. A constant current loop is simply postulated, and the force is calculated, leaving the energy budget unclosed.

The notable exception is the analysis by Robert Wald \cite{Wald_AdvancedEM}. Wald rigorously confronts this problem, analyzing the mechanics of a magnetic dipole moving in a field gradient (where $\dot{\mathbf{B}}$ is experienced in the comoving frame). He concludes that because the magnetic field itself performs no work, the energy balance can \textit{only} be satisfied if the work done by the dipole comes from its \textbf{internal energy}. For an elementary particle, he explicitly identifies this internal reservoir as the particle's \textbf{Rest Mass} ($mc^2$).

\subsection{The Solution: The Microscopic Active Host Interface}

To incorporate spin into a consistent classical framework without violating the conservation of energy, the mechanism that enforces the constant constraint is explicitly modeled. We propose that the electron is defined not as a passive loop (like a superconducting ring), but as an \textbf{Active Superconducting Loop}.

It is proposed that the loop is stabilized by the \textbf{Microscopic Host Interface}. Unlike the passive lattice of a wire, the electron possesses an \textbf{Active Interface} coupled to its \textbf{Intrinsic Energy Reservoir} ($mc^2$).

\begin{itemize}
    \item \textbf{The Reservoir:} The electron spin maintains its magnitude by drawing upon its internal rest mass energy. This reservoir acts as the microscopic realization of the external voltage source in our macroscopic model.
    \item \textbf{The Mechanism:} When the spin interacts with an external field, the \textbf{Active Host Interface} instantaneously supplies or absorbs the exact amount of power required to cancel the induced EMF and maintain $|\mathbf{j}|$ constant.
    \item \textbf{The Energy Balance:} The work done by the field on the current ($\mathbf{j} \cdot \mathbf{E}$) is transferred directly into the \textbf{intrinsic reservoir} of the particle via the host interface ($P_{\text{host}}$).
\end{itemize}

This restores mechanical consistency. The Lorentz force $\mathbf{j} \times \mathbf{B}$ remains the valid mechanism of interaction, provided we account for the cost paid by the \textbf{Active Host Interface} to maintain the structural integrity of the loop.

\subsection{Alternative Formulations}
\label{sec:Magnetic_Charge_Alternative}

It is crucial to note that this energetic problem is not resolved by adopting alternative energy-momentum tensors such as Minkowski or Abraham. We will consider those tensors in more detail in Part VI.

Moreover, it is acknowledged that this energetic anomaly is an artifact of the Amperian (current loop) model. One could construct an alternative theory based on \textbf{Magnetic Charges} (Gilbert Dipoles).
\begin{itemize}
    \item A magnetic charge $q_m$ in a $\mathbf{B}$-field experiences a force $\mathbf{F} = q_m \mathbf{H}$ (analogous to $q\mathbf{E}$).
    \item Crucially, magnetic charges are not subject to Faraday induction. A constant magnetic dipole composed of separated magnetic charges requires no internal energy source to maintain its strength in a changing field.
\end{itemize}

While this would resolve the energy deficit elegantly, it contradicts the empirical reality of quantum matter. As definitively analyzed by Jackson \cite{Jackson1977}, the two models predict fundamentally different magnetic field topologies \textit{inside} the particle (the singular contact term).
\begin{itemize}
    \item The Gilbert model predicts an internal field antiparallel to the moment ($\int \mathbf{B}_{\text{int}} dV \propto -\mathbf{m}$).
    \item The Amperian model predicts an internal field parallel to the moment ($\int \mathbf{B}_{\text{int}} dV \propto +2\mathbf{m}$).
\end{itemize}
Experimental evidence from the \textbf{Hyperfine Structure} of atomic spectra confirms the Amperian prediction. The electron behaves physically as a current loop.

\textbf{Scope of this Work:} Consequently, the Amperian model is accepted as the baseline. The thermodynamic cost that comes with it is therefore accepted. It is claimed that the \textbf{Active Host Interface} is the necessary consistency condition to make this empirically correct model mechanically consistent with the conservation of energy.

\section{Conclusion: The Necessity of the Active Host}
\label{sec:Spin_Conclusion}

A fundamental energetic deficit has been identified in the classical description of magnetism. By analyzing the mechanics of the electron, a condition has been uncovered that is often overlooked in standard texts: \textbf{Intrinsic Spin is not a passive property; it is an energetically active constraint.}

\subsection{Empirical Confirmation}
The choice of model is not arbitrary. As definitively shown by Jackson \cite{Jackson1977}, the hyperfine structure of atomic spectra empirically contradicts the Magnetic Charge (Gilbert) model in favor of the Amperian Current Loop. We are therefore bound by physical reality to model magnetization as a circulation of charge.

However, accepting the Amperian model imposes a thermodynamic condition. To satisfy the quantum requirement of constant spin magnitude ($\dot{\mathbf{m}}=0$) in a classical framework, we must accept that the electron performs work against induced electromotive forces.

\subsection{The Consistency Condition}
This cost is resolved by formalizing the \textbf{Microscopic Host Interface}.
\begin{itemize}
    \item It is acknowledged that the energy required to maintain the current is drawn from the particle's \textbf{Intrinsic Energy Reservoir} (as derived by Wald).
    \item This exchange is classified as the \textbf{microscopic host power} ($P_{\text{host}}$).
\end{itemize}

This constitutes the necessary "Consistency Condition" that renders the standard model consistent. It allows us to retain the Lorentz force ($\mathbf{j} \times \mathbf{B}$) as the universal mechanism of interaction, while ensuring that the Conservation of Energy is satisfied locally.

\subsection{Scope and Outlook}
While alternative theories based on magnetic charges might offer a simpler energy balance, they are inconsistent with the internal topology of matter. This book is dedicated to deriving macroscopic laws from the established microscopic baseline. Therefore, we accept the "Active Loop" model as the structural basis for our analysis.

With this microscopic foundation secured, the analysis can now proceed to the macroscopic scale. In the next chapter, these active loops are averaged to derive the macroscopic force density on magnetized matter, showing how the \textbf{Intrinsic Reservoirs} aggregate to form the bulk energetic properties of magnetic materials.

\chapter{Magnetics: Binding Energy}
\label{chap:Magnetic_Binding_Energy}
\label{chap:Magnetic_Host}

The investigation of the Host Interface now shifts to the physics of bound currents. The energetics of high-dynamic field variance are translated to the magnetic domain. The detailed derivation performed for the dielectric testbed is not repeated; the conceptual logic mirrors the electric case: the macroscopic potential energy of a material is physically located in the variance of the microscopic fields filtered out by the spectral filtering process.

\section{The Taxonomy of Magnetic Response}

\subsection{Diamagnetism: The Inductive Opposition}
Diamagnetism is a universal effect present in all matter, arising from the classical orbital motion of electrons. It is the microscopic analog of the macroscopic superconductor.

\begin{itemize}
    \item \textbf{Microscopic Mechanism (Lenz's Law):} When an external magnetic field is applied, the changing flux induces a microscopic electric field ($\mathbf{e}_{\text{ind}}$) within each electron orbit via Faraday's Law.
    \item \textbf{Response:} This field exerts a tangential force that accelerates or decelerates the orbital current. The resulting change in the magnetic moment $\Delta \mathbf{m}$ is directed to \textbf{oppose} the applied field.
    \item \textbf{Energetics:} Analogous to the rigid superconductor (Case Study 1), this is a mechanism of \textbf{Field Energy Redistribution}. Microscopically, there is only electromagnetic field energy being redistributed in space; no non-electromagnetic potential energy is created within the electron orbital itself.
\end{itemize}

\subsection[Paramagnetism and Ferromagnetism]{Paramagnetism and Ferromagnetism: \\ The Alignment of Variance}

In contrast to the induced moments of diamagnetism, paramagnetism and ferromagnetism arise from the re-orientation of \textit{pre-existing} intrinsic magnetic moments (\textbf{Active Superconducting Loops}, as derived in Chapter \ref{chap:Spin}).

\subsubsection{The Initial State: Microscopic Potential Energy}
Consider a ferromagnetic material in its unmagnetized state. Microscopically, the material is populated by intense magnetic fields generated by the intrinsic spins.
\begin{itemize}
    \item \textbf{Microscopic Reality:} High-frequency field variance is immense ($\langle \mathbf{b}^2 \rangle \gg 0$).
    \item \textbf{Macroscopic Appearance:} Due to thermal randomization (paramagnetism) or domain cancellation (ferromagnetism), these microscopic dipoles are oriented such that their fields vectorially sum to zero over the spectral filtering volume.
    \item \textbf{Spectral Status:} The magnetic energy exists, but it resides entirely in the "Noise Band" (High-Frequency spatial variance). It is filtered out of the macroscopic description ($\mathbf{B}=0, \mathbf{M}=0$).
\end{itemize}

\subsubsection{The Process: Spectral Transfer via Torque}
When an external field is applied, it exerts a torque ($\boldsymbol{\tau} = \mathbf{m} \times \mathbf{b}$) on the individual dipoles.
The external field performs work to rotate the dipoles into the direction of the macroscopic low-frequency external field. This external torque is, however, strictly counterbalanced by internal forces.

To understand the nuance, recall the dielectric case (Part III). There, when an external field separated the charges of a dipole, the counterbalancing microscopic force arose from the \textbf{local} microscopic electric field of the dipole itself, as well as the quantum boundary forces responsible for stability. The energy was stored effectively "inside" the stretched dipole field.

In the magnetic case, the situation is distinctly different. The quantum boundary conditions are indeed responsible for the stability of the dipole itself (holding the current locally inside the loop). However, the counterbalancing microscopic electromagnetic torque that opposes rotation comes \textbf{not from the dipole itself}, but from the \textbf{high-dynamic fields present in the surrounding matter}. It is the high-frequency interaction with the fields of neighboring dipoles (or the crystal field) that resists alignment.

Crucially, the \textbf{total electromagnetic torque is always zero} microscopically; the dipole is in static equilibrium at every instant ($\boldsymbol{\tau}_{\text{ext}} + \boldsymbol{\tau}_{\text{int}} = 0$). 

\paragraph{The Filtering of Torque.}
It must be emphasized that this microscopic rotational balance \textbf{does not translate} directly to the macroscopic domain.
The averaging process acts as a filter. The force from the external low-frequency part of the field is counterbalanced by internal high-frequency electromagnetic forces. When the system is filtered, the external low-frequency torque remains visible, while the counterbalancing high-frequency internal interaction is filtered out (as it belongs to the "Noise Band").

This missing counter-torque then emerges as a new parameter within the \textbf{Host Interface} of the macroscopic domain. Macroscopically, there is no torque; there are only circulating currents. The microscopic rotation of the spins is filtered out during the transition from micro to macro. The low-frequency macroscopic observer cannot "see" that the dipoles are rotating internally. They observe only the macroscopic current response ($\mathbf{J}_M = \nabla \times \mathbf{M}$).
Imagine a block where the macroscopic magnetization changes; macroscopically, this corresponds exactly to a changing current density. One cannot discern that the physical origin is the orientational rotation of microscopic dipoles. The torque and the rotation have been filtered out, replaced by a linear \textbf{Binding Force Density} acting on the current fluid.

Moreover, it must be kept in mind that while the magnetic dipoles rotate, the magnetic flux through the current loops making up the dipoles might change. Within the Active Spin model (Chapter \ref{chap:Spin}), this implies a further microscopic energy exchange, originating from the \textbf{Intrinsic Energy Reservoir} of the spin model itself.

\subsection{Summary on Binding Energy in Magnetism}

The conceptual idea of binding energy and the emergent macroscopic force parallels the dielectric case. Microscopically, there are only electromagnetic field energies present. The quantum forces are responsible for stability.

The internal microscopic field interactions are filtered out when translating into the macroscopic domain and emerge as part of the Host Interface. When microscopically dipoles change in magnitude due to induction or microscopically dipoles rotate into the direction of the external applied field, there are only microscopic fields present describing the interactions. When filtering the system and emerging to the macroscopic domain, the high-frequency field interactions are filtered out and emerge as the macroscopic force density of the \textbf{Host Force}. Paralleling the dielectric case, we call this force the \textbf{Binding Force} and the corresponding energy \textbf{Binding Energy}.

In the following we will consider some further case studies to illustrate the concept in more detail.

\section{Case Study 1: The Ideal Diamagnet (The Inductive Sink)}
\label{sec:IdealDiamagnet}

The ideal baseline for bound currents is considered: a rigid, lossless Diamagnetic material. To derive the macroscopic forces, the \textbf{Lattice of Isolated Superconducting Loops} is utilized as the conceptual testbed (structurally isomorphic to the "Lattice of Conductors" in the dielectric case).

\subsection{Phase I: The Transient Response}

Consider the material exposed to a time-varying external magnetic field. By Faraday's Law, this generates a rotational macroscopic electric field $\mathbf{E}$ throughout the volume.

\subsubsection{Microscopic Reality}
Microscopically, the material consists of isolated, atomic-scale superconducting loops (orbitals).
\begin{itemize}
    \item \textbf{The Driver:} The local electric field accelerates the electron fluid within each loop, inducing a microscopic current change ($\Delta \mathbf{j}$) to oppose the flux change.
    \item \textbf{The Constraint:} The Quantum Mechanical boundary conditions act as a rigid geometric constraint. They hold the radius of the current loop constant, but they do \textit{not} oppose the tangential acceleration of the current itself.
\end{itemize}

\subsection{Macroscopic Force Balance: The Emergence of the Host Force}

The spectral filtering operator is now applied to the momentum balance of the charge fluid.
Just as in the dielectric case, the filtering of the non-linear microscopic interactions generates a residual force term. This emergent term is defined as the \textbf{Host Force} ($\mathbf{f}_{\text{host}}$). It compresses the microscopic field interactions of the microscopic current loop into a new emergent macroscopic low-frequency force density.

In the quasi-static limit (where the electron inertia is negligible on the macroscopic scale), the macroscopic force balance for the magnetization current fluid ($\mathbf{J}_M$) is:
\begin{equation}
    \underbrace{ \rho_- \mathbf{E} }_{\text{Driving Force}} + \underbrace{ \mathbf{f}_{\text{host}} }_{\text{Host Force}} \approx \mathbf{0}
\end{equation}

\textbf{Physical Interpretation:}
Macroscopically, in a perfect conductor, the external field is completely counterbalanced by internal currents. The internally induced electric field cancels the external driver. In diamagnetism, the externally induced electric field is only \textit{partially} counterbalanced by the Host Interface, specifically by the \textbf{Binding Force} emergent from the microscopic field interactions.

The Binding Force ($\mathbf{f}_{\text{binding}}$) stops the external field from being completely screened. This behavior parallels the dielectric case, where the external field is only partly counterbalanced by the induced electric field of the dielectric itself.

Physically, this force is the macroscopic manifestation of the \textbf{Spectral Transfer}. The work done against the Binding Force transfers energy into the low-frequency fields of the material, into \textbf{Binding Energy}.

\subsection{Power Balance: Reversible Storage}

The energetic role of this force is revealed by the Macroscopic Energy Balance. The electric field performs work on the induced magnetization currents:
\begin{equation}
    P_{\text{in}} = \mathbf{J}_M \cdot \mathbf{E} =\underbrace{ - \left( \frac{\partial u_{\text{EM}}}{\partial t} + \nabla \cdot \mathbf{S}_{\text{EM}} \right) }_{\text{Field Dynamics}}
\end{equation}

When the external magnetic field increases, work is done against the binding force. Energy leaves the macroscopic field domain ($u_{\text{EM}}$) and is stored in Binding Energy. There is a local sink of macroscopic field energy. The moving charges perform work while moving with velocity into the direction of the induced electric field.

Just as in the dielectric testbed, this energy is not lost. It is stored in the increased variance of the \textbf{microscopic magnetic fields} (the intense near-fields of the induced dipoles). In the macroscopic framework, we call this Binding Energy.

\section{Case Study 2: The Ideal Paramagnet/Ferromagnet (Spectral Alignment)}
\label{sec:IdealParamagnet}

Next we consider para- and ferromagnetic materials. Here a different macroscopic effect arises: the magnetic field is not counterbalanced, but \textbf{enhanced}. As the internal microscopic persisting dipoles align with the external field, the macroscopic magnetic field grows.
This covers both paramagnetism (thermal randomization) and ferromagnetism (domain randomization). The ideal, lossless limit is considered in the following.

\subsection{Phase I: The Transient Response (The Alignment)}

Consider the material in an initial unmagnetized state.
\begin{itemize}
    \item \textbf{Macroscopically ($\mathbf{B} \approx 0$):} The material appears inert.
    \item \textbf{Microscopically ($\langle \mathbf{b}^2 \rangle \gg 0$):} The volume is filled with the intense magnetic near-fields of the electron spins and microscopic magnetic dipoles. However, due to their internal orientation, these fields destructively interfere over any macroscopic averaging volume. The energy exists, but it is locked in the "Noise Band".
\end{itemize}

When an external field is applied, the external low-frequency field exerts a torque on the dipoles and tries to rotate them into the direction of the external field.
This torque is counterbalanced by internal high-frequency microscopic fields that stop the dipoles from rotating completely—the high-frequency field generated by the surrounding dipoles inside matter.
The total microscopic electromagnetic torque is always zero. These high-dynamic field interactions then emerge within the \textbf{Host Interface} within the macroscopic domain.

\subsubsection{Macroscopic Force Balance: The Emergent Generator}

Macroscopically, the complex individual rotations of the domains are filtered out. The only observable remnant of this microscopic activity is the emergence of a smooth magnetization current density $\mathbf{J}_M$ (where $\mathbf{J}_M = \nabla \times \mathbf{M}$).

To understand the force balance, the fields must be tracked.
\begin{enumerate}
    \item \textbf{The Trigger:} The time-varying external flux induces a macroscopic electric field $\mathbf{E}_{\text{ext}}$.
    \item \textbf{The Enhancement:} In a ferro- and paramagnet, the material responds by creating a magnetization current $\mathbf{J}_M$ that generates a \textit{parallel} self-field, enhancing the total magnetic flux.
\end{enumerate}

Standard induction (Lenz's Law) would try to oppose the change with a Back-EMF. The binding force here acts as an \textbf{Energy Source}, which generates a macroscopic current in the direction of the external electric field. 
The spectral filtering of the microscopic alignment torques yields an \textbf{Active Binding Force} ($\mathbf{f}_{\text{bind}}$) that is part of the Host Interface. 
The force balance for the magnetization current fluid is:
\begin{equation}
    \underbrace{ \rho_- \mathbf{E}_{\text{total}} }_{\text{Field Force (Back-EMF)}} + \underbrace{ \mathbf{f}_{\text{bind}} }_{\text{Binding Force}} \approx \mathbf{0}
\end{equation}

\textbf{Physical Interpretation:}
Here, $\mathbf{f}_{\text{bind}}$ acts as a \textbf{Generator}. It represents the macroscopic "push" resulting from the collective unlocking of the microscopic dipoles.
Physically, this force is the mechanism that injects energy into the current fluid, driving $\mathbf{J}_M$ to create a magnetic field stronger than the external stimulus. It describes the energy brought into the macroscopic system from the internal, previously hidden microscopic field reservoirs.

\subsubsection{Power Balance: The Source of Field Energy}

The energetic description here is fundamentally different from the diamagnetic case. In diamagnetism, the material acted as a sink of macroscopic field energy, as the magnetization increased. Here, the system \textbf{enhances} the field and acts as an \textbf{enhancing source} of electromagnetic field energy as the magnetization increases.

As the dipoles align, their microscopic fields superimpose constructively. The macroscopic magnetic energy density $u_{\text{EM}}$ rises beyond what the external source alone provides.
To balance the local field energy equation, the material must act as a supplier. This is explicitly shown via the Host Power $P_{\text{host}}$:
\begin{equation}
    \underbrace{ \frac{\partial u_{\text{EM}}}{\partial t} + \nabla \cdot \mathbf{S}_{\text{EM}} }_{\text{Field Growth}} = \underbrace{ - (\mathbf{J}_M \cdot \mathbf{E}) }_{\text{Injection via Gateway}} = \underbrace{ P_{\text{host}} }_{\text{Input from Host}} > 0
\end{equation}

As the external field decreases, and the magnetization decreases, the process reverses. The macroscopic material acts as a sink of macroscopic field energy, returning the energy to the hidden noise band.

\textbf{The Origin of the Energy:}
This extra energy is not created ex nihilo. It is \textbf{transferred} from the microstructure to the macrostructure.
\begin{itemize}
    \item \textbf{Microscopic Reality:} The magnetic energy was always present, stored in the intense microscopic field inherent in the microscopic structure of matter.
    \item \textbf{The Macroscopic Representation:} In the macroscopic system, discrete rotating dipoles are not observed. Instead, the spectral filtering process yields an \textbf{Emergent Host Force} ($\mathbf{f}_{\text{host}}$). This force acts as a driver in the momentum equation.
    \item \textbf{The Process:} The Host Force performs positive work on the magnetization current fluid ($P_{\text{host}} > 0$). This macroscopic work represents the effective energy transfer resulting from the microscopic alignment. It allows the energy to flow from microscopic variance into the macroscopic field.
\end{itemize}

\section{The General Case: Dissipation and Hysteresis}
\label{sec:Magnetic_Dissipation}

In real materials, the reorientation of magnetic moments is rarely a frictionless process. Whether the material is diamagnetic, paramagnetic, or ferromagnetic, the dynamic change of the magnetization state ($\dot{\mathbf{M}} \neq 0$) often triggers irreversible energy losses.

\subsection{Microscopic Mechanisms: The Internal Friction}

Microscopically, the "smooth" change of magnetization observed macroscopically is often a discontinuous process involving mechanical and electrical friction at the atomic scale. Common mechanisms include:
\begin{itemize}
    \item \textbf{Domain Wall Friction:} In ferromagnets, the expansion of domains involves the physical movement of domain walls (Bloch/Néel walls). These walls can "snap" past crystal impurities or grain boundaries (Barkhausen effect), generating acoustic noise and heat.
    \item \textbf{Micro-Eddy Currents:} Even in insulators, the rapid reorientation of local dipoles can induce microscopic circulating currents in the surrounding lattice, dissipating energy via Ohmic heating at the unit-cell level.
    \item \textbf{Magnetostriction:} The rotation of dipoles often deforms the electron orbitals, which in turn exerts stress on the crystal lattice (micro-elastic deformation). As the material cycles, this internal mechanical rubbing generates vibrations (heat).
\end{itemize}

\subsection{Macroscopic Representation: The Dissipative Host Force}

Just as the individual rotations are averaged out, these specific microscopic loss mechanisms are filtered out of the macroscopic electromagnetic description. They manifest solely as an \textbf{Irreversible Energy Sink} within the Host Interface.

To account for this, the Host Force $\mathbf{f}_{\text{host}}$ is generalized to include a dissipative component:
\begin{equation}
    \mathbf{f}_{\text{host}} = \underbrace{ \mathbf{f}_{\text{binding}} }_{\text{Storage/Injection}} + \underbrace{ \mathbf{f}_{\text{diss}} }_{\text{Friction}}
\end{equation}

\begin{itemize}
    \item \textbf{The Storage/Injection Component:} This is the force discussed in Case Studies 1 and 2. It is conservative (reversible). It represents energy stored in the binding potential or injected from the \textbf{Intrinsic Reservoir}.
    \item \textbf{The Dissipative Component:} This force opposes the rate of change of magnetization ($\mathbf{f}_{\text{diss}} \propto - \dot{\mathbf{M}}$). It represents the "drag" forces acting on the microstructure.
\end{itemize}

\subsection{Power Balance: Spectral Degradation}

The energy balance now includes a terminal sink. The total work done by the field on the magnetization current is split:
\begin{equation}
    \underbrace{ \mathbf{J}_M \cdot \mathbf{E} }_{\text{Input}} = \underbrace{ P_{\text{stored}} }_{\text{Reversible}} + \underbrace{ P_{\text{diss}} }_{\text{Irreversible}}
\end{equation}

This dissipative Host Power is the physical origin of the \textbf{Hysteresis Loop} observed in macroscopic measurements ($\oint \mathbf{H} \cdot d\mathbf{B}$). The area of the loop corresponds exactly to the work done against the dissipative component of the Host Force ($\oint \mathbf{f}_{\text{diss}} \cdot d\mathbf{l}$) over a complete cycle.

\paragraph{Conclusion.}
The unified framework successfully elucidates the physics of magnetic energy. For free charges and currents on conductors, two energy gateways were identified: work and dissipation. This chapter introduced the third: \textbf{Storage}. The Binding Force, as part of the Host Interface, counterbalances the microscopic forces and stores energy in a separate Binding Energy reservoir. It physically stores the electrodynamic field energy that has effectively left the macroscopic domain.

Force density in magnetic materials is the subject of the subsequent chapter. Having identified the dissipation and storage energy gateways, the analysis will extend to the gateway of \textbf{Work}—the physical coupling of the macroscopic electromagnetic domain to the Continuum Mechanics domain. This will again parallel the analysis of dielectrics.

\chapter{Force Density in Magnetics: The Ampere Topology}
\label{chap:MagneticForce}

\section{The Universal Framework: Recapitulation}
\label{sec:MagForce_Framework}

In the dielectric part (Part III), a fundamental limit regarding the mechanical forces in dielectric media was established. It was demonstrated that the question of deformation is \textbf{macroscopically indeterminate}—one cannot calculate the local elastic stress solely from the macroscopic field variables.

As the transition to the magnetic domain is made, three foundational principles are carried forward that constitute the universal framework for force density:

\subsection{1. The Invariance of the Vacuum Tensor}

We propose, just like in the electric case, that the total macroscopic force density budget is rigorously given by the averaged Lorentz force acting on the bound sources, microscopically as well as macroscopically.

Microscopically, the total electromagnetic force $\mathbf{f}_{\text{total}}$ is distributed between two distinct mechanical targets:
\begin{itemize}
    \item \textbf{Internal Stress ($\mathbf{f}_{\text{internal}}$):} Forces acting \textit{within} the atomic current loops (e.g., expanding the loop radius against quantum binding forces). These sum to zero over a single dipole and translate not into the macroscopic deformation of matter but into the continuum mechanics domain.
    \item \textbf{Deformation Force ($\mathbf{f}_{\text{def}}$):} Total forces acting on the center of mass of the dipoles, which are transmitted to the crystal lattice. This is the component responsible for macroscopic actuation (Magnetostriction).
\end{itemize}

The filtering or averaging process yields a new macroscopic system, with separate macroscopic physical laws. It is a compressed, separate emergent system, distinct from the microscopic reality—a lower resolution of the same system.

Macroscopically, the force budget is split into 3 emergent macroscopic pathways. For a magnetic material defined by magnetization $\mathbf{M}$, the bound current density is $\mathbf{J}_b = \nabla \times \mathbf{M}$. The governing equation for the total momentum budget is:

\begin{equation}
\boxed{
\begin{aligned}
    \underbrace{ \left( \frac{\partial \mathbf{G}_{\text{mech}}}{\partial t} + \nabla\cdot\mathbf{T}_{\text{mech}} \right) }_{\text{Total Material Response}}
    \quad &\overset{\text{Coupling}}{\textbf{=}} \quad
    \underbrace{(\nabla \times \mathbf{M} )\times\mathbf{B} }_{\text{Lorentz Force}} \\
    &\overset{\text{Identity}}{=} \quad
    \underbrace{ -\left(\frac{\partial \mathbf{G}_{\text{EM}}}{\partial t} + \nabla\cdot \mathbf{T}_{\text{EM}}\right) }_{\text{Vacuum Tensor Source}}
\end{aligned}
}
\end{equation}

The momentum budget is split into 3 emergent macroscopic pathways:
\begin{enumerate}
    \item \textbf{Storage:} This corresponds to the emergent Binding Force from averaging, representing energy exchanging with the binding storage (micro-field storage).
    \item \textbf{Dissipation:} This component is active only in a dynamic process. It is counterbalanced by friction on moving charges (electrons and ions) and dissipates energy into the thermodynamic domain.
    \item \textbf{Work:} This pathway enters the continuum mechanics domain. It performs macroscopic work on matter and is responsible for deformation and the internal stress field in matter.
\end{enumerate}

In the following, we aim to identify the corresponding force density for the magnetic domain that enters the continuum mechanics equation and is responsible for deformation of matter. We proceed parallel to the dielectric chapter \ref{chap:ForceDensity}.

\section{Microscopic Foundation: The Dipole Force}
\label{sec:Mag_Micro_Foundation}

To determine how the bulk material deforms, the force acting on a single constituent unit must first be determined. Here, a distinct asymmetry is encountered between the electric and magnetic domains.

An electric dipole is mechanically simple: two point charges separated by a distance. The force is simply the sum of the Lorentz forces on the static charges.
A magnetic dipole, however, is mechanically complex: it is a \textbf{Current Loop} (specifically, an \textbf{Active Superconducting Loop}, as detailed in Chapter \ref{chap:Spin}). It contains moving charges with internal mechanical momentum. Consequently, the calculation of the force on its center of mass requires a relativistic treatment that accounts for the coupling between the internal motion of the carriers and the external electric field.

\subsection{The Necessity of Hidden Momentum}

Consider a stationary magnetic dipole defined by a microscopic magnetization density $\mathbf{m}$ placed in a static external electric field $\mathbf{e}$.
Electromagnetically, this system possesses a microscopic field momentum density $\mathbf{g} = \varepsilon_0 (\mathbf{e} \times \mathbf{b})$. To find the net electromagnetic momentum, we integrate over all space. To calculate the total electromagnetic momentum, we reformulate the fields mathematically. Let us consider Maxwell's equations in a static scenario, where only the current representing the magnetization is present:
\begin{equation}
    \nabla \times \mathbf{b} = \mu_0 \nabla \times \mathbf{m}
\end{equation}
We can introduce the vector field $\mathbf{h}= \mathbf{b}-\mu_0 \mathbf{m}$ as the curl-free auxiliary field. 
The integral splits into two terms:
\begin{equation}
    \mathbf{P}_{\text{field}} = \varepsilon_0 \mu_0 \int_{V_{\infty}} (\mathbf{e} \times \mathbf{h}) \, dV + \varepsilon_0 \mu_0 \int_{V_{\text{matter}}} (\mathbf{e} \times \mathbf{m}) \, dV.
\end{equation}

For static fields with no free current, both $\mathbf{e}$ and $\mathbf{h}$ are curl-free gradients ($\mathbf{e}=-\nabla \phi, \mathbf{h}=-\nabla \psi$). A fundamental vector identity dictates that the volume integral of the cross product of two gradients over unbounded space is identically zero. Thus, the first term vanishes.

This leaves a non-zero net field momentum localized entirely within the material volume. Defining the total dipole moment as $\mathbf{m}_{\text{dp}} = \int \mathbf{m} \, dV$:
\begin{equation}
    \mathbf{P}_{\text{field}} = \varepsilon_0 \mu_0 \int (\mathbf{e} \times \mathbf{m}) \, dV \approx \varepsilon_0 \mu_0 (\mathbf{e} \times \mathbf{m}_{\text{dp}}).
\end{equation}

This creates a discrepancy: the system is static (nothing is moving), yet it possesses net linear electromagnetic momentum. According to the Relativistic Center-of-Energy Theorem \cite{Babson2009}, the total momentum of a static closed system must be zero ($\mathbf{P}_{\text{total}} = 0$).

Therefore, the matter composing the current loop must possess an equal and opposite \textbf{Hidden Mechanical Momentum}:
\begin{equation}
    \mathbf{P}_{\text{hidden}} = - \mathbf{P}_{\text{field}} = \varepsilon_0 \mu_0 (\mathbf{m}_{\text{dp}} \times \mathbf{e}).
\end{equation}
Physically, this momentum resides in the relativistic mass variation of the charge carriers as they accelerate and decelerate around the loop in the presence of the electric potential gradient, as shown in more detail in for example \cite{Vaidman1990}.

\subsection{Derivation of the Center-of-Mass Force}
The force relevant for material deformation is the acceleration of the dipole's \textbf{Center of Mass}. The change of the center of mass momentum of the dipole is what translates into the movement of the dipole and ultimately into the deformation of the matter.

The Lorentz force represents the total change of electromagnetic momentum, by identity:
\begin{equation}
     \rho_{\text{micro}}\mathbf{e} + \mathbf{j}_{\text{micro}}\times\mathbf{b} =-\frac{\partial \mathbf{g}_{\text{em}}}{\partial t} - \nabla\cdot \mathbf{t}_{\text{em}}.
\end{equation}
This equals the total change of all mechanical momentum inside an isolated system, which is the hidden mechanical momentum and the center of mass momentum. For the isolated magnetic dipole it holds:
\begin{equation}
    \frac{d\mathbf{P}_{\text{hidden}}}{dt} + \frac{d\mathbf{P}_{\text{CM}}}{dt} =\int \rho_{\text{micro}}\mathbf{e} + \mathbf{j}_{\text{micro}}\times\mathbf{b} \, dV.
\end{equation}

For the center of mass momentum it thus holds: 
\begin{equation}
    \frac{d\mathbf{P}_{\text{CM}}}{dt} =\int (\rho_{\text{micro}}\mathbf{e} + \mathbf{j}_{\text{micro}}\times\mathbf{b}) \, dV - \frac{d\mathbf{P}_{\text{hidden}}}{dt} 
\end{equation}

The two terms are computed explicitly in the following.

\subsubsection{1. The Total Lorentz Force}
The Lorentz force on the bound current density $\mathbf{j} = \nabla \times \mathbf{m}$ is:
\begin{equation}
    \mathbf{f}_{\text{Lorentz}} = \int (\nabla \times \mathbf{m}) \times \mathbf{b} \, dV.
\end{equation}
Using standard vector identities and the vacuum Maxwell relation $\nabla \times \mathbf{b} = \varepsilon_0 \mu_0 \dot{\mathbf{e}}$, this expands to:
\begin{equation}
    \mathbf{f}_{\text{Lorentz}} = (\mathbf{m} \cdot \nabla)\mathbf{b} + \varepsilon_0 \mu_0 (\mathbf{m} \times \dot{\mathbf{e}}).
\end{equation}

\subsubsection{2. The Hidden Momentum Rate}
The time derivative of the hidden momentum applies to both the changing field and the changing dipole moment:
\begin{equation}
    \frac{d \mathbf{P}_{\text{hidden}}}{dt} = \frac{d}{dt} \left[ \varepsilon_0 \mu_0 (\mathbf{m} \times \mathbf{e}) \right] = \varepsilon_0 \mu_0 (\dot{\mathbf{m}} \times \mathbf{e}) + \varepsilon_0 \mu_0 (\mathbf{m} \times \dot{\mathbf{e}}).
\end{equation}

\subsubsection{3. The Cancellation and Result}
Subtracting the two terms reveals the simplification. The term involving the time-varying electric field $\mathbf{m} \times \dot{\mathbf{e}}$ appears in both expressions and cancels exactly:
\begin{align}
    \frac{d\mathbf{P}_{\text{CM}}}{dt} &= \left[ (\mathbf{m} \cdot \nabla)\mathbf{b} + \underline{\varepsilon_0 \mu_0 (\mathbf{m} \times \dot{\mathbf{e}})} \right] - \left[ \varepsilon_0 \mu_0 (\dot{\mathbf{m}} \times \mathbf{e}) + \underline{\varepsilon_0 \mu_0 (\mathbf{m} \times \dot{\mathbf{e}})} \right] \nonumber \\
    &= (\mathbf{m} \cdot \nabla)\mathbf{b} - \varepsilon_0 \mu_0 (\dot{\mathbf{m}} \times \mathbf{e}).
\end{align}
Using the anti-symmetry of the cross product, we arrive at the final microscopic force law:
\begin{equation}
    \boxed{ \frac{d\mathbf{P}_{\text{CM}}}{dt} = (\mathbf{m} \cdot \nabla)\mathbf{b} + \frac{1}{c^2} \left( \mathbf{e} \times \frac{d\mathbf{m}}{dt} \right) }
    \label{eq:Micro_Mag_Force}
\end{equation}
This represents the change of the center of mass momentum of the magnetic dipole, which is the force responsible for material deformation. The final force terms have a symmetric structure to the dielectric force on the electric dipole.

\section{Macroscopic Transition: The Ampere Adaptation}
\label{sec:Mag_Macro_Adaptation}

Having established the force on a single isolated dipole, the macroscopic transition is addressed. The challenge is to sum these forces to find the bulk deformation density $\mathbf{F}_{\text{def}}$.

As with the dielectric case, we propose a \textbf{Lattice Adaptation Term} for the macroscopic force density. 
The magnetic dipole effectively resides in the interstitial vacuum regions between neighboring magnetic dipoles. It "blindly misses" the field singularities hidden within the bodies of its neighbors. Consequently, it \textbf{does not encounter} the extremely high field singularities (the ``inner'' fields) that exist inside massive dipoles.

\subsection{The Topological Sign Inversion}

The sign of the local field correction depends entirely on the topology of the field lines inside the source.

\begin{enumerate}
    \item \textbf{The Electric Case (Separation):} An electric dipole consists of separated charges. The internal field lines run from positive to negative, \textit{opposing} the direction of the dipole moment $\mathbf{P}$. This creates a \textit{depolarizing} singularity:
    \begin{equation}
        \mathbf{E}_{\text{inner}} = -\frac{\mathbf{P}}{3\varepsilon_0} \quad \implies \quad \text{Correction is Positive.}
    \end{equation}

    \item \textbf{The Magnetic Case (Circulation):} A magnetic dipole consists of a current loop. The field lines do not terminate; they flow continuously \textit{through} the loop, \textit{aligned} with the direction of the moment $\mathbf{M}$. This creates a \textit{magnetizing} singularity.
\end{enumerate}

For a spherical exclusion volume (representing a standard cubic lattice), the average field inside a uniformly magnetized sphere is given by \cite{Jackson1999}:
\begin{equation}
    \mathbf{B}_{\text{inner}} = +\frac{2\mu_0}{3}\mathbf{M}.
\end{equation}
Note the positive sign and the factor of 2. The magnetic self-field reinforces the moment rather than opposing it.

\subsection{The Effective Field and Force Density}

The lattice dipoles experience the total average field \textit{minus} this excluded self-field. Thus, the effective magnetizing field is:
\begin{equation}
    \mathbf{B}_{\text{eff}} = \mathbf{B} - \mathbf{B}_{\text{inner}} = \mathbf{B} - \frac{2\mu_0}{3}\mathbf{M}.
\end{equation}

This effective field is substituted into the microscopic force law derived in Eq.~\eqref{eq:Micro_Mag_Force}. Assuming the relativistic term is driven by the macroscopic electric field (which does not suffer the same exclusion singularity in a neutral medium), the \textbf{hypothesis of microstructure} yields the following force density:

\begin{align}
    \mathbf{F}_{\text{def}} &= n \langle \mathbf{f}_{\text{CM}} \rangle \nonumber \\
    &\approx (\mathbf{M} \cdot \nabla)\mathbf{B}_{\text{eff}} + \varepsilon_0 \mu_0 \left( \mathbf{E} \times \frac{\partial \mathbf{M}}{\partial t} \right) \nonumber \\
    &= (\mathbf{M} \cdot \nabla)\left( \mathbf{B} - \frac{2\mu_0}{3}\mathbf{M} \right) + \frac{1}{c^2} \left( \mathbf{E} \times \frac{\partial \mathbf{M}}{\partial t} \right).
\end{align}

If one ignores our proposed lattice adaptation term, the remaining force density is identified as the magnetic equivalent of the \textbf{Einstein-Laub Force Density}—the standard formulation obtained by naively applying the macroscopic field to the dipole moment.

\subsection{Limitations: The Hypothesis of Microstructure}
\label{sec:Mag_Limitations}

The correction term derived above—$-\frac{2\mu_0}{3}(\mathbf{M} \cdot \nabla)\mathbf{M}$—is explicitly framed as a \textbf{Hypothesis of Microstructure} rather than a universal constitutive law. Like its dielectric counterpart, its specific magnitude relies on structural symmetries that do not hold for all magnetic materials.

The derivation relied on the ``Test Dipole'' argument with a spherical exclusion volume, yielding the factor of $2/3$. This assumes a statistical isotropy characteristic of amorphous materials.

In real ferromagnetic crystals (e.g., hexagonal cobalt or tetragonal martensite), the magnetic dipoles are located at fixed sites with anisotropic symmetry. The scalar factor might transform into a complex \textbf{Interaction Tensor} dependent on the specific lattice constants. Consequently, two magnetic materials could possess identical macroscopic magnetization $\mathbf{M}$ but experience different internal deformation forces depending on their microscopic packing.

\section{Summary of Part IV: The Magnetic Response}

In Part IV of this book, we have rigorously analyzed all interactions of the \textbf{Rotational Currents} ($\mathbf{J}_{\perp}$). This is the component of the current density that possesses curl ($\nabla \cdot \mathbf{J} \neq 0$), leading to the generation of solenoid fields ($\nabla \times \mathbf{B} \neq 0$).

We have demonstrated that a single unified framework—the \textbf{Vacuum Lorentz Framework}—rigorously describes the mechanics of both free charges (conductors) and bound charges (magnets).

\subsection{The Three Mechanical Gateways}
By enforcing the \textbf{Force-Energy Consistency Criterion (FECC)}, we identified that the Host Interface of any material partitions the total electromagnetic work ($P = \mathbf{J} \cdot \mathbf{E}$) into three distinct physical channels:

\begin{enumerate}
    \item \textbf{Dissipation (The Thermal Sink):} Energy is irreversibly transferred to the random motion of the lattice (heat). This is the physics of Resistance and hysteretic losses.
    \item \textbf{Work (The Continuum Sink):} Energy is orderedly transferred to the bulk motion of the material (kinetic energy and deformation). This is the physics of Actuation and Magnetostriction.
    \item \textbf{Storage (The Potential Sink):} Energy is reversibly transferred to the microscopic degrees of freedom (potential energy). This is the physics of Magnetic Binding.
\end{enumerate}

\subsection{The Universality of the Vacuum Tensor}
In the last chapter, we proved the \textbf{Theorem of Macroscopic Indeterminacy}, showing that local stress cannot be derived from macroscopic fields alone without microstructural side-channel information.

However, we concluded that the \textbf{Total Momentum Budget} is structurally invariant. The vacuum energy-momentum tensor correctly describes the conservation of momentum for the entire system:
\begin{equation}
\boxed{
\begin{aligned}
    \underbrace{ \left( \frac{\partial \mathbf{G}_{\text{mech}}}{\partial t} + \nabla\cdot\mathbf{T}_{\text{mech}} \right) }_{\text{Total Material Response}}
    \quad &\overset{\text{Coupling}}{\textbf{=}} \quad
    \underbrace{\rho \mathbf{E} + \mathbf{J}\times\mathbf{B} }_{\text{Lorentz Force}} \\
    &\overset{\text{Identity}}{=} \quad
    \underbrace{ -\left(\frac{\partial \mathbf{G}_{\text{EM}}}{\partial t} + \nabla\cdot \mathbf{T}_{\text{EM}}\right) }_{\text{Vacuum Tensor Source}}
\end{aligned}
}
\end{equation}
This equation is the master key to the electromagnetic response. It gives the momentum budget (or electromagnetic force budget) that the Host Interface splits into the three gateways (Dissipative Force, Binding Force, and Work Force). It ties the force $\mathbf{f}$ irrevocably to the energy flux $\mathbf{S}$ and the mass carrier $\mathbf{v}$.

With the magnetic and electric response complete, we now turn our attention to the electrodynamics of moving matter.

\part{Part V: Moving Matter and Relativistic Sources}
\label{part:MovingMatter}

\chapter{The Physics of Moving Sources}
\label{chap:Moving_Matter}

\section{Introduction: The Geometry of Sources}

In the preceding parts of this work, the "vacuum framework" has been established as the rigorous baseline for macroscopic electrodynamics. However, a final frontier remains: the relativistic consideration of \textbf{moving matter}.

Historically, the electrodynamics of moving media has been a source of conceptual ambiguity. Conventional approaches often introduce complex "constitutive relations" for moving bodies.
However, we have demonstrated that independent of whether charges are bound or free, or whether the material is magnetic or dielectric, all interactions are fundamentally described by the vacuum tensor. Matter consists solely of charges and currents. 

Building on the "Two-Fluid" model (Part I) and the "Lattice Models" (Parts III and IV), it is established that $\mathbf{P}$ and $\mathbf{M}$ are not independent physical fluids. They are merely macroscopic mathematical variables quantifying \textbf{Bound Charge} ($\rho_b$) and \textbf{Bound Current} ($\mathbf{J}_b$). There is no "moving medium" in the fundamental equations; there are only moving charges. Once the sources are correctly identified via the relativistic source tensor $M^{\mu\nu}$, the "moving media problem" vanishes, leaving behind only standard vacuum electrodynamics.

By strictly adhering to the covariant transformation of sources, it is demonstrated that the ``moving medium'' requires no independent physical laws—its behavior is entirely determined by the relativistic kinematics of its underlying currents. It mirrors the microscopic transformation of sources exactly.

Therefore, we propose that the physics of "Moving Matter" is simply the physics of \textbf{Moving Sources}. The Vacuum Stress Tensor requires no modification. The force on a moving dielectric is rigorously predicted by applying the standard Lorentz force to the relativistically transformed sources.

\section{Relativistic Consistency}
\label{sec:ProposedCovariant}

The framework's physical consistency is best displayed in four-dimensional covariant notation. We define the state of the system using the standard vacuum field tensor $F^{\mu\nu}$ and the total source density 4-vector $J^{\nu}_{\text{total}}$.

The bound sources are contained within the antisymmetric \textbf{polarization-magnetization tensor} $M^{\mu\nu}$, which combines the spatial vectors $\mathbf{P}$ and $\mathbf{M}$ into a single relativistic geometric object. The bound current is simply the four-divergence of this tensor ($J^{\nu}_b = \partial_\mu M^{\mu\nu}$).

The entire system is described by two fundamental equations:
\begin{align}
    \partial_\mu F^{\mu\nu} &= \mu_0 J^{\nu}_{\text{total}} = \mu_0 \left( J^{\nu}_{f} + \partial_\sigma M^{\sigma\nu} \right), \label{eq:CovariantMaxwell_Final} \\
    \partial_\nu T^{\mu\nu}_{\text{vac}} &= -F^{\mu\lambda} J_{\lambda, \text{total}}. \label{eq:TensorDivergence_Final}
\end{align}

This structural form is \textbf{identical} to that of free charges in a vacuum.
\begin{itemize}
    \item The field is described universally by the standard field-strength tensor $F^{\mu\nu}$.
    \item The field energy-momentum is described universally by the standard vacuum tensor $T^{\mu\nu}_{\text{vac}}$.
    \item The interaction is strictly local, occurring between the field and the \textit{total} current $J^{\nu}_{\text{total}}$.
\end{itemize}

There is no need for modified energy-momentum tensors or complex auxiliary field tensors. By embedding the bound sources into the geometry of the source current $J^\nu_{total}$, the need for complex, ad-hoc ``constitutive relations'' for moving media is avoided. The physics is not about how the ``medium'' moves, but simply how the source coordinates transform. 

The governing equations are inherently covariant and apply to any inertial frame without modification. The motion of the medium is naturally accounted for by applying the standard Lorentz transformation to the source tensor $M^{\mu\nu}$. The well-known phenomenological fact—that a moving magnet appears polarized, or a moving dielectric appears magnetized—is an automatic consequence of the tensor transformation properties:
\begin{equation}
    M'^{\mu\nu} = \Lambda^\mu_\alpha \Lambda^\nu_\beta M^{\alpha\beta}.
\end{equation}

This mixing of polarization and magnetization is mathematically analogous to the familiar mixing of currents and charges under a boost. For instance, a purely magnetized body ($\mathbf{P}=0, \mathbf{M}\neq 0$ in its rest frame) will, in a frame where it moves with velocity $\mathbf{v}$, naturally acquire an electric polarization component $\mathbf{P}' \approx (\mathbf{v} \times \mathbf{M})/c^2$. This arises purely from the coordinate transformation of the source tensor. $\mathbf{M}$ and $\mathbf{P}$ simply represent the corresponding source projections in the observing frame. There is no need for separate "moving media" postulates; the physics is contained entirely within the geometry of spacetime.

\subsection{Unpacking the Covariant Physics: From Tensor to 3D Sources}
\label{sec:Covariant_Unpacking}

While the 4-vector notation $\partial_\mu M^{\mu\nu}$ provides mathematical compactness, it is essential to unpack it to reveal the internal physical mechanism. The polarization-magnetization tensor $M^{\mu\nu}$ is simply a matrix that organizes the dipole densities into temporal and spatial components.

Using the standard Minkowski metric convention $(+,-,-,-)$, the antisymmetric source tensor is explicitly:
\begin{equation}
    M^{\mu\nu} =
    \begin{pmatrix}
    0 & cP_x & cP_y & cP_z \\
    -cP_x & 0 & -M_z & M_y \\
    -cP_y & M_z & 0 & -M_x \\
    -cP_z & -M_y & M_x & 0
    \end{pmatrix}
\end{equation}

Physically, this structure partitions the bound sources by their spacetime geometry:
\begin{itemize}
    \item \textbf{Polarization ($\mathbf{P}$):} Resides in the \textbf{Time-Space} components ($M^{0i}$). It represents the separation of charge. In terms of currents, its time-derivative ($\partial_t \mathbf{P}$) corresponds to the \textbf{Polarization Current}, which describes the linear motion of bound charges.
    \item \textbf{Magnetization ($\mathbf{M}$):} Resides in the \textbf{Space-Space} components ($M^{ij}$). It represents the circulation of charge. Its spatial curl ($\nabla \times \mathbf{M}$) corresponds to the \textbf{Magnetization Current}.
\end{itemize}

\subsubsection{Recovering the Source Equations}
The bound source currents are derived by taking the 4-divergence of this tensor ($J^\nu_b = \partial_\mu M^{\mu\nu}$). This operation naturally splits the relativistic definition into the two familiar 3D equations utilized throughout this book.

\textbf{1. The Bound Charge Density ($\nu = 0$ component):}
The time component of the divergence yields the accumulation of charge:
\begin{equation}
    \rho_b = \frac{1}{c} J^0_b = \partial_i M^{i0} = -\nabla \cdot \mathbf{P}.
\end{equation}
This confirms that the time-component of the divergence extracts the edges of the polarization field (the charge accumulation).

\textbf{2. The Bound Current Density ($\nu = 1,2,3$ components):}
The spatial components of the divergence yield the total bound flux:
\begin{equation}
    \mathbf{J}_b = \underbrace{\frac{\partial \mathbf{P}}{\partial t}}_{\substack{\text{Time variation} \\ \text{of } M^{0i}}} + \underbrace{\nabla \times \mathbf{M}}_{\substack{\text{Spatial curl} \\ \text{of } M^{ij}}}.
\end{equation}
This confirms that the macroscopic bound current is simply the sum of the linear change in dipole moment (Polarization Current) and the circulation of the magnetic moment (Magnetization Current).

\subsection{The Physics of Moving Media: Lorentz Transformation}
\label{sec:Moving_Media_Transform}

Because $\mathbf{P}$ and $\mathbf{M}$ are components of a single tensor $M^{\mu\nu}$, they must mix under coordinate transformations just as time and space mix. If a material has polarization $\mathbf{P}$ and magnetization $\mathbf{M}$ in its rest frame, an observer seeing it move with velocity $\mathbf{v}$ will observe new effective fields $\mathbf{P}'$ and $\mathbf{M}'$:
\begin{align}
    \mathbf{P}' &= \gamma \left( \mathbf{P} + \frac{\mathbf{v} \times \mathbf{M}}{c^2} \right), \label{eq:P_transform} \\
    \mathbf{M}' &= \gamma \left( \mathbf{M} - \mathbf{v} \times \mathbf{P} \right). \label{eq:M_transform}
\end{align}

\subsubsection{Physical Interpretation}
Equations (\ref{eq:P_transform}) and (\ref{eq:M_transform}) reveal the physical mechanism of moving media without requiring any new postulates or auxiliary fields:

\begin{itemize}
    \item \textbf{Motional Polarization ($\mathbf{v} \times \mathbf{M}$):} A moving magnet ($\mathbf{M}$) naturally exhibits an electric polarization. 
    \textit{Physical Mechanism:} The magnetization consists of circulating current loops. Due to the \textbf{relativity of simultaneity}, the charge density in the leading and trailing legs of the loop appears different to a moving observer, creating a net electric dipole moment where none existed in the rest frame.
    
    \item \textbf{Motional Magnetization ($-\mathbf{v} \times \mathbf{P}$):} A moving dielectric ($\mathbf{P}$) naturally exhibits a magnetization. 
    \textit{Physical Mechanism:} The dielectric consists of separated positive and negative charges. When the material moves, these separated charges constitute two parallel convection currents. Since the charges are spatially offset, their currents do not cancel perfectly, creating a net magnetic moment.
\end{itemize}

By strictly defining $\mathbf{P}$ and $\mathbf{M}$ as source representations, the complex phenomenology of moving media emerges automatically as a simple coordinate transformation of the sources.

\subsection{Example: The Moving Polarized Square}
\label{sec:Example_MovingSquare}

To demonstrate that $\mathbf{P}$ and $\mathbf{M}$ are not independent physical entities but rather coupled components of the total current, we consider a simplified thought experiment: a rigid dielectric square with uniform polarization $\mathbf{P} = P_0 \hat{y}$ (pointing upwards).

Physically, the bulk is neutral. The only charge carriers are the bound surface charges $\sigma_b$ located at the top and bottom boundaries, where the polarization terminates ($\sigma_b = \mathbf{P} \cdot \hat{n}$). The vertical sides are electrically neutral.

\subsubsection{The Physical Picture (Microscopic Baseline)}
Assume this object moves to the right with velocity $\mathbf{v} = v\hat{x}$. From the perspective of the underlying sources (the "Microscopic Baseline"), the physics is elementary. The current is simply the mechanical convection of the existing bound charges:

\begin{itemize}
    \item \textbf{Top Surface:} Positive charges moving right form a current sheet $\mathbf{K}_{\text{top}} = (+\sigma_b) \mathbf{v} \propto +\hat{x}$.
    \item \textbf{Bottom Surface:} Negative charges moving right form a current sheet $\mathbf{K}_{\text{bottom}} = (-\sigma_b) \mathbf{v} \propto -\hat{x}$.
    \item \textbf{Vertical Sides:} There is no charge ($\sigma_b = 0$). Therefore, there is \textbf{no current}.
\end{itemize}
Total Physical State: Two parallel currents flowing in opposite directions.

\subsubsection{The Mathematical Reformulation}
Now, this same system is analyzed using the standard macroscopic formalism, treating $\mathbf{P}$ and $\mathbf{M}$ as separate fields. The motion of the polarized medium creates an effective "motional magnetization":
\begin{equation}
    \mathbf{M} = \mathbf{P} \times \mathbf{v} = (P_0 \hat{y}) \times (v \hat{x}) = -P_0 v \hat{z}.
\end{equation}
The formalism defines the total bound current as the sum of two distinct mathematical terms:
\begin{equation}
    \mathbf{J}_{\text{bound}} = \underbrace{ \frac{\partial \mathbf{P}}{\partial t} }_{\text{Polarization Current}} + \underbrace{ \nabla \times \mathbf{M} }_{\text{Magnetization Current}}
\end{equation}

When analyzed individually, these terms predict "Artifactual Currents" where physically none exist:

\begin{enumerate}
    \item \textbf{The Polarization Artifact ($\partial \mathbf{P}/\partial t$):}
    As the square moves, the local polarization at a fixed point in space changes abruptly at the vertical edges.
    \begin{itemize}
        \item \textbf{Leading Edge:} Vacuum is replaced by matter. $\mathbf{P}$ jumps from 0 to $P_0$. This is mathematically equivalent to a vertical current pulse pointing \textbf{Up} ($+\hat{y}$).
        \item \textbf{Trailing Edge:} Matter is replaced by vacuum. $\mathbf{P}$ drops from $P_0$ to 0. This is mathematically equivalent to a vertical current pulse pointing \textbf{Down} ($-\hat{y}$).
    \end{itemize}

    \item \textbf{The Magnetization Artifact ($\nabla \times \mathbf{M}$):}
    The effective magnetization $\mathbf{M}$ is uniform inside the square and points into the page ($-\hat{z}$). This creates a circulating solenoidal surface current $\mathbf{K}_{\text{mag}} = \mathbf{M} \times \mathbf{n}$ around the \textit{entire} perimeter.
    \begin{itemize}
        \item \textbf{Vertical Sides:} This predicts vertical currents flowing \textbf{Down} at the leading edge and \textbf{Up} at the trailing edge.
    \end{itemize}
\end{enumerate}

\subsubsection{The Synthesis (Cancellation of Artifacts)}
The complexity introduced by splitting the sources is resolved only when the terms are summed to find the total current:

\begin{itemize}
    \item \textbf{On the Vertical Sides:} The "Polarization Current" (Up) and the "Magnetization Current" (Down) are exactly equal and opposite.
    \begin{equation}
        \mathbf{J}_{\text{vertical}} = \mathbf{J}_P + \mathbf{J}_M = 0.
    \end{equation}
    The artifacts cancel perfectly, recovering the physical truth that there is no current on the uncharged sides.
    
    \item \textbf{On the Horizontal Surfaces:} The magnetization current is parallel to the velocity and matches the convection current, recovering the physical transport of the surface charge.
\end{itemize}

\textbf{Conclusion:} The decomposition of the source into $\mathbf{P}$ and $\mathbf{M}$ generates "artifactual currents" on the vertical sides—mathematical artifacts that exist in the separate terms but vanish in the sum.
This confirms that $\mathbf{P}$ and $\mathbf{M}$ are not independent physical fluids. They are merely coupled components of the single source tensor $M^{\mu\nu}$. The vacuum formalism, by focusing on the total source $\mathbf{J}_{\text{total}}$, avoids this complexity.

\subsection{Example: The Moving Magnetized Square}
\label{sec:Example_MovingMagnet}

To complete the symmetry, we consider the inverse case: a constant, non-moving magnetic square with uniform magnetization $\mathbf{M} = M_0 \hat{z}$ (pointing out of the page).

Physically, this is not a block of "magnetic fluid." It is a loop of current flowing counter-clockwise around the perimeter. The bulk is neutral and carries no current.

\subsubsection{The Physical Picture (Microscopic Baseline)}
Now, assume this current loop moves to the right with velocity $\mathbf{v} = v\hat{x}$.
From the perspective of the underlying sources, Special Relativity must be applied to the charge densities in the wires.
\begin{itemize}
    \item \textbf{The Ionic Background:} The positive lattice ions move at velocity $v$. They undergo standard Lorentz contraction, increasing their linear charge density by $\gamma$.
    \item \textbf{The Bottom Wire:} The macroscopic current flows right ($+\hat{x}$). Thus, the electrons flow \textbf{left} ($-\hat{x}$), opposite to the loop's velocity. Their speed in the lab frame is lowest ($v - u_{\text{drift}}$). They experience \textit{minimum} Lorentz contraction.
    \item \textbf{The Top Wire:} The macroscopic current flows left ($-\hat{x}$). Thus, the electrons flow \textbf{right} ($+\hat{x}$), parallel to the loop's velocity. Their speed in the lab frame is highest ($v + u_{\text{drift}}$). They experience \textit{maximum} Lorentz contraction.
\end{itemize}

\textbf{Result:} The bottom wire acquires a net \textbf{Positive Charge}. The top wire acquires a net \textbf{Negative Charge}. The loop has physically become an electric dipole.

\subsubsection{The Mathematical Reformulation}
This system is now analyzed using the macroscopic formalism. The motion of the magnetized medium creates a "motional polarization":
\begin{equation}
    \mathbf{P} = \frac{\mathbf{v} \times \mathbf{M}}{c^2}
\end{equation}
Given $\mathbf{v} = v\hat{x}$ and $\mathbf{M} = M_0 \hat{z}$, the polarization points downwards:
\begin{equation}
    \mathbf{P} \propto -\hat{y}
\end{equation}

Interpreting this field in terms of bound sources ($\rho_b = -\nabla \cdot \mathbf{P}$):
\begin{itemize}
    \item \textbf{Inside the Bulk:} $\mathbf{P}$ is uniform, so $\rho_b = 0$. (Matches the neutral vacuum inside the loop).
    \item \textbf{Top Surface:} $\mathbf{P}$ starts, creating a positive surface charge $+\sigma_b$.
    \item \textbf{Bottom Surface:} $\mathbf{P}$ terminates, creating a negative surface charge $-\sigma_b$.
\end{itemize}

\subsubsection{The Synthesis (Identity)}
The mathematical prediction of "Motional Polarization" matches the physical reality of "Relativistic Charge Imbalance" perfectly.

\textbf{Conclusion:}
There is no "Dielectric Fluid" being created inside the magnet. The $\mathbf{P}$-field is simply a macroscopic bookkeeping tool that describes the relativistic modification of the current loop's charge density.
Just as the moving dielectric demonstrated that $\mathbf{M}$ is a current, the moving magnet validates that $\mathbf{P}$ is a charge distribution. They are not independent fields; they are the spatial and temporal projections of the single, unified source tensor $M^{\mu\nu}$.

\section{Relativistic Verification: The Moving Dielectric}
\label{sec:MovingDielectric_Relativistic}

The covariant framework is now applied to a specific, illustrative case: a rigid piece of dielectric matter with constant polarization moving with velocity $\mathbf{v}$.

This serves as the consistency verification. It demonstrates that the "vacuum framework" handles the conversion of electromagnetic energy into macroscopic mechanical work (the motor/generator effect) purely through the transformation of sources, without requiring ad-hoc modifications to the stress tensor or "effective" field definitions.

\subsection{Momentum Balance: The Transformed Source Method}

First, the sources are defined. In the rest frame of the dielectric, the system is simple:
\begin{equation}
    \rho = - \nabla \cdot \mathbf{P}, \quad \mathbf{J} = \frac{\partial \mathbf{P}}{\partial t} = 0 \quad (\text{assuming static P}).
\end{equation}
To describe the moving system, the laws of electrodynamics are not modified. Instead, the new effective sources ($\rho', \mathbf{J}'$) and fields ($\mathbf{E}', \mathbf{B}'$) are simply calculated using the standard Lorentz transformations derived in Sec. \ref{sec:ProposedCovariant}.

The macroscopic momentum balance retains the exact same architectural structure as the stationary case. We simply insert the transformed sources:

\begin{equation}
\boxed{
\begin{aligned}
    &\underbrace{ \frac{\partial \mathbf{G}_{\text{mech}}}{\partial t} + \nabla \cdot \mathbf{T}_{\text{kin}} }_{\substack{\text{Bulk Mechanical} \\ \text{Response}}}
    - \underbrace{ \mathbf{F}_{\text{host}} }_{\substack{\text{Internal Material} \\ \text{Response}}} \\
    &\quad \overset{\text{Physical Coupling}}{=} 
    \underbrace{ \rho' \mathbf{E}' + \mathbf{J}' \times\mathbf{B}' }_{\substack{\textbf{The Macroscopic Gateway} \\ (\mathbf{F}_{\text{Lorentz}})}} \\
    &\overset{\text{Math. Identity}}{=}
    \underbrace{ -\left(\frac{\partial \mathbf{G}'_{\text{vac}}}{\partial t} + \nabla\cdot \mathbf{T}'_{\text{vac}}\right) }_{\substack{\text{Macroscopic EM} \\ \text{Response}}}
\end{aligned}
}
    \label{eq:MacroMomentumDomainSeparation}
\end{equation}

\subsubsection{Interpretation of the Host Force}
The interpretation of $\mathbf{F}_{\text{host}}$ depends on the internal state of the matter. In Parts III and IV, the Host Interface was seen to act in the three presented ways: \textbf{Storage} (Binding force), \textbf{Dissipation}, and \textbf{Work}.

In this example, a \textbf{rigid body} with constant polarization is considered; thus, the gateways for dissipation and storage are closed. The bound charges are locked to the lattice.
There is no relative motion of charges (no current flow relative to the lattice), so there is no resistive drag ($\mathbf{F}_{\text{res}} = 0$). There is no deformation of the dipoles.

Therefore, the Host Force represents pure force transmission for macroscopic bulk motion. The host interface transmits the Lorentz force acting on the transformed bound charges directly to the bulk mass of the rigid body:

\subsection{Energy Balance: The Origin of Mechanical Work}

The corresponding energy equation follows via the FECC. Again, the structure is invariant; the work done by the transformed electric field $\mathbf{E}'$ on the transformed total current $\mathbf{J}'$ is calculated. Crucially, $\mathbf{J}'$ now contains the \textbf{Convection Current} and the \textbf{Röntgen Current} automatically:

\begin{equation}
\boxed{
\begin{aligned}
    &\underbrace{ \frac{\partial U_{\text{mech}}}{\partial t} + \nabla \cdot \mathbf{S}_{\text{mech}} }_{\substack{\text{Carrier Kinetic} \\ \text{Response}}}
    - \underbrace{ P_{\text{host}} }_{\substack{\text{Internal Material} \\ \text{Power}}} \\
    &\quad \overset{\text{Physical Coupling}}{=} 
    \underbrace{ \mathbf{J}' \cdot \mathbf{E}' }_{\substack{\textbf{The Macroscopic Gateway} \\ (\text{Work Rate})}} \\
    &\overset{\text{Math. Identity}}{=}
    \underbrace{ -\left(\frac{\partial U'_{\text{vac}}}{\partial t} + \nabla \cdot \mathbf{S}'_{\text{vac}}\right) }_{\substack{\text{Macroscopic EM} \\ \text{Response}}}
\end{aligned}
}
    \label{eq:MacroEnergyDomainSeparation}
\end{equation}

\section{Summary: Moving Matter}
\label{sec:Moving_Matter_Summary}

In this chapter, the Vacuum Framework has been subjected to its most rigorous test: the domain of relativistic motion. The results confirm that the mechanical consistency established in the static domain extends seamlessly to dynamic systems without the addition of any new postulates.
All of moving matter is described by a single geometric principle: the \textbf{polarization-magnetization tensor} ($M^{\mu\nu}$).
It has been demonstrated that $\mathbf{P}$ and $\mathbf{M}$ are not distinct physical entities but are merely the temporal and spatial projections of the same underlying source distribution.
\begin{itemize}
    \item \textbf{The Moving Dielectric:} It was shown that "Motional Magnetization" is simply the convection current of bound surface charges.
    \item \textbf{The Moving Magnet:} It was shown that "Motional Polarization" is simply the relativistic charge accumulation arising from Lorentz contraction differences in current loops.
\end{itemize}
These are not "effective" fields; they are the literal physical consequences of observing sources from a moving frame. Because this transformation is purely kinematic, it guarantees that the \textbf{Force-Energy Consistency Criterion (FECC)} is satisfied automatically. Since the sources transform geometrically, the energy and momentum transfer must track the mass exactly, leaving no room for a "Constitutive Slip" or ambiguous hidden reservoirs.

\part{Part VI: The Analysis of Macroscopic Models}
\label{part:NewPart6}

\chapter{The Vacuum Framework: The Standard of Judgment}
\label{ch:review_vacuum_framework}

Part I through Part V of this monograph have been dedicated to a constructive synthesis. By strictly adhering to the "Microscopic Baseline"---the deterministic reality of discrete charges and vacuum fields---a comprehensive macroscopic framework has been derived. It has been demonstrated that the \textbf{Vacuum (Amperian) Formulation} is not merely a "microscopic approximation" but a fully robust macroscopic field theory capable of describing the complete phenomenology of ponderable media, from the electromechanics of dielectrics and conductors to the relativistic dynamics of moving matter.

In this sixth part of the work, the analytical direction is reversed. The argument shifts from \textit{bottom-up construction} to \textit{top-down critique}.

For over a century, the literature of electrodynamics has been characterized by a proliferation of competing energy-momentum tensors---Minkowski, Abraham, Einstein-Laub, Chu, and others. The modern consensus suggests that the choice between these formulations is a matter of convention, resolvable only by the arbitrary assignment of a "material counterpart" to balance the conservation laws.

Before these historical formulations can be subjected to a rigorous mechanical audit, the standard against which they will be judged must be explicitly defined. This chapter summarizes the findings of the preceding parts to establish the \textbf{Mechanically Consistent Framework}. It formalizes the criteria of validity that will identify the physical distinction between a "mathematically balanced equation" and a "mechanically consistent reality."

\section{The Criterion of Consistency}
\label{sec:mechanically_consistent_framework}

The central conclusion of the derivation in Part II was that the macroscopic equations of motion are not independent postulates; they are the \textit{spectral projections} of the underlying microscopic reality. The application of this spectral filter revealed a system that is strictly over-determined by its mechanical constraints.

Specifically, the "Vacuum Framework" was shown to satisfy a rigorous closure condition that links the electromagnetic and mechanical sectors. This link is not a flexible coupling variable; it is a rigid identity derived from Newton's Second Law. We designate this the \textbf{Force-Energy Consistency Criterion (FECC)}.

\subsection{The Kinematic Lock}

In any consistent physical theory, "Momentum" and "Energy" are not independent currencies; they are indivisible facets of the same interaction.
\begin{itemize}
    \item \textbf{Force} ($\mathbf{f}$) is the rate of momentum transfer to a mass.
    \item \textbf{Power} ($P$) is the rate of work performed by that force.
\end{itemize}
These two descriptions are strictly coupled by the \textbf{kinematic constraint}: the velocity $\mathbf{v}$ of the specific mass target acting as the sink. The power density $P$ is, by axiomatic definition, the scalar product of the force and its kinematic target:
\begin{equation}
    P \equiv \mathbf{f} \cdot \mathbf{v}.
    \label{eq:mech_power_def_kinematic}
\end{equation}
This identity is fundamental. It asserts that a physical theory cannot define $\mathbf{f}$ and $P$ in isolation. If a theory postulates a power density (a source/sink of field energy), it has implicitly defined the velocity field to which that power couples.

\textbf{The Kinematic Lock:} \textit{Energy absorption requires a moving carrier.} Wherever there is a sink of electromagnetic energy ($-\nabla \cdot \mathbf{S} - \dot{u} \neq 0$), there must be a massive charge-carrier population moving at that exact coordinate to accept the work. To "sink" energy is to "accelerate" mass.

The derivation of the Vacuum Framework demonstrated that the electromagnetic interaction intrinsically satisfies this condition. The macroscopic Lorentz force ($\rho \mathbf{E} + \mathbf{J}\times\mathbf{B}$) acts upon specific fluid elements (electrons and ions) with defined velocities. The resulting energy transfer is exactly the work done on these fluids:
\begin{equation}
\boxed{
\begin{aligned}
    \underbrace{ \left( \frac{\partial U_{\text{mech}}}{\partial t} + \nabla\cdot\mathbf{S}_{\text{mech}} \right) - P_{\text{diss}} }_{\substack{\text{Total Mechanical Response}}}
    \quad &\overset{\text{Coupling}}{\textbf{=}} \quad
    \underbrace{ \mathbf{J}\cdot\mathbf{E} }_{\substack{\text{The Energy Gateway} \\ (P_{\text{Lorentz}})}} \\
    &\overset{\text{Identity}}{=} \quad
    \underbrace{ -\left(\frac{\partial U_{\text{EM}}}{\partial t} + \nabla\cdot \mathbf{S}_{\text{EM}}\right) }_{\substack{\text{Macroscopic EM Source}}}
\end{aligned}
}
\end{equation}
This framework is "Mechanically Consistent by Construction." The energy leaving the field matches exactly the work accepted by the mass.

\section{The Topology of Interaction}

The constructive synthesis further clarified the specific "Gateways" through which this energy passes. By resolving the material into its kinematic components (Ordered, Relative, and Unordered motion), the analysis identified three distinct pathways for electromagnetic work:

\begin{enumerate}
    \item \textbf{Dissipation (The Thermal Sink):}
    Energy transferred to the unordered, random motion of the charge carriers. Physically, this corresponds to the irreversible "friction" of the Host Interface (e.g., Ohmic heating, dielectric loss).
    
    \item \textbf{Work (The Continuum Sink):}
    Energy transferred to the ordered, coherent motion of the material bulk. This is the gateway to Continuum Mechanics, driving the forces of actuation, electrostriction, and magnetostriction.
    
    \item \textbf{Storage (The Potential Sink):}
    Energy transferred to the internal potential energy of the material lattice. As derived in Part III, this encompasses the energy stored in the "Binding Reservoir"---the intense, high-frequency microscopic fields that maintain the structural integrity of the dielectric or magnetic dipoles.
\end{enumerate}

The Vacuum Framework distinguishes these pathways precisely. Crucially, it identifies the "Binding Energy" (Category 3) as a \textit{mechanical} storage term---energy stored in the spring-like constraints of the matter---rather than limiting it to the definition of the macroscopic electromagnetic field.

\section{The Protocol of the Audit}

With this standard established, the following chapters will proceed with the critique of the historical formulations (Minkowski, Abraham, Einstein-Laub).

The analysis will reject the hypothesis of "formal equivalence." Instead, the \textbf{Principle of Physical Exclusivity} is invoked: distinct mathematical formulations of the energy-momentum tensor are treated as mutually exclusive physical hypotheses regarding the nature of the interaction.

The audit protocol is forensic:
\begin{enumerate}
    \item \textbf{Assume Validity:} We will assume, arguendo, that the proposed tensor (e.g., Minkowski) is the correct description of reality.
    \item \textbf{Identify the Variables:} We will attempt to identify the mechanical state variables (Mass Target, Velocity Field) implied by that tensor's force and energy definitions.
    \item \textbf{Test Compatibility:} We will apply the \textbf{FECC} ($P \stackrel{?}{=} \mathbf{f} \cdot \mathbf{v}$) to determine if the proposed force density and energy flux are kinematically compatible.
\end{enumerate}

If a formulation predicts a flux of energy ($P \neq 0$) in a system where it defines no moving Mechanical Target ($\mathbf{v} = 0$), it will be identified as \textbf{mechanically inconsistent}. It will be argued that such theories rely on "Phantom Work"---algebraic balancing terms that have no correspondence to the motion of mass.

The Vacuum Framework, having been proven to possess a closed and consistent topology, serves as the control against which these deviations are measured.

\chapter{The Mechanical Audit of the Minkowski Tensor}
\label{chap:Macro_Critique}

A mechanically consistent macroscopic description of electrodynamics---the "Vacuum Framework"---has been established. In the previous chapter, this framework was sharpened into a diagnostic tool: the \textbf{Force-Energy Consistency Criterion (FECC)}.

The analytical focus now shifts to the investigation of alternative formulations. We begin with the most prominent and mathematically elegant candidate: the Minkowski Energy-Momentum Tensor.

\section{The Hypothesis of Macroscopic Reality}

In Part I, Chapter \ref{chap:Critique_ArbitrarySplit}, the analysis identified the Minkowski formulation as being topologically distinct at the microscopic level. The "expanding dipole" test case revealed that the Minkowski tensor implies the existence of a force density within the vacuum gap between discrete charges---a prediction that requires the postulate of \textbf{microscopic vacuum inertia}.

In this chapter, we extend the \textit{arguendo} assumption of validity to the macroscopic scale. The Minkowski tensor is treated not as a convenient approximation, but as a candidate for a \textbf{wholesale macroscopic hypothesis} that claims to replace the microscopic Lorentz description.

The question is posed: "If the Minkowski hypothesis were true, what mechanical reality would be required to sustain it?" The analysis assumes the algebraic validity of the tensor and attempts to identify the mechanical variables (mass, velocity) that would satisfy the \textbf{Axiom of Actuation}.

The investigation is forensic. The "thermodynamic divergence" that arises is not presented as an accusation, but as the inevitable mathematical consequence of the equation's own balance.

\section{The Analytical Strategy}

Microscopically, the distinct topology of the Minkowski framework was relatively demonstrable. Macroscopically, however, the divergence is more subtle. In bulk matter, polarization and magnetization are always co-located with mass. One will not find "inertia where there is no mass" in a continuum. Therefore, the systematic analysis must be more precise.

\subsection{The Indeterminacy of Deformation}
Experimentally distinguishing tensors by measuring bulk deformation is not merely difficult; it is a category error.
As proven in Part III, Chapter \ref{chap:ForceDensity} (\textit{The Theorem of Macroscopic Indeterminacy}), the force responsible for mechanical deformation is not a single-valued function of the macroscopic fields ($\mathbf{P}, \mathbf{E}$) alone. It depends explicitly on the microscopic topology of the material lattice (``Side-Channel Information'').
Two materials with identical field states ($\mathbf{P}, \mathbf{E}$) can experience radically different internal stress distributions depending on whether their microstructure is continuous or discrete.
Therefore, no single energy-momentum tensor---whether Minkowski, Abraham, or Lorentz---can claim to represent the \textit{universal} ``Deformation Force Density.'' The total momentum budget is invariant, but its partition into ``Deformation'' (Work) versus ``Internal Binding'' (Storage) is material-dependent. Consequently, deformation measurements test the \textit{material model}, not the fundamental \textit{field theory}.
To rigorously audit the Minkowski tensor, we must therefore ignore deformation and focus on the invariant \textbf{Total Energy-Momentum Budget}. We test whether the tensor provides a consistent description of the total exchange.

\subsection{The Protocol: Internal Mechanical Consistency}
Instead, the focus is placed entirely on \textbf{Internal Mechanical Consistency}. For a theory to be mechanically valid, it must strictly satisfy the kinematic constraints of its own definitions. This applies particularly to the description of non-linear and hysteretic behavior.

Mechanical consistency requires that the fundamental elements of the interactions be clearly identified:
\begin{enumerate}
    \item The \textbf{Mass Target} of the force.
    \item The \textbf{Velocity} of that target.
    \item The corresponding \textbf{Power Equation}, which must satisfy the identity $P = \mathbf{f} \cdot \mathbf{v}$.
\end{enumerate}

The investigation proceeds in three rigorous stages:
\begin{enumerate}
    \item \textbf{Linear Matter:} Testing the tensor on linear dielectrics (The Null Test).
    \item \textbf{Non-linear, Non-dissipative Matter:} Testing the tensor against saturation (The "Free Energy Paradox").
    \item \textbf{Dissipative Matter:} Testing the tensor against friction and heating (The "Missing Mechanism").
\end{enumerate}

\textbf{Changing fields} are the primary focus, as the interest lies in the process of \textit{mechanical work} ($P > 0$). If the Minkowski tensor predicts a force on a target that has zero velocity ($\dot{\mathbf{P}} = 0$), yet the material heats up, the tensor will be identified as mechanically inconsistent.

\section[The Divergence of Energy-Momentum]{The Divergence of the\\ Energy-Momentum Tensor}

The mathematical divergence between the two realities must be explicitly defined.

\subsection[The Divergence of Momentum]{The Divergence of Macroscopic\\ Momentum and Force}
\label{sec:Macro_Momentum_Divergence}

The Vacuum Framework adheres to the sum of Lorentz forces on constituents:
\begin{equation}
\boxed{
\begin{aligned}
    &\underbrace{ (\rho - \nabla \cdot \mathbf{P})\mathbf{E} + \left(\mathbf{J} + \frac{\partial \mathbf{P}}{\partial t} + \nabla \times \mathbf{M}\right)\times\mathbf{B} }_{\substack{\textbf{The Lorentz Definition} \\ \text{(Total Force on Constituents)}}} \\
    &\quad \overset{\text{Identity}}{\textbf{=}} \quad
    \underbrace{ -\left(\frac{\partial \mathbf{G}_{\text{vac}}}{\partial t} + \nabla\cdot \mathbf{T}_{\text{vac}}\right) }_{\substack{\textbf{The Vacuum Definition} \\ \text{(Momentum Supply)}}}
\end{aligned}
}
    \label{eq:Macro_Momentum_Architecture_Split}
\end{equation}

Minkowski's formulation proposes a substantial redefinition. By algebraically absorbing the terms involving bound sources ($\mathbf{P}, \mathbf{M}$) into the field tensor, the formulation implicitly \textbf{redefines} the concept of field momentum. Consequently, to maintain the equation's balance, the definition of "Mechanical Force" must also change.

\begin{equation}
\boxed{
\begin{aligned}
    \underbrace{ (\rho \mathbf{E} + \mathbf{J} \times \mathbf{B}) + \mathbf{F}_{\text{matter}, M} }_{\substack{\textbf{The Minkowski Postulate} \\ \text{(Free Force + Proposed Matter Force)}}}
    \quad &\overset{\text{Definitional Identity}}{\textbf{=}} \quad
    \underbrace{ -\left( \frac{\partial \mathbf{G}_M}{\partial t} + \nabla \cdot \mathbf{T}_M \right) }_{\substack{\textbf{The Minkowski Definition} \\ \text{(Field + Conflated Momentum)}}}
\end{aligned}
}
    \label{eq:Macro_Minkowski_Momentum_Architecture}
\end{equation}

Here, the macroscopic variables are defined analogously to their microscopic counterparts:
\begin{align}
    \mathbf{G}_M &\equiv \mathbf{D} \times \mathbf{B} \\
    \mathbf{T}_M &\equiv \mathbf{E} \otimes \mathbf{D} + \mathbf{H} \otimes \mathbf{B} - \frac{1}{2}(\mathbf{E} \cdot \mathbf{D} + \mathbf{B} \cdot \mathbf{H})\mathbf{I}
\end{align}

The Minkowski force density ($\mathbf{F}_{\text{matter}, M}$) is a new mathematical object, distinct from the Lorentz force:
\begin{equation}
    (\mathbf{F}_{\text{matter}, M})_k \equiv \frac{1}{2} \sum_{j=1}^{3} \left[ P_j (\partial_k E_j) - E_j (\partial_k P_j) + M_j (\partial_k B_j) - B_j (\partial_k M_j) \right].
    \label{eq:MinkowskiForce_Macro}
\end{equation}

The \textbf{physical discrepancy} ($\Delta \mathbf{F}$) between the proposed Minkowski force and the actual Lorentz force is non-zero. It corresponds exactly to the time rate of change of the conflated momentum components:
\begin{equation}
    \Delta \mathbf{F} = \mathbf{F}_{\text{matter}, M} - \mathbf{F}_{\text{Lorentz}} = \frac{\partial \mathbf{G}_\Delta}{\partial t} + \nabla \cdot \mathbf{T}_\Delta = \mathbf{\Phi}
\end{equation}
The discrepancy is explicitly identified as the anomalous vector $\mathbf{\Phi}$:
\begin{equation}
    \mathbf{\Phi} = \frac{\partial (\mathbf{P} \times \mathbf{B})}{\partial t} + \nabla \cdot \left[ (\mathbf{P} \cdot \mathbf{E} + \mathbf{M} \cdot \mathbf{B})\mathbf{I} - (\mathbf{E}\mathbf{P} + \mathbf{B}\mathbf{M}) \right]
\end{equation}
This $\mathbf{\Phi}$ represents the force that the Minkowski formulation effectively "reallocates" from the material lattice (the Host Interface) and reassigns to the field description.

\subsection[The Divergence of Energy]{The Divergence of Macroscopic\\ Energy and Power}
\label{sec:Macro_Energy_Divergence}

A similar divergence occurs in the energy domain. The derived ground truth asserts that energy transfer occurs strictly via the electric work on the macroscopic currents:
\begin{equation}
\boxed{
\begin{aligned}
    \underbrace{ \left( \mathbf{J} + \frac{\partial \mathbf{P}}{\partial t} + \nabla \times \mathbf{M} \right) \cdot \mathbf{E} }_{\substack{\textbf{The Lorentz Definition} \\ \text{(Work on All Matter)}}}
    \quad &\overset{\text{Identity}}{\textbf{=}} \quad
    \underbrace{ -\left(\frac{\partial U_{\text{vac}}}{\partial t} + \nabla \cdot \mathbf{S}_{\text{vac}}\right) }_{\substack{\textbf{The Vacuum Definition} \\ \text{(Field Energy Supply)}}}
\end{aligned}
}
    \label{eq:Macro_Energy_Architecture_Split}
\end{equation}

The Minkowski formulation, however, \textbf{postulates} a new interaction term $P_{\text{matter}, M}$. By redefining the field energy flux (Poynting vector) to include the transport of bound energy, it necessitates a redefinition of the work term:
\begin{equation}
\boxed{
\begin{aligned}
    \underbrace{ \mathbf{J} \cdot \mathbf{E} + P_{\text{matter}, M} }_{\substack{\textbf{The Minkowski Postulate} \\ \text{(Free Work + Proposed Matter Work)}}}
    \quad &\overset{\text{Definitional Identity}}{\textbf{=}} \quad
    \underbrace{ -\left( \frac{\partial U_M}{\partial t} + \nabla \cdot \mathbf{S}_M \right) }_{\substack{\textbf{The Minkowski Definition} \\ \text{(Field + Conflated Energy)}}}
\end{aligned}
}
    \label{eq:Minkowski_Energy_Architecture}
\end{equation}

Here, the macroscopic energy variables are defined as:
\begin{align}
    U_M &\equiv \frac{1}{2}(\mathbf{E} \cdot \mathbf{D} + \mathbf{B} \cdot \mathbf{H}) \\
    \mathbf{S}_M &\equiv \mathbf{E} \times \mathbf{H}
\end{align}

The \textbf{energy discrepancy} ($\Delta P$) is given by the evolution of the "conflated" energy density and flux:
\begin{equation}
    \Delta P = P_{\text{matter}, M} - P_{\text{Lorentz}} = \frac{\partial U_\Delta}{\partial t} + \nabla \cdot \mathbf{S}_\Delta = \Phi_E
\end{equation}
with the Minkowski power transfer density given by:
\begin{equation}
    P_{\text{matter}, M} = \frac{1}{2} \left( \frac{\partial \mathbf{E}}{\partial t} \cdot \mathbf{D} - \mathbf{E} \cdot \frac{\partial \mathbf{D}}{\partial t} + \frac{\partial \mathbf{B}}{\partial t} \cdot \mathbf{H} - \mathbf{B} \cdot \frac{\partial \mathbf{H}}{\partial t} \right).
    \label{eq:MinkowskiPowerVector_Discrepancy_Audit}
\end{equation}
Explicitly, the energy-domain anomalous vector is:
\begin{equation}
    \Phi_E = \nabla \cdot (\mathbf{E} \times \mathbf{M}) - \frac{1}{2} \frac{\partial}{\partial t}(\mathbf{E} \cdot \mathbf{P} - \mathbf{M} \cdot \mathbf{B})
\end{equation}
This term $\Phi_E$ represents the \textbf{Energy Reallocation}---the portion of mechanical work (stretching the Host Interface) that the Minkowski framework reclassifies as stored field energy.

\section{The Principle of Mutually Exclusive Hypotheses}

Before dissecting the specific components of the Minkowski tensor, the \textbf{Principle of Physical Exclusivity} is reaffirmed.

As established in Part I-V, the Vacuum Framework (represented by $\mathbf{F}_{\text{Lorentz}}$ and $P_{\text{Lorentz}}$) is mechanically closed. The Minkowski formulation implies a completely different set of Force and Power equations. The difference between this proposed reality and the Vacuum Reality matches exactly the difference terms $\mathbf{\Phi}$ and $\Phi_E$:

\begin{equation}
    \mathbf{F}_{\text{mech, M}} - \mathbf{\Phi} \quad \equiv \quad \mathbf{F}_{\text{Lorentz}} - \mathbf{\Phi}
\end{equation}
\begin{equation}
    P_{\text{mech, M}} - \Phi_E \quad \equiv \quad (\mathbf{f} \cdot \mathbf{v})_{\text{Lorentz}} - \Phi_E
\end{equation}

Physically, the Minkowski formulation stands as a distinct, self-contained hypothesis about how fields and matter interact. By changing the definition of field momentum density, it fundamentally alters the physics of the system. It changes the force density, redefines the mass target, and severs the rigorous connection between energy transfer and mechanical work found in the Lorentz theory.

\textbf{It abandons the kinematic constraints of the Lorentz theory in favor of a new, mutually exclusive physical reality.}

Whether these new definitions preserve internal mechanical consistency ($P = \mathbf{f} \cdot \mathbf{v}$) is not guaranteed by their algebraic balance; it is the specific hypothesis we must test.

\section[The Mechanical Consistency]{The Mechanical Efficiency of\\ the Minkowski Tensor}

The detailed forensic analysis now proceeds. To investigate the validity of the Minkowski proposal, the strict protocol of \textbf{Internal Mechanical Consistency} (FECC) is adhered to. The Minkowski tensor is scrutinized not as a field description, but as a mechanical hypothesis---a "Wholesale Reality" that replaces the microscopic Lorentz description with a mutually exclusive macroscopic framework.

The audit procedure is deductive: it attempts to verify if the tensor can be mechanically consistent with its own definitions.
\begin{enumerate}
    \item \textbf{Identify Variables:} We strictly identify the mass target and velocity variable defined \textit{within} the Minkowski framework itself.
    \item \textbf{Evaluate Couplings:} We calculate the Force $\mathbf{f}_M$ and the Energy Divergence $\mathcal{D}_M$ independently.
    \item \textbf{Test the FECC:} We check if the fundamental identity holds:
    \begin{equation}
        -\nabla \cdot \mathbf{S}_M - \dot{u}_M \overset{?}{=} \mathbf{f}_M \cdot \mathbf{v}_{\text{target}}
    \end{equation}
\end{enumerate}

\subsection{Case Study 1: The Linear Dielectric (The Null Test)}

The investigation begins with the simplest possible material system: a rigid, stationary, non-dissipative dielectric that responds linearly to the applied field.

\paragraph{The Physical Setup.}
We define the system state rigorously:
\begin{enumerate}
    \item \textbf{Kinematics:} The material bulk is held stationary by external constraints ($\mathbf{v}_{\text{bulk}} = 0$).
    \item \textbf{Response:} The material is linear and isotropic, such that $\mathbf{D} = \varepsilon \mathbf{E}$ (where $\varepsilon$ is a time-independent scalar constant).
    \item \textbf{Dynamics:} We apply a time-varying external field ($\frac{\partial \mathbf{E}}{\partial t} \neq 0$) to drive the interaction.
\end{enumerate}

\paragraph{Step 1: The Energy Audit (The Balance Sheet).}
First, the energy transfer predicted by the Minkowski Poynting Theorem is calculated. The interaction term $\mathcal{D}_M$ (the divergence of the energy flux) describes the rate at which energy leaves the electromagnetic sector to enter the matter.

Using the Minkowski definition of energy density $U_M = \frac{1}{2}\mathbf{E}\cdot\mathbf{D}$, the power transfer density is derived from the continuity equation:
\begin{equation}
    \mathcal{D}_M = -\left( \frac{\partial U_M}{\partial t} + \nabla \cdot \mathbf{S}_M \right) =\frac{1}{2} \left( \frac{\partial \mathbf{E}}{\partial t} \cdot \mathbf{D} - \mathbf{E} \cdot \frac{\partial \mathbf{D}}{\partial t} \right).
\end{equation}
Substituting the linear constitutive relation $\mathbf{D} = \varepsilon \mathbf{E}$:
\begin{equation}
    \mathcal{D}_M = \mathbf{E} \cdot (\varepsilon \dot{\mathbf{E}}) - \frac{1}{2} \frac{\partial}{\partial t} (\varepsilon |\mathbf{E}|^2) = \varepsilon (\mathbf{E} \cdot \dot{\mathbf{E}}) - \varepsilon (\mathbf{E} \cdot \dot{\mathbf{E}}) \equiv 0.
\end{equation}
\textbf{Result:} For a linear medium, the Minkowski formulation predicts \textbf{zero energy transfer}. The theory asserts that no work is performed on the matter; rather, the energy is entirely retained within the definition of the macroscopic field energy $U_M$.

\paragraph{Step 2: The Mechanical Audit (The Search for the Mechanism).}
Next, the mechanical side of the ledger is systematically inspected. 
Because the Minkowski framework claims to be a fundamental, mutually exclusive reality, the "microscopic dipoles" of the Lorentz theory cannot be invoked to explain it. The tensor must be audited strictly on its own terms. It is treated as a "Black Box" proposal, asking: \textit{What is moving inside?}

\begin{itemize}
    \item \textbf{The Force Definition:} The Minkowski force density involves terms such as $P_j \partial_k E_j$. This force acts on the field variable $\mathbf{P}$. 
    \item \textbf{The Search for the Target:} To satisfy Newton's Second Law ($\mathbf{f} = \rho \mathbf{a}$), there must be a mass density $\rho_m$ associated with $\mathbf{P}$. We identify this simply as the "Dielectric Matter," the macroscopic carrier of the polarization state.
    
    \item \textbf{The Search for Velocity (The Investigation):} 
    What is the velocity $\mathbf{v}$ of this matter? 
    \begin{enumerate}
        \item \textbf{Bulk Motion:} The material bulk is constrained to be stationary ($\mathbf{v}_{\text{bulk}} = 0$).
        \item \textbf{Internal Motion (The Internal Motion Hypothesis):} Could there be hidden \textit{internal} motions of the mass—a "swirling fluid" of polarization—that we cannot see?
    \end{enumerate}
    
    \item \textbf{The Status of Internal Motion:}
    Even though the bulk is at rest ($\mathbf{v}_{\text{bulk}}=0$), there exists an internal motion of massive constituents, $\mathbf{v}_{\text{int}}(\mathbf{x},t)\neq 0$, on which the Minkowski force could act.
    
    The energy audit above, however, constrains what such a postulate would have to accomplish. In this linear case, the Minkowski balance sheet yields a \textbf{null interaction density} ($\mathcal{D}_M \equiv 0$). Therefore, any admissible internal motion must either:
    \begin{enumerate}
        \item be everywhere \textbf{power-orthogonal} to the Minkowski force density (so that $\mathbf{f}_M \cdot \mathbf{v}_{\text{int}} \equiv 0$ pointwise), or
        \item involve \textbf{exact internal cancellations} between multiple colocated subsystems (positive work on one component and negative work on another at the same coordinate), such that the net Minkowski interaction remains identically zero.
    \end{enumerate}
    
    Crucially, the Minkowski framework does not supply the kinematic structure required to \textit{define} such an internal velocity field, nor does it provide a dynamical law that would \textit{enforce} the required orthogonality/cancellation. This hypothesis therefore does not yield a predictive mechanical model; it introduces an arbitrary, unconstrained degree of freedom. In other words, the Minkowski tensor can be made to \textit{avoid} work in the linear case only by postulating \textbf{zero internal motion} or by leaving the internal mechanics \textbf{undefined}.
    
    \item \textbf{Conclusion (for the Linear Case):} The Minkowski ``medium'' is therefore treated as a stationary macroscopic target with no mechanically defined internal velocity field available. In this null test, the framework remains \textbf{mechanically null}: it is consistent because its own bookkeeping asserts that no net energy transfer occurs, while no internal motion is doing work.
\end{itemize}

Crucially, in this linear case, the debate is moot. Because the energy audit yielded zero ($\mathcal{D}_M \equiv 0$), the mechanical audit requires zero work ($P_{\text{mech}} \equiv 0$).
The force pushes against a stationary wall:
\begin{equation}
    P_{\text{mech}} = \mathbf{f}_M \cdot \mathbf{v}_{\text{target}} = \mathbf{f}_M \cdot \mathbf{0} \equiv 0.
\end{equation}

\paragraph{The Verdict: Mechanically Consistent (But Static).}
The Minkowski tensor passes the audit for the linear case. However, it does so naturally only because the net energy transfer is zero. Comparing the two frameworks reveals a distinct philosophical divergence:

\begin{itemize}
    \item \textbf{The Lorentz Reality (The Engine):} Even in the linear case, the material is treated as an engine of moving fluids ($\mathbf{J}_P = \partial_t \mathbf{P}$). Energy is actively transferred ($P = \mathbf{J} \cdot \mathbf{E}$) and stored in the "binding energy" of the Host Interface.
    \item \textbf{The Minkowski Reality (The Block):} The material is treated as a passive block with changing parameters. There is no motion, no work, and no distinct "storage" mechanism other than the re-labeling of the field energy itself.
\end{itemize}

While the Minkowski approach is mathematically consistent in this static limit, it achieves this consistency by postulating no internal mechanism of the interactions.

\subsection{The Thermodynamic Isomorphism (The Trap of Linearity)}
\label{sec:Linearity_Constraint}

Before identifying the failure modes, it is essential to understand why the Minkowski formulation has survived for a century. The definition of the energy density $u_M = \frac{1}{2}\mathbf{E}\cdot\mathbf{D}$ is not merely an "error"; it is a compellingly reasonable hypothesis.

In the vast majority of engineering applications (linear, non-dissipative dielectrics), the term $\frac{1}{2}\mathbf{E}\cdot\mathbf{D}$ correctly captures the \textit{Total Thermodynamic Work} required to assemble the field in the material. It represents the sum of the energy stored in the vacuum and the potential energy stored in the material lattice.

\begin{equation}
    u_M = \frac{1}{2} \mathbf{E} \cdot \mathbf{D} = \underbrace{\frac{1}{2}\varepsilon_0 E^2}_{\text{Vacuum Energy}} + \underbrace{\frac{1}{2}\mathbf{E} \cdot \mathbf{P}}_{\text{Binding Energy}}.
\end{equation}

Because the Minkowski functional correctly accounts for the total bill in this limit, it is easy to assume it also correctly identifies the location of the storage. The error lies not in the \textit{amount} of energy, but in its \textit{attribution}. The Minkowski Force-Energy architecture implicitly assumes that this entire sum resides "in the field," whereas the mechanical reality is that the second term resides "in the springs" (the binding potential of the matter).

This distinction is subtle in the linear regime. As long as the "springs" are linear (Hookean), the energy stored in them ($u \propto P^2 \propto E^2$) scales exactly like field energy. The two reservoirs are indistinguishable by external observation. This algebraic isomorphism is the origin of the historical persistence of the Minkowski tensor. It is only when we step outside the linear limit that the "mask" slips and the misattribution becomes a divergence.

\paragraph{The Macroscopic Perspective.}
We propose that this misidentification is not an arbitrary mistake; it is the logical consequence of the \textbf{macroscopic viewpoint}. The macroscopic electric field $\mathbf{E}$ is, by definition, a low-pass filtered quantity. It structurally filters out the intense microscopic fields that hold the material together. 
Consequently, a physicist working comfortably within the macroscopic domain cannot "see" the internal binding mechanism; they see only that energy enters the volume and remains there. Without access to the microscopic baseline, it is entirely rational—though arguably methodologically limited—to attribute this stored energy to the macroscopic field definition itself ($u \propto \mathbf{E}\cdot\mathbf{D}$). 

This is the "Constraint of Linearity." Because the energy bookkeeping works in this limit, it is assumed that the \textit{mechanism} (Field Energy) is also correct. The distinction between "Energy stored in the macroscopic field" and "Energy stored in the hidden microscopic constraint" is invisible in the linear regime.
The failure only becomes visible when we push the system out of this "Safe Zone"---into saturation or dissipation. Once the linear isomorphism breaks, the decision to assign the energy to the field ($u_M$) rather than the work ($P_{mech}$) renders the theory mechanically inconsistent.

\subsection{Case Study 2: Saturation (The Failure of the Block)}
\label{sec:saturation_critique}

The limitations of the Minkowski "Macroscopic Hypothesis" are exposed as soon as the material response separates from the field drive. We consider the case of \textbf{Saturation}.

\paragraph{The Physical Setup.}
We modify the previous setup:
\begin{enumerate}
    \item \textbf{Kinematics:} The material bulk remains rigid and stationary ($\mathbf{v}_{\text{bulk}} = 0$).
    \item \textbf{Response:} The material is non-linear and has reached saturation, such that the polarization is clamped at a maximum value $\mathbf{P}_{\text{sat}}$ and can no longer increase ($\dot{\mathbf{P}} = 0$).
    \item \textbf{Dynamics:} The external field continues to increase ($\dot{\mathbf{E}} \neq 0$).
\end{enumerate}

\paragraph{Step 1: The Energy Audit (The Effective Sink).}
The Minkowski energy transfer is recalculated via the divergence term:
\begin{equation}
    \mathcal{D}_M = \frac{1}{2} \left( \frac{\partial \mathbf{E}}{\partial t} \cdot \mathbf{D} - \mathbf{E} \cdot \frac{\partial \mathbf{D}}{\partial t} \right).
\end{equation}
Using the definition $\mathbf{D} = \varepsilon_0 \mathbf{E} + \mathbf{P}_{\text{sat}}$ and evaluating with $\dot{\mathbf{P}} = 0$, the vacuum terms cancel, but the cross-term leaves a non-zero residual:
\begin{equation}
    \mathcal{D}_M =  \frac{1}{2} \mathbf{P}_{\text{sat}} \cdot \frac{\partial \mathbf{E}}{\partial t} \neq 0.
\end{equation}
\textbf{Result:} The Minkowski formulation predicts a \textbf{Non-Zero Energy Transfer}. The theory asserts that energy is being extracted from the total electromagnetic field budget ($U_M$).

\paragraph{Step 2: The Mechanical Audit (The Broken Bridge).}
Now, the mechanical destination of this energy is sought.
\begin{itemize}
    \item \textbf{Target:} As established in Case 1, the force acts on the "Dielectric Matter" carrying $\mathbf{P}$.
    \item \textbf{Velocity:} The bulk velocity is zero by construction ($\mathbf{v}_{\text{bulk}}=0$). A defender may again attempt to rescue the theory by postulating an internal motion of massive constituents, $\mathbf{v}_{\text{int}}(\mathbf{x},t)\neq 0$. However, the Minkowski framework does not provide a kinematic law that defines such a velocity field, nor does it supply a constitutive mechanism that would tie it uniquely to the macroscopic state variable $\mathbf{P}$ (which is clamped here).
    \item \textbf{Property Check:} The polarization $\mathbf{P}$ itself is constant ($\dot{\mathbf{P}}=0$). Even the "internal state" is not changing.
\end{itemize}
Consequently, within the closed Minkowski framework, the mechanical work channel is strictly closed:
\begin{equation}
    P_{\text{mech}} = \mathbf{f}_M \cdot \mathbf{v} \equiv 0.
\end{equation}

\paragraph{The Verdict: Mechanically Inconsistent (FECC Divergence).}
Comparing the two sides, we find a structural divergence from the force-energy consistency criterion (FECC):
\begin{equation}
    \underbrace{ \mathcal{D}_M }_{\text{Energy In}} \left( \neq 0 \right) \quad \neq \quad \underbrace{ \mathbf{f}_M \cdot \mathbf{v} }_{\text{Work Out}} (0).
\end{equation}
The Minkowski tensor predicts that energy disappears from the field but provides no mechanically defined work channel in which that energy can arrive. To avoid the inconsistency, one must postulate an internal moving mass target. But since the framework does not define such a target or its velocity, this postulate is external to the theory.

The logical chain of argument appears to be complete:
\begin{enumerate}
    \item The Minkowski Force definition implies $\mathbf{P}$ is a state of a stationary block $\implies \mathbf{v}=0$.
    \item The Minkowski Energy definition implies energy leaves the field $\implies$ it needs a sink.
    \item A mechanical sink requires kinetic motion to accept work ($\mathbf{v} \neq 0$).
    \item \textbf{Contradiction:} The target cannot be both stationary ($v=0$) and moving ($v \neq 0$) at the same time.
\end{enumerate}

\subsubsection{Comparative Forensic Check: The Lorentz Solution}

In contrast, the Vacuum Framework handles this scenario trivially.

\paragraph{1. The Lorentz Reality (The Locked Engine).}
\begin{itemize}
    \item \textbf{Mechanism:} Since the material is saturated ($\dot{\mathbf{P}}=0$), the polarization current is physically zero ($\mathbf{J}_P = \dot{\mathbf{P}} = 0$).
    \item \textbf{Work:} The power delivered is practically zero: $P = \mathbf{J}_P \cdot \mathbf{E} = 0$.
    \item \textbf{Consistency:} No current flows, no work is done. The "springs" are fully stretched and locked against the stops. All limits are respected.
\end{itemize}

\paragraph{Conclusion.}
The Minkowski tensor fails because it relies on an algebraic definition of energy ($U_M$) that does not correspond to the physical state of the matter. It is \textbf{internally inconsistent}: it cannot define a mass target and velocity that simultaneously satisfy the force equation and the energy budget. The force implies a stationary target, while the energy residual demands a moving target. This contradiction presents a structural inconsistency of the hypothesis.

\subsection{Case Study 3: Dissipation (The Missing Heat Sink)}
\label{sec:dissipation_critique}

Finally, the definitive test is considered: \textbf{Dissipation} (Hysteresis). This is the "observable smoking gun" of the energy analysis, as it represents an irreversible transfer of energy (Heat) that cannot be swept under the rug of reversible storage.

\paragraph{The Physical Setup.}
\begin{enumerate}
    \item \textbf{Kinematics:} The material bulk is rigid and stationary ($\mathbf{v}_{\text{bulk}} = 0$).
    \item \textbf{Response:} The material exhibits hysteresis. The relationship between $\mathbf{P}$ and $\mathbf{E}$ forms an open loop area $\oint \mathbf{E} \cdot d\mathbf{P} > 0$.
    \item \textbf{Dynamics:} We drive the system through a full cycle of the external field.
\end{enumerate}

\paragraph{Step 1: The Energy Audit (The Heat Generation).}
The Minkowski power transfer is integrated over one full cycle:
\begin{equation}
    Q_M = \oint \mathcal{D}_M \, dt = \oint \left( \dot{U}_M \right) dt= \oint \dot{\left( \frac{1}{2} \mathbf{E} \cdot \mathbf{D}  \right)} dt.
\end{equation}
Because the $\mathbf{D}$-$\mathbf{E}$ relationship is hysteretic, the integral is non-zero ($Q_M > 0$).
\textbf{Result:} The Minkowski formulation correctly predicts that energy is extracted from the field per cycle. It successfully anticipates the thermodynamic reality that energy leaves the electromagnetic domain.

\paragraph{Step 2: The Mechanical Audit (The Missing Mechanism).}
Where did this energy go? In a physical system, "Heat" is the random kinetic energy of microscopic particles. To heat a material, one must perform work on its internal constituents (e.g., vibrating the lattice).
\begin{itemize}
    \item \textbf{Target:} The "Dielectric Matter."
    \item \textbf{Velocity:} The bulk is stationary by construction ($\mathbf{v}_{\text{bulk}}=0$). Minkowski's macroscopic field variables do not define a kinematic internal velocity field for the matter state $\mathbf{P}$. Any "internal motion" would be an ad-hoc postulate external to the theory.
    \item \textbf{Mechanism:} Therefore, within the Minkowski macroscopic reality itself, there is no mechanically defined channel ($\mathbf{f}\cdot\mathbf{v}$ on a specified mass target) that can deliver the heat $Q_M$.
\end{itemize}
Accordingly, the mechanical work calculation within the theory remains null:
\begin{equation}
    W_{\text{mech}} = \oint (\mathbf{f}_M \cdot \mathbf{v}) \, dt \equiv 0.
\end{equation}

\paragraph{The Verdict: The "Force Without Work" Divergence.}
The audit identifies the mechanism of the divergence:
\begin{itemize}
    \item The Minkowski force vector $\mathbf{f}_M$ exists and is mathematically non-zero. It "touches" the matter.
    \item However, the theory provides no mechanically defined velocity field for a massive target that could receive work. Without a velocity $\mathbf{v}$, the projection of force onto motion is null ($\mathbf{f}_M \cdot \mathbf{v} = 0$).
\end{itemize}
Thus, the force acts, but performs no work. It acts as a "Force of Constraint" rather than a "Force of Actuation." This leads to the structural violation of the \textbf{Axiom of Actuation}:
\begin{equation}
    \underbrace{ Q_M }_{\text{Heat In}} (>0) \quad \neq \quad \underbrace{ W_{\text{mech}} }_{\text{Work Out}} (0).
\end{equation}
The Minkowski tensor predicts that the material heats up, yet provides no mechanical pathway to deliver that heat. It implies the deposit of energy into the matter without performing the necessary work to transfer it.
Heat is kinetic energy. To create heat, one must actuate a mass. The Minkowski tensor implies force on a stationary target. In the context of mechanics, this is an inconsistency.

\subsubsection{Comparative Forensic Check: The Lorentz Explanation}

\paragraph{1. The Lorentz Reality (The Friction Engine).}
The Vacuum Framework identifies the mechanism immediately.
\begin{enumerate}
    \item \textbf{Mechanism:} The polarization current $\mathbf{J}_P$ represents real charge motion against a "friction" force (the damping in the Host Interface).
    \item \textbf{Work:} The field performs real positive work on the current: $P = \mathbf{J}_P \cdot \mathbf{E} > 0$.
    \item \textbf{Destination:} This work is dissipated as heat in the Host Interface (Conceptually: $I^2 R$ losses).
\end{enumerate}
The energy leaves the field, is transferred via mechanical work ($\mathbf{J} \cdot \mathbf{E}$), and becomes heat. The audit trail is unbroken.

\paragraph{2. The Minkowski Reality (Action Without a Carrier).}
The Minkowski framework is mutually exclusive. It claims the energy leaves the field ($Q_M > 0$) but does not define the moving mechanical carrier required to receive it within its own macroscopic variables. It is therefore a theory of \textbf{Action without a defined kinematic sink}.

\paragraph{Conclusion.}
The Minkowski tensor fails the consistency check. It describes an "Abstract Framework" that maintains consistency primarily in the static limit. As soon as the material is required to act (Saturate) or feel (Dissipate), the tensor's internal inconsistency is exposed.

\subsection{The Acknowledged Thermodynamic Limits}
\label{sec:acknowledged_limits}

It is crucial to acknowledge that the failure of the conventional time-domain formulation to account for dissipation is an explicitly recognized limitation in authoritative literature.
Standard texts (e.g., Landau \& Lifshitz \cite{LandauLifshitzVol8}, Jackson \cite{Jackson1999}) admit that for dispersive or lossy media, the quantity $u_{\text{conv}}$ "cannot be rationally defined as a thermodynamic quantity." They recognize that the work done becomes path-dependent. 
Rather than resolving the mechanical deficit (by restoring the work term $\mathbf{f}\cdot\mathbf{v}$), the standard recourse is to abandon the time-domain description in favor of frequency-domain averaging. This tacitly accepts the structural failure of the Minkowski tensor to provide a \textbf{mechanically consistent description of instantaneous energy transfer}.

\section{Verdict: The Forced Conclusion}

The analysis is complete. By rigorously attempting to validate the "Minkowski Reality" against the Axiom of Actuation, we have arrived at an impasse.

\paragraph{1. The Failure of the Constructive Search.}
The tensor was treated as a "Black Box" hypothesis and the population of strictly mechanical variables that would satisfy the energy conservation laws was attempted.
\begin{enumerate}
    \item \textbf{The Linear Constraint:} In the simplest case, the tensor passed the audit solely because both sides of the ledger were zero. It works only when "nothing happens" (Sterile Reality).
    \item \textbf{The Fatal Divergence:} As soon as real physical processes occur---Saturation (Force Limit) or Dissipation (Heat)---the energy budget diverges from the mechanical capacity. The tensor predicts energy transfer ($\mathcal{D} \neq 0$) while simultaneously predicting zero mechanical work ($P_{\text{mech}} = 0$).
\end{enumerate}

\paragraph{The Implication: Non-Local Energy Transfer.}
Because it is difficult to identify a velocity field that satisfies the energy budget, the Minkowski framework implicitly relies on a non-mechanical mechanism: \textbf{Non-Local Energy Transfer}. Energy vanishes from the field flux at point $\mathbf{x}$ and appears as heat at point $\mathbf{x}$ without traversing the necessary kinetic bridge of mechanical work. In a mechanical universe, this implies a bypass of the local work condition.

\paragraph{2. The Persistence of the Divergence (The Averaging Limit).}
This result answers the fundamental question of the audit: "Can averaging resolve the topology?"
In Part I, the Minkowski theory was found to be topologically distinct at the microscopic scale (Vacuum Inertia). In Part VI, it is found to be thermodynamically distinct at the macroscopic scale (Non-Local Transfer).
This observation confirms that the averaging process does not "erase" the topological features of the underlying definition. 
Therefore, the defense that the Minkowski tensor is a "valid macroscopic approximation" is challenged. A valid approximation should preserve the conservation laws of the underlying reality. The Minkowski formulation requires a \textbf{Mutually Exclusive Reality} where energy can move without a mechanical carrier.

The conclusion is that the Minkowski tensor is \textbf{mechanically inconsistent}. It is limited in describing the physical connection of how electromagnetic fields exchange energy with matter.

\chapter{The Focus on Symmetry: The Abraham Audit}
\label{chap:Abraham_Audit}

In the previous chapter, the Minkowski energy-momentum tensor was identified as a \textbf{kinematically decoupled state} in the context of the macroscopic limit. It was demonstrated that while it accounts for the energy lost by the field during dissipation, it does not explicitly define a mechanical mechanism (Work) to transfer that energy to the matter in the absence of bulk motion. It implies "heat without work," verifying the diagnosis of thermodynamic inconsistency.

The analysis now turns to its primary historical alternative: the \textbf{Abraham Energy-Momentum Tensor}.

For over a century, the discussion between the Abraham and Minkowski formulations has been a central theme in macroscopic electrodynamics. However, the application of the \textbf{Force-Energy Consistency Criterion (FECC)} suggests that this distinction does not resolve the underlying thermodynamic divergence.
The analysis indicates that the Abraham tensor involves a modification of the Minkowski tensor that addresses the \textit{symmetry} of the momentum density but retains the \textit{thermodynamic} structure. Consequently, it is subject to the same consistency challenges identified in Chapter \ref{chap:Macro_Critique}.

\section{The Motivation: The Quest for Symmetry}

The historical dissatisfaction with the Minkowski tensor was not driven by the "thermodynamic inconsistency" we identified (which has largely gone unnoticed), but by an "angular momentum error."
The Minkowski tensor $\mathbf{T}_M$ is not symmetric. Specifically, the simple relationship between momentum density and energy flux is broken:
\begin{equation}
    \mathbf{g}_M \neq \frac{1}{c^2} \mathbf{S}_M
\end{equation}
(Recall: $\mathbf{g}_M = \mathbf{D} \times \mathbf{B}$ while $\mathbf{S}_M = \mathbf{E} \times \mathbf{H}$).

To a relativistic physicist, this asymmetry is disturbing because it complicates the conservation of angular momentum ($\mathbf{L} = \mathbf{r} \times \mathbf{p}$). Max Abraham (1909) proposed a modification to restore this symmetry, defining a new momentum density:
\begin{equation}
    \mathbf{g}_A = \frac{1}{c^2} \mathbf{S}_A = \frac{1}{c^2} (\mathbf{E} \times \mathbf{H})
\end{equation}
For a century, the argument has been framed as: "Which momentum density is correct?"
This question prioritizes the algebraic properties of the field tensor over the mechanical consistency of the coupling.

\section{The Shared Indictment: The Energy Domain}

To apply the FECC, the \textbf{energy domain} is inspected first.
The question is asked: \textit{Does the Abraham formulation change the definition of Energy Transfer?}

The Abraham formulation defines the scalar Energy Density and Vector Flux as:
\begin{itemize}
    \item \textbf{Energy Density:} $u_A = \frac{1}{2} (\mathbf{E} \cdot \mathbf{D} + \mathbf{B} \cdot \mathbf{H})$
    \item \textbf{Energy Flux:} $\mathbf{S}_A = \mathbf{E} \times \mathbf{H}$
\end{itemize}
These definitions are \textbf{identical} to the Minkowski definitions.
\begin{equation}
    u_A \equiv u_M, \quad \quad \mathbf{S}_A \equiv \mathbf{S}_M
\end{equation}

Consequently, the \textbf{energy balance sheet} of the Abraham tensor is structurally identical to that of the Minkowski tensor.
\begin{equation}
    \nabla \cdot \mathbf{S}_A + \frac{\partial u_A}{\partial t} \equiv \nabla \cdot \mathbf{S}_M + \frac{\partial u_M}{\partial t}
\end{equation}

This identity has verified consequences for the Abraham viability. It means that the \textbf{thermodynamic inconsistency} identified in Chapter \ref{chap:Macro_Critique} applies to the Abraham tensor \textbf{verbatim}.

Recall the "Dissipation Case" (Section \ref{sec:dissipation_critique}): A stationary material with hysteresis absorbs energy from the field.
\begin{itemize}
    \item Minkowski predicted a non-zero energy consumption: $Q > 0$.
    \item Therefore, Abraham predicts the \textbf{exact same} energy consumption: $Q > 0$.
\end{itemize}

Unlike the Minkowski tensor, the Abraham formulation introduces a symmetric stress tensor. However, the thermodynamic implication remains: energy leaves the electromagnetic field and enters the material sector. To be mechanically consistent, this energy must be accounted for as Mechanical Work.

\section{The Momentum Domain Proposal}

The proponent of the Abraham tensor will argue: \textit{"But the Force is different! Abraham added a term to the force definition. Maybe this extra force does the work?"}
We now audit this defense.

To achieve this symmetry, the Abraham formulation requires a reconstruction of the entire stress tensor. The theory cannot simply "patch" the momentum; it must re-engineer the stress to maintain conservation.
The Abraham formulation defines the symmetric stress tensor $\mathbf{T}_A$ by symmetrizing the field products:
\begin{equation}
    \mathbf{T}_A = \frac{1}{2} \left[ (\mathbf{E} \otimes \mathbf{D} + \mathbf{D} \otimes \mathbf{E}) + (\mathbf{H} \otimes \mathbf{B} + \mathbf{B} \otimes \mathbf{H}) \right] - u_A \mathbf{I}
\end{equation}
The \textbf{Abraham Force Density} $\mathbf{f}_A$ is obtained from the divergence of this new tensor, minus the time evolution of the new momentum ($\mathbf{f} = -\nabla \cdot \mathbf{T} - \dot{\mathbf{g}}$):
\begin{equation}
    \mathbf{f}_A =  -\nabla \cdot \mathbf{T}_A - \dot{\mathbf{g}}_A
\end{equation}

\subsection{The Work Audit: The Proof of Inconsistency}
We now perform the critical check: \textbf{Does this complex force structure perform the missing work in dissipative media?}

To answer this, the velocity ($\mathbf{v}$) of the mass target must be determined. We rely on the baseline established in Chapter \ref{chap:Macro_Critique}, applying a strict \textbf{Black Box Audit} of the Abraham framework variables.

\paragraph{Step 1: The Linear Baseline (The Null State).}
In the analysis of the linear dielectric (Section \ref{sec:Macro_Momentum_Divergence}), an energy audit of the Minkowski/Abraham framework was performed. It was found that for a linear material, the energy transfer is identically zero ($Q=0$).
\begin{itemize}
    \item Since no energy enters the mechanical system, there is no source for kinetic energy.
    \item Consequently, the framework provides no mechanically required actuation channel. Any attempt to postulate internal motion in this null case must either be everywhere power-orthogonal to the force or involve exact internal cancellations; but the Abraham/Minkowski variables provide no kinematic law that defines such a velocity field or enforces those constraints. The ``internal motion'' rescue is therefore mechanically \textit{underdetermined}.
\end{itemize}
This is the same internal-motion issue identified in the Minkowski audit of the linear dielectric (Chapter \ref{chap:Macro_Critique}, Case Study 1): it requires extra microscopic kinematics not defined by the macroscopic tensor.
In the linear regime, the force targets a bound charge distribution that is mechanically locked to the stationary lattice. The framework treats $\mathbf{P}$ as a static property of the stationary medium, not as a moving fluid.

\paragraph{Step 2: The Dissipative Extension.}
Now hysteresis (dissipation) or saturation is introduced. The Energy Audit reveals a change: the framework now predicts a net energy loss from the field ($Q > 0$).
However, the question must be asked: \textbf{Does the mechanical nature of the force target change?}

The Abraham force density $\mathbf{f}_A$ acts on the macroscopic field variable $\mathbf{P}$. 
Crucially, within the Abraham/Minkowski definition, $\mathbf{P}$ is a \textbf{state variable}, not a kinematic object. The theory defines its magnitude and direction, but it does \textbf{not} define a velocity field $\mathbf{v}_P$ associated with its "motion." The binding between mass and the state variable $\mathbf{P}$ remains identical to the linear case. The introduction of a thermodynamic phase lag (hysteresis) does not sever the mechanical bond between the charge and the mass. It does not suddenly "unlock" a velocity vector for the force to ride on.
Unlike the Lorentz framework, which explicitly models $\mathbf{P}$ as a current of moving charges ($\mathbf{J}_P = \partial_t \mathbf{P}$), the Abraham framework provides no kinematic definition for the variable it exerts force upon.

Therefore, the mechanical state of the target remains unchanged:
\begin{equation}
    \mathbf{v}_{target} \equiv \mathbf{v}_{bulk} = 0
\end{equation}

We calculate the work done by the Abraham Force in this dissipative scenario:
\begin{equation}
    P_{mech} = \mathbf{f}_A \cdot \mathbf{v}_{target} = \mathbf{f}_A \cdot \mathbf{0} \equiv 0
\end{equation}

\paragraph{The Verdict: The Same Contradiction.}
This contradiction reveals that the structural divergence persists:
\begin{itemize}
    \item \textbf{Thermodynamics:} The theory predicts energy enters the matter ($Q > 0$).
    \item \textbf{Mechanics:} The theory predicts zero work is done ($W = 0$).
\end{itemize}

The Abraham formulation asserts that energy is transferred (Heat) without a mechanically defined kinematic sink (Work). Adding the "Abraham Force" term fails to resolve the issue because it adds a force vector without supplying the required velocity field of a massive carrier on which that force could do work within the macroscopic theory itself.

The "Abraham Correction" is purely algebraic. It rearranges the momentum variables to satisfy a relativistic symmetry preference, but it fails to address the fundamental disconnect between the Field Energy Sink ($Q>0$) and the Mechanical Work Sink ($W=0$).

\section{Conclusion: The Symmetry Artifact}

The debate between Abraham and Minkowski represents a significant historical divergence in the philosophy of electrodynamics. For a century, the discourse has focused on which "Force" is correct, while the mechanical audit suggests an alternative priority: \textbf{calculating the work done reveals a null result for both formulations.}

The Abraham tensor represents an \textbf{Incomplete Resolution}.
\begin{enumerate}
    \item \textbf{It inherits the Thermodynamic Inconsistency:} Because it uses the Minkowski energy flux ($\mathbf{E} \times \mathbf{H}$), it predicts energy loss without a mechanical carrier.
    \item \textbf{It offers a Null Force:} Because the "Abraham Term" is a time-derivative of a field vector, it exerts force, but that force calculates to zero work ($\mathbf{f} \cdot \mathbf{v} = 0$) in stationary media.
\end{enumerate}

This analysis suggests that the "Symmetry Debate" is secondary to the mechanical stability of the theory. Whether the tensor is symmetric or not is irrelevant if the tensor is mechanically inconsistent. A ledger that balances its "assets" (symmetry) but fails to balance its "energy budget" (work) is architecturally incomplete.

\paragraph{The Structural Contrast.}
It is worth noting the contrast in this historical development. Abraham sought to "restore" symmetry to the electromagnetic description. Yet, as established in the Vacuum Framework, the fundamental Vacuum Tensor $\mathbf{T}_{vac}$ is \textbf{already inherently symmetric}:
\begin{equation}
    \mathbf{g}_{vac} = \varepsilon_0 (\mathbf{E} \times \mathbf{B}) = \frac{1}{c^2} \left( \mathbf{E} \times \frac{\mathbf{B}}{\mu_0} \right) = \frac{1}{c^2} \mathbf{S}_{vac}
\end{equation}
The asymmetry that Abraham sought to correct was not a property of nature; it was a property of the \textit{Minkowski conflation}. By incorporating material momentum (bound current) into the field definition, Minkowski broke the symmetry. Abraham attempted to address the symptom (asymmetry) without resolving the underlying divergence (conflation). Had the analysis adhered to the Microscopic Baseline (Lorentz), the symmetry would have been self-evident without the need for additional force terms.

The \textbf{Vacuum Tensor} resolves both. The conclusion is that the Abraham tensor is \textbf{mechanically inconsistent} just as the Minkowski tensor. It is limited in describing the physical connection of how electromagnetic fields exchange energy with matter.

\chapter{The Focus on Deformation: The Point-Dipole Audit}
\label{chap:PointDipole_Audit}

The forensic audit of the macroscopic continuum has thus far yielded two negative results regarding consistency:
\begin{enumerate}
    \item \textbf{Minkowski:} The attempt to define a self-contained macroscopic reality, limited by the "Force without Work" contradiction in the dissipative regime.
    \item \textbf{Abraham:} The attempt to symmetrize the tensor, limited by the same thermodynamic inheritance.
\end{enumerate}

Even though we have already considered the formulations in \ref{chap:ForceDensity}, we will now concisely re-examine the \textbf{Point-Dipole Formulations} through the specific lens of the Vacuum Framework's Force-Energy Consistency Criterion (FECC).

Represented by \textbf{Kelvin} in electrostatics and \textbf{Einstein and Laub} in electrodynamics, this approach does not strive for the abstract elegance of relativistic symmetry. Instead, it offers a \textbf{pragmatic slice} of the interaction. It builds the force density from the bottom up, summing the forces on discrete dipoles. This provides a description that is thermodynamically incomplete but mechanically precise regarding a specific domain: the deformation of the bulk lattice and the coupling towards the continuum mechanics domain.

\section{The Pragmatic Approach: Defining the Scope}

The Einstein-Laub force density was derived not from a `top-down` imposition of symmetry, but from a `bottom-up` integration of the forces acting on dipoles.
Their result for the ponderomotive force density is:
\begin{equation}
    \mathbf{f}_{EL} = (\mathbf{P} \cdot \nabla)\mathbf{E} + (\mathbf{M} \cdot \nabla)\mathbf{H} + \frac{\partial \mathbf{P}}{\partial t} \times \mathbf{B} - \varepsilon_0 \frac{\partial \mathbf{M}}{\partial t} \times \mathbf{E}
\end{equation}
For a century, this expression has been presented as a competitor to the Lorentz force. However, viewed through the criterion of the \textbf{Force-Energy Consistency Criterion}, it is not a competitor but a \textbf{subset}.

The key to understanding the Einstein-Laub formulation lies in its derivation. It is effectively the relativistic generalization of the \textbf{Kelvin polarization force} known from electrostatics ($\mathbf{f}_{Kelvin} = (\mathbf{P} \cdot \nabla)\mathbf{E}$). Both formulations share a common basis: they assume a \textbf{point-dipole model} for the constituents of matter.

As analyzed in Part III (Chapter \ref{chap:ForceDensity}), the Point-Dipole model acts as a specific type of low-pass filter. By mathematically collapsing the distributed charge structure of an atom into a singularity, the model effectively "deletes" the internal geometry.
\begin{itemize}
    \item \textbf{Input:} It correctly captures the Net Force on the center of mass (Deformation).
    \item \textbf{Loss:} It discards the Internal Stress required to hold the dipole together (Binding).
\end{itemize}

\subsection{The Thermodynamic Consequence}
This filtering has a profound thermodynamic consequence. By removing the internal moving parts of the dipole ($ \mathbf{v}_{int} $), the model removes the gateways into the \textbf{Storage Domain} (binding energy) as well as the gateways into the \textbf{Thermodynamic Domain} (heat generation).
\begin{equation}
    \mathbf{v}_{int} \to 0 \quad \implies \quad P_{dissipation} \to 0 \quad \text{and} \quad P_{storage} \to 0
\end{equation}
Therefore, the Einstein-Laub formulation is \textbf{structurally decoupled} from the storage and thermodynamic domains. It cannot describe dielectric loss or magnetic hysteresis because it has filtered out the machinery required to generate heat.

This is not a failure of the theory; it is a strict definition of its \textbf{scope}. The Einstein-Laub force is the force density acting on the \textit{centers} of the dipoles—on the lattice itself. It is the \textbf{Force of Deformation}, separate from the Force of Binding.

\section{The Geometrical Adaptation}

However, while the \textit{intent} of the Einstein-Laub formulation (to describe deformation) is accepted, we proposed in Part III that an adaptation is required to correctly represent the lattice structure of matter. 
As demonstrated in the "Test Dipole Insertion" thought experiment (Chapter \ref{chap:ForceDensity}), the \textbf{Sampling Bias} creates a discrepancy between the macroscopic average field $\mathbf{E}$ and the effective local field $\mathbf{E}_{eff}$ actually experienced by the discrete lattice sites. Since the test dipole never visits the singular source regions of its neighbors, it effectively subtracts their contribution from the average:
\begin{equation}
    \mathbf{E}_{eff} = \mathbf{E}_{macro} - \langle \mathbf{E}_{inner} \rangle_{excluded}
\end{equation}
We thus proposed the \textbf{Lattice Adaptation}, presented in detail in \ref{chap:ForceDensity}.

\section{Assessment: The Valid Subset}

The audit of the Einstein-Laub (Point-Dipole) formulation concludes with a qualified endorsement.

\begin{enumerate}
    \item \textbf{Is it Fundamental?} \textbf{No.} It is a filtered approximation. It violates the FECC for general thermodynamic processes (Heat/Storage) because it lacks the internal degrees of freedom.
    \item \textbf{Is it Useful?} \textbf{Yes.} It successfully isolates the \textbf{continuum mechanical sector}. It filters out the "internal noise" (Lorentz forces balanced by binding constraints) to provide a clear signal for the bulk deformation stress.
\end{enumerate}

With the geometrical adaptation applied, we propose that the Einstein-Laub force density stands as the most accurate available description for the \textbf{deformation slice} of the macroscopic reality. It is the "engineering baseline," distinct from but derived from the "universal baseline" of the Lorentz force.

\chapter{The Variational Circularity: Force, Energy, and Mass}
\label{chap:VariationalMethod}

\section{Introduction: The Geometry of Interaction}

In the preceding parts of this work, the competing formulations of electrodynamics—Minkowski, Abraham, Lorentz—have been rigorously examined and tested against the laws of mechanics. The forensic derivation has demonstrated, through specific examples, that the Vacuum (Lorentz) Framework is the only one that maintains the \textbf{Force-Energy Consistency Criterion (FECC)}.

This chapter serves as a foundational synthesis of that conclusion. Rather than introducing new proofs, the analysis examines the tool that has been at the center of the historical controversy: the \textbf{Variational Method}.

In the theoretical literature, the Principle of Virtual Work ($\mathbf{F} = -\nabla U$) is often presented as a primary source—a mathematical procedure that generates force laws from energy scalars, seemingly bypassing the need for detailed mechanical models. In Chapter \ref{chap:ForceDensity} (The Theorem of Indeterminacy), it was shown that the \textbf{Korteweg-Helmholtz force density} is physically incomplete for describing local stress, as it is limited by its topological input. However, having established \textit{that} it yields physically distinct results from the Lorentz baseline, the task is now to explain \textit{how} it appears to function effectively in limited pragmatical cases.

In this chapter, a detailed examination of the \textbf{variational mechanism} itself is conducted. It is demonstrated that this force is not a fundamental discovery, but a mathematical \textbf{consequence} of the specific energy functional postulated by Minkowski.

The objective is to demonstrate that the Variational Method is not a separate law of physics, but a direct application of the FECC ($P = \mathbf{f} \cdot \mathbf{v}$). The algebraic link between force and energy exists \textit{only} because of the kinematic link between position and motion.

The central thesis of this chapter is simple: \textbf{The Variational Method is a spatial projection of the power balance.} It works perfectly, but only if the variational coordinate is rigorously identified with the position of a mass. By understanding this geometry, it becomes possible to explain not just \textit{that} the historical formulations are limited, but exactly \textit{why} the mathematical procedure produced them.

\section{The Physical Meaning of the Gradient}
\label{sec:Variational_Meaning}

Mathematically, the gradient operator $\nabla$ is purely geometric. It interrogates how a scalar field changes as one moves in an abstract coordinate space. Physically, however, in the context of mechanics, the gradient acquires a specific, constrained meaning defined by the \textbf{Force-Energy Consistency Criterion (FECC)}.

\subsection{The Fundamental Identity}

Mechanics is built on the definition of Work. Energy ($U$) is defined as the capacity to do work ($W$), and work is defined as the action of a force ($\mathbf{F}$) moving a mass over a distance ($d\mathbf{x}$):
\begin{equation}
    dW = -dU = \mathbf{F} \cdot d\mathbf{x}_{\text{mass}}
\end{equation}
This identity is the "fundamental link" that explicitly connects the three concepts. It implies that Force is the spatial derivative of Energy:
\begin{equation}
    \mathbf{F} = -\frac{dU}{d\mathbf{x}_{\text{mass}}}
\end{equation}

\subsection{The Target of the Variable}

Crucially, the variable $d\mathbf{x}$ in the denominator is not an arbitrary coordinate. It is the \textbf{position vector of the mass}.

When the Variational Method is applied, the implied physical question is:
\begin{center}
    \textit{"How much does the system's energy change if one displaces this specific element?"}
\end{center}

The resulting answer (the "force") acts specifically on the element displaced.
\begin{itemize}
    \item If one varies the position of an electron ($d\mathbf{x}_e$), the gradient $\nabla_{x_e} U$ yields the force on the electron.
    \item If one varies the position of a bulk boundary ($d\mathbf{x}_{bound}$), the gradient yields the force on the boundary.
\end{itemize}
We thus propose that the force density of the Variational Method is completely dependent on the kinematic identity of the variational coordinate and the predefined energy functional.

\section{The Assumption of Energy Primacy}
\label{sec:EnergyMyth}

Before diagnosing the procedural specifics, it is necessary to contextualize the philosophical foundation of this approach. In the literature, the Variational Method is often presented as a primary or "more fundamental" path to truth than the direct calculation of forces from fields. Stratton, for example, argues that the energy principle is the primary definition from which force must be derived \cite{Stratton1941}. This perspective reflects a \textbf{philosophical priority}: the belief that Energy is the primary scalar invariant of nature.

However, this priority introduces a \textbf{definition dependency}. The energy method does not discover forces independent of the definitions provided to it; it reflects the physics encoded in the energy functional.

\subsection{The Circularity of Variation}
Mechanically, Potential Energy $U$ is defined as the work done against a force $\mathbf{F}$.
\begin{equation}
    U(\mathbf{x}) \equiv - \int_{\infty}^{\mathbf{x}} \mathbf{F}(\mathbf{x}') \cdot d\mathbf{x}'
\end{equation}
The Variational Method then "derives" the force by differentiating this energy:
\begin{equation}
    \mathbf{F} = - \nabla U = - \nabla \left( - \int \mathbf{F} \cdot d\mathbf{x} \right) = \mathbf{F}
\end{equation}
This operation is theoretically self-consistent, but it offers no new physical insight beyond what was assumed in the integral. The validity of the output $\mathbf{F}$ is strictly dependent on the validity of the input integral $U$. If the energy functional $U$ was constructed by filtering out certain degrees of freedom (e.g., the independent motion of bound charges), the differentiation will faithfully report that those forces do not exist in the model.

\subsection{Definition by Coordinate Selection}
The utility of the method lies not in the differentiation, but in the \textbf{choice of the variational coordinate}.

Standard texts often obscure this by the variation of a generic spatial coordinate $\mathbf{x}$. But as established by the FECC (kinematic lock), the gradient operator $\nabla_{\mathbf{x}}$ is physically an instruction to "test the work done when mass located at $\mathbf{x}$ moves."
\begin{itemize}
    \item If $U$ is defined in terms of the position of the macroscopic boundary, $\nabla_{\text{boundary}} U$ yields the force on the boundary.
    \item If $U$ is defined in terms of the position of atomic dipoles, $\nabla_{\text{atom}} U$ yields the force on the atoms.
\end{itemize}

Therefore, the result of the energy method is \textbf{completely dependent on the specific mass defined to be varied}. It is not a privileged orbital view; it is a mechanism that enforces the initial hypothesis about where the mass is located. The inability of the Korteweg-Helmholtz method to find bulk forces is simply a consequence of the fact that its energy functional $U_{\text{total}}$ is defined globally, effectively associating all energy changes with the movement of the boundaries.

The Variational Method is not a higher valid definition of force, but a projection of the power balance. \textbf{Force and energy are two different views on the same physical quantity, fundamentally connected by the FECC.}

\section{The Mechanics of Variation: A Derivation from Power}
\label{sec:Mechanics_of_Variation}

The connection between the Variational Method (virtual work) and the fundamental laws of mechanics is clarified here by deriving the variational principle directly from the rigorous power balance established in this book.

It is shown that the equation $\mathbf{F} = -\nabla U$ is not an axiom; it is the time-integral of the \textbf{Force-Energy Consistency Criterion (FECC)}.

\subsection{The Bridge between Worlds: Mass and Charge}

The unification of Electrodynamics and Mechanics relies on a single, concrete entity: the \textbf{charged particle} (or the continuum fluid element). This entity possesses two distinct attributes that serve as the "coupling hooks" for the two theories:
\begin{itemize}
    \item \textbf{Charge ($q$):} The handle by which the Electromagnetic Field grips the particle.
    \item \textbf{Mass ($m$):} The handle by which Inertial Mechanics grips the particle.
\end{itemize}
Because these two attributes reside on the same physical object, they share the same kinematic state: a single position $\mathbf{x}(t)$ and a single velocity $\mathbf{v}(t)$. This kinematic lock is the bridge that allows energy to flow between the two worlds.

\subsection{The Chain Rule Derivation}

We begin with the rigorous Poynting theorem derived in the vacuum framework. For a closed system (or a local control volume in the quasistatic limit where radiative flux $\nabla \cdot \mathbf{S} \to 0$), the rate of change of field energy must equal the mechanical power delivered to the matter:

\begin{equation}
    P_{\text{mech}} = -\frac{\partial U_{\text{EM}}}{\partial t}
    \label{eq:Power_Balance_Base}
\end{equation}

Applying the FECC, mechanical power is strictly defined as the dot product of the Force acting on the mass and the velocity of that mass:
\begin{equation}
    P_{\text{mech}} \equiv \mathbf{F} \cdot \mathbf{v}
\end{equation}

Substituting this into the balance equation:
\begin{equation}
    \mathbf{F} \cdot \mathbf{v} = -\frac{\partial U_{\text{EM}}}{\partial t}
\end{equation}

The definitive step is the application of the \textbf{chain rule}. If the energy $U_{\text{EM}}$ changes, it is because the configuration of the system ($\mathbf{x}$) is changing in time. The time derivative is decomposed:
\begin{equation}
    \frac{\partial U_{\text{EM}}}{\partial t} = \frac{\partial U_{\text{EM}}}{\partial \mathbf{x}} \cdot \underbrace{\frac{d\mathbf{x}}{dt}}_{\mathbf{v}}
\end{equation}

Substituting this back into the power balance:
\begin{equation}
    \mathbf{F} \cdot \mathbf{v} = -\left( \nabla U_{\text{EM}} \right) \cdot \mathbf{v}
\end{equation}

For this identity to hold for any arbitrary motion $\mathbf{v}$, the vector coefficients must be equal:
\begin{equation}
    \boxed{ \mathbf{F} = -\nabla U_{\text{EM}} }
\end{equation}

\subsection{The Crucial Insight: The Identity of \texorpdfstring{$\mathbf{x}$}{x}}
This derivation reveals the hidden constraint of the variational method.
The gradient $\nabla U_{\text{EM}}$ yields the force $\mathbf{F}$ \textit{if and only if} the coordinate $\mathbf{x}$ in the gradient $\frac{\partial}{\partial \mathbf{x}}$ is the same coordinate that defines the velocity $\mathbf{v} = \frac{d\mathbf{x}}{dt}$ in the power equation.

Since $\mathbf{v}$ is the \textbf{velocity of mass}, $\mathbf{x}$ must be the \textbf{position of mass}.

\begin{itemize}
    \item \textbf{Valid variation:} If the position of a charge carrier ($\mathbf{x}_{\text{charge}}$) is varied, motion ($\mathbf{v}$) is simulated. The resulting gradient is the physical Lorentz force acting on that carrier.
    \item \textbf{Invalid variation:} If a parameter that effectively moves the "field coordinates" or a constitutive property is varied without moving the mass, the chain rule is technically misapplied relative to the mechanical definition. A mathematical gradient is calculated that has force units, but does not correspond to a rate of mechanical work on a massive body. The result is a "virtual force"—a quantity that clearly has units of force, but does not correspond to any rate of mechanical work on a massive body.
\end{itemize}

This is why the Force-Energy connection is fundamental. The Variational Method is nothing more than the spatial projection of the power balance. It works exclusively because the "particle" binds the electrical world (Energy source) and the mechanical world (Power sink) to a single geometric point.

\section{The Divergence of Energy Functionals: A Comparative Case Study}
\label{sec:Energy_Functionals}

In the preceding parts of this book, detailed demonstrations established the physical inconsistencies of the Minkowski and Abraham tensors. These critiques are not repeated here. Instead, these historical formulations are used as \textbf{comparative inputs} for the meta-analysis of the Variational Method.

To understand exactly how the variational "machine" works—and to see the force-energy consistency criterion in action—it is fed different energy definitions and the mechanical outputs are observed. This experiment demonstrates that the Variational Method is a neutral algorithm: it converts an energy hypothesis into a force conclusion.

Two distinct formulations of Poynting's Theorem are examined, often conflated in the literature. While they appear identical in static linearity, they diverge fundamentally in the context of the variational variation (motion).

\subsection{1. The Conventional Formulation}
Presented in authoritative textbooks such as Jackson \cite{Jackson1999}, Stratton \cite{Stratton1941}, and Zangwill \cite{Zangwill2013}, this formulation is derived directly from Maxwell's macroscopic equations without assuming a specific constitutive relation. It relies on the auxiliary fields $\mathbf{D}$ and $\mathbf{H}$ to produce the following balance equation:
\begin{equation}
    \mathbf{j}_f \cdot \mathbf{E} = -\nabla \cdot (\mathbf{E} \times \mathbf{H}) - \left( \mathbf{E} \cdot \frac{\partial \mathbf{D}}{\partial t} + \mathbf{H} \cdot \frac{\partial \mathbf{B}}{\partial t} \right).
    \label{eq:intro_conventional_poynting}
\end{equation}
Here, the Poynting vector is defined as $\mathbf{S}_{\text{conv}} = \mathbf{E} \times \mathbf{H}$. Crucially, the term in parentheses is a \textbf{rate of work}, not necessarily the derivative of a state function. It represents the total electromagnetic power fed into the volume, some of which increases the field energy and some of which performs work on bound charges.

\subsection{2. The Minkowski Formulation}
The Minkowski formulation (and the Abraham tensor) takes a different approach. It \textbf{postulates} a specific state function for the energy density, symmetric in form:
\begin{equation}
    u_M = \frac{1}{2} (\mathbf{E} \cdot \mathbf{D} + \mathbf{B} \cdot \mathbf{H})
\end{equation}
and fixes the energy flux as $\mathbf{S}_M = \mathbf{E} \times \mathbf{H}$.
To satisfy energy conservation with this specific choice of $u_M$, the power balance equation must be modified. It requires an additional term $P_{\text{matter}, M}$ to account for the discrepancy between the rate of change of the chosen state function and the actual power input:

\begin{equation}
    \frac{\partial u_M}{\partial t} + \nabla \cdot \mathbf{S}_M = -(\mathbf{j}_f \cdot \mathbf{E} + P_{\text{matter}, M}).
    \label{eq:Minkowski_Energy_Balance_Final}
\end{equation}

By expanding the time derivative of $u_M$ and comparing it to Eq. \ref{eq:intro_conventional_poynting}, the explicit form of this "Minkowski power" is found:

\begin{equation}
    P_{\text{matter}, M} = \frac{1}{2} \left( \frac{\partial \mathbf{E}}{\partial t} \cdot \mathbf{D} - \mathbf{E} \cdot \frac{\partial \mathbf{D}}{\partial t} + \frac{\partial \mathbf{B}}{\partial t} \cdot \mathbf{H} - \mathbf{B} \cdot \frac{\partial \mathbf{H}}{\partial t} \right).
    \label{eq:MinkowskiPowerVector_Explicit_General_Relativistic}
\end{equation}

\subsection{The Critical Divergence}
The two formulations are NOT physically equivalent.
\begin{itemize}
    \item \textbf{Linear, stationary media:} If $\mathbf{D} = \varepsilon \mathbf{E}$ with constant $\varepsilon$ (time-invariant), then $\mathbf{E} \cdot \dot{\mathbf{D}} = \frac{1}{2}\partial_t (\mathbf{E} \cdot \mathbf{D})$. In this limited case, $P_{\text{matter}, M} = 0$, and the formulations coincide.
    \item \textbf{Non-linear or moving media:} If the medium moves (simulating the variational displacement $\delta \mathbf{x}$), $\varepsilon$ becomes time-dependent at a fixed point. In this case, $P_{\text{matter}, M} \neq 0$. The Minkowski formulation assigns a different energy budget to the field than the Conventional formulation.
\end{itemize}
Since the Variational Method relies on calculating $\delta U$ under a virtual displacement (which is kinematically equivalent to motion), the choice of functional $U$ dictates the resulting force. It is now shown that applying the method to these different functionals leads to mechanically distinct results, highlighting the danger of selecting $U$ based on algebraic convenience rather than the FECC.

\section{Case Study 1: The Null Prediction of Conventional Theory}
\label{sec:CaseStudy_Conventional}

Applying the Variational Method to the conventional Poynting formulation, as established in Eq. \ref{eq:intro_conventional_poynting}, the work term is explicitly:
\begin{equation}
    P_{\text{mech}} = \mathbf{J}_f \cdot \mathbf{E}
\end{equation}
This term represents the power delivered by the electromagnetic field to \textbf{free charge carriers}.

Consider a neutral material object (whether dielectric or magnetic). By definition, the free charge density $\rho_f$ is zero, and consequently, the free current density $\mathbf{J}_f$ (assuming no conduction) is zero.
\begin{equation}
    \rho_f = 0 \implies P_{\text{mech}} = 0
\end{equation}
The conventional formulation asserts, unequivocally, that the electromagnetic field performs \textbf{zero mechanical work} on neutral matter, regardless of its polarization $\mathbf{P}$ or magnetization $\mathbf{M}$.

\subsection{The Variational Outcome}
If the Variational Method is now applied, the question becomes: "How much work is done if one displaces the material by $\delta \mathbf{x}$?"
Since the power integral is identically zero for all time, the work functional $W = \int P dt$ is also zero.
\begin{equation}
    \delta W = \mathbf{f} \cdot \delta \mathbf{x} = 0 \implies \mathbf{f} = 0
\end{equation}

\subsection{The Result}
A mechanically consistent application of the conventional energy formulation predicts \textbf{zero force} on neutral media. This result differs from observation (glass is attracted to capacitors, iron to magnets), but it is \textbf{mathematically consistent} with the input premise. The conventional formulation simply does not account for the work done on bound charges or currents; therefore, its variational derivative correctly reports that no such work exists.

The fact that texts derive non-zero forces (like $\mathbf{f} \sim \nabla \varepsilon E^2$ or $\mathbf{f} \sim \nabla \mu H^2$) from this starting point is not a feature of the theory, but a demonstration of the \textbf{procedural adaptation}. They obtain a result only by modifying the kinematic lock—varying a parameter ($\varepsilon$ or $\mu$) that was assumed constant during the definition of the energy, effectively injecting "work" that was never in the energy balance to begin with.

\section{Case Study 2: The Minkowski Postulate}
\label{sec:CaseStudy_Minkowski}

Attention now turns to the Minkowski formulation. Unlike the null-prediction of the conventional theory, Minkowski \textit{postulates} a specific form for the total energy density:
\begin{equation}
    u_M = \frac{1}{2} (\mathbf{E} \cdot \mathbf{D} + \mathbf{B} \cdot \mathbf{H})
\end{equation}
This postulate is not derived from microscopic physics but is a mathematical construction rooted in symmetry arguments. Its mechanical inconsistency has been presented in the previous parts of this book.
Still, we will derive the force corresponding to this proposed energy functional in the following.

\subsection{Deriving Force from Power}
As derived in Eq. \ref{eq:MinkowskiPowerVector_Explicit_General_Relativistic}, the enforcement of energy conservation with this specific functional requires the existence of a "matter power" term:
\begin{equation}
    P_{\text{matter}, M} = \frac{1}{2} \left( \frac{\partial \mathbf{E}}{\partial t} \cdot \mathbf{D} - \mathbf{E} \cdot \frac{\partial \mathbf{D}}{\partial t} + \frac{\partial \mathbf{B}}{\partial t} \cdot \mathbf{H} - \mathbf{B} \cdot \frac{\partial \mathbf{H}}{\partial t} \right).
\end{equation}
Considering a linear, isotropic medium where $\mathbf{D} = \varepsilon \mathbf{E}$ and $\mathbf{B} = \mu \mathbf{H}$, this expression simplifies significantly.
Using the product rule $\mathbf{E} \cdot \partial_t (\varepsilon \mathbf{E}) = \varepsilon \mathbf{E} \cdot \partial_t \mathbf{E} + E^2 \partial_t \varepsilon$, the terms involving field derivatives cancel, leaving only the time-derivatives of the material parameters:

\begin{equation}
    P_{\text{matter}, M} = -\frac{1}{2} E^2 \frac{\partial \varepsilon}{\partial t} - \frac{1}{2} H^2 \frac{\partial \mu}{\partial t}
\end{equation}

To find the force, it is necessary to link this power to the velocity of the mass $\mathbf{v}$.
If the material properties move with the mass (advection), the local time derivative is related to the spatial gradient by the definition of the convective derivative:
\begin{equation}
    \frac{d\varepsilon}{dt} = \frac{\partial \varepsilon}{\partial t} + (\mathbf{v} \cdot \nabla)\varepsilon = 0 \implies \frac{\partial \varepsilon}{\partial t} = -(\mathbf{v} \cdot \nabla)\varepsilon
\end{equation}
Substituting this kinematic identity into the power equation:
\begin{equation}
    P_{\text{matter}, M} = \frac{1}{2} E^2 (\mathbf{v} \cdot \nabla \varepsilon) + \frac{1}{2} H^2 (\mathbf{v} \cdot \nabla \mu)
\end{equation}
Finally, applying the FECC ($P = \mathbf{f} \cdot \mathbf{v}$), one can factor out the velocity vector:
\begin{equation}
    \mathbf{f}_M \cdot \mathbf{v} = \left( \frac{1}{2} E^2 \nabla \varepsilon + \frac{1}{2} H^2 \nabla \mu \right) \cdot \mathbf{v}
\end{equation}
Since this must hold for any velocity $\mathbf{v}$, the implied force density yields:
\begin{equation}
    \boxed{ \mathbf{f}_M = \frac{1}{2} E^2 \nabla \varepsilon + \frac{1}{2} H^2 \nabla \mu }
\end{equation}

\subsection{The Conclusion of the Experiment}
This result is identically the \textbf{Korteweg-Helmholtz force}.
The derivation demonstrates a crucial point: The Minkowski formulation yields a non-zero force not because it discovered a physical interaction, but because it \textbf{postulated an energy functional} ($U_M$) that explicitly includes the material parameters $\varepsilon$ and $\mu$.

By assigning an arbitrary energy to the field-matter system and using the FECC, one may derive a force. However, this force contains no more physical insight than the energy postulate itself. Force, Mass, and Energy are not independent variables; treating them as such is a \textbf{category error}. If one postulates the energy, one has already postulated the force.

\section{The Pragmatic Success: The Thermodynamic Black Box}
\label{sec:Black_Box}

If the derivation in Case Study 2 reveals that the Minkowski force density relies on a physically inconsistent energy functional for moving media, a critical question remains: Why does the engineering community successfully use this Korteweg-Helmholtz force to design actuators?

\subsection{Global Truth vs. Local Ambiguity}
The answer lies in the nature of thermodynamics. State functions (like Total Energy) are \textbf{path-independent}.
\begin{itemize}
    \item The total work required to assemble a capacitor with a dielectric slab inside is a fixed quantity, regardless of whether strict limits were held constant or varied during the process.
    \item Therefore, the \textbf{total energy difference} $\Delta U_{\text{total}}$ between "Slab Out" and "Slab In" is correct.
\end{itemize}

Because the \textbf{net} energy change is correct, the Principle of Virtual Work guarantees that the gradient of this total—the \textbf{net force}—must be correct.
\begin{equation}
    \mathbf{F}_{\text{net}} = -\nabla U_{\text{total}} \quad \text{(Correct)}
\end{equation}

The Korteweg-Helmholtz method functions as a \textbf{thermodynamic black box}. It correctly maps the input (total energy) to the output (net force). Let us consider the energy functional in more detail.

\subsection{The Conflated Potential as a Total Budget}
It is crucial to emphasize that the selection of the Minkowski functional ($u_M = \frac{1}{2}\mathbf{D}\cdot\mathbf{E}$) was not an arbitrary error. In the context of linear, stationary media, it represents a \textbf{profoundly efficient compression} of the physics.

As established in the mechanical audit (Chapter \ref{chap:Macro_Critique}), the total energy density of a linear dielectric is the algebraic sum of the fundamental vacuum field energy and the mechanical potential energy stored in the bound lattice:
\begin{equation}
    u_{\text{Minkowski}} = u_{\text{vacuum}} + u_{\text{bound}}.
\end{equation}
By bundling the energy that leaves the macroscopic domain—transferring through the \textbf{host interface} to store as elastic potential in the microscopic lattice—directly into the field term, the Minkowski formulation creates a powerful "effective energy functional" for the system.

For \textbf{non-dissipative} (conservative) linear dielectrics, this bundling is numerically exact. The energy calculated by the Minkowski functional is identical to the sum of the field and mechanical energy changes derived from the rigorous Lorentz formulation.
\begin{equation}
    \Delta U_{\text{total, Minkowski}} \equiv \Delta U_{\text{field}} + \Delta U_{\text{mechanical}}
\end{equation}
Because this equivalence holds for the static limit, the choice of $u_M$ was a rational and pragmatically sound hypothesis. The structural divergence is subtle: it only becomes visible when one demands a local description of the \textbf{mechanism} of transfer. Discovering the error requires the resolution of a \textbf{two-fluid model} (Lorentz), which explicitly distinguishes between the "field space" and the "matter coordinates." Without this high-resolution separation, the Minkowski functional stands as the most logical definition of the system's total potential.
It thus calculates the correct \textbf{net force}.

\subsection{The Hydrodynamic Equivalence (Incompressible Limit)}
There is a second, equally practical reason for the engineering success of the Korteweg-Helmholtz force in fluid dynamics. As demonstrated by Haus and Melcher in their seminal analysis of electrohydrodynamics \cite{HausMelcher1989}, the distinction between the microscopic (Kelvin) force density and the macroscopic (Korteweg-Helmholtz) force density often vanishes in the context of \textbf{incompressible fluids}.

The difference between the two force densities is typically a gradient term (e.g., electrostriction gradients). In the Navier-Stokes equation governing the fluid motion:
\begin{equation}
    \rho \frac{D\mathbf{v}}{dt} = -\nabla p + \mathbf{f},
\end{equation}
any force term that can be expressed as a perfect gradient ($\nabla \psi$) can be mathematically absorbed into the pressure term ($p' = p - \psi$). Since incompressible flow is determined solely by the curl of the driving forces (vorticity dynamics), gradient forces do not affect the flow pattern; they only redefine the static pressure field.

Consequently, Haus and Melcher show that for constant-permittivity in incompressible fluids, the Kelvin force (based on point dipoles) and the Korteweg-Helmholtz force (based on energy variation) predict \textbf{identical kinematics}. The pragmatic engineer, observing the same fluid motion, naturally concludes the force formulations are equivalent. The error is hidden in the pressure definition, which is rarely a primary observable in flow visualization.

\subsection{The Local Limitation: Single-Fluid vs. Two-Fluid Realities}
The limitation of the local force distribution is a direct result of the \textbf{model of matter} employed.

The Minkowski formulation effectively treats the dielectric as a \textbf{single fluid}—a unitary mass distribution characterized by a parameter $\varepsilon$. In this model, there are no internal degrees of freedom (like polarization displacement); there is only the bulk parameter. Consequently, when the energy functional is varied, the math generates a force $\mathbf{f} \sim \nabla \varepsilon$ wherever the material property changes.

Crucially, there is \textbf{no microscopic physical origin} for this force. Unlike the Lorentz force, which acts on specific charges or currents, the Minkowski force acts on the "gradient of the parameter." Unless there is a real accumulation of charge at that gradient (which is not guaranteed), this force is effective. It is a \textbf{mathematical consequence} of the bundled energy functional: by assigning the elastic energy to the field term and coupling it to a single mass coordinate, the variational calculus generates a "effective force" to account for the energy change, despite the absence of any physical mechanism (like charge separation) at that location.

Lorentz, in contrast, employs a \textbf{two-fluid model}. It explicitly recognizes that the "mass" is composed of two distinct populations (ions and electrons) that move relative to each other. Because it tracks this internal coordinate (Polarization), it correctly assigns force to the internal constituents. The force is not just on the surface of the block; it is distributed throughout the deep bulk, acting on every displaced charge pair.

\section{Conclusion: The Variational Synthesis}

In this synthesis, it has been argued that the Variational Method is not an independent source, but an implicit application of the \textbf{Force-Energy Consistency Criterion (FECC)}:
\begin{equation}
    P = \mathbf{f} \cdot \mathbf{v} \implies \mathbf{f} = -\nabla U
\end{equation}
The central insight is that \textbf{energy and force are not independent variables}: they are two sides of the same coin. The Variational Method is simply the mathematical procedure that enforces this connection. If one postulates an energy functional, one has implicitly postulated the force; if one postulates a force law, one has implicitly defined the energy.

By examining the Korteweg-Helmholtz (KH) force through this lens, its nature is re-evaluated. It appears not as a fundamental discovery of a new interaction, but as a logical mathematical consequence of the Minkowski energy budget. Because Minkowski postulates that the field energy depends on the bulk parameter $\varepsilon$ (at least for linear media), the variational calculus inevitably yields a force proportional to $\nabla \varepsilon$.

As shown, this force density can successfully calculate the \textbf{total net force} in conservative, non-dissipative systems (the thermodynamic black box). Its widespread success is a testament to the utility of this macroscopic compression.

The limitation, however, lies in the information loss. As detailed widely in \textbf{Part III} and \textbf{Part IV}, the Minkowski/Abraham formulation is mechanically constrained in the general case because it relies on a single-fluid model. This was not an oversight, but a \textbf{necessary constraint} of the purely macroscopic era. Without the microscopic resolution to track the "Two-Fluid" reality of the lattice (Option A), the single-fluid effective energy (Option B) was the only rational path available.

True mechanical consistency requires the abandonment of these bundled functionals and a return to the rigorous, microscopic power balance of the \textbf{two-fluid model}. Only by respecting the kinematic lock at the microscopic scale can the inconsistencies of the macroscopic world be resolved.

\chapter{Conclusion: The Resolution of the Paradox}
\label{chap:Part6_Conclusion}

The forensic audit of the electromagnetic stress tensor is now complete. The century-long debate, often framed as a binary conflict between alternative formulations (Minkowski vs. Abraham), has been shown to be a symptom of a deeper methodological challenge: the attempt to describe the mechanical behavior of matter using only macroscopic field variables.

Our proposed resolution of this "Century of Divergence" was not a trivial task. It was not a matter of simply spotting a mathematical error in the derivations of Minkowski or Abraham—both of whom were mathematically rigorous within their starting assumptions. Rather, the path to the \textbf{Vacuum Tensor} required the simultaneous integration of five consistent physical insights. Each of these pillars represents a subtle departure from the standard macroscopic approximation, and it is only when they are assembled that the paradox is resolved.

\section{The Five Pillars of Resolution}

The reason the controversy persisted for so long is that the \textbf{Vacuum Tensor} (the Lorentz force) appears, at first glance, to fail in predicting the correct deformation of dielectrics. To see why it is actually the correct fundamental description, one must peel back layers of approximation that are usually baked into the macroscopic theory.

\subsection[Pillar I: The Necessity of Microstructure]{Pillar I: The Necessity of\\ Microstructure (Opening the Black Box)}
The first hurdle was the assumption of the "Opaque System." Historical formulations were constrained by the premise that the force density must be expressible as a function of the macroscopic fields ($\mathbf{E}, \mathbf{H}, \mathbf{D}, \mathbf{B}$) alone.
\begin{itemize}
    \item As proven in the \textbf{Laminated Conductor} experiment (Part III), the macroscopic fields are topology-limited. They cannot distinguish between a state of internal tension and a state of zero stress.
    \item The resolution required "opening the black box"—acknowledging that the \textbf{Deformation Force} is not a single-valued function of the fields, but depends on the explicit connectivity of the lattice.
\end{itemize}
By accepting that "Side-Channel Information" (the hypothesis of microstructure) is required, we resolve the indeterminacy that plagued the pure field theories.

\subsection{Pillar II: The Redefinition of Energy (The High-Field Storage)}
The second pillar is the subtle distinction between "Field Energy" and "Binding Energy."
\begin{itemize}
    \item \textbf{Minkowski's Rational Choice:} Minkowski postulated that the total work done on the system ($ \int \mathbf{E} \cdot d\mathbf{D} $) should be counted as "Field Energy." For linear media, this is a pragmatic and efficient compression.
    \item \textbf{The Vacuum Distinction:} The audit revealed, however, that this "Total Energy" is actually a sum of the true Vacuum Energy ($\varepsilon_0 E^2/2$) and the Mechanical Potential Energy stored in the atomic binding fields.
\end{itemize}

\subsection{Pillar III: The Two-Fluid Model (The Kinematic Split)}
The third insight concerns the model of matter itself. The single-fluid continuum model (used implicitly by Minkowski and Abraham) treats the dielectric as a unitary block.
\begin{itemize}
    \item This model hides the internal mechanism of polarization. It cannot represent the relative motion between the electron cloud and the ion lattice.
    \item By adopting the \textbf{Two-Fluid Model} (Lorentz), we introduced the necessary kinematic degree of freedom: the internal velocity $\mathbf{v}_{\text{internal}}$. This allows the theory to rigorously define the "Power Gateway"---distin\-guish\-ing between work done to accelerate the bulk (deformation) and work done into internal high-frequency energy storage.
\end{itemize}

\subsection{Pillar IV: The Partitioning of Force (Resolving Indeterminacy)}
The fourth pillar addresses the primary critique of the Vacuum formulation: "It predicts the wrong force."
\begin{itemize}
    \item It is true that the Vacuum Tensor ($\rho \mathbf{E} + \mathbf{J} \times \mathbf{B}$) yields a total force density that includes components balanced by quantum constraints—forces that do not deform the macroscopic lattice.
    \item The resolution lies in the \textbf{Partitioning Theorem}: The Total Force is not the Deformation Force. The total budget $\mathbf{f}_{\text{total}}$ splits into mechanically distinct roles:
    \begin{equation}
        \mathbf{f}_{\text{total}} = \mathbf{f}_{\text{deformation}} + \mathbf{f}_{\text{binding}}+ \mathbf{f}_{\text{dissipation}}
    \end{equation}
\end{itemize}
Standard theories attempted to find a Level 2 tensor that yielded $\mathbf{f}_{\text{deformation}}$ (a Level 3 quantity) directly. They failed because they demanded that the field description alone solve the allocation problem. The Vacuum Tensor yields the \textbf{total force budget} $\mathbf{f}_{\text{total}}$, which is the physically conserved quantity.

\subsection{Pillar V: Spectral Independence (The Orthogonality of Scales)}
Finally, the validity of separating the "Vacuum" from the "Material" relies on the concept of \textbf{Spectral Filtering}.
\begin{itemize}
    \item One might argue that inside a dense solid, the "vacuum field" is a fiction.
    \item However, as shown in Part II, the microscopic field can be rigorously decomposed into a low-frequency (macroscopic) component and a high-frequency (atomic) component. Even though the energy density is quadratic ($E^2$), the total energy separates cleanly because the cross-terms between widely separated frequencies vanish (orthogonality).
\end{itemize}
This justifies the mathematical separation of the interaction into "Macroscopic Work" (Low-Frequency) and "Binding Energy" (High-Frequency), validating the binding energy framework of the Vacuum Tensor.

These insights help to propose that the Vacuum energy momentum tensor is the exclusive mechanically consistent description of the electromagnetic interaction.

\section{Implications for Part VII: Toward a Unified View}

Once the Vacuum Tensor is accepted as the fundamental baseline, an \textbf{isomorphism of classical physics} emerges. This mechanical synthesis, which unifies the thermodynamic, continuum mechanical, and electrodynamic views, will be demonstrated in the following chapter.

This concludes the forensic audit of the 20th-century electrodynamic formulations. In Part VII, it will be argued that the disciplines of Thermodynamics, Continuum Mechanics, and Electrodynamics may be viewed not as separate magisteria, but as three \textbf{spectral views} of a single, unified mechanical reality.

\part{Part VII: The Mechanical Synthesis}
\label{part:Synthesis}

\chapter{Review of Parts I-VI}
\label{chap:ReviewPart1to6}
This analysis commenced with the recognition of a \textbf{divergence in interpretation}. A survey of the modern landscape of the energy-momentum stress tensor revealed a conflicting array of definitions for the theory's core elements: the electromagnetic stress (momentum flux), momentum density, energy density, and energy flux. The tensors of Abraham, Minkowski, and the Amperian vacuum formulation stand as the primary exemplars of this divergence.
To resolve this ambiguity, the criterion of \textbf{mechanical consistency} was advanced: a valid tensor formulation must not only balance algebraically but must also predict a force distribution that aligns strictly with its own energy budget.
\paragraph{The Analysis of Equivalence}
The conventional perspective---exemplified by Penfield, Haus, and the modern consensus---posits that all tensor formulations are equally valid conventions, provided they sum to the correct total divergence. This analysis suggests a different conclusion.
The \textbf{principle of physical exclusivity} demonstrates that physics requires more than the balancing of an algebraic equation ($LHS=RHS$); it posits the \textit{ontological identity} of the terms. Arbitrary reassignment of terms between "field" and "matter" obscures the physical mechanism of the host interface.
\paragraph{The Microscopic Resolution}
At the microscopic level, the ambiguity is clarified. The microscopic Lorentz theory was identified as the universally valid baseline. Consequently, alternative microscopic formulations—such as the Abraham or Minkowski tensors—are shown to be mechanically inconsistent. They present a reality mutually exclusive to the Lorentz framework.
\paragraph{The Macroscopic Synthesis}
The macroscopic field and force equations were derived strictly as the low-frequency limit of this microscopic baseline. In Parts III through V, it was demonstrated that the vacuum tensor framework explains all force and energy interactions of electromagnetic matter in a mechanically consistent manner.
The unified coupling for all interactions is rigorously described by the \textbf{gateway equations}:
\begin{equation}
\boxed{
\begin{aligned}
    \underbrace{ \left( \frac{\partial \mathbf{G}_{\text{mech}}}{\partial t} + \nabla\cdot\mathbf{T}_{\text{mech}} \right) - \mathbf{F}_{\text{host}} }_{\substack{\text{Total Mechanical Response}}}
    \quad &\overset{\text{Physical Coupling}}{\textbf{=}} \quad
    \underbrace{\rho\mathbf{E} + \mathbf{J}\times\mathbf{B} }_{\substack{\textbf{The Momentum Gateway} \\ (\mathbf{F}_{\text{Lorentz}})}} \\
    &\overset{\text{Math. Identity}}{=} \quad
    \underbrace{ -\left(\frac{\partial \mathbf{G}_{\text{EM}}}{\partial t} + \nabla\cdot \mathbf{T}_{\text{EM}}\right) }_{\substack{\text{Macroscopic EM Response}}}
\end{aligned}
}
    \label{eq:Macro_Momentum_Architecture_Unification}
\end{equation}
And the corresponding \textbf{energy balance}, which identifies the source/sink nature of the interaction:
\begin{equation}
\boxed{
\begin{aligned}
    \underbrace{ \left( \frac{\partial U_{\text{mech}}}{\partial t} + \nabla\cdot\mathbf{S}_{\text{mech}} \right) - P_{\text{host}} }_{\substack{\text{Total Mechanical Response}}}
    \quad &\overset{\text{Physical Coupling}}{\textbf{=}} \quad
    \underbrace{ \mathbf{J}\cdot\mathbf{E} }_{\substack{\textbf{The Energy Gateway} \\ (P_{\text{Lorentz}})}} \\
    &\overset{\text{Math. Identity}}{=} \quad
    \underbrace{   -\left( \frac{\partial u_{\text{EM}}}{\partial t} + \nabla \cdot \mathbf{S}_{\text{EM}} \right)   }_{\substack{\text{Macroscopic EM Response}}}
\end{aligned}
}
    \label{eq:Macro_Energy_Architecture_Unification}
\end{equation}
The vacuum energy-momentum stress tensor is defined unambiguously by:
\begin{itemize}
    \item \textbf{Energy Density:} $u_0 = \frac{1}{2} (\varepsilon_0 E^2 + \mu_0^{-1} B^2)$
    \item \textbf{Momentum Density:} $\mathbf{g}_0 = \varepsilon_0(\mathbf{E} \times \mathbf{B})$
    \item \textbf{Energy Flux (Poynting Vector):} $\mathbf{S}_0 = \mu_0^{-1}(\mathbf{E} \times \mathbf{B})$
    \item \textbf{Stress Tensor:} $\mathbf{T}_0 = \varepsilon_0 \mathbf{E} \otimes \mathbf{E} + \mu_0^{-1} \mathbf{B} \otimes \mathbf{B} - u_0 \mathbf{I}$
\end{itemize}
The derived host interaction gateways act as the universal routing mechanism, directing this raw electromagnetic power to one of three mechanical destinations:
\begin{enumerate}
    \item \textbf{Work (Connecting to the Continuum Mechanics Domain):} The fluid moves the bulk lattice $\to$ \textbf{kinetic energy}. (\textit{Action: Actuation, Acceleration})
    \item \textbf{Storage (Connecting to the Binding Energy Domain):} The fluid performs work against a microscopic restoring force $\to$ \textbf{binding energy}. (\textit{Action: Binding})
    \item \textbf{Dissipation (Connecting to the Thermodynamic Domain):} The fluid rubs against the lattice $\to$ \textbf{heat}. (\textit{Action: Resistance, Hysteresis})
\end{enumerate}
This framework couples the macroscopic electrodynamic domain into thermodynamics and continuum dynamics in a mechanically consistent manner.
\paragraph{Coupling to Continuum Mechanics}
It was further derived that the total macroscopic Lorentz force density, which determines the divergence of the vacuum stress tensor, implies a subdivision of forces:
\begin{enumerate}
    \item \textbf{Internal Structural Forces:} Forces counterbalanced by microscopic constraints (e.g., dipole binding forces).
    \item \textbf{Continuum Forces:} Forces that translate into the continuum domain (e.g., bulk deformation).
\end{enumerate}
The point dipole approach of Kelvin, Einstein, and Laub was identified as the correct method to derive the force density component coupling to the continuum mechanics domain. However, we propose that an adaptation term is required. By utilizing the "test dipole" thought experiment—inserting a dipole into a statistically isotropic bulk material—we proposed the specific structural form:
\begin{equation}
\begin{split}
    \mathbf{F}_{\text{deformation}} &\approx \underbrace{\left[ (\mathbf{P} \cdot \nabla)\mathbf{E} + \frac{\partial \mathbf{P}}{\partial t} \times \mathbf{B} \right] + \left[ (\mathbf{M} \cdot \nabla)\mathbf{B} + \frac{1}{c^2} \mathbf{E} \times \frac{\partial \mathbf{M}}{\partial t} \right]}_{\text{Standard Einstein-Laub}}\\
    &\quad + \underbrace{(\mathbf{P} \cdot \nabla)\left( \frac{\mathbf{P}}{3\varepsilon_0} \right) - \frac{2\mu_0}{3} (\mathbf{M} \cdot \nabla)\mathbf{M}}_{\text{Geometrical Stress}}
\end{split}
\end{equation}
This derivation acknowledges that the force density is explicitly dependent on the exact microscopic structure and effective geometry (e.g., the spherical exclusion volume leading to the factor $1/3$) of the material.
\paragraph{The Relativistic Verification (Part V)}
The framework was rigorously tested against the phenomenology of moving matter. By utilizing the covariant source transformation, it was proven that "Moving Media" electrodynamics requires no new postulates. The complex effects of motional polarization ($\mathbf{P}'$) and magnetization ($\mathbf{M}'$) are shown to be simply the coordinate transformation of the single, unified source tensor $M^{\mu\nu}$. There is no "medium"; there are only moving sources.
\paragraph{Assessment of Alternative Formulations}
Other macroscopic tensor formulations were audited for mechanical consistency. The Minkowski and Abraham formulations were found to be \textbf{mechanically inconsistent}. Specifically, a well-defined mass target for their proposed force densities could not be isolated, nor could the energy budget be reconciled with the mechanical work performed. The physical mechanism of dissipation remains opaque in these frameworks.
Furthermore, the laminated dielectric example demonstrated that it is generally not possible to derive a definite force density from the macroscopic tensor formalism alone. Energy-based approaches such as the Korteweg-Helmholtz method, which rely exclusively on macroscopic field variables and neglect microscopic structure, are thus structurally underdetermined regarding local force density.
Regarding alternative formulations, such as that proposed by Chu: while it is theoretically possible to construct a mechanically consistent tensor based on magnetic charges, this work intentionally adheres to the empirical constraint of their absence. As the provided framework explains all interaction phenomena consistently without magnetic monopoles, the introduction of additional macroscopic tensors is deemed unnecessary.
\paragraph{Conclusion: The Return to Simplicity}
When mechanical consistency is enforced, the ambiguity of the energy-momentum stress tensor is resolved. This framework suggests the vacuum tensor alone suffices to describe all interactions, from magnetics to dielectrics to moving matter. The theory returns to the simplicity of its microscopic origins, scaled precisely to the macroscopic domain.

\chapter{The Mechanical Unification}
\label{chap:MechanicalUnification}

The preceding parts of the book have systematically analyzed electrodynamic interactions and their coupling to the continuum mechanical and thermodynamic domains. Throughout this analysis, the distinction between the microscopic baseline and the macroscopic approximation has been central.

With the derivation complete, the framework now allows for the presentation of the integrating matrix of classical physics proposed in Part \ref{part:MacroscopicFilter}. This chapter synthesizes the spectral projection of the underlying microscopic reality, detailing the architecture that unifies the domains. This suggests that the perceived separations between mechanics, thermodynamics, and electrodynamics are consequences of spectral filtering rather than fundamental physical distinctions.

\section{Parallel Phenomena in Classical Physics}
\label{sec:Unification_Parallels}

A comparative analysis of the continuum mechanical, thermodynamic, and electrodynamic domains reveals striking structural parallels. These are not merely analogies but identical mechanical processes viewed through different spectral filters.

\subsection{The Domain of Binding Energy (The Complementary Domain)}
\label{sec:Unification_Potential}

The electrodynamic analysis identified a specific pathway for energy storage in the macroscopic domain: potential energy stored in high-frequency fields, which are filtered out of the macroscopic field representation. This was designated as \textbf{binding energy}.

The microscopic origin of this hidden potential energy lies in the relative motion of charges and currents—specifically, the displacement of charges in dielectrics or the orientation of currents in magnetic materials. The principle consistently manifests as energy stored in the high-frequency component of the electromagnetic fields, sustained by the relative configuration of ions and electrons inside matter.

A direct parallel exists within the continuum mechanics domain. The conceptual process is identical, though the domain definition shifts from the relative motion of ions and electrons to the relative motion of neutral atoms. Macroscopic elasticity is, microscopically, the reorientation of neighboring atoms, which alters the microscopic electromagnetic high-frequency fields. These fields, having been filtered out of the electrodynamic domain, are compressed into a macroscopic material parameter known as \textbf{elasticity}.

Thus, what is termed "binding energy" in the macroscopic electrodynamic domain (arising from electron-ion relative motion) is mechanically identical to "elastic energy" in the continuum mechanics domain (arising from atom-atom relative motion). Both represent the high-frequency storage of electromagnetic energy, hidden from the low-frequency macroscopic view.

\subsection{Dissipation}
\label{sec:Unification_Dissipation}

The second parallel is the process of \textbf{dissipation}, defined as the degradation of ordered particle motion into unordered motion. Here again, the continuum mechanical and electrodynamic domains exhibit a shared mechanism.

In the continuum domain, the analysis considers the movement of neutral atoms. Friction arises from the relative motion of neighboring layers having an ordered velocity gradient ($\nabla \mathbf{v}$). This relative, ordered movement of adjacent atomic layers excites random motion in the bulk of the material, which manifests as heat. The macroscopic filtered effect is captured by the material parameter \textbf{viscosity}.

The same conceptual process occurs in electrodynamics. The distinction lies only in the participants: rather than neighboring neutral atoms moving past each other, it is the relative motion of ions and electrons. Whether manifesting as resistance in a conductor or dissipation in magnetization and polarization types, the process is mechanically identical: the relative ordered motion of ions and electrons excites unordered particle motion, which manifests as heat.

Consequently, the inputs from both the continuum mechanics domain and the macroscopic electrodynamics domain to the thermodynamics domain represent the same physical process: the irreversible transfer of kinetic energy from the ordered spectral modes to the unordered spectral mode.

\section{The Architecture of Classical Physics}
\label{sec:Unification_Architecture}

A key finding of this analysis is that classical physics is a single, unified subject. The historical categorization into three distinct disciplines --- thermodynamics, continuum mechanics, and electrodynamics --- represents three distinct \textbf{analytical perspectives} (or spectral views) of the same underlying reality.

Before presenting the unified matrix, the conceptual characteristics of these three domains are reviewed.

\subsection{The Unordered Domain (Thermodynamics)}
\label{sec:Unification_Thermodynamics}

This domain captures the unordered motion of particles. It is defined by randomness ($\langle \mathbf{v} \rangle = 0$) and the irreversible loss of information.

The primary state variable is \textbf{temperature} ($U_K$), representing the mechanical, kinetic energy of random molecular impacts ($u \propto k_B T$). Furthermore, this domain inherits \textbf{thermal radiation}, representing the electromagnetic fields corresponding to the unordered motion of particles.

The dynamics of heat flow within the macroscopic thermodynamic domain are governed by the heat diffusion equation. This domain couples to the continuum domain through \textbf{pressure}. Pressure performs work to accelerate bulk matter within the continuum domain. The inverse effect is \textbf{adiabatic cooling}, where energy is extracted from the thermodynamics domain to perform mechanical work, causing the temperature to drop. Additionally, as detailed above, the thermodynamic domain couples to both the continuum mechanical and electrodynamic domains through dissipation.

\subsection{The Ordered Domain (Continuum Mechanics)}
\label{sec:Unification_Continuum}

The continuum mechanics domain is the domain of ordered neutral particle motion (principally the ions). It is defined by coherent bulk motion ($\mathbf{v}_{\text{bulk}}$).

The primary variable of interest is \textbf{macroscopic momentum} ($\mathbf{G} = \rho_m \mathbf{v}$). The equation of motion is the Cauchy Momentum Equation, which effectively represents the "physics of the skeleton."

Coupling to the thermodynamic domain occurs via pressure and dissipation. Coupling to the electrodynamic domain is mediated by the electromagnetic force density. As derived in the preceding parts of the book, this force density is structurally determined by the material's microstructure.

\subsection{The Relative Domain (Electrodynamics)}
\label{sec:Unification_Electrodynamics_Domain}

The domain of \textbf{electrodynamics} is fundamentally characterized by the requirement of a \textbf{two-fluid model}. It explicitly tracks the \textbf{relative motion} between the positive ion fluid and the negative electron fluid.

This domain describes the motion of the light components (electrons) relative to the heavy skeleton (ions). The ordered relative motions generate macroscopic charges, currents, and electromagnetic fields. The state variables are the Electromagnetic Fields ($\mathbf{E}, \mathbf{B}$), along with charge density ($\rho$) and current density ($\mathbf{J}$).

The electromagnetic field performs work directly on these charge carriers. This interaction couples the electromagnetic domain to the continuum domain (via force density on the bulk) and to the thermodynamic domain (via dissipation), serving as the input of work and heat, respectively.

\section{The Unified Matrix of Classical Physics}
\label{sec:Unified_Matrix}

The microscopic Lorentz theory is completely described by the microscopic laws of electromagnetism and mechanics. The process of filtering from micro to macro gives rise to the emergent domains of classical physics.

The framework proposes that separating these motions into three spectral modes yields the origin of the three domains. The \textbf{spectral mode} criterion asks: Is the motion Ordered, Relative, or Unordered? Furthermore, each domain is analyzed in terms of its \textbf{frequency component}: high-frequency versus low-frequency signal.

This derivation results in a $2 \times 3 \times 2$ unified matrix:
(Mechanical vs. Electromagnetic) $\times$ (Ordered vs. Relative vs. Unordered) $\times$ (Low Freq vs. High Freq).

A crucial distinction must be made regarding the nature of these variables:
\begin{itemize}
    \item \textbf{Low-Frequency Variables (Surviving Signal):} These are physically identical to their microscopic counterparts (mass momentum, charges, currents, fields). They are merely the filtered, smoothed version of the microscopic reality. No new physics emerges here; information is merely filtered.
    \item \textbf{High-Frequency Variables (Emergent):} These are the \textbf{emergent properties} of the macroscopic domain. They arise entirely from the process of \textbf{information compression}. The complex, high-bandwidth details of microscopic motion (e.g., atomic vibration, orbital deformation) are compressed into singular new macroscopic low frequency state parameters (temperature, elasticity, binding mechanics) that did not exist as distinct entities in the microscopic domain.
\end{itemize}

The high-frequency fields, containing too much bandwidth to be retained in the macroscopic description, are filtered out and compressed into these new emergent macroscopic variables. The overarching objective of macroscopic physics is precisely this \textbf{information compression}: reducing the unmanageable bandwidth of the microscopic reality into a computable, effective description.

The proposed unified table of classical physics is presented below:

\begin{table}[h]
\centering
\footnotesize
\caption{\textbf{The unified matrix of classical physics.} Macroscopic physics acts as information compression: retaining the low-frequency signal while compressing the high-frequency storage.}
\label{tab:UnifiedMatrix}
\begin{tabular}{|c|c||c|c|c|}
\hline
\multicolumn{2}{|c||}{} & \textbf{\shortstack{Thermo-\\dynamics}} & \textbf{\shortstack{Continuum\\Mechanics}} & \textbf{\shortstack{Electro-\\dynamics}} \\
\multicolumn{2}{|c||}{} & \textbf{1. Unordered} & \textbf{2. Ordered} & \textbf{3. Relative} \\
\multicolumn{2}{|c||}{} & (Fluctuation) & (Common Motion) & (Diff. Motion) \\
\hline \hline
\multirow{4}{*}{\textbf{\shortstack{Low\\frequency\\signal}}} & \textbf{Mech.} & \textbf{---} & \textbf{\shortstack{Bulk\\momentum}} & \textbf{\shortstack{Carrier\\momentum}} \\
 & & & $\mathbf{p} = m \mathbf{v}_{\text{bulk}}$ & $\mathbf{p}_e = m_e \mathbf{v}_e$ \\
\cline{2-5}
 & \textbf{EM} & \textbf{---} & \textbf{---} & \textbf{\shortstack{Electro-\\dynamics}} \\
 & & & & $\mathbf{J}, \rho$ \\
\hline \hline
\multirow{4}{*}{\textbf{\shortstack{Compressed\\high\\frequency}}} & $\langle \textbf{Mech.} \rangle$ & \textbf{Temperature} & \textbf{---} & \textbf{---} \\
 & & T & & \\
\cline{2-5}
 & $\langle \textbf{EM} \rangle$ & \textbf{\shortstack{Thermal\\radiation}} & \textbf{\shortstack{Elastic\\energy}} & \textbf{\shortstack{Binding\\energy}} \\
 & & & & \\
\hline
\end{tabular}
\end{table}

This table serves as the "systematic framework" of macroscopic variables.

\paragraph{Thermodynamics (Column 1)}
The first column represents the thermodynamic domain. As this domain describes unordered motion, there is neither a low-frequency macroscopic momentum nor a low-frequency electromagnetic field response that constitutes a signal. The describing macroscopic variables are Temperature and Thermal Radiation, representing the compressed high-frequency responses—the "Complementary Domain"—of the mechanical and electromagnetic parts of the microscopic reality.

\paragraph{Continuum Mechanics (Column 2)}
The second column represents the continuum mechanics domain, defined by the ordered motion of the bulk mass (ions). Consider the mechanical aspect: the ordered motion of point particles is averaged into a single macroscopic fluid, represented by continuum mechanical momentum. The difference between the micro-state and the average (the high-frequency mass fluctuation) offers no physical insight as it stores no energy in this context; gravity is negligible at this scale, and the averaging is a valid simplification.

Consider the electromagnetic aspect of the ordered ions. As there is no relative motion between ions and electrons in this mode, the macroscopic electromagnetic field is by definition zero. However, a high-frequency part exists which is responsible for energy storage. This is \textbf{Elastic Energy}—energy stored in the high-frequency deformations of the electromagnetic fields between atoms—conceptually identical to binding energy.

\paragraph{Macroscopic Electrodynamics (Column 3)}
The third column is defined by the relative motion of electrons and ions. 

Mechanically, this involves the additional kinetic energy of the electron cloud moving relative to the ion lattice. It represents \textbf{carrier inertia} ($m_{\text{eff}}\mathbf{J}$). As noted in previous chapters, this small but finite mass is critical for the FECC: the Lorentz force must have a massive target to accelerate. Similar to the continuum domain, this relative motion is averaged, and only the low-frequency fluid representation is physically meaningful for the signal.

Electromagnetically, the low-frequency parts define the domain. The relative motions of ions and electrons generate macroscopic fields ($\mathbf{E}, \mathbf{B}$) and macroscopic sources ($\rho, \mathbf{J}$). However, the relative motions also induce further low-frequency parts that are filtered out of the macroscopic field domain. This corresponds to the \textbf{bound energy} (polarization and magnetization energy), which parallels the elastic energy of the continuum domain.

\paragraph{Interactions}
The state variables of the $2 \times 3 \times 2$ matrix have now been established. 
Beyond these state variables, specific macroscopic properties define the \textbf{interactions} between the domains. These variables have also been described throughout the audit: pressure, electromagnetic force density, resistance, viscosity, and hysteresis dissipation govern the exchange of energy and momentum across the spectral boundaries.

\section{Conclusion: The Mechanical Synthesis}
\label{sec:Unification_Conclusion}

Classical physics is typically presented as a collection of disparate disciplines: thermodynamics for heat, continuum mechanics for deformation, and electrodynamics for fields. The central proposal of this work is that these are not three separate subjects.

The analysis has revealed deep inherent symmetries between the continuum mechanical and electromagnetic domains: dissipation (viscosity and resistance) and storage (elasticity and binding energy) are conceptually identical processes. The difference lies merely in the spectral view.

They are three \textbf{spectral views} of a single, unified mechanical reality.

\chapter{Conclusion: The Mechanical Synthesis}
\label{chap:Conclusion}

The present inquiry concludes here. This work has attempted a systematic mechanical analysis of the structure of macroscopic electrodynamics.

The objective was not to introduce new physics, but to clarify the existing structure. The central question posed at the outset was whether a mechanically consistent description of macroscopic electrodynamics can be established without resorting to the customary "arbitrary split."

The findings suggest that the \textbf{Force-Energy Consistency Criterion (FECC)} and the filtering perspective provide a mechanism for such a resolution.

\section{Recapitulation: The Logic of the Argument}
\label{sec:Conclusion_Journey}

This work began with the recognition of a structural divergence in the standard consensus. The foundations of macroscopic electrodynamics have long been characterized by competing formulations, with no agreement on the most basic entities of the theory: energy, force, and momentum in matter. The modern response to this confusion—the "hypothesis of equivalence"—safeguards the theory by asserting that all tensor formulations are valid conventions, provided they sum to the correct total.

This analysis has explored an alternative view. By proposing the \textbf{principle of physical exclusivity}, it is suggested that physics, unlike accounting, may require a unique definition for its terms. The argument posits that arbitrarily shifting terms between "field" and "matter" masks the physical insight of the differential equation.

The argument treats electrodynamics not as a self-contained algebraic system, but as a closely intertwined discipline with general mechanics. By enforcing the strict requirements of mechanical work ($P=\mathbf{f} \cdot \mathbf{v}$) and applying rigorous signal processing (filtering) to the vacuum, the ambiguity is clarified. The question shifts from "which tensor balances the books?" to "which force moves the mass?"

This shift—from algebraic accounting to mechanical analysis—reduces the complexity. It reveals that the variety of macroscopic tensors arises from a failure to track the exact mass target of a force. Once the \textbf{host interface} is understood as the mechanical mediator, the need for complex, asymmetric tensors vanishes. The journey leads not to a new exotic theory, but to the rigorous re-validation of the simplest possible one: the microscopic Lorentz force, averaged.

\section{Core Contributions}
\label{sec:Conclusion_Contributions}

This work summarizes its findings as eight structural observations regarding the standard interpretation of electrodynamics.

\subsection[The FECC]{1. The Force-Energy Consistency\\ Criterion (FECC)}
We proposed a rigorous standard for a mechanically consistent theory: the \textbf{force-energy consistency criterion}.
Force and energy are not independent choices; they are two sides of the same coin, rigidly locked by the definition of mechanical work ($P = \mathbf{f} \cdot \mathbf{v}$).
This criterion acts as a "Razor" that excludes historical tensors (Minkowski, Abraham) because they algebraically balance the books but fail to provide a specific \textbf{mass target} for the force. It addresses the foundations of standard variational methods (Korteweg-Helmholtz), demanding that every interaction be defined by a force acting on a specific mass velocity.

\subsection{2. The Theorem of "Non-Local Transfer"}
A structural inconsistency was identified in the Minkowski and Abraham tensors when applied to dissipative media.
By demonstrating that these tensors predict energy leaving the field ($Q > 0$) without a mechanically defined carrier to receive it ($W=0$), they are shown to be mechanically inconsistent.
This phenomenon implies they describe a kinematically decoupled geometry.

\subsection{3. The Spectral Unification of Classical Physics}
A theoretical unification was proposed: \textbf{thermodynamics}, \textbf{continuum mechanics}, and \textbf{electrodynamics} are presented not as separate systems, but as three \textbf{spectral filters} of the same underlying reality.
\begin{itemize}
    \item \textbf{Unordered motion} $\to$ Thermodynamics.
    \item \textbf{Ordered motion} $\to$ Continuum mechanics.
    \item \textbf{Relative motion} $\to$ Electrodynamics.
\end{itemize}
Table \ref{tab:UnifiedMatrix} maps the emergent macroscopic variables of each domain. Parallels between binding energy and elastic energy, as well as between viscosity, resistance, friction, and hysteresis, have been proposed.

\subsection{4. The Unity of Sources}
The analysis removes the artificial distinction between "free" and "bound" matter. Polarization ($\mathbf{P}$) and Magnetization ($\mathbf{M}$) are presented strictly as \textbf{Charges} and \textbf{Currents} ($\rho_b, \mathbf{J}_b$).
This demystification resolves the "Moving Media" paradoxes by reducing them to simple relativistic coordinate transformations of the source current tensor. The auxiliary fields $\mathbf{D}$ and $\mathbf{H}$ are viewed as bookkeeping artifacts rather than fundamental entities.

\subsection{5. The Gateway Architecture}
The map of electromagnetic energy routing was derived. The \textbf{host interface} is modeled as a router with exactly three outputs:
\begin{enumerate}
    \item \textbf{Work} (Continuum motion).
    \item \textbf{Dissipation} (Heat/scattering).
    \item \textbf{Elasticity/Binding} (High-frequency storage).
\end{enumerate}
This clarifies that "potential energy" is not an abstract scalar but high-frequency field energy mechanically stored in the variance of the lattice.

\subsection{6. The "Magnetic Pivot" Mechanism}
The apparent paradox of magnetostatic work—how a magnetic field drives a motor despite doing no work ($\mathbf{F} \perp \mathbf{v}$)—was explicated.
The mechanical model of the B-field was defined as a \textbf{Lossless Fulcrum}. It does not \textbf{do} the work; it \textbf{steers} the electrical work ($\mathbf{J} \cdot \mathbf{E}$) into the mechanical axis via the host interface. This redefines permanent magnets from "passive potentials" to "active energy reservoirs" powered by quantum stability.

\subsection{7. The Theorem of Macroscopic Indeterminacy}
It was demonstrated that it is \textbf{impossible} to derive a universal local force density from macroscopic fields alone.
Because macroscopic fields are averages, they delete the geometric information (topology) needed to calculate local stress. The analysis showed that to obtain the correct deformation force, this information must be re-injected via the \textbf{Hypothesis of Microstructure} (The Lattice Adaptation Term). Standard tensors and variational methods (like Korteweg-Helmholtz) filter out this topology.

\subsection{8. The Principle of Physical Exclusivity}
Finally, it has been proposed that Physics is not merely algebra.
The analysis suggests that adding arbitrary vector terms to the laws of motion is algebraically valid but physically inaccurate. Definitions cannot be arbitrarily mixed; mechanical consistency requires a unique path for the flow of energy and momentum.

\section{The Architecture of Verification}
\label{sec:Conclusion_Architecture}

It is important to distinguish the logical structure of this work from its narrative presentation. While the chapters follow a linear path, the validity of the arguments draws from five methodologically orthogonal pillars. These lines of inquiry are independent. They mutually enhance one another, converging on a similar verdict.

\subsection{1. The Microscopic Critique}
\textbf{Part I.} It assumes only the validity of the microscopic Lorentz force and analyzes the definitions of the competing tensors (Minkowski, Abraham) microscopically.
\begin{itemize}
    \item \textbf{Methodology:} The Theorem of Mutual Exclusivity.
    \item \textbf{Independence:} It argues that the competitors are flawed at the root (definitions) by demonstrating distinct, mutually exclusive energy pathways resulting in "Artifactual Work" and "Vacuum Inertia". This critique holds true even if the macroscopic derivation is ignored entirely.
\end{itemize}

\subsection{2. The Macroscopic Audit}
\textbf{Part VI.} It applies the \textbf{Force-Energy Consistency Criterion (FECC)} to the macroscopic continuum, auditing the internal books of the competing historical models.
\begin{itemize}
    \item \textbf{Methodology:} The macroscopic tensors have been tested for internal mechanical consistency.
    \item \textbf{Independence:} It demonstrates that the Minkowski force density does not couple to the Minkowski energy flux. Furthermore, it explicitly links Energy and Force as "two sides of the same coin," equating the specific force density to the defined energy functional. This exposes that variational solutions (like Korteweg-Helmholtz) are not independent physical discoveries, but strictly dependent consequences of their initial energy postulates.
\end{itemize}

\subsection{3. The Constructive Verification}
\textbf{Parts III--V.} It demonstrates that the Vacuum Framework (coupled via the Host Interface) successfully explains all experimental observations of free charges and conductors as well as dielectrics and magnetics in a mechanically consistent manner.
\begin{itemize}
    \item \textbf{Methodology:} The resolution of paradoxes (e.g., The Magnetic Pivot, Relativistic Moving Media).
    \item \textbf{Independence:} It explains the experimental reality—from the behavior of dielectrics to the mechanics of magnets—without requiring ad-hoc modifications or "hidden" variables.
\end{itemize}

\subsection{4. The Structural Derivation (Mathematical Inevitability)}
\textbf{Parts II \& VII.} This is an architectural argument. It derives the macroscopic laws directly from the microscopic baseline using rigorous signal processing.
\begin{itemize}
    \item \textbf{Methodology:} Spatial Filtering and Symmetry Analysis (The Unified Matrix).
    \item \textbf{Independence:} It indicates that the Vacuum framework is \textit{natural}. It shows that the "Unified Matrix" and the symmetry with Continuum Mechanics are not inventions, but inescapable mathematical consequences of filtering the vacuum.
\end{itemize}

\subsection{5. The Theorem of Indeterminacy}
\textbf{Part III.} This is a \textit{protective} argument. It establishes the theoretical limits of the continuum assumption, immunizing the theory against falsification by material-specific deformation anomalies.
\begin{itemize}
    \item \textbf{Methodology:} The "laminated media" counter-example (the proof of non-uniqueness).
    \item \textbf{Independence:} It proves that force density is partially \textit{indeterminate} from macroscopic fields alone. By demonstrating that the force depends on the internal microscopic structure, it invalidates the claim that "continuum thermodynamics" (Korteweg-Helmholtz) represents a universal truth, reinforcing the necessity of specific dipole models.
\end{itemize}

\subsection{The Convergence}
The coincidence of these five independent vectors on a single solution—the universal validity of the Lorentz law—provides, we believe, robust evidence for the theory. The convergence is not circular; it is mutually reinforcing.

\section{The Resolution of the Ambiguity}
\label{sec:Resolution_Ambiguity}

This analysis returns to the "Landscape of Ambiguity" with a clarified view. If one accepts the fundamental kinematic axiom ($P=\mathbf{f}\cdot\mathbf{v}$), the "structural indeterminacy" that has characterized the discipline is significantly reduced.

\subsection{A Systematic Return to Simplicity}
\label{subsec:Conclusion_Simplicity}

The "rigorous analysis" suggests a simplification of the conceptual landscape. It is concluded that while macroscopic electrodynamics constitutes a distinct emergent theory, it operates as the low-frequency projection of microscopic electrodynamics.

Consequently, there is one set of fundamental field equations:
\begin{align}
    \nabla \cdot \mathbf{E} &= \frac{\rho_{\text{total}}}{\varepsilon_0}\\
    \nabla \cdot \mathbf{B} &= 0\\
    \nabla \times \mathbf{E} &= -\frac{\partial \mathbf{B}}{\partial t}\\
    \nabla \times \mathbf{B} &= \mu_0 \mathbf{J}_{\text{total}} + \mu_0 \varepsilon_0 \frac{\partial \mathbf{E}}{\partial t}
\end{align}
And arguably only one fundamental interaction mechanism, the Lorentz Force:
\begin{equation}
    \mathbf{f} = \rho_{\text{total}} \mathbf{E} + \mathbf{J}_{\text{total}} \times \mathbf{B}
\end{equation}

The distinction between "free" and "bound" sources is exclusively kinematic, not fundamental. Physically, they act as identical sources of the field and experience identical forces from the field.

\subsection{The "Arbitrary Split" Revisited}

The conventional resolution—that the choice of stress tensor is a matter of convention—is challenged here not as a matter of preference, but as a \textbf{category error}.

The "Arbitrary Split" hypothesis rests on the assumption that global conservation is the only constraint. But mechanics imposes a local constraint: force requires a mass target. Because the mass of the vacuum and the mass of the dielectric are distinct, the split of momentum between them is strictly determined by source topology.

By strictly enforcing the Force-Energy Consistency Criterion, the vacuum tensor is identified as the description consistent with the proposed FECC.

This resolution points toward a broader epistemological structure, which we now address.

\subsection{The Hierarchy of Uniqueness: Separation of Powers}
\label{subsec:Conclusion_Hierarchy}

The resolution of the paradox lies in a strict "Separation of Powers." We have established a hierarchy of uniqueness that distinguishes between the \textit{impressed} budget and the \textit{absorbed} stress.

\paragraph{Level 1: Microscopic Uniqueness (Ground Truth).}
At the carrier level, the coupling is absolute. The Lorentz force is unique.

\paragraph{Level 2: Macroscopic Uniqueness of Input.}
The \textbf{Vacuum Tensor} is the unique master equation for the input ledger. It determines the \textit{total} force and energy supplied to the material system. It is robust because it is invariant under filtering.

\paragraph{Level 3: Macroscopic Indeterminacy of Allocation.}
The \textbf{Material Response} determines the internal split (Deformation vs. Heat vs. Binding). This is \textit{indeterminate}; it depends on the microscopic topology. 

Thus, the vacuum tensor is the unique answer to "What is supplied?", while the force density is the indeterminate answer to "Where does it go?".

\section{The Hierarchy of Resolutions: A Generalization}

\subsection{The Three Stages of Analysis}

The specific resolution of the electrodynamic problem developed above suggests that physical reality must be described at distinct, yet interconnected, levels of resolution. These are not independent ontological layers piled on top of one another, but rather different optical resolutions of the same underlying reality.

We distinguish three fundamental distinct stages of analysis:

\begin{enumerate}
    \item \textbf{Stage 1 (The Quantum Baseline):} Quantum Electrodynamics (QED). The domain of probabilistic wavefunctions, Heisenberg uncertainty, and fundamental quantised interactions.
    \item \textbf{Stage 2 (The Classical Microscopic Baseline):} The Lorentz Theory. The domain of deterministic trajectories, continuous fields, and discrete charge carriers interacting in a vacuum. This is the "high-fidelity" classical limit.
    \item \textbf{Stage 3 (The Macroscopic Model):} The Continuum Theory. The domain of spatially averaged fields, smooth continuous media, and observable engineering variables.
\end{enumerate}

Throughout this work, we have navigated the transitions between these stages, specifically analyzing the descent from Stage 2 to Stage 3. We have characterized this transition not merely as an approximation, but conceptually as a \textit{high-frequency information filtering process}. 

As demonstrated in the preceding parts of this book, this filter is a destructive operation. By spatially and temporally averaging the microscopic reality, the process effectively deletes the "high-frequency" information—the granular details of particle position and local force distribution. Moving down the ladder of resolution reduces complexity by compressing information, but it simultaneously imposes a fundamental epistemological limit on what the resulting theory can predict. 

\subsection{Application: The Epistemological Boundaries of Macroscopic Electrodynamics}

This perspective led directly to the demonstration of \textit{Macroscopic Indeterminacy}. We showed that the force density required to calculate material deformation cannot generally be expressed using Stage 3 variables alone. This limitation is a direct consequence of the information loss inherent in the filtering process.

We established this argument by constructing two distinct physical scenarios that exhibit divergent behavior in Stage 2 (microscopic reality) yet remain mathematically indistinguishable in Stage 3 (macroscopic observable). This finding allowed us to delineate the epistemological boundaries of any macroscopic theory operating at Stage 3: strictly speaking, a Stage 3 theory cannot uniquely determine local mechanical forces without re-importing information from Stage 2.

\subsection{Case Study: The Navier-Stokes Singularity}

The epistemological framework developed here is not limited to electrodynamics. The method of distinguishing between "Stage 2 reality" and "Stage 3 models" can be applied to any continuum field theory to identify its fundamental limits. A prominent example is the Navier-Stokes equation for fluids.

Microscopically, a fluid is governed by the same fundamental "ground truth" presented in Part 1. It consists of discrete particles (Stage 2) obeying deterministic equations of motion. A macroscopic theory is derived by applying a high-frequency filter (averaging procedure) to this reality.

\paragraph{The Microscopic Reality (Stage 2).}

At Stage 2, the fluid is composed of discrete ions and molecules. The system is described by the microscopic equations of motion (Newton's Laws applied to particles):

\begin{equation}
    \frac{d \mathbf{p}_i}{dt} = \mathbf{F}_{i} 
\end{equation}

In this domain, the mass density $\rho$ and velocity $\mathbf{v}$ are highly granular. They are defined by delta functions located at the particle positions. There are extreme gradients: the density jumps from "atomic density" to zero (vacuum) over the scale of an electron orbital. There is no continuous field; there is only a collection of discrete, rigid bodies with no internal velocity gradients.

\paragraph{The Filtering Operation (Stage 2 $\rightarrow$ Stage 3).}

The transition to the continuum perspective involves a "low-pass" filtering operation. This mathematical filter averages the discrete granular fields over a control volume, deleting all "high-frequency" spatial variations (the positions of individual molecules) and generating smooth, continuous fields for density $\rho_{\text{macro}}$ and velocity $\mathbf{u}$.

Crucially, this process fundamentally alters the system. The filter deletes the true microscopic velocity fields—the rapid, chaotic thermal motion of individual particles—and replaces them with a single mean drift velocity.

\paragraph{The Macroscopic Model (Stage 3).}

The result is a smooth, continuous description of the fluid governed by the Navier-Stokes equation:

\begin{equation}
    \rho \left( \frac{\partial \mathbf{u}}{\partial t} + \mathbf{u} \cdot \nabla \mathbf{u} \right) = -\nabla p + \mu \nabla^2 \mathbf{u} + \mathbf{f}
    \label{eq:NavierStokes}
\end{equation}

This equation provides an excellent engineering description of fluid flow. However, it is an approximation that assumes the fluid is a continuous medium down to infinitesimal scales. From a signal processing perspective, it is a bandwidth-limited representation of reality; it contains no information above a certain spatial frequency.

\paragraph{A Perspective on the "Millennium Problem".}

One of the greatest open problems in mathematics is the existence of "blow-up" solutions (singularities) to the Navier-Stokes equation. The question is whether, for smooth initial conditions, the velocity or energy density can become infinite in finite time.

While we do not purport to address the mathematical intricacies of this problem, the filtering perspective offered here provides a relevant physical context.

From the viewpoint of spectral analysis, a mathematical singularity (a blow-up) corresponds to a structure requiring infinite bandwidth. However, as established in this framework, the macroscopic theory (Stage 3) is inherently bandwidth-limited by the averaging process used to derive it.

This suggests that even if the Navier-Stokes equations were found to admit a mathematical singularity, such a solution would lie outside the physical domain of validity of the model. Long before a singularity forms, the gradients would exceed the validity of the averaging filter (the continuum hypothesis would break down), and the system would revert to its discrete, granular nature. Thus, in this specific physical view, a blow-up in the Stage 3 model would capture an artifact of the filtered approximation rather than a prediction of the underlying Stage 2 reality.

\subsection{The Spectral Epistemology of Continuum Physics}

The conclusion of this work is therefore a call for epistemological precision. Macroscopic field theories—whether Electrodynamics, Fluid Mechanics, or Thermodynamics—are indispensable tools. However, as Stage 3 models, they are constructed by filtering out the microscopic "noise" of reality.

The specific contribution of this framework is to operationalize this common understanding using the language of signal processing. By treating the transition from microscopic to macroscopic not just as an "averaging" but as a \textit{bandwidth-limiting filter}, we gain a precise diagnostic tool.

When we push these models to their limits—searching for singularities or probing behavior at the atomic scale—we are, in spectral terms, trying to reconstruct high-frequency data from a low-pass filtered signal. The resolution of the paradoxes in macroscopic electrodynamics, as shown in this book, lies not in inventing new macroscopic mathematics, but in recognizing the cut-off frequency of our own lens.

\section{The Engineering Frontier: The Epistemological Wall}
\label{sec:Conclusion_Frontier}

Classical Electromagnetism is the invisible infrastructure of the modern world. By understanding these laws, we unlock the operating manual for nearly every technology we rely on.

This presents the "Paradox of Competence": \textit{How is the standard engineering consensus so highly effective if its theoretical foundations regarding force and stress are ambiguous?}

The resolution lies in the distinction between the "Black Box" (Performance) and the "White Box" (Structure). Engineering applications typically depend on the \textit{integrated} total force or energy transfer. As shown in Chapter 1, the global integrals of the competing tensors (Minkowski, Abraham, Lorentz) are often identical for isolated rigid bodies. The engineer is protected by the integral.

However, as engineering pushes toward the limits of material science---high-energy density dielectrics, optical tweezers, and nano-electromechanical systems (NEMS)---the "black box" approach fails. When the structural integrity of the material itself is the limiting factor, the internal stress distribution becomes the primary observable. At this frontier, we encounter the \textbf{epistemological limit} of the macroscopic theory.

\subsection{The Observability Wall}
All engineering is ultimately based on the mechanical response. An electric motor is not driven by the $\mathbf{H}$-field; it is driven by the force $\mathbf{f}$ that the field exerts on the rotor's mass.

It must be recognized that macroscopic physics—whether Thermodynamics, Continuum Mechanics, or Electrodynamics—is always a simplification. Defining a material property, such as the dielectric constant $\varepsilon_r$ or magnetic permeability $\mu_r$, is a \textbf{spectral filtering operation}.

The act of measurement itself defines the cutoff frequency. It physically filters out the high-frequency spatial information (the precise position of ions, domain walls, lattice defects) and compresses the result into a scalar value. Consequently, a macroscopic theory based on these variables is \textit{structurally incapable} of predicting the exact local stress state in general, because it has discarded the necessary geometric data.

\subsection{The Risk of the Continuum: Safety Factors as Topological Insurance}
This insight reframes standard engineering practice. Engineers routinely apply large "Safety Factors" (often $2\times$ or $3\times$) when designing insulation systems or structural magnetic components.

Physical Exclusivity reveals that these safety factors are not merely guards against manufacturing defects; they essentially function as unintentional acknowledgments of the \textbf{Theorem of Indeterminacy}. The macroscopic field calculates the \textit{average} stress. However, material failure is initiated by the \textit{peak} microscopic stress. Because the continuum theory filters out the peaks, it systematically underestimates the risk. The "Safety Factor" is the engineer's manual re-introduction of the topological variance that the theory deleted.

\subsection{The Way Forward: Computational Topology}
Therefore, for materials with complex unit cells or high packing densities, analytical approximations are insufficient. The engineering workflow must likely evolve to:
\begin{enumerate}
    \item \textbf{Simulate} the exact lattice geometry in a CAD environment.
    \item \textbf{Calculate} the Lorentz force on the distributed charge clouds.
    \item \textbf{Average} these discrete forces to obtain the effective macroscopic stress tensor for that specific material.
\end{enumerate} 

This workflow bypasses the "Problem of the Energy-Momentum Tensor" entirely. It acknowledges that the "stress of the medium" is an emergent property of the specific arrangement of its constituents.

\section{Outlook: Limitations and Future Directions}
\label{sec:Conclusion_Outlook}

The synthesis presented here attempts a closed mechanical system, yet it opens new avenues for rigorous inquiry.

\paragraph{Force Density Confirmation via Computational Topology.}
Future research could move beyond the search for a universal analytical force density and instead focus on developing standardized CAD workflows for calculating the \textbf{Lattice Adaptation Term} of common crystal structures. By simulating the exact internal field gradients, engineers could build a library of "correction factors" that bridge the gap between the simple Lorentz force and the complex reality of material deformation.

\paragraph{The Alternative Magnetic Basis.}
This work relied exclusively on the Amperian model---defining magnetization via microscopic electric currents---to construct a mechanically consistent theory. However, as proposed by \cite{FanoChuAdler1960}, it is possible to formulate a self-consistent macroscopic theory based on magnetic charges (the Gilbertian model).

While experimental evidence regarding the hyperfine structure of the neutron favors the current-loop model for intrinsic spin (\cite{Jackson1977}), a rigorous "structured examination" of the Gilbertian formulation serves as a valuable counter-investigation. Future theoretical work could apply the Force-Energy Consistency Criterion to the magnetic charge model to definitively map the divergences in mechanical stress predictions between the two approaches. 

\paragraph{The Closing of the Ledger.}
Ultimately, the "Problem of the Energy-Momentum Tensor" appears to be resolved by a single kinematic requirement: identifying the exact mass target of any proposed force. The theoretical ledger balances naturally once the currency is tracked correctly. By recognizing that \textbf{Force requires Mass} and \textbf{Energy requires Work}, the Lorentz force is revealed as the precise mechanical accounting of the vacuum's interaction with matter.

\bibliographystyle{unsrt}
\bibliography{references}

\end{document}